\pdfoutput=1
\documentclass[12pt]{JHEP3}
\input epsf.tex
\epsfclipon

\usepackage{color}
\let\normalcolor\relax
\usepackage{epsfig}
\usepackage{epstopdf}
\usepackage{epsf}
\input{epsf.sty}
\usepackage{graphicx,amsmath,amssymb}
\usepackage{epsfig,multicol}
\usepackage{multirow}
\allowdisplaybreaks

\usepackage{bbm,bm,amsmath,amssymb}

\usepackage[normalem]{ulem}

\usepackage{comment}

\def\figs/B{B}
\def\ba{\begin{eqnarray}}
\def\ea{\end{eqnarray}}
\def\bg{\begin{eqnarray}}
\def\nd{\end{eqnarray}}
\def\sin{{\rm sin}}
\def\cos{{\rm cos}}

\def\log{{\rm log}}
\def\ln{{\rm log}}

\title{Bulk Viscosity at Extreme Limits: From Kinetic Theory to Strings}
\author{Alina Czajka$^{1, 2}$, Keshav Dasgupta$^{1}$, Charles Gale$^{1}$, Sangyong Jeon$^{1}$,
Aalok Misra$^{3}$, Michael Richard$^{4}$, Karunava Sil$^{5}$\\
\vskip.03in
${}^1$ Department of  Physics, McGill University \\
~~3600 rue University, Montr\'{e}al, Qu\'{e}bec, Canada H3A 2T8\\
${}^2$ Institute of Physics, Jan Kochanowski University\\
~~Swietokrzyska 15 street, 25-406 Kielce,  Poland\\
${}^3$  Department of Physics, Indian Institute of Technology Roorkee\\
~~Uttarakhand-247667, India\\
${}^4$ John Abbott College, 21275 Lakeshore Dr\\
~~ Sainte-Anne-de-Bellevue, Qu\'{e}bec, Canada H9X 3L9\\
${}^5$ Department of Physics, Indian Institute of Technology Ropar\\
~~Nangal Road, Rupnagar, Punjab 140001, India\\
{\tt aczajka, keshav, gale, jeon@hep.physics.mcgill.ca}
~~{\tt micheal.richard@mail.mcgill.ca, aalokfph@iitr.ac.in, karunavasil@gmail.com}}

\date{\today}
\maketitle

\abstract{In this paper we study bulk viscosity in a thermal QCD model with large number of colors at two extreme limits: the very weak and the very strong 't~Hooft couplings. The weak coupling scenario is based on kinetic theory, and one may go to the very strong coupling dynamics via an intermediate coupling regime. Although the former has a clear description in terms of kinetic theory, the intermediate coupling regime, which uses lattice results, suffers from usual technical challenges that render an explicit determination of bulk viscosity somewhat difficult. On the other hand, the very strong 't~Hooft coupling dynamics may be studied using string theories at both weak and strong string couplings using gravity duals in type IIB as well as M-theory respectively. In type IIB we provide the precise fluctuation modes of the metric in the gravity dual responsible for bulk viscosity, compute the speed of sound in the medium and analyze the ratio of the bulk to shear viscosities. In M-theory, where we uplift the type IIA mirror dual of the UV complete type IIB model, we study and compare both the bulk viscosity and the sound speed by analyzing the quasi-normal modes in the system at strong IIA string coupling. By deriving the spectral function, we show the consistency of our results both for the actual values of the parameters involved as well for the bound on the ratio of bulk to shear viscosities.}

\begin{document}

\section{Introduction and summary}

The wide and thoroughgoing experimental programs pioneered at the Relativistic Heavy Ion Collider (RHIC) and  pursued at the Large Hadron Collider (LHC) offer a unique opportunity to study properties of a most exotic state of matter: the quark-gluon plasma (QGP). Although there is a common agreement that droplets of QGP are produced in heavy ion collisions in the pursued experiments, a unequivocal and quantitative determination of the properties of such a state is still the topic of much research. The time evolution of the plasma, its transport properties, the parameters of the transition to the confined phase, are some of the features that are currently being addressed along with many others. The difficulties in extraction of QGP properties owe much to the fact that the excited nuclear matter produced by colliding heavy ions at currently achievable energy scales is strongly coupled. Accordingly, the applicability of known fundamental methods and approaches to study the system in this regime is very limited, and hence all obtained findings in this limit have to be examined critically. On the other hand, this situation may also provide an opportunity to explore new technical facets of known tools and to explore new directions.

One of the methods that have proven useful to study the properties of QGP in the domain accessible experimentally is viscous hydrodynamics - a low-frequency long-wavelength effective theory. The application of the hydrodynamic framework to heavy ion collisions \cite{Teaney:2000cw,Huovinen:2001cy,Kolb:2001qz,Hirano:2002ds,Kolb:2002ve,Romatschke:2007mq,Luzum:2008cw,Dusling:2007gi,Song:2007ux} and its use in the interpretation of a wide range of experimental observables \cite{Adcox:2004mh,Back:2004je,Arsene:2004fa,Adams:2005dq} allowed to conclude that the experimentally produced QGP is a strongly coupled system. In particular, the studies on the hadronic flow and the emergence of other collective phenomena in the hydrodynamic description of QGP were taken as an indication of its fluid-like nature. Moreover, the success of hydrodynamics seemed to necessitate a fast near-thermalization of the QGP. All these arguments led to the conclusion that the created quark gluon plasma must be strongly coupled~\cite{Shuryak:2008eq}. For  reviews on hydrodynamic applications and formulations see \cite{Gale:2013da,Heinz:2013th,Romatschke:2009im,Florkowski:2017olj,Jeon:2015dfa}.

Another powerful tool to study systems in the limit of strong 't~Hooft coupling originated with the discovery of the AdS/CFT correspondence \cite{malda}. Even though it was hydrodynamic predictions and analyses that provided the empirical evidence that the shear viscosity to entropy density ratio is small \cite{Teaney:2009qa}, it was the AdS/CFT conjecture that established an analytical bound of $\eta/s=1/4\pi$ \cite{Kovtun:2004de}\footnote{Violation of this bound is seen in the presence of higher derivative terms, discussed first in \cite{kazpet}. In the absence of these terms, the KSS \cite{Kovtun:2004de} bound continues to hold at strong 't~Hooft coupling.}.

Transport coefficients are valuable elements of the hydrodynamic description as they carry information about the microscopic properties of a medium. In the case of the shear viscosity of strongly interacting matter, numerous phenomenological studies, the AdS/CFT result,  kinetic theory calculations in the high-temperature weakly coupled regime of QGP $\eta \propto 1/(g_{YM}^4 \log(1/g_{YM}))$ \cite{Arnold:2000dr}, and  non-perturbative estimates \cite{Christiansen:2014ypa} allow a schematic global understanding of the shear viscosity to entropy density ratio. It is known that shear viscosity is large in the perturbative, high-temperature,  limit, smaller  near the phase transition temperature \cite{Nakamura:2004sy,Csernai:2006zz}, and large again in the confined, pion gas domain \cite{Lang:2012tt}. However, the physics of the bulk viscosity is less satisfactorily understood. There are strong indications that bulk viscosity behaviour follows a trend opposite to that of the shear viscosity. In the limit of high-temperature QCD, bulk viscosity was found to have a very small value \cite{Arnold:2006fz}: this is to be expected, as  the coefficient of bulk viscosity can be written as a correlator of the trace anomaly (see Section \ref{alina1}), and QCD is known to be approximately conformal at high temperatures. Although it may seem that, in the very large coupling regime, a direct application of AdS/CFT techniques to bulk viscosity exploration is not possible as the conjecture relies on the ${\cal N} = 4$ super Yang-Mills theory which is perfectly conformal: the bulk viscosity vanishes identically, this is not quite true. Approaches based on holography in fact have proven useful by providing a lower bound on the ratio of bulk to shear viscosities: $\zeta/\eta \geq 2(1/3-c_s^2)$ \cite{Buchel-bound}\footnote{The non-conformal string theory studied in \cite{Buchel-bound} is different from what we consider here. In \cite{Buchel-bound} it's the ${\cal N} = 2^{\ast}$ supersymmetric gauge theory obtained by a mass deformation of ${\cal N} =4$ Super Yang Mills theory. See also \cite{Jeon:1994if} and \cite{benimadav} for an even earlier studies on bulk viscosity from first principles.}. In the vicinity of the transition from QGP to hadronic degrees of freedom, the bulk viscosity should, in principle, be calculated from the equation of state extracted the lattice data \cite{Borsanyi:2014rza,Bazavov:2017dsy}. It is expected to be proportional to the trace anomaly $(\epsilon -3P)/T^4$ and hence be notably peaked. Various investigations, both formal and phenomenological \cite{Kharzeev:2007wb,Karsch:2007jc,Moore:2008ws,Gubser:2008yx,NoronhaHostler:2008ju,Denicol:2009am}, confirm this expectation. Recently, it was demonstrated that the presence of a coefficient of bulk viscosity is important in hydrodynamical simulations, as it has a significant impact on the elliptic flow coefficients \cite{Denicol:2009am,Song:2008hj,Denicol:2010tr} and other heavy-ion observables, strongly interacting and otherwise \cite{Ryu:2015vwa,Ryu:2017qzn,Paquet:2015lta,Bozek:2017kxo,Monnai:2016kud}. However, it is fair to write that the precise behavior of the bulk viscosity for systems in extreme conditions of temperature and density  is not yet firmly established and therefore needs further studies.

Understanding the behavior of bulk viscosity and knowing how it changes when the coupling strength varies is important for several reasons. First, bulk viscosity is an inherent property of nonconformal systems, and finite-temperature systems obeying QCD are good examples of such environments. The behavior of the bulk viscosity is fixed by  parameters that break conformal symmetry. These include, at least at the perturbative region, finite masses of plasma constituents and the Callan-Symanzik $\beta$-function which expresses the coupling constant as a function of an energy scale \cite{Arnold:2006fz,Jeon:1994if}. Equivalently, these parameters enter the definition of the speed of sound, and bulk viscosity can be conveniently expressed as some function of $1/3-c_s^2$ as well. From the phenomenological point of view, expressing bulk viscosity via the speed of sound is  practical as this enables a direct connection with the lattice QCD equation of state.  Second, bulk viscosity plays an essential role in the hydrodynamical description and modelling of hot and dense strongly-interacting matter. One could attempt to compute the coefficient within a theory which captures the microscopic interactions, and then insert it into hydrodynamic equations. Alternatively, fluid dynamics may be viewed as an effective theory of the long wavelength behaviour, and its transport coefficients are to be extracted empirically. Either way, viscous hydrodynamics serves as a powerful tool to investigate the strongly coupled nuclear medium produced in RHIC and LHC experiments. It provides information on the dynamics of the plasma, informs how the plasma evolves and also helps to extract, or at least constrain, other plasma characteristics. In addition, bulk viscosity studies have the potential to further develop new theoretical methods to study the conformal anomaly of QCD.

Because of different system dynamics at different coupling regimes one may expect a different dependence of the bulk viscosity on the factor $1/3-c_s^2$. This is what was observed by comparing the bulk to shear viscosity ratios at perturbative and very strong-coupling limits: $\zeta/\eta \propto (1/3-c_s^2)^2$ \cite{Arnold:2006fz} and $\zeta/\eta \propto (1/3-c_s^2)$ \cite{Buchel-bound}, respectively. Analyzing this difference is the main objectives of our studies. What is more, we focus on the intrinsic physics of bulk viscosity which is a unique measure of a system's departure from the conformal symmetry independently of the differences in the dynamics at different energy scale. Importantly, we also examine methods employed for the viscosity calculation. Our ultimate goal is to provide a detailed picture of bulk viscosity as a physical quantity of the SU($M$) gauge theory. Due to complex nature of the theory we examine analytically accessible limits of the quantity, which clearly requires us to use very different methods.

The system under consideration is governed by the SU($M$) theory with the interaction strength determined by the 't~Hooft coupling $\lambda=g_{YM}^2 M$, where $g_{YM}$ is the gauge coupling and $M$ is the number of colors. The 't~Hooft coupling may be thought as an effective coupling of QGP. We distinguish here 3 regions of the 't~Hooft coupling: the weak coupling region, the intermediate coupling region (near the phase crossover  temperature), and the strong (infinite) coupling one. In each region a different microscopic approach is applicable. The extreme limits are discussed comprehensively while the intermediate coupling part includes a brief summary and discussion on conceptual difficulties preventing one from determining bulk viscosity in this domain.

As already mentioned, the weak-coupling studies on bulk viscosity for QCD were done within kinetic theory in \cite{Arnold:2006fz}. In our work we intend to adjust the kinetic theory result to the 't~Hooft coupling. In this way we provide the form of bulk viscosity which can be directly confronted with its strong-coupling counterpart discussed via string theory methods. In this approach quark contribution is always suppressed by a factor $1/M$ and may be neglected in the leading order analysis. Kinetic theory \cite{Arnold:2000dr,Arnold:2002zm,Arnold:2003zc} is an effective theory which is commonly and successfully used to compute transport coefficients. Its correspondence to fundamental microscopic theory was directly shown for scalar theory in \cite{Jeon:1994if} and then also for QED \cite{Gagnon:2006hi,Gagnon:2007qt}. In this manuscript we undertake the task to justify the validity of kinetic theory for SU$(M)$ theory by providing power counting of microscopic processes contributing to the collision kernel of the Boltzmann equation. Since a derivation of the transport equation from diagrammatic representation of any non-Abelian theory is highly non-trivial, we intend to present a procedure on how to justify the collision kernel diagrammatically and we do not intend to derive the Boltzmann equation. We discuss how the pinching poles and nearly pinching poles control power counting of elastic and inelastic processes, respectively. The consequences of soft physics on power counting are emphasized. We also show how the integral equations emerge by discussing all topological structures of planar diagrams contributing to them. We believe that this examination may provide solid arguments to prove an equivalence of the Boltzmann equation with the analysis based on the loop expansion.

The intermediate coupling region is considered mostly to summarize the status of studies of bulk viscosity done with microscopic analyses, that is, which need usage of lattice QCD. The bulk viscosity in this regime can be obtained if one can extract the low frequency behavior of the corresponding spectral density \cite{Kharzeev:2007wb,Karsch:2007jc,Moore:2008ws,Romatschke:2009ng,Meyer:2007dy,Meyer:2010ii,Astrakhantsev:2018oue}. Although these approaches provide some constrains, they do not yet allow definite conclusions on the behaviour of the bulk viscosity to be drawn. We briefly discuss the difficulties.


On the other hand, the strong 't~Hooft coupling behavior of bulk viscosity is an interesting playground to study sting theory and gauge theory because of the use of gauge/gravity duality. In fact since the bulk viscosity should truly be studied in a theory with running couplings, the famed AdS/CFT duality is not very useful, as discussed above. Going beyond CFT will require us to find the right gravity dual to answer any questions related to running couplings, and especially questions related to bulk viscosity. The gravity dual that we seek has been first proposed in \cite{KS, ouyang} and the full UV completion was given from the type IIB side in \cite{metrics, 3reg, chenchen} and more recently from the type IIA side in \cite{UVcom}.

At this stage one might ask as to how a gravitational background, which has hitherto no connection to gauge theory, could in principle enter the picture to help us solve strongly coupled system like the one that we concentrate on here. There are two ways to answer this question, but none are completely satisfactory. The first one is to relegate this to the magic of duality. However this duality is special because all dualities studied so far have either been between two different gauge theories or between two different supergravity theories. There has never been a duality between a gauge theory and a gravitational theory before AdS/CFT \cite{malda}.

The second one is to view the gauge theory as to be somehow {\it contained} inside a gravitational background. To elaborate this viewpoint, let us consider a four-dimensional Minkowski spacetime. This serves as an arena for gauge theory interactions, and for simplicity we decouple all gravitational interactions by tuning the Newton constant. The gauge theory interactions can happen at various energy scales, and we can assume that a specific slice of four-dimensional Minkowski spacetime is associated with a  specific energy configuration. We can stack up all the slices together such that the low energy slices are at the bottom and the high energy slices are kept on top of one another in an increasing fashion. Clearly the topmost slice will be at infinite energy.

The above construction immediate provides a {\it five}-dimensional space and if we assume the energy direction to be parametrized by a radial coordinate $r$, then at $r = 0$ we have IR physics and at $r \to \infty$ we have UV physics. This is also by construction a five-dimensional {\it gravitational} background, and by this simple assumption we seem to have got a five-dimensional gravity theory that captures the dynamics of the four-dimensional gauge theory from IR to UV! Of course this is a very simple construct and does not answer all questions related to gauge/gravity duality but it is instructive to see how two seemingly unrelated physics, one of gauge theory and other of gravity, may be united in a framework like above.

A few quick checks may be easily performed at this stage. If the gauge theory is a CFT, i.e scale independent, then the slicing idea will tell us that we need not worry too much of the physics at any scale $r$, and instead study the dynamics of the corresponding gauge theory from the boundary at $r \to \infty$. Of course this is what AdS/CFT dictionary tells us, but what is lacking in our simple construct is the justification that the gravitational background is indeed an $AdS_5$ space. Maybe the idea of scale invariance, combined with decoupling and the supergravity EOMs could uniquely fix that, but this has not been checked.

On the other hand if we are dealing with a gauge theory that is not scale invariant, then every point on the slice matters. At every $r$ we have the corresponding gauge theory dynamics at {\it that} scale\footnote{This argument entails the fact that if we keep $r$ fixed and move along the remaining four-dimensions, nothing should change. However we can envision more generic scenario where the energy scale is mapped to a certain {\it combination} of $r$ and the other three directions. In this case the Wilsonian effective action will be sensitive to where we are on a given slice. Of course it should be possible to redefine coordinates in such a way to find a new {\it radial} coordinate that will again correspond to the energy scale. For this paper we will however stick to the simplest case where $r$ is mapped to the energy scale, $r_c$ to the UV cut-off, and $r_h$, the horizon radius, to the temperature.}.
Indeed in the Wilsonian sense at this scale all high energy degrees of freedom are integrated out and we are left with a set of relevant, marginal and irrelevant operators. This is of course the premise of our construction in \cite{metrics}, and the UV completion in \cite{3reg, chenchen} is done by introducing new degrees of freedom from the so-called Region 2
of \cite{3reg} onwards.

Other checks, that include the exact mapping of the gauge theory operators to supergravity states, are much harder to perform and in fact the dictionary for gauge/gravity duality for the non-conformal case is not yet fully developed compared to the conformal case. Nevertheless one thing is for sure: to have any control on the computations on the supergravity side we need small $g_s$. For a background with a constant dilaton $-$ an example would be the Klebanov-Strassler background \cite{KS} $-$ a little bit of numerology can tell us that $g_s$ may be made arbitrarily small\footnote{For example if we take $\varphi = -45$, then $g_s \equiv e^{\varphi} = 2.86 \times 10^{-20}$ which is a very small number. The minus sign can be easily accommodated by assuming that the background appears from S-dualizing a NS5 background, i.e a torsional background (see for example \cite{DEM, UVcom}).}.

There is also an additional requirement of large number of colors. For a SU$(M)$ gauge theory, the corresponding supergravity theory will make sense if $\lambda \equiv g_sM$ is very large. In this limit all computations can be restricted to classical supergravity alone, and stringy corrections can be entirely ignored. However if we want to study an actual large $M$ QCD we will have to explore string coupling $g_s = {\cal O}(1)$. How can we ignore stringy corrections now and restrict ourselves to supergravity alone?

A way out of this conundrum was first proposed in \cite{MQGP} by performing a sequence of two stringy dualities: mirror transformation and M-theory uplift. The mirror transformation is a special kind of duality that takes a type IIB background to a type IIA background by simply interchanging the K\"ahler and the complex structures of the internal manifolds on both sides of the duality. In \cite{MQGP} this was implemented by performing three T-dualities along the isometry directions of the internal manifold in the type IIB side \cite{syz}. Being T-dualities they do not change the behavior of the dilaton too much, and therefore takes a weakly coupled background into another weakly coupled one.

The second duality is when we increase the type IIA coupling. At strong coupling a new internal direction opens up and the theory goes to eleven-dimensional M-theory where the dynamics is now miraculously governed by eleven-dimensional {\it supergravity}. All the type IIA stringy corrections are now captured succinctly by classical supergravity analysis in M-theory \cite{witten11}, and therefore $g_s = {\cal O}(1)$ can again be studied using supergravity, albeit from eleven-dimensions. Such a dual description was termed as the MQGP limit of thermal QCD with large number of colors in \cite{MQGP}.

The above considerations tell us that the strong 't~Hooft coupling regime may be studied from the perspectives of both weak and strong string couplings. In the presence of $N_f$ flavors, it means we are exploring both $g_sN_f \to 0$ as well as $g_s N_f = {\cal O}(1)$ limits\footnote{By strong string coupling or by ${\cal O}(1)$ coupling we will mean $g_s$ close to 1 but slightly less. This is because we always want to keep the combination $(g_sN_f)^k \left({g_sM^2\over N}\right)^m << 1$ even for $m = 1$ and $k \in \mathbb{Z}$. See also footnote \ref{7khoon}.}. This in turn boils down to saying that we can have analytic control on the transport coefficients $-$ and here we will concentrate only on bulk viscosities $-$  for pure glue as well as for flavored large $M$ thermal QCD. Section \ref{nf0} of the paper will therefore be dedicated to studying the bulk viscosity using weak string coupling and with vanishing number of flavors, whereas section \ref{nfn0} will be dedicated to studying the bulk viscosity using the other limit, namely strong string coupling and non-vanishing number of flavors.

There is yet another limit where we can remain at weak string coupling, but explore strong YM coupling. In the type IIB side such a scenario becomes possible once $N_f$ flavor degrees of freedom are switched on. That this could happen is a consequence of two conspiracies: one, the dilaton picks up ${\cal O}(g_sN_f)$ corrections forcing it away from being a constant, and two, the NS 2-form field, through the vanishing two-cycle on which we have the $M$ wrapped D5-branes, also picks up ${\cal O}(g_sN_f)$ corrections. These corrections provide additional structure to the already non-constant field, but more importantly they add to the dilaton factor constructively to provide the full structure of the YM coupling.

Interestingly, from either of these limits at strong 't~Hooft coupling, the ratio of the bulk to shear viscosities remains proportional to linear power of $\left({1\over 3} - c_s^2\right)$. The difference however lies in the precise coefficients that control the lower bounds at weak and strong string couplings. For example at strong string coupling, the lower bound is almost 9 times bigger than the Buchel bound \cite{Buchel-bound} as we will discuss in section \ref{nfn0}. Of course nowhere we see any violation of the Buchel bound, so  presumably the violation can only occur once we dimensionally reduce the four-dimensional theory to two-dimensions. This is much like the scenario presented in \cite{violation}, but we will not discuss it any further here.

What we will discuss however is the appearance of the linear power of the deviation factor, $\left({1\over 3} - c_s^2\right)$, when we study spectral function using the weakly coupled type IIA theory. The spectral function is an important aspect in the study of QGP, and its derivation is rather complicated at weak 't~Hooft coupling. At strong 't~Hooft coupling there is a way to derive it from the gauge/gravity duality, but the derivation is technical and involves various manipulations of the background. Nevertheless an answer can be found in the present set-up and the final result shows a linear dependence on the deviation factor. In the limit of vanishing frequency, the result matches well with actual QGP, despite the presence of a large number of colors. Such a success points towards some inherent universality, and it will be interesting to explore this further.

\subsection{Organization of the paper}

The paper is organized as follows.
In section \ref{weak} we study bulk viscosity at weak 't~Hooft coupling. After short introductory remarks on the definition of bulk viscosity and the applied microscopic theory, in section~\ref{alina1} we discuss the Kubo formula which provides a general and first-principle method of the viscosity computation. In section \ref{kinetic} we briefly summarize results on the leading order bulk viscosity calculation performed within kinetic theory by solving the Boltzmann equation. Sections~\ref{alina2} $-$ \ref{alina6} contain a diagrammatic, but qualitative only, analysis which is to justify validity of the effective kinetic theory formulation for studies of transport coefficients. In section~\ref{alina2} we consider the one-loop diagram to find a typical size of the bulk viscosity. This step shows also that fermionic contributions are subleading in favor of the gluonic ones. Then, in section~\ref{alina3}, the power counting of the relevant self-energies is done. Section~\ref{sec-just} is devoted to an evaluation of typical sizes of multi-loop diagrams which represent scattering processes. Both particle number conserving and particle number changing processes are studied and the role of the soft physics is emphasized in subsections~\ref{alina4} and~\ref{alina5}. In section~\ref{alina6} a schematic form of the relevant integral equations needed for a diagrammatic bulk viscosity computation is presented. In section~\ref{alina7} we briefy refer to other diagrammatic methods which can be employed to real-time dynamics investigations.

In section~\ref{int} an intermediate coupling regime is discussed. The section consists of a brief overview of literature on the approaches aiming at an extraction of bulk viscosity from lattice QCD results by studying mostly QCD sum rules and finding constraints on the spectral density. The difficulties in the quantitative determination of the bulk viscosity are pointed out.

The strong coupling results are discussed in sections~\ref{nf0}, \ref{nfn0} and \ref{spectral}. In section~\ref{nf0}, the weak string but strong 't~Hooft coupling regime is discussed. We start by giving a detailed description as to where the string theory techniques would fit in in the study of bulk viscosity. The various domains of compatibility as well as the UV completion are emphasized and the consistency of the background is shown from both type IIB and its dual type IIA pictures. In section~\ref{2bbg}, a slightly simplified background is taken to quantify various parameters associated with the computation of bulk viscosity. For example, one of the important parameter is the fluctuation associated with the vielbeins. This is elaborated in section~\ref{bvcomp}. The fluctuation modes can be divided into positive and negative frequencies, and we show that there are pieces of the fluctuations, called  $p_{nk}$, that are related to certain sources $\Delta^{(n)}_{ab}$ in the gravity dual picture. The analysis of the sources is rather complicated and therefore in section~\ref{amandas} we first take a toy example to study the equations connecting $p_{nk}$ fluctuations with the $\Delta^{(n)}_{ab}$ sources. The toy example is based on a simplifying constraint, and using this the simplest zero and the non-zero modes of the fluctuations are shown to satisfy equations that relate them to the sources. In section~\ref{galgad} we go beyond the simple toy example by studying the equations governing the fluctuating modes in a generic setting. As before, the zero and the non-zero modes satisfy equations relating them to certain sources.

Once we have the fluctuations, we can use them to compute the transport coefficients. In section \ref{sound} we perform two important computations: one, the sound speed, and two, the ratio of the bulk to the shear viscosities. The former is given by an equation which takes into account not only the scale dependence of the temperature, but also the background fluctuations. Needless to say, the ratio of the bulk to the shear viscosities should depend on the sound speed, and we elucidate this by first computing the precise ratio and then showing that the ratio is indeed bounded below by the deviation of the sound speed from its conformal value and more interestingly, is independent of the cut-off.

The remaining two sections are devoted to studying bulk viscosity at strong string and strong 't~Hooft couplings. The first, i.e section~\ref{nfn0}, has to do with obtaining a Buchel-like bulk-viscosity-to-shear-viscosity bound by looking at scalar modes of metric perturbations and the associated quasi-normal modes. The second, i.e section~\ref{spectral}, has to do with obtaining the same result from spectral functions. Here is a more detailed plan of these two sections.

 In section~\ref{nfn0}, we first briefly review the Strominger-Yau-Zaslow (SYZ) type IIA mirror of \cite{metrics}'s top-down type IIB holographic dual of large-$N$ thermal QCD, as well as its M-theory uplift as constructed in \cite{MQGP}. This is followed by a discussion on obtaining the EOM for a linear combination of scalar modes of metric perturbations gauge invariant under infinitesimal diffeomorphisms and obtaining the associated quasi-normal modes in section~\ref{lostgls}; it is noted that with a non-zero bare resolution parameter, the horizon turns out to be an irregular singular point, a fact that proves in fact to be quite helpful in obtaining the aforementioned quasi-normal modes. In section~\ref{zerbare}, we show that one cannot avoid non-normalizable modes if one were to turn off the bare resolution parameter resonating well with similar non-normalizable perturbation modes obtained in section~\ref{nf0}. A Buchel-like bound for the ratio of the bulk and shear viscosities in terms of the linear power of the deviation of the square of the speed of sound from its conformal value is finally obtained, both for $N_f=0$ and $N_f\neq0$ in section~\ref{suhaag}.

In section~\ref{spectral}, we follow a different route $-$ that of spectral function involving correlation function of gauge fluctuations in background value of gauge fields on the world volume of the flavor D6-branes of the aforementioned SYZ type IIA mirror. In section~\ref{spec2}, we obtain the background value of a D6-brane world volume gauge field ${\cal A}_t(r)$, $r$ being the radial coordinate and set up the EOM for fluctuations about the same. We obtain and explicitly solve the EOM $-$ there turns out to be only one linearly independent EOM $-$ in the zero-momentum limit in section~\ref{spec3}. From the on-shell action, the gauge-fluctuation correlation function and hence the spectral function per unit frequency in the vanishing frequency limit, is worked out in section~\ref{onshala} and it is explicitly seen that the difference between the same at non-zero and zero temperatures is precisely proportional to the deviation of the square of the speed of sound from its conformal value. In section~\ref{gsnf1}, we argue that unlike sections~\ref{spec2} $-$ \ref{onshala} wherein one had considered weak-string-coupling strong-'t-Hooft-coupling limit, the result of section~\ref{onshala} goes through even for the strong string and strong 't~Hooft couplings' limit. We argue therein that the $g_s \to 0$ limit alongwith non-trivial $B$-field along the vanishing two-cycle conspires to produce a $g^2_{YM}$ in the gauge theory side that is no longer a small number.

Finally in the appendices we discuss three topics. The first one is on a gauge invariant combination of the scalar modes of metric fluctuations. Such a combination is useful to study the quasi-normal modes.The second one is on the derivation of the on-shell action and Green's function required to study the spectral function. The third one is on an estimation of the horizon radius.

\section{Bulk viscosity at weak 't~Hooft coupling \label{weak}}

When the system exhibits a small deviation from thermal equilibrium, its evolution is well described by the equations of hydrodynamics. These are given in terms of conservation laws of currents accompanied by the equation of state. Here, we focus only on the energy and momentum currents which are encoded in the stress-energy tensor $T^{\mu\nu}$. Its spatial part is:
\bg
\label{aline}
T^{ij}=T^{ij}_{\rm eq} + \eta(\nabla^i u^j+ \nabla^j u^i- 2/3
g^{ij} \nabla \cdot u)+\zeta g^{ij} \nabla \cdot u,
\nd
where
$\zeta$ and $\eta$ are the bulk and shear viscosities, $u^i$ is the fluid flow velocity and the metric is mostly negative.\footnote{ In Secs.~\ref{weak} and \ref{int} we use the $(+,-,-,-)$ signature for the metric, which is commonly used in diagrammatic approaches, while in Secs.~\ref{nf0}-\ref{spectral} we use $(-,+,+,+)$, commonly applied for calculations in string theories. The choice is made for convenience since we refer to many known results on both sides throughout the manuscript.} A many-body system can be driven out of its equilibrium state through uniform compression or rarefaction and both processes lead to changes in the energy density $\epsilon$, the increase or decrease, respectively. The pressure $\mathcal{P}$ also changes but its change is different than that provided by the equation of state $P(\epsilon)$. The trace of the stress tensor carries information on changes in pressure. The deviation from the equilibrium pressure when the system is expanding or contracting is characterized by the bulk viscosity $\zeta$:
\ba
\label{press-bulk}
\mathcal{P}=P - \zeta \nabla \cdot u,
\ea
where $\nabla \cdot u$ is the expansion parameter. Bulk viscosity, as well as other transport coefficients, is determined by microscopic dynamics. Here we discuss how bulk viscosity emerges when the system is governed by the non-Abelian SU($M$) gauge theory with the Lagrangian:
\ba
\label{L-SUN}
{\cal L} = -\frac{1}{4}F^{\mu \nu}_a F_{\mu \nu}^a
+i \bar \psi \gamma_\mu D^\mu\psi.
\ea
Here $\psi$ is the quark field with $M \times N_f$ degrees of freedom, where $M$ is the number of colors and $N_f$ is the number of flavors, $D_\mu = \partial_\mu + ig_{YM}A_\mu$ is the covariant derivative with the gluon field $A_\mu$, which has $M^2 -1$ degrees of freedom, and $F_{\mu\nu} = {1\over ig_{YM}}[D_\mu, D_\nu]$ is the field strength tensor. The strength of interaction is fixed by the gauge coupling~$g_{YM}$.

Classically, this theory has conformal symmetry as long as the quarks are massless. Quantum mechanically, renormalization breaks the conformal symmetry since the Callan-Symanzyk $\beta$-function is non-zero. Therefore, it is expected that the bulk viscosity of the massless SU($M$) gauge theory is directly related to the $\beta$-function. This is manifestly shown within the effective kinetic theory analysis in Ref.~\cite{Arnold:2006fz}. In the rest of our analysis, we mainly consider the large $M$ limit. In this limit, the relevant interaction strength turns out to be the 't~Hooft coupling $\lambda = g_{YM}^2 M$ and then $\beta \sim \lambda^2/M$ \cite{Gross:1973ju}.

In principle, to study bulk viscosity comprehensively one should consider massive fermion fields, since a constant mass is a parameter that breaks conformal symmetry as well. In the light of the forthcoming discussion it is, however, not necessary here as the quark contribution will be $M$ suppressed compared
to the gluon contribution.

\subsection{Kubo formula for bulk viscosity \label{alina1}}

The first-principles prescription to compute bulk viscosity is given by the Kubo formula \cite{Kubo57}:
\ba
\label{bulk-zeta}
\zeta = \frac{1}{2}\lim_{\omega \to 0}\lim_{{\bf k}\to 0}
\frac{\rho_{PP}(\omega, {\bf k})}{\omega} ,
\ea
where $\rho_{PP}(\omega, {\bf k})$ is the spectral function of the pressure-pressure correlator and $\omega$ is the frequency of the hydrodynamic mode. In the following discussion, we will often omit the ${\bf k}$ dependence from the correlation functions and spectral densities. The common ${\bf k}\to 0$ limit should be understood in those cases.

The spectral function is related to the imaginary part of the pressure-pressure retarded correlation function:
\ba
\label{spec}
\rho_{PP}(\omega) = 2{\text{Im}} G^{PP}_R(\omega)\equiv i(G^{PP}_R(\omega)-G^{PP}_A(\omega)),
\ea
where we used $G_A=G_R^*$.  In the rest frame of the fluid cell, the pressure operator is given by the trace of the stress-tensor $\hat{P} = -{1\over 3} \hat{T}^i_i$. Because of the energy-momentum conservation, one can easily show that the spectral functions $\rho_{\epsilon P}(\omega, {\bf k})$ and $\rho_{\epsilon\epsilon}(\omega, {\bf k})$ must vanish in the same limit~\cite{Jeon:2015dfa}, where $\hat{\epsilon} = \hat{T}^{00}$ is the energy density operator. For theoretical analysis, it is often more advantageous to use the trace of the full stress-energy tensor $\hat{\Theta}/3 = \hat{T}^{\mu}_{\mu}/3 = {1\over 3}\hat{\epsilon} - \hat{P}$ which is Lorentz invariant, or the more 
kinetic-theory-friendly combination $\hat{P'} = \hat{P} - c_s^2\hat{\epsilon}$ 
({\it c.f.} Eq.(\ref{q}) below and also see Refs.~\cite{Jeon:1994if,Czajka:2017wdo})
which reduces to $-\hat{\Theta}/3$ in the conformal limit.
Here $c_s^2 = \partial P/\partial \epsilon$ is the speed of sound squared.
Similar combination also arises naturally in calculations of bulk viscosity in gauge-gravity
theories \cite{Buchel-bound}. 
Therefore,
\ba
\label{bulk-zeta1}
\zeta = \lim_{\omega \to 0}\frac{{\text{Im}} G_R^{OO}(\omega)}{\omega}
=  \lim_{\omega \to 0} \partial_\omega {\text{Im}} G_R^{OO}(\omega)
\ea
with the retarded correlation function given in coordinate space as
\ba
\label{gr}
G^{OO}_R(x) =
-i\theta(t)  \left\langle \left[\hat{\mathcal{O}}(t,x), \hat{\mathcal{O}}
(0,0)\right]  \right\rangle.
\ea
Here the operator $\hat{\mathcal{O}}$ can be $\hat{P}$, $\hat{\Theta}/3$ or $\hat{P}'$. This correlation function contains all the essential information about the physics of bulk viscosity and their structures are fixed by the Lagrangian (\ref{L-SUN}) and  thermal medium effects.

Although the Kubo formula (\ref{bulk-zeta1}) is general, in this section we focus on the regime of the sufficiently high energy scale, where the expansion in the small 't~Hooft coupling $\lambda$ may be applied. We consider here the limit where the 't~Hooft coupling $\lambda = g_{YM}^2 M$ remains small and the number of flavours $N_f$ is fixed while $M\to\infty$. In this limit one should, in principle, be able to calculate bulk viscosity perturbatively. Due to very complex multi-scale nature of the non-Abelian theory, a comprehensive quantitative computation of bulk viscosity using field theoretical tools is not an easy task. To date, a complete diagrammatic analysis of the bulk viscosity in QCD has not been carried out (for other transport coefficients of QED, see \cite{Gagnon:2006hi,Gagnon:2007qt}). However, an equivalent approach to compute the coefficient is offered by using  effective kinetic theory.

\subsection{Bulk viscosity from kinetic theory \label{kinetic}}

The foundations of the effective kinetic theory of the SU($M$) theory were formulated in Refs. \cite{Arnold:2000dr,Arnold:2002zm,Arnold:2003zc}. The scattering processes governing transport properties of the medium are embedded in the collision kernel of the Boltzmann equation and their sizes in terms of the gauge coupling $g_{YM}$, the numbers of degrees of freedom and the Casimir operators are explicitly shown in Ref. \cite{Arnold:2002zm}. Further this formulation was used in Ref. \cite{Arnold:2006fz} to calculate bulk viscosity of QCD. Here we briefly summarize these results in the leading order in the 't~Hooft coupling $\lambda= g_{YM}^2M$.

In the large-$M$ limit, the bulk viscosity in the leading order in $\lambda$ is governed only by the pure gluodynamics since quarks are suppressed by at least a factor of $M$. This can be clearly seen from the following analysis. Bulk viscosity depends on two factors. First, it must be proportional to the nonconformality parameter reflecting the incompressibility of the system. Second, it is controlled by the mean free path carrying the information on the microscopic properties of the medium, in particular, on the nature of interaction, and relevant degrees of freedom. From Ref. \cite{Arnold:2006fz} one observes that the same dependence of bulk viscosity on the nonconformality parameter is obtained either for quark and for gluon contributions. The mean free path of the quark contribution and of the gluon one is parametrically the same but it is associated with the corresponding numbers of degrees of freedom, which are different. While the number of gluons scales as $M^2$, the number of quarks scales as $M$. This dependence occurs for both the number conserving and number changing processes and can be extracted when analyzing all matrix elements and associated degrees of freedom shown explicitly in \cite{Arnold:2002zm}. Hence we ignore the quark contribution at every step of the forthcoming analysis.

In kinetic theory one focuses on the evolution of the distribution function of relevant quasiparticles. The evolution of the gluon distribution function $f(p,x)$ is governed by the Boltzmann equation of the form:
\ba
(\partial_t + {\bf v} \cdot \nabla_{\bf x}) f({\bf p},{\bf x},t) = - \mathcal{C} [f].
\ea
Since $f({\bf p},{\bf x},t)$ is slightly out of equilibrium it can be expressed as $f= f_{\rm eq} + f_1$, where $f_{\rm eq}$ is of the form $f_{\rm eq}({\bf p},{\bf x},t) = (e^{\beta(t) \gamma_u (E_p (x)-{\bf p}\cdot {\bf u}({\bf x}))} - 1)^{-1}$, with $\gamma_u=(1-u^2)^{-1/2}$. $f_{\rm eq}$ is therefore a function of time-space dependent quantities: $\beta(t)$ being the inverse of temperature $T(t)$ and $E_p(x)=\sqrt{{\bf p}^2 + m^2_{\rm th}(x)}$ - the energy of a gluon where the $x$ dependence appears through the thermally fluctuating mass $m_{\rm th}(x)$. $f_1$ is the nonequilibrium correction, which includes both the action of hydrodynamic forces and the correction due to thermally fluctuating mass. $\mathcal{C} [f]$ is a collision term, that contains processes involving only gluons, namely, the number conserving $gg \to gg$ scatterings and the number changing $g\to gg$ splittings. Its explicit form can be found in \cite{Arnold:2002zm}. The left-hand side of the Boltzmann equation at the linearized order is then:
\ba
\label{LHS}
(\partial_t + {\bf v} \cdot \nabla_{\bf x}) f_{\rm eq}({\bf p},{\bf x},t)
= -\beta^2(t) S({\bf p}) \nabla \cdot {\bf u}({\bf x})\Big|_{\beta(t)=\beta, {\bf u}({\bf x})=0},
\ea
where $S({\bf p}) = -T q({\bf p}) f_0(E_p)(1+f_0(E_p))$ and $f_0$ is the Bose-Einstein distribution function $(e^{\beta E_p } - 1)^{-1}$. The form of the quantity $q({\bf p})$ is most essential as it establishes the final parametric dependence of bulk viscosity on the nonconformality parameter. It reads:
\ba
\label{q}
q({\bf p}) =\frac{1}{E_p}\bigg[\left(\frac{1}{3} -  c_s^2\right) {\bf p}^2 - c_s^2 \tilde m^2 \bigg].
\ea
The quantity $\tilde m^2$ is of the form:
\begin{eqnarray}
\label{m-til}
\tilde m^2=m^2_{\rm th}- \frac{d(m^2_{\rm th})}{d(\ln T^2)} = -\frac{M T^2}{6} \beta_\lambda.
\end{eqnarray}
The formula (\ref{q}) is obtained by taking into account the stress-energy conservation law, thermodynamic relations and space dependence of the quasiparticle energy. Note that as the consequence of the temperature dependence of the quasiparticle mass, given by $m^2_{\rm th}=g_{YM}^2(T)MT^2/6$, the beta function of SU$(M)$ theory $\beta_\lambda=-11\lambda^2/(48\pi^2M)$ arises in the formula (\ref{m-til}) and, consequently, in Eq. (\ref{q}). The $\beta_\lambda$-function is just the parameter that breaks conformal symmetry in the system and the factor $1/3-c_s^2$, with the speed of sound squared $c_s^2=\partial P/\partial \epsilon$, is equivalent to it through the relation:
\ba
\label{spd}
\frac{1}{3} - c_s^2= -\frac{5}{72\pi^2}M \beta_\lambda =
\frac{55}{3456\pi^4} \lambda^2.
\ea
Due to such a dependence, $q({\bf p})$ can be expressed in a simple form:
\ba
\label{q1}
q({\bf p}) =\left(\frac{1}{3} -  c_s^2\right) \left[|{\bf p}|- \frac{4\pi^2}{5}
\frac{T^2}{|{\bf p}|}\right].
\ea
In all formulas, terms which are suppressed by any powers of $M$ were omitted. The form of left-hand side of the Boltzmann equation, Eq.~(\ref{LHS}), dictates also the form of the correction $f_1$ which, in turn, fixes the form of the linearized collision kernel. The correction is $f_1=\beta^2 f_0(1+f_0) \chi \nabla \cdot {\bf u}$, so that both sides of the Boltzmann equation are proportional to $ \nabla \cdot{\bf u}$. By dropping this scalar factor, the Boltzmann equation can be expressed in a convenient form $S({\bf p}) = [\mathcal{C} \chi] ({\bf p})$. Bulk viscosity may be then found as:
\ba
\label{i-eq}
\zeta = \tilde S_{m} \tilde C^{-1}_{mn} \tilde S_{n},
\ea
where the matrix is $\tilde C_{mn} = 2M^2 \int_p \phi_m(p) [\mathcal{C}\phi_n] (p)$ and the column vector is $\tilde S_{m}=2M^2 \int_p \phi_m(p) S(p)$, with the basis functions $\phi_m(p) = p^m T^{K-m-1}/(T+p)^{K-2}$ and $m=1,...,K$. The numerical procedure relies on the variational method. Since $\zeta \propto S^2 \propto q^2$, the bulk viscosity is clearly expressed by the nonconformality parameter squared, $(1/3-c_s^2)^2$ or equivalently $\beta^2_\lambda$ and the inverted collision kernel introduces the mean free path. The final expression then scales as:
\ba
\label{zet}
\zeta = \frac{a T^3 M^2}{\lambda^2 \ln(b/\lambda)}
\left(1/3-c_s^2\right)^2,
\ea
where $a$ and $b$ should be obtained by solving the integral equation~(\ref{i-eq}). The whole procedure of finding bulk viscosity coefficient of QCD is comprehensively discussed in \cite{Arnold:2006fz} for different values of the number of flavors $N_f$. One can then reproduce the dependence of bulk viscosity of the SU($M$) theory on the coupling constant $\lambda$ from Fig. 1 of Ref.~\cite{Arnold:2006fz} by setting all quark masses to 0, taking $N_f=0$, and rescaling the coupling $4\pi M \alpha_s \to \lambda$. Due to the same sizes of the nonconformality parameter and the 't~Hooft coupling constant squared, given by the relation~(\ref{spd}), one can write:
\ba
\label{bulk-s}
\frac{\zeta}{s} \propto \frac{\lambda^2}{\ln(b/\lambda)} \propto
\frac{\left(1/3-c_s^2\right)}{\ln(b/\lambda)} ,
\ea
where we used the entropy density $s=(P+\epsilon)/T \propto M^2 T^3$. The formula (\ref{bulk-s}) shows that in the very weak coupling regime the leading order bulk viscosity over entropy density ratio is a linear function of the nonconformality parameter $1/3-c_s^2$, up to the logarithm. This occurs due to the fact that $\beta_\lambda$ function is of the same order as the inverse of the mean free path. This behavior is characteristic for the theories when the conformal symmetry is broken only by the $\beta_\lambda$ function. These are, for example, SU$(M)$ in the large $M$ limit or massless QCD. Also, the shear viscosity coefficient of QCD with the effective coupling $\lambda$ was studied in \cite{Huot:2006ys} and the
result is:
\ba
\label{shear}
\frac{\eta}{s} = \frac{A}{\lambda^2 \ln(B/\lambda)}
\ea
with $A$ and $B$ being numerical constants. Combining Eqs. (\ref{shear}) and (\ref{zet}), one finds that the ratio of $\zeta/\eta$ is characterized by the quadratic dependence of the nonconformality parameter:
\ba
\label{sellroth}
\frac{\zeta}{\eta} \propto \left(1/3-c_s^2\right)^2.
\ea

\subsection{One-loop diagram and power counting \label{alina2}}

So far kinetic theory has been the only utilizable method allowing for a quantitative computation of transport coefficients of non-Abelian weakly coupled gauge theories. However, it is an effective description of quasiparticle dynamics and its equivalence to quantum field theoretical approach has not been fully shown for the SU($M$) theory. In particular, a rigorous diagrammatic derivation of the Boltzmann equation of that theory, and consequently bulk viscosity, has not yet been presented. Given that, transport coefficients of QED were analyzed using standard diagrammatic techniques in Refs.~\cite{Gagnon:2006hi,Gagnon:2007qt}. In this manuscript we strongly rely on procedures shown in Refs.~\cite{Gagnon:2006hi,Gagnon:2007qt}, which are similar here, but due to different nature of interaction in the non-Abelian gauge theory there are additional diagrams contributing to the kernel and the effective vertices have more complicated structure. All these subtleties will be discussed in the forthcoming parts in some detail.

As was shown in Refs.~\cite{Jeon:1994if,Arnold:2001ba,Arnold:2001ms,Arnold:2002ja}, the equivalence of the diagrammatic method and the kinetic theory description can be established when the ladder diagram resummation dominates the leading order result. To carry out a qualitative analysis of the weakly coupled large $M$ Yang-Mills theory, it will be therefore enough to confirm that the planar ladder diagrams dominate in the viscosity calculations. The goal of the forthcoming subsections is to do just that for the purpose of establishing the qualitative behavior of the bulk viscosity in the weakly coupled theory. We will not, however, attempt to carry out the full analytical computation necessary for the quantitative analysis as this is beyond the scope of this work.

To perform the qualitative analysis we need to establish the necessary basic ingredients dictated by the Kubo formula (\ref{bulk-zeta1}). The full stress-energy tensor of the SU($M$) gauge theory is given by:
\ba
\label{Tens}
T^{\mu\nu} =
F_a^{\mu\alpha} F_{\alpha}^{a\,\nu}
- g^{\mu\nu}{\cal L}_g.
\ea
To have some insight into the parametric form of the bulk viscosity and to establish a starting point for evaluating the size of microscopic processes governing its behavior, it is illuminating to consider only the kinetic terms of the stress-energy tensor, that is, the first term in Eq. (\ref{Tens}). Since quarks are subleading we focus only on the gluonic contribution to the stress-energy tensor; we briefly comment on this issue later.

Power counting of the gluon one-loop diagram is most conveniently accomplished using the $(r,a)$ basis of the thermal field theory. This was shown for the scalar field theory in \cite{Wang:1998wg,Wang:2002nba} and also for gauge theories in~\cite{Arnold:2001ba}. In this basis, the elementary gluon propagators are the retarded propagator $G_{ra}$, advanced one $G_{ar}$ and the auto-correlation function, which introduces information on the medium momentum distribution, $G_{rr}=(1+2n_B)(G_{ra}-G_{ar})$, where $n_B$ is the Bose-Einstein statistics. These propagators carry indices related to color, spin or the Lorentz structure, but within this analysis we will not explicitly show them.

Since all these propagators describe a propagation of a given particle in a thermal medium they are dressed with self-energies. The retarded self-energy is given by $\Pi=\textrm{Re}\,\Pi -i \textrm{Im}\,\Pi$ and the retarded propagator is:
\ba
G_{ra}(p) = \frac{A^g(p)}{p^2  -  \Pi (p)},
\ea
where $A^g(p)$ carries the necessary color and tensor indices. The advanced propagator then is given by $G_{ar}=G_{ra}^*$.

In the weakly coupled limit, the retarded propagator has poles at $p^0 \approx \pm E_p - i\Gamma_p^g$ where the quasi-particle energy is given by $E_p=\sqrt{{\bf p}^2 + m_{{\rm th}}^2}$ with the thermal mass $m_{{\rm th}}^2 = {\rm Re} \Pi(p) $. The thermal width is given by the imaginary part of the self-energy at the on-shell momentum $\Gamma_p={\textrm{Im}\,\Pi(E_p, |{\bf p}|)}/{2E_p}$. The resummed propagator can be then expressed as:
\ba
\label{propagators-dressed-ra}
G_{ra}(p) &=& \frac{A^g}{(p_0+i\Gamma_p)^2 - (E_p)^2}.
\ea
In using the propagators in the $(r,a)$ basis to evaluate the Kubo formula, we encounter two different types of singularities: The pinching pole singularity and the collinear singularity. Both are regulated by the thermal self-energies but they complicate the power counting. In this section, we discuss the pinching pole singularity and its ramification. The effect of collinear singularity is discussed in the later section.

Using the operator $\hat P'$ defined below Eq. (\ref{spec}) one finds the gluonic one-loop contribution to the bulk viscosity in the pinching pole approximation as:
\ba
\label{exp}
\zeta &\propto&
\int \frac{d^4p}{(2\pi)^4}\big[c_s^2 p_0^2-1/3 {\bf p}^2\big]^2 n_B(p_0)(1+n_B(p_0))
G_{ra}(p)G_{ar}(p)\;\;\;\;\;
\ea
Note that the propagator part is written in a symbolic way as all internal indices and traces over them are not written explicitly. The retarded propagator has poles at $p^0 = \pm E_p - i\Gamma^g$ and the advanced one at $p^0 = \pm E_p + i\Gamma^g$. Hence the two poles at ${\rm Re}\,p^0 = E_p$, for instance, are separated by $2i\Gamma_p^g$ in the imaginary direction on the opposite side of the integration contour. When integrated, these ``pinching poles'' result in a large $1/\Gamma^g$ factor leading to the following power counting:
\ba
\label{gl}
&&\int dp^0  n_B(p_0)(1+n_B(p_0)) \bigg[c_s^2 p_0^2-\frac{1}{3} {\bf p}^2\bigg]^2
G_{ra}(p)G_{ar}(p) \\
&& \qquad\qquad\qquad\qquad \to  n_B(E_p)(1+n_B(E_p))
\bigg[\frac{(1/3-c_s^2) {\bf p}^2
- c_s^2 m_{\rm th}^2}{E_p}\bigg]^2 \frac{M^2-1}{\Gamma_p^g}. \nonumber
\ea
This expression requires a few comments. First, the factor in the square bracket has analogous form to the quantity $q({\bf p})$ found within the kinetic theory and given by Eq. (\ref{q}), up to the thermal mass term. The expression (\ref{gl}) is obtained, however, only from the one-loop analysis and it does not include all effects. We expect that when the Lagrangian part and the interaction terms of the stress-energy tensor operator are included in the computation, the term $d(m_{\rm th}^2)/d(\ln T^2)$ will emerge in Eq. (\ref{gl}). This term, when subtracted from $m^2_{\rm th}$ in (\ref{gl}) will be analogous to the expression (\ref{m-til}) and therefore will result in the $\beta_\lambda$ function emergence, or equivalently $(1/3-c_s^2)$, analogously to what was obtained within the kinetic theory. The inclusion of the temperature dependence of the thermal mass was justified in Ref. \cite{Jeon:1994if} and explicitly incorporated to formulate fluid dynamic equations in Ref. \cite{Czajka:2017wdo}, but for scalar theories only. We expect that performing full analysis of the spectral function of the SU($M$) theory will result in this dependence of the nonconformality parameter, but we do not intend to derive it. We do focus on discussing the consequences of the presence of $1/\Gamma^g$ factor in the formula (\ref{gl}), which governs the mean free path behavior. Before doing that let us point out that $M^2-1$ in (\ref{gl}) reflects the number of degrees of freedom and since the number of colors is large we will be neglecting further the constant ``$-1$''.

\begin{figure}[b]
\centering
\begin{tabular}{ccc}
a)&&b)\\
\includegraphics[width=0.3\textwidth]{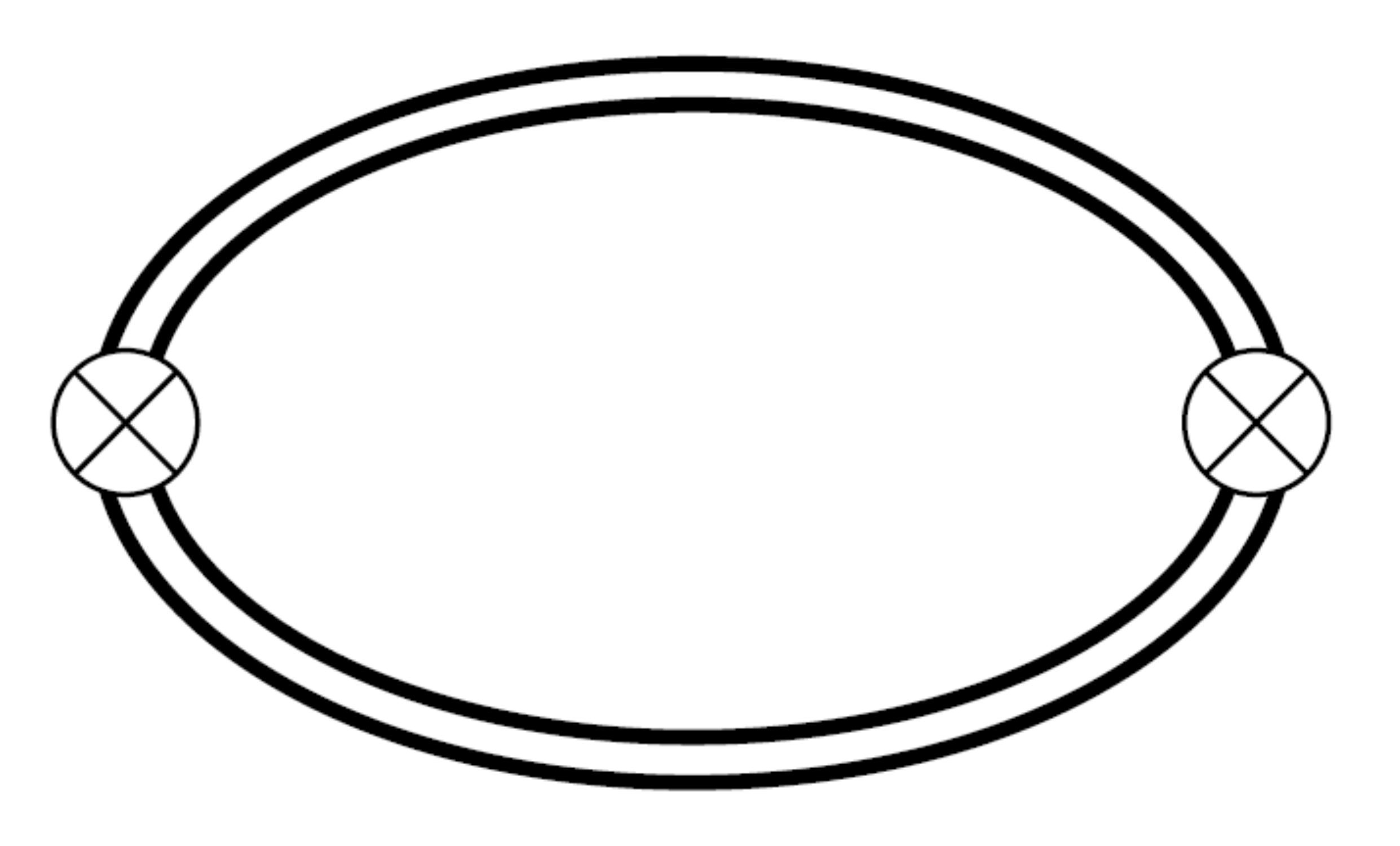} &&
\includegraphics[width=0.3\textwidth]{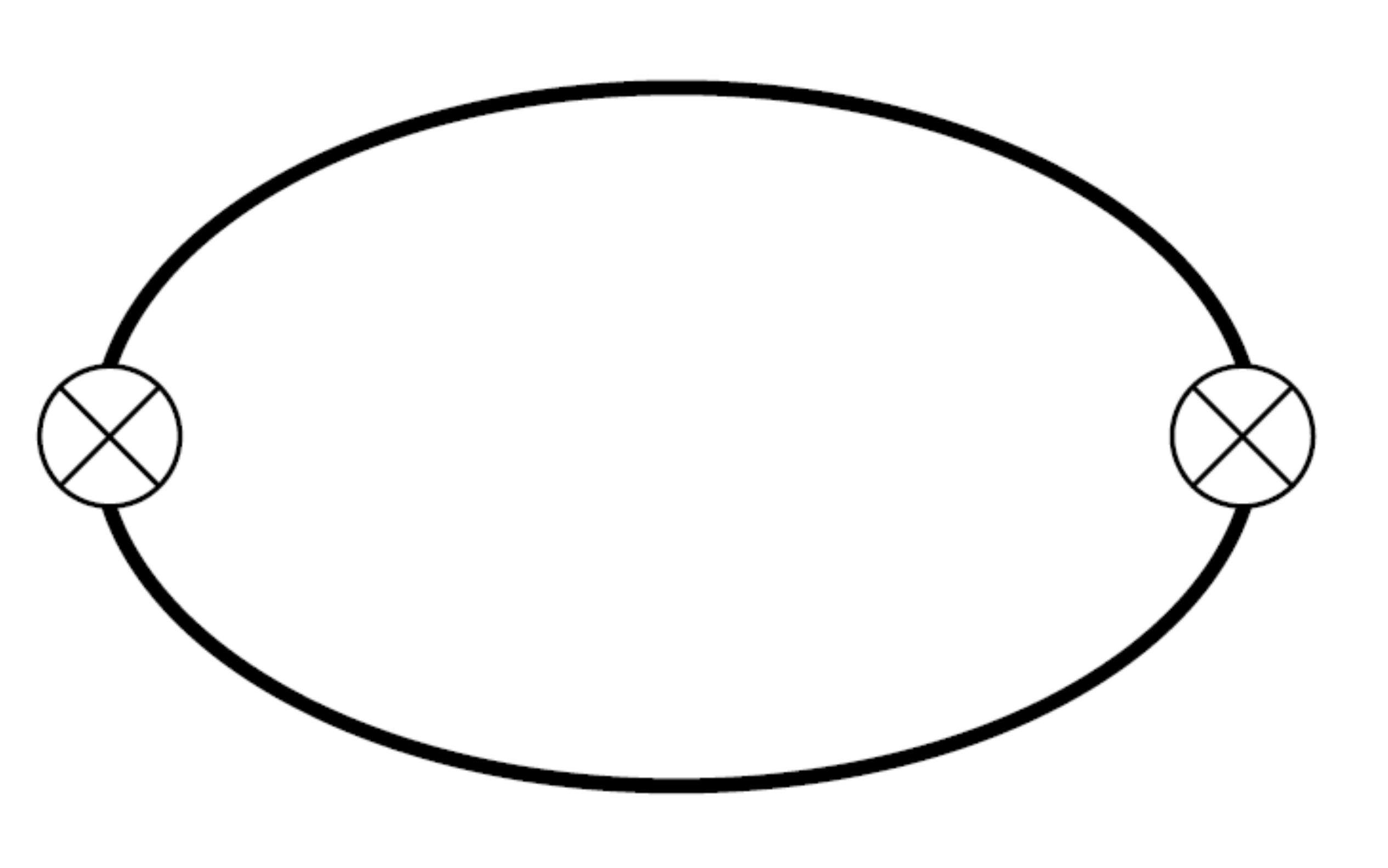}  \\
$\propto M^2/\Gamma^g$ && $\propto M N_f/\Gamma^f$
\end{tabular}
\caption{One-loop contribution to the spectral function. a) gauge boson
loop, b) fermion loop. The crossed vertices denote insertions of the full trace of
the stress-energy operator.}
\label{fig-one}
\end{figure}

To represent the expression (\ref{gl}) diagrammatically, it is convenient to use the 't~Hooft notation \cite{tHooft:2002ufq} so that a double line corresponds to a gluon propagator and any fermion propagator is represented by a single line. In this representation power counting relies on the simple formula~\cite{tHooft:2002ufq}:
\ba
\label{N}
g_{YM}^{V_3+2V_4}M^{L}N_f^{L_f},
\ea
where $L$ is the number of closed loops, $V_3$ is the number of the 3-body interaction vertices and $V_4$ is the number of 4-body interaction vertices. In case of a fermion occurrence there is an extra factor of $N_f$ and $L_f$ is the number of fermion loops. Using the 't~Hooft notation, the one-loop diagram corresponding to the expression~(\ref{gl}), together with its typical size, is depicted in Fig.~\ref{fig-one}a), where the crossed vertices stand for the insertion of the renormalized operator of the trace of the stress-energy tensor. For a comparison, in Fig. ~\ref{fig-one}b) we also present the fermionic one loop with its typical size given in terms of the corresponding degrees of freedom and the fermionic thermal width $\Gamma^f$ being given by the imaginary part of the fermionic self-energy. Therefore, the gauge boson contribution to the correlation function at the leading order scales as $M^2/\Gamma^g$, since the diagram is made of two closed loops and the fermionic one scales as $MN_f/\Gamma^f$. As may be implied from Fig.~\ref{fig-one} each factor of the thermal width is associated with the presence of a pair of propagators. Thus one observes that the fermion contribution is subleading by a factor of $N_f/M$ comparing to the gluonic one as long as the same parametric dependence in the parameter $1/3-c_s^2$ holds and $\Gamma^f$ is of the same order as $\Gamma^g$. From the kinetic theory findings in Refs. \cite{Arnold:2006fz,Arnold:2002zm} one finds the parameter $(1/3-c_s^2)^2$ being common for gluons and fermions and we rely on this result. Given that, the estimates of the sizes of the fermionic and gluonic thermal widths are still needed. What is more, to fully estimate the leading order 2-point correlation function for bulk viscosity, one also needs to know typical sizes of the corresponding thermal masses, which are essential for number changing processes.

\subsection{Self-Energy Power Counting \label{alina3}}

Both the real part and the imaginary part of the self energy plays a key role in the calculation of transport coefficients. The role of the imaginary part (the thermal width) as a regulator for the pinching pole singularity has already been discussed in the previous section. The role of the real part (the thermal mass) is to regulate the infrared and collinear singularities that occur at finite temperature. Hence, the size of the thermal mass defines the soft scale while the temperature itself defines the hard scale. In QCD, we know that the thermal mass is of $O(g_{YM}T)$ while the thermal width is of $O(g_{YM}^4T)$ when the particle momentum is hard (For instance, see \cite{Gagnon:2006hi}). In the large $M$ limit, these need to be re-expressed using the 't~Hooft coupling.

\begin{figure}[h]
\centering
\begin{tabular}{ccccc}
a)&& b)&& c)  \\
\includegraphics[width=0.22\textwidth]{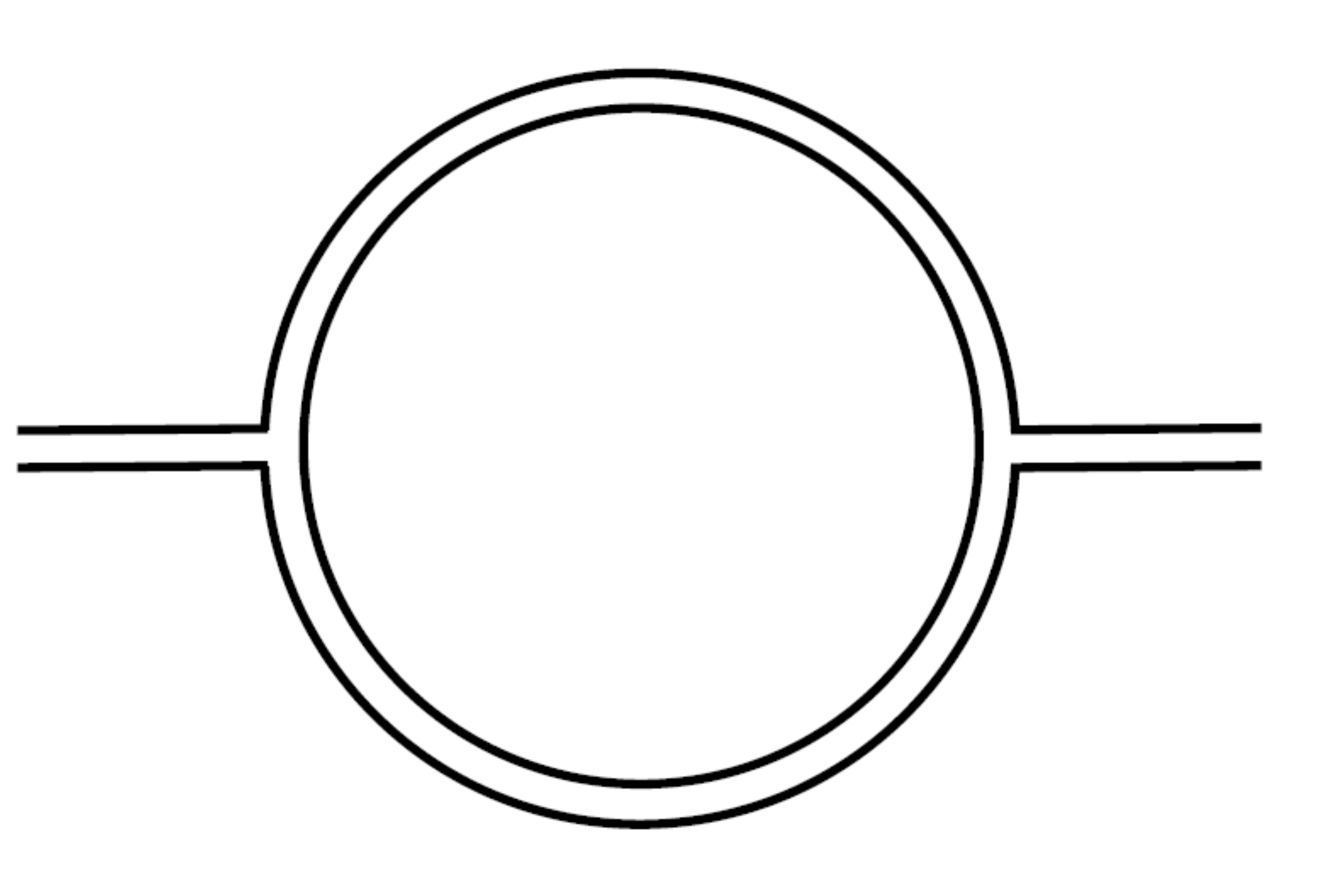} &&
\includegraphics[width=0.18\textwidth]{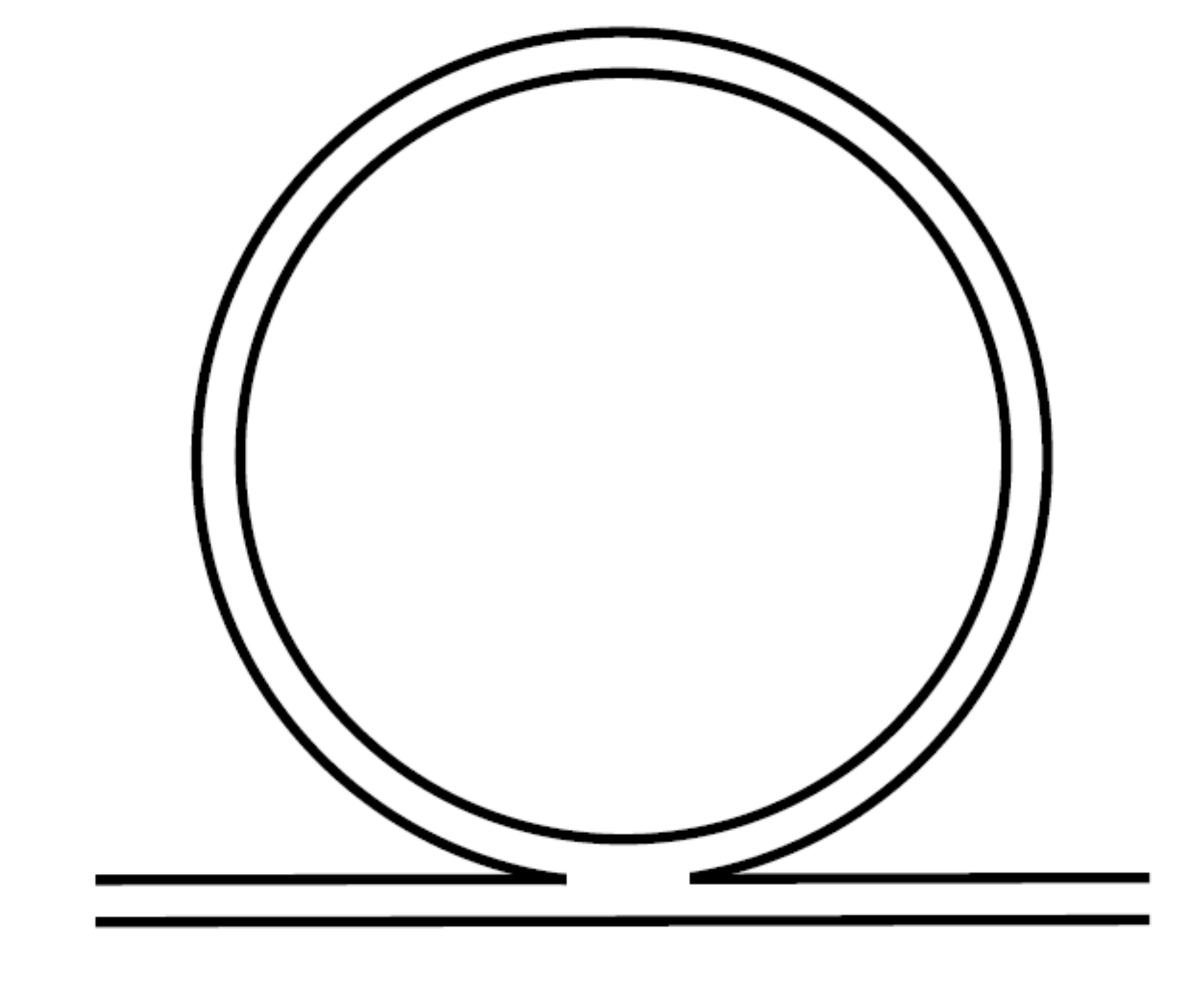}  &&
\includegraphics[width=0.22\textwidth]{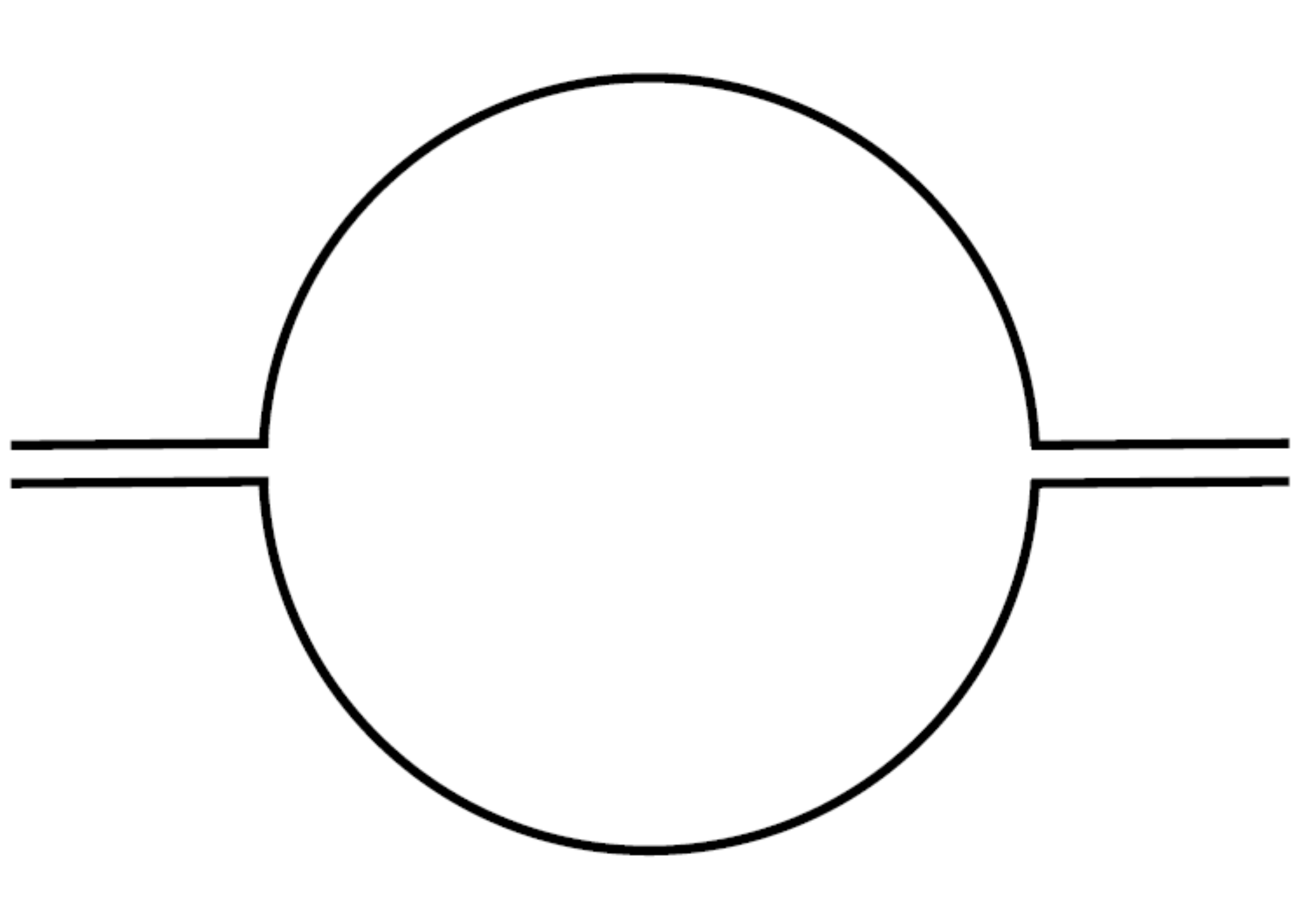} \\
$\propto g_{YM}^2M$ &&
$\propto g_{YM}^2M$ &&
$\propto g_{YM}^2N_f$
\end{tabular}
\caption{1-loop diagrams contributing to the real part of the gluon polarization tensor in a non-Abelian gauge theory together with their relative sizes given in terms of the gauge coupling $g_{YM}$, number of colors $M$ and the number of flavors $N_f$.  a) gluon loop, b) gluon tadpole, c) fermion loop.}
\label{fig-two}
\end{figure}

\begin{figure}[h]
\centering
\begin{tabular}{c}
\includegraphics[width=0.24\textwidth]{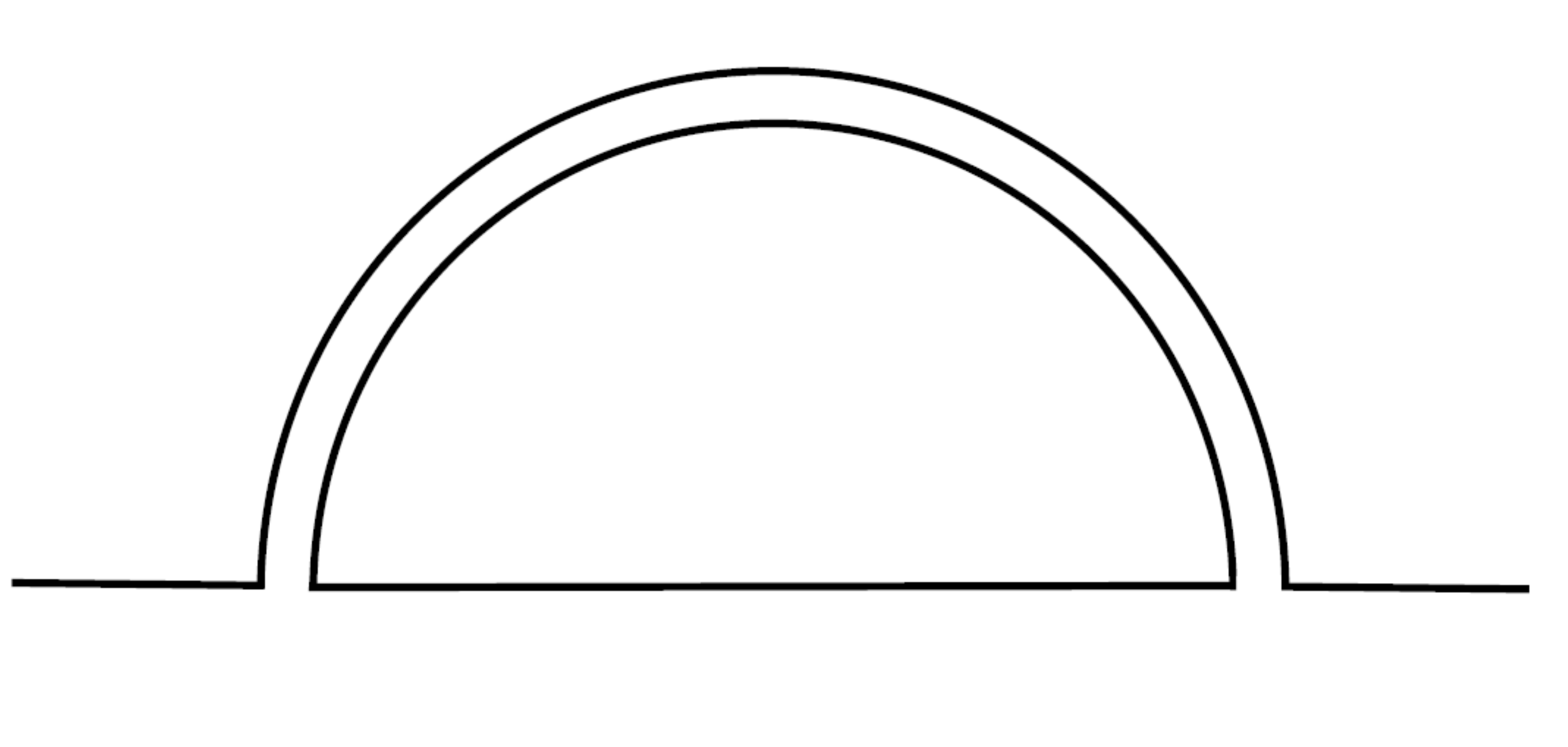} \\
$\propto g_{YM}^2M$
\end{tabular}
\caption{1-loop contribution to the real part of the quark self-energy.}
\label{fig-three}
\end{figure}

The thermal mass is determined by the real part of the self-energy of a particle. In our case, the leading order contribution comes from one-loop diagrams. The corresponding diagrams contributing to the gluon thermal mass in the double-line representation are shown in Fig.~\ref{fig-two}. For a systematic comparison, the fermion leading contribution to the real part of self-energy is shown in Fig.~\ref{fig-three}.

The coupling dependence comes from counting the interaction vertices and number of degrees of freedom using the formula (\ref{N}). As in case of one-loop diagrams contributing to the spectral density, when $M \to \infty$ the gluon loops, Figs.~\ref{fig-two}a) and ~\ref{fig-two}b), dominate over the fermion one by a factor of $M/N_f$. Thus, the leading order of the gluon thermal mass as well as the fermion one (Fig. \ref{fig-three}) in the large $M$ limit is:
\ba
\label{re}
{\rm Re}\Pi^{\rm HTL} \propto {\rm Re}\Sigma^{\rm HTL} \propto \lambda T^2.
\ea
For explicit expressions, see~\cite{Kalashnikov:1980tk,Klimov:1981ka,Braaten:1989mz}.

\begin{figure}[h]
\centering
\begin{tabular}{ccccccccc}
a)&& b)&& c) && d) && e)\\
\includegraphics[width=0.15\textwidth]{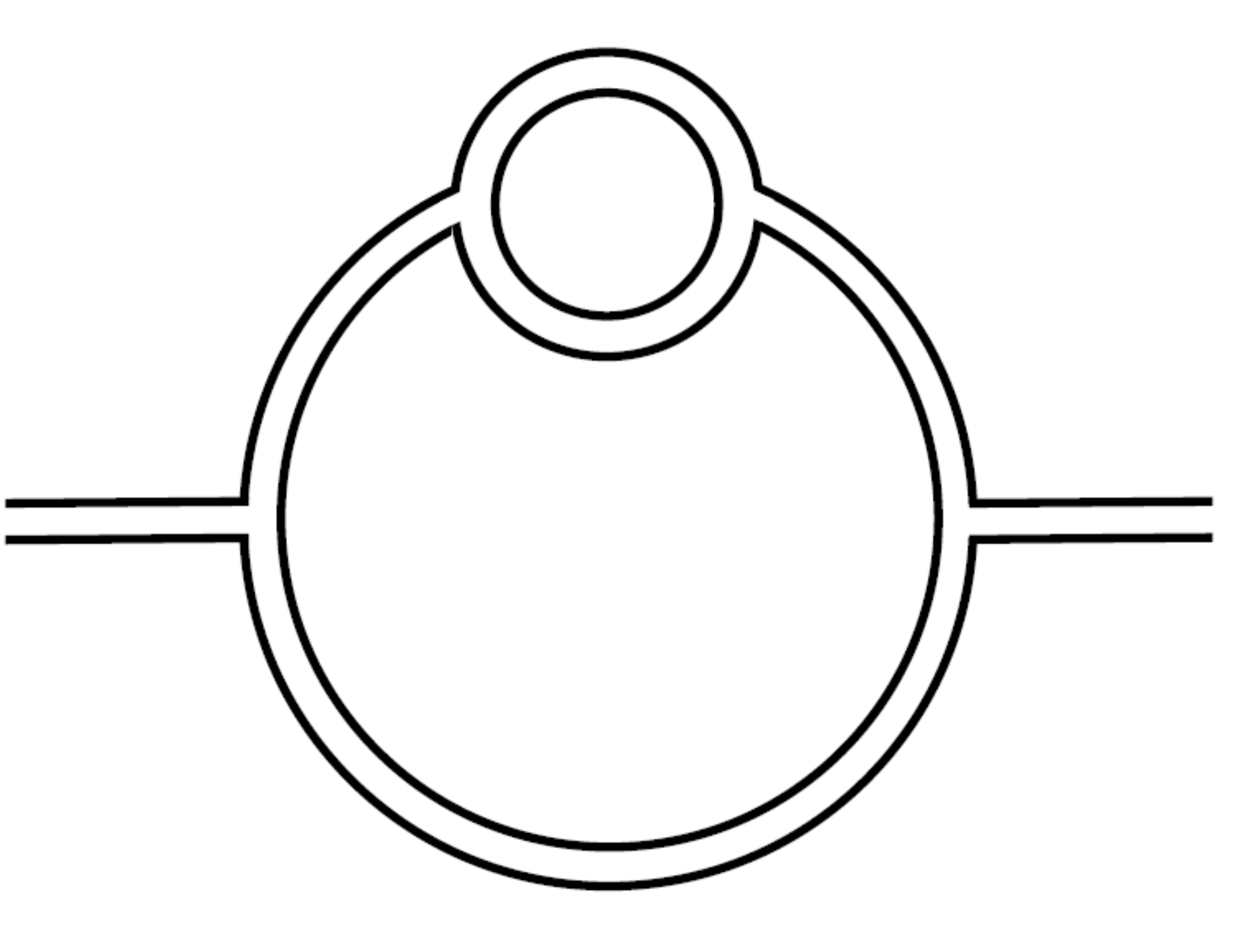} &&
\includegraphics[width=0.15\textwidth]{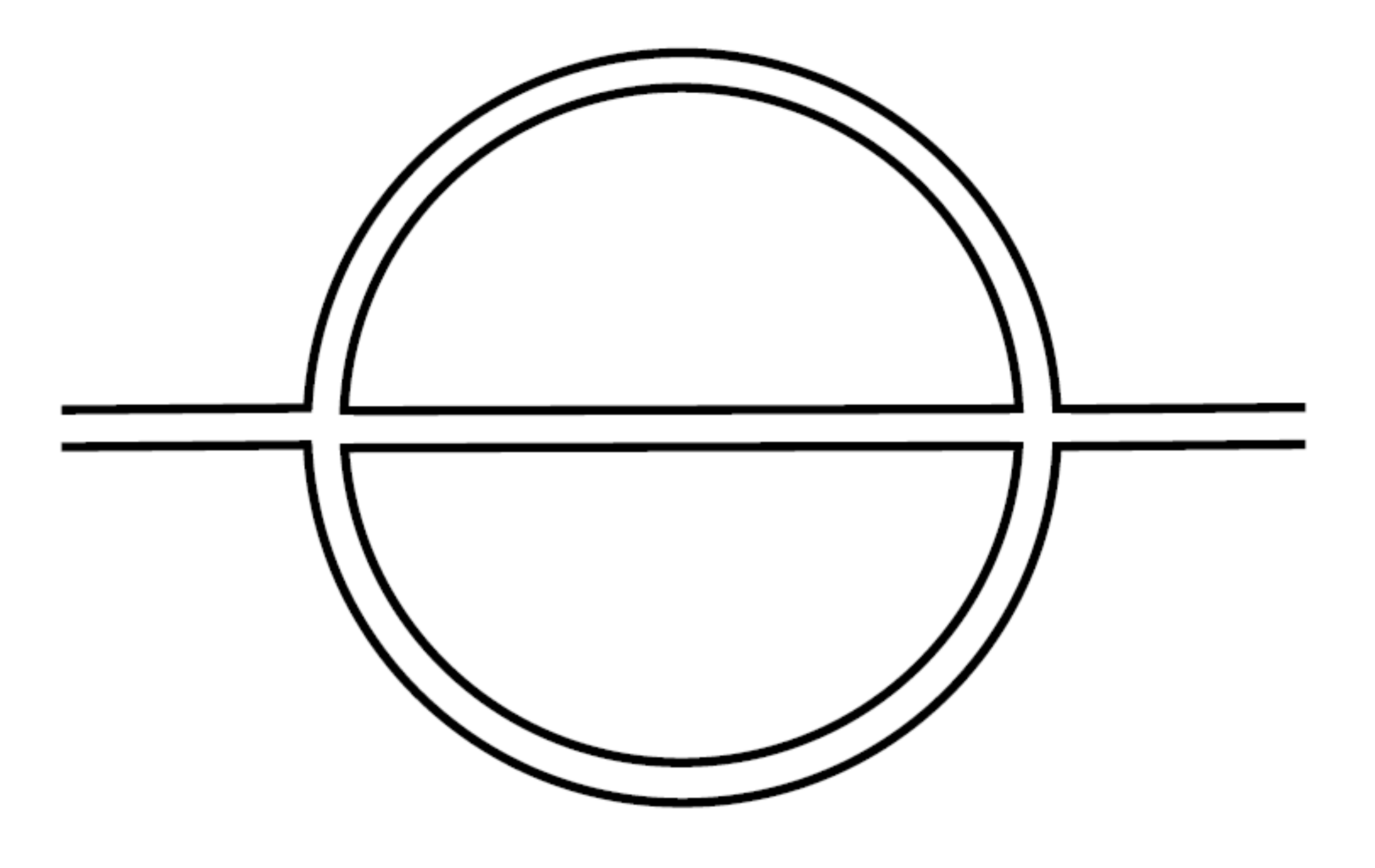} &&
\includegraphics[width=0.15\textwidth]{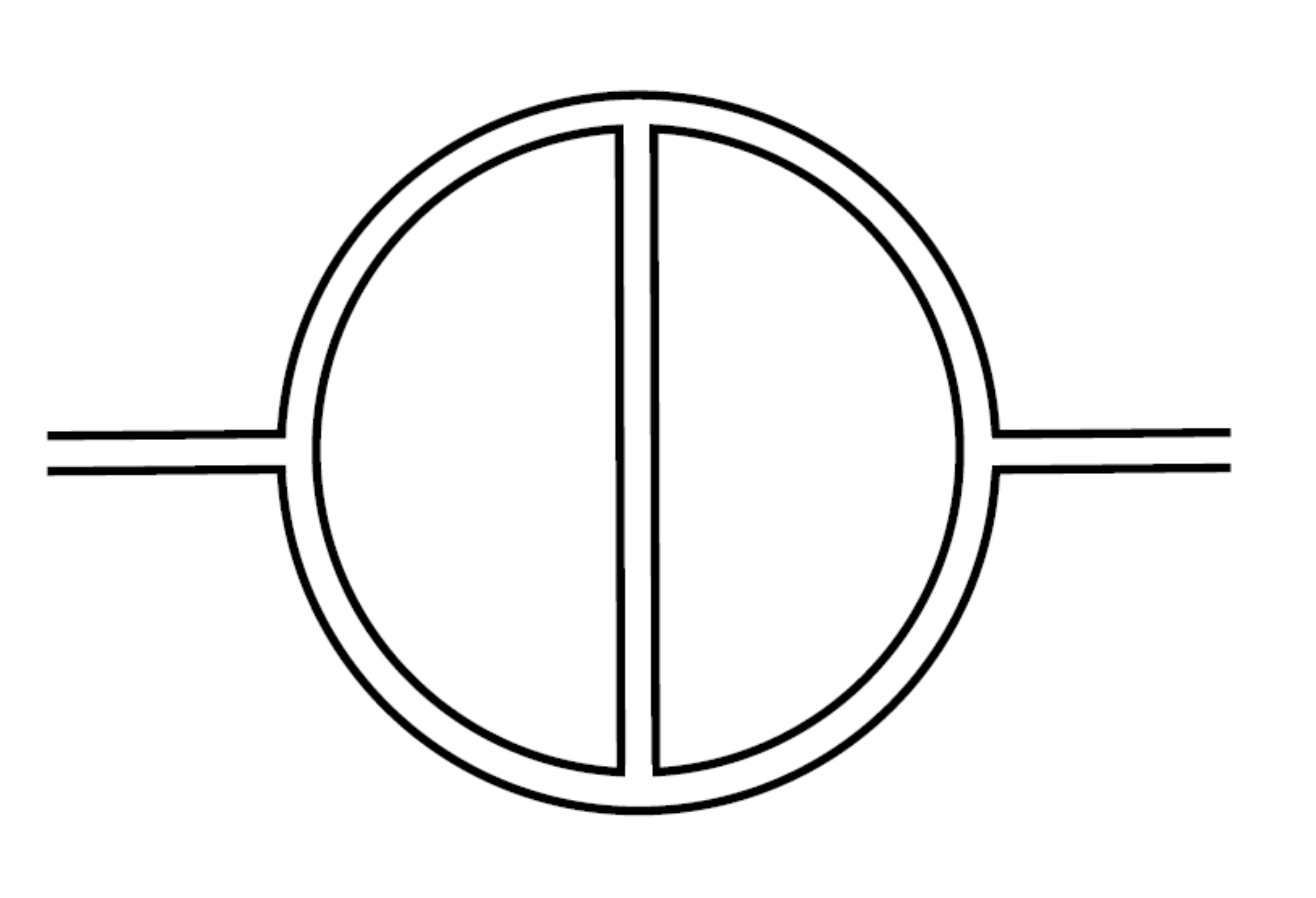}  &&
\includegraphics[width=0.15\textwidth]{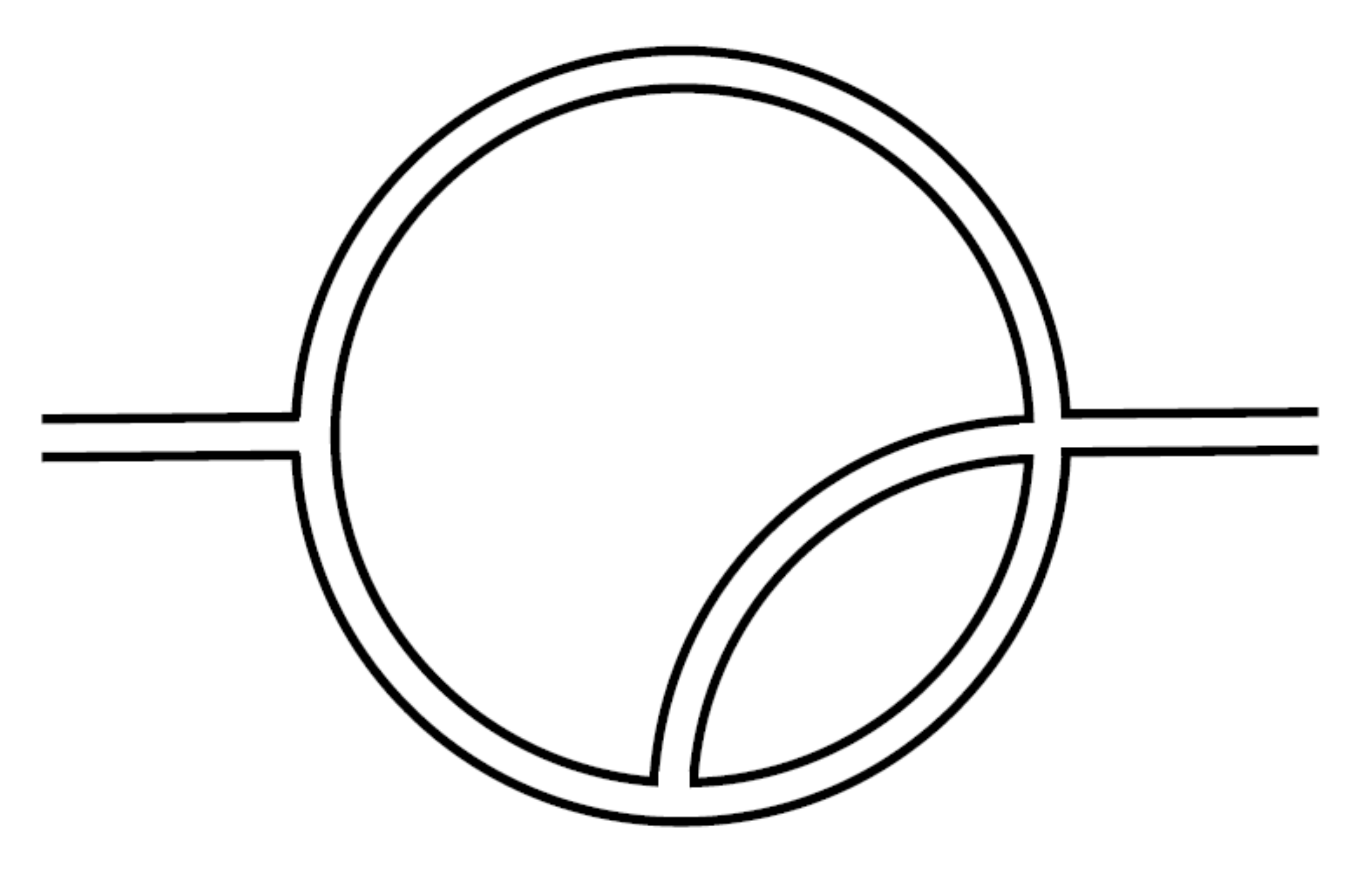} &&
\includegraphics[width=0.15\textwidth]{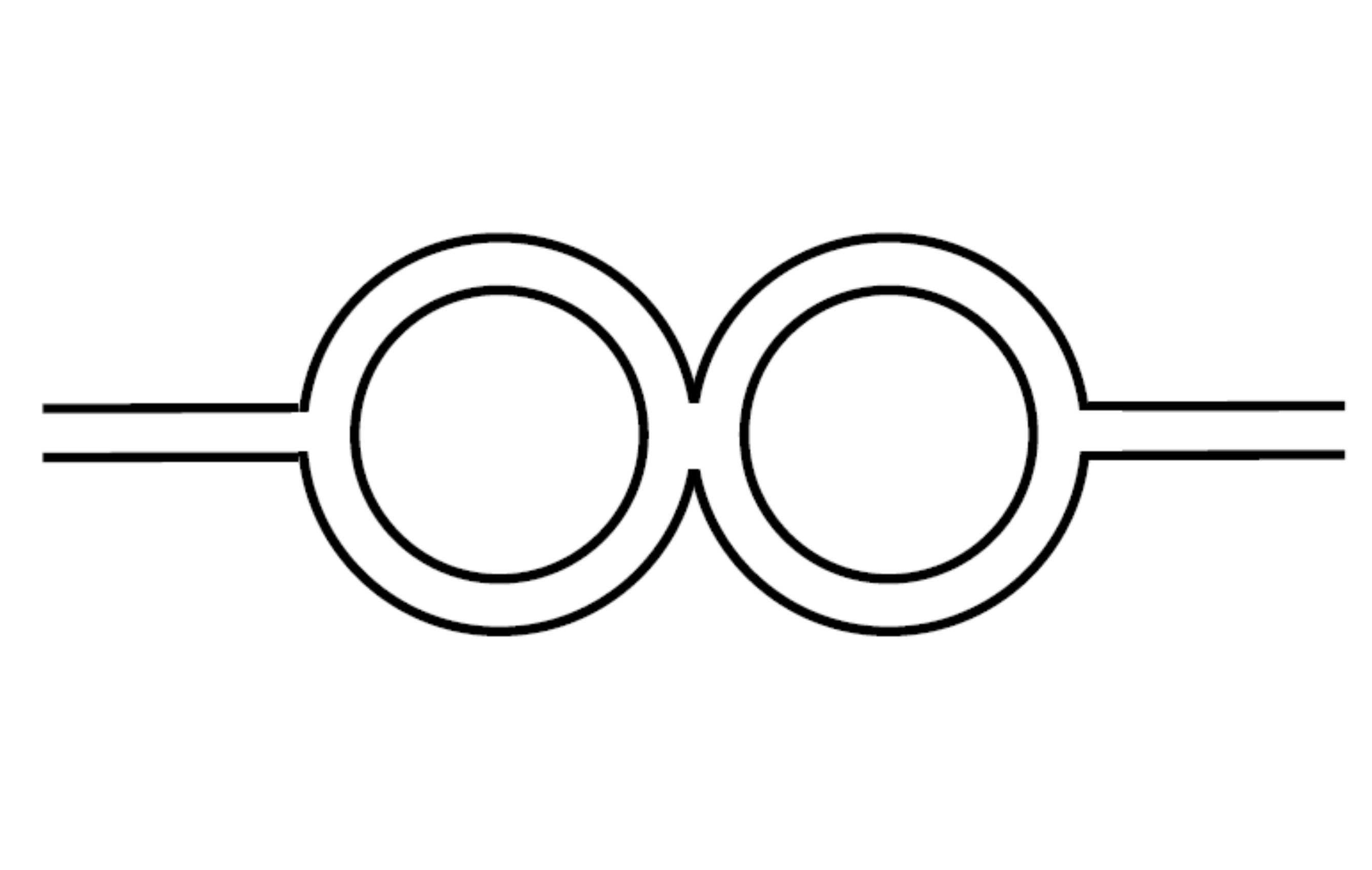} \\
$\propto g_{YM}^4M^2$ &&
$\propto g_{YM}^4M^2$ &&
$\propto g_{YM}^4M^2$ &&
$\propto g_{YM}^4M^2$ &&
$\propto g_{YM}^4M^2$
\\&&&&&&&& \\
&& f) && g) && h) && \\
&&
\includegraphics[width=0.15\textwidth]{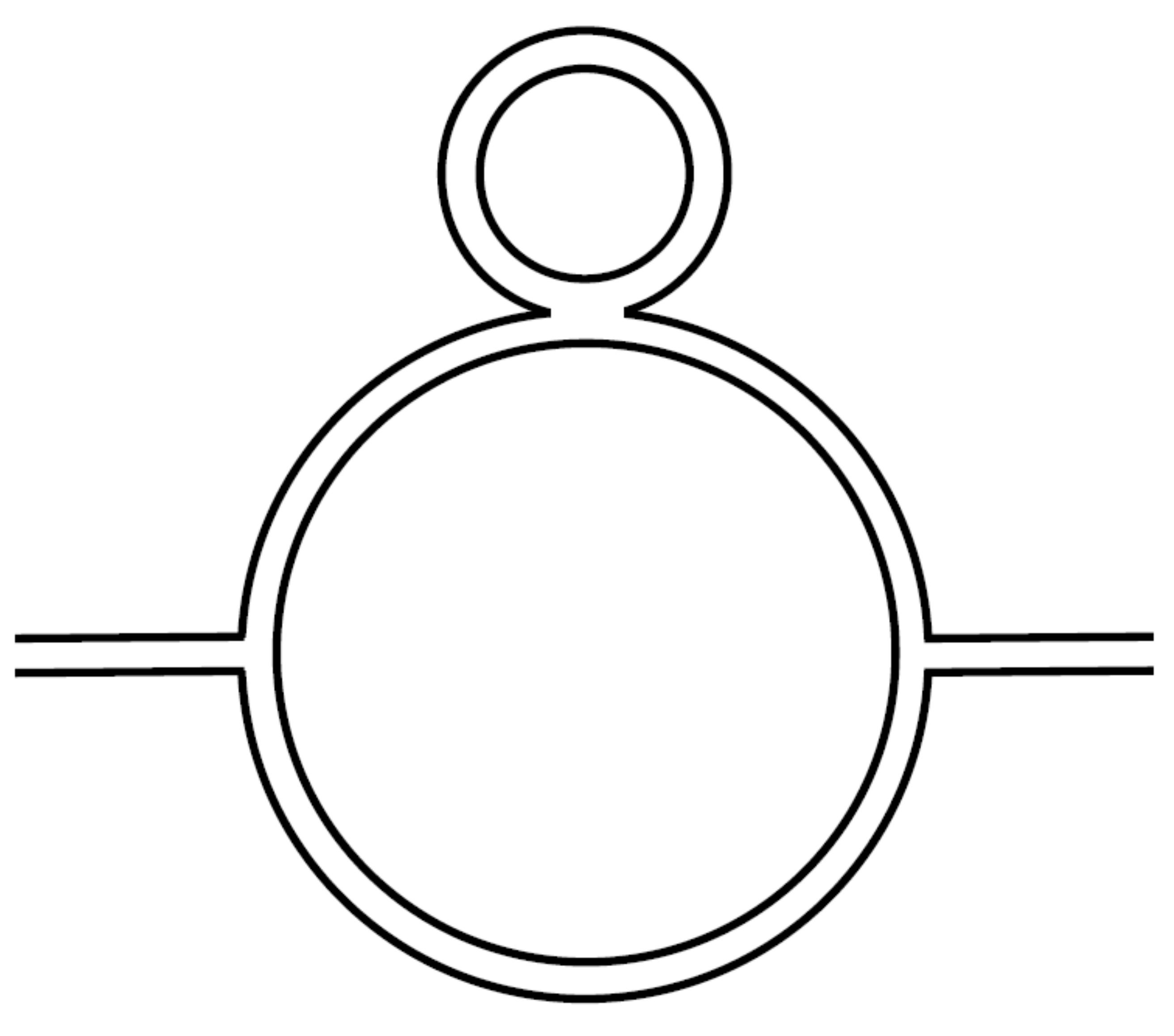}  &&
\includegraphics[width=0.15\textwidth]{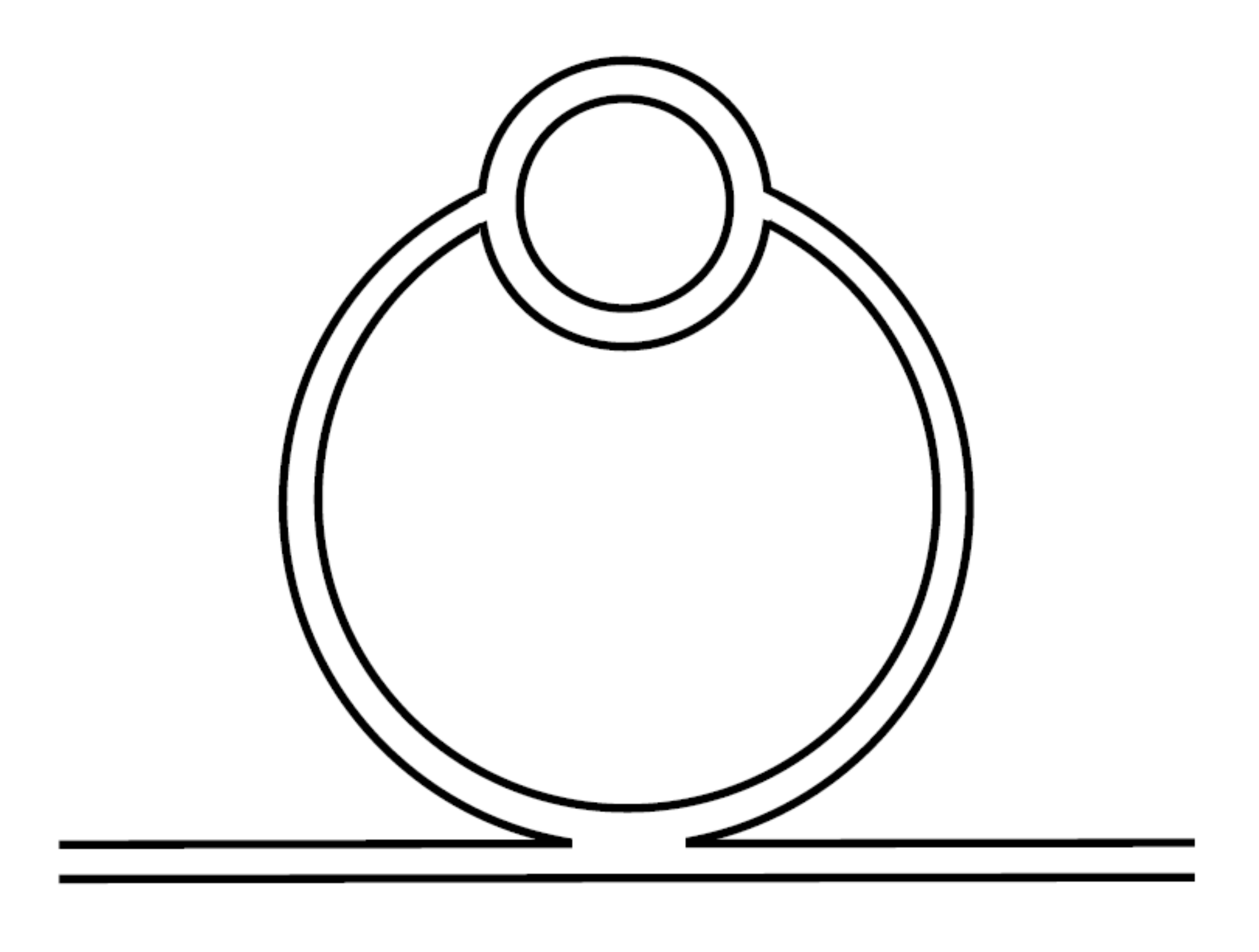} &&
\includegraphics[width=0.15\textwidth]{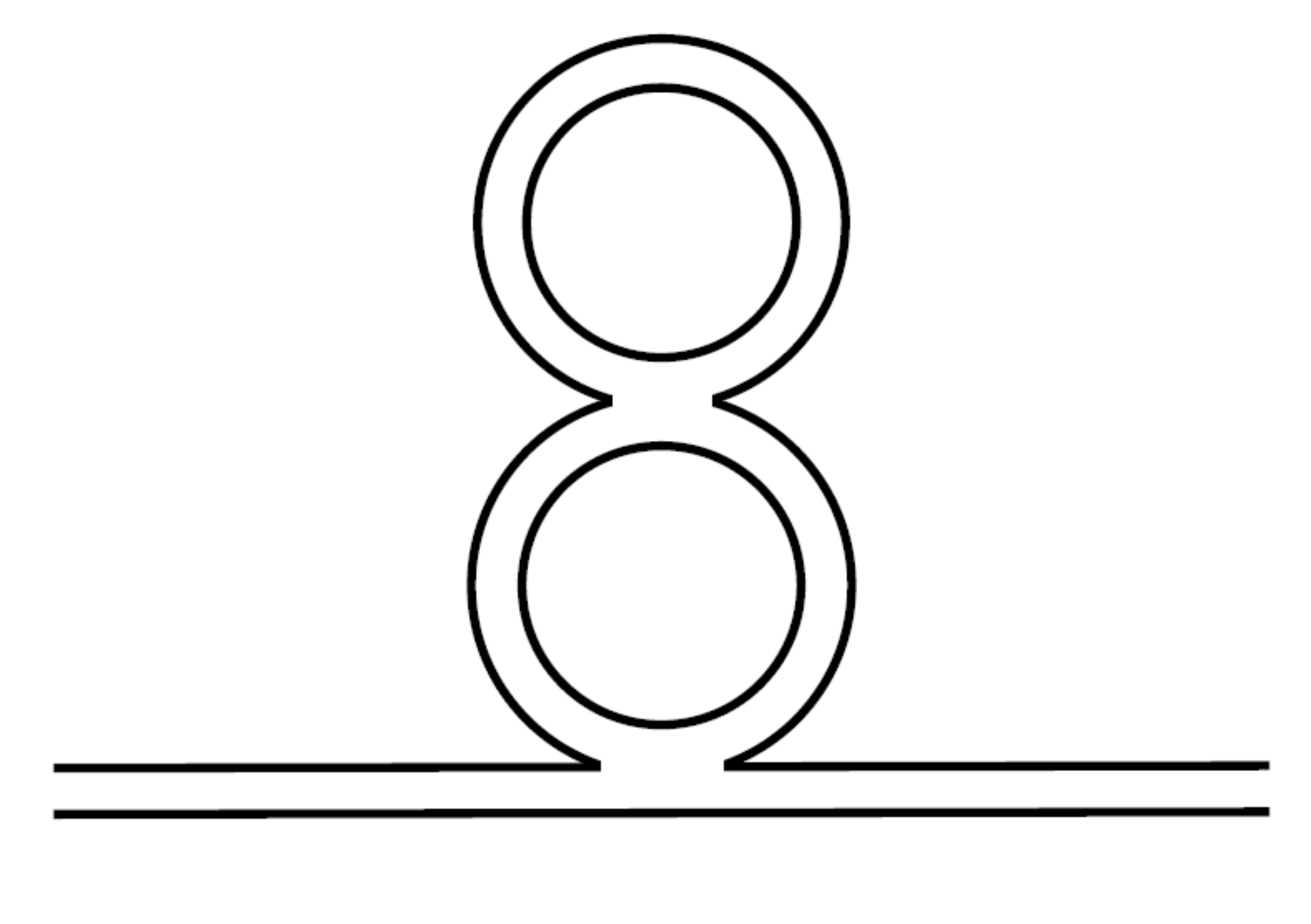} &&  \\
&&
$\propto g_{YM}^4M^2$ &&
$\propto g_{YM}^4M^2$ &&
$\propto g_{YM}^4M^2$ &&
\\&&&&&&&& \\
i) && j) && k) && l) && m) \\
\includegraphics[width=0.15\textwidth]{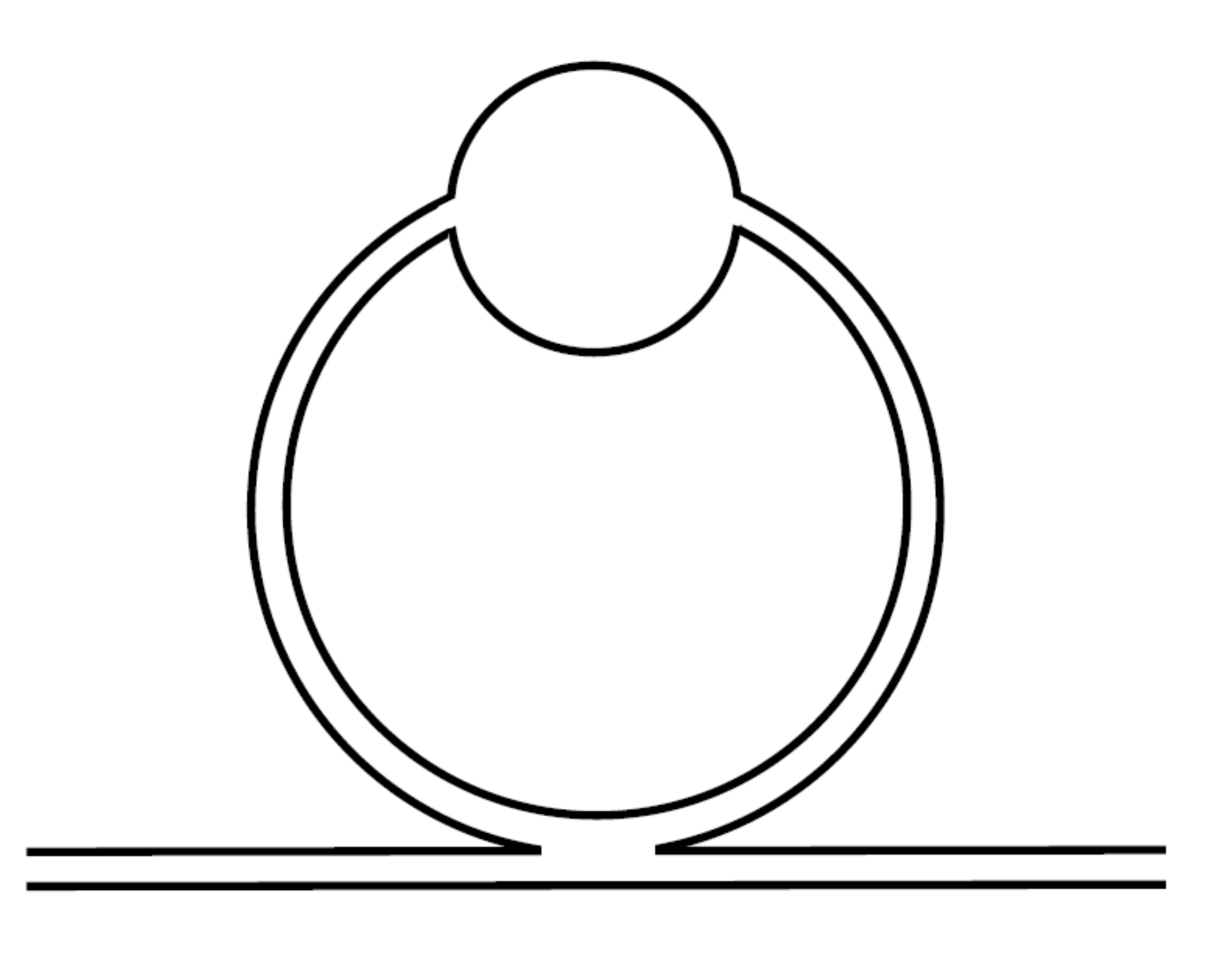} &&
\includegraphics[width=0.15\textwidth]{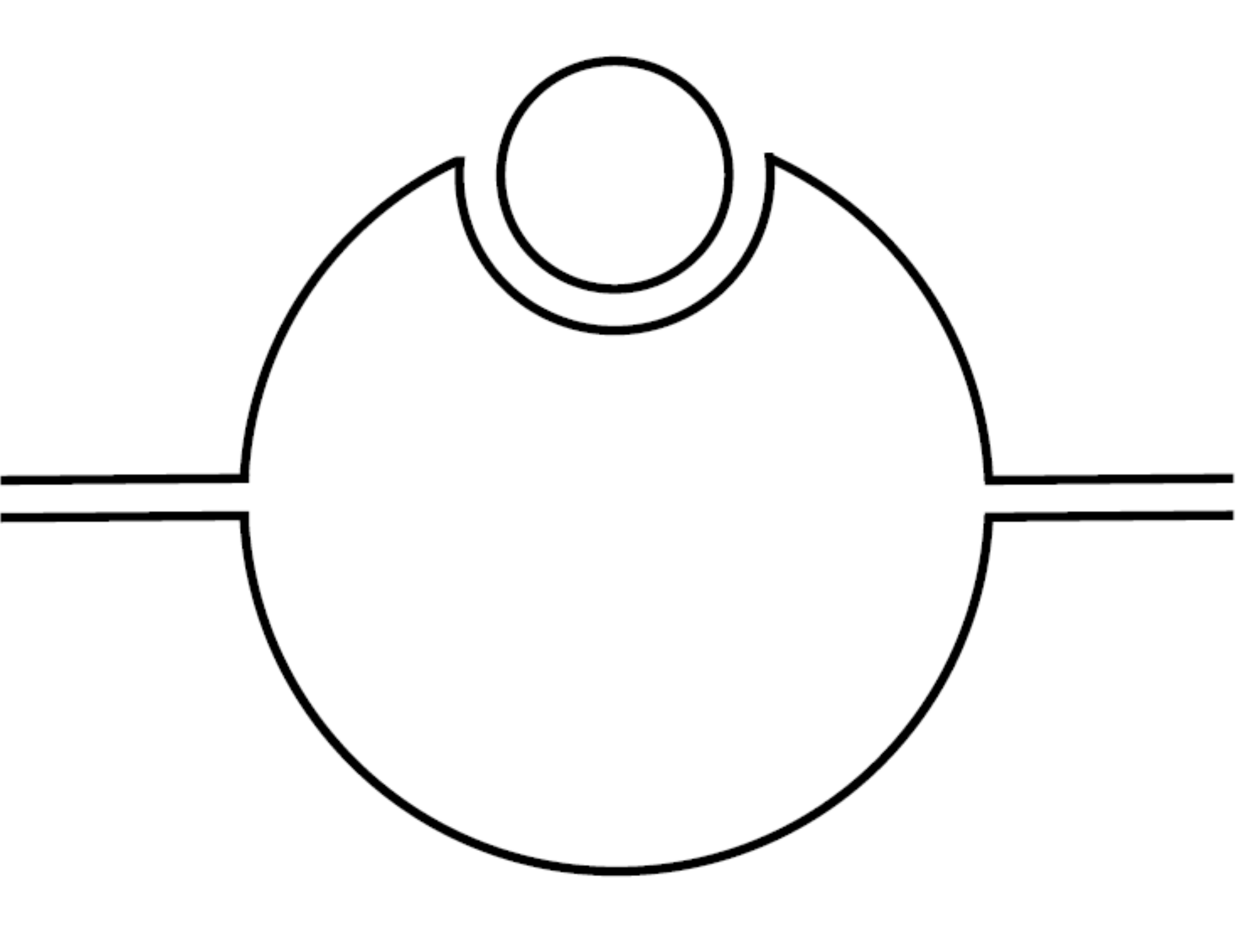}  &&
\includegraphics[width=0.15\textwidth]{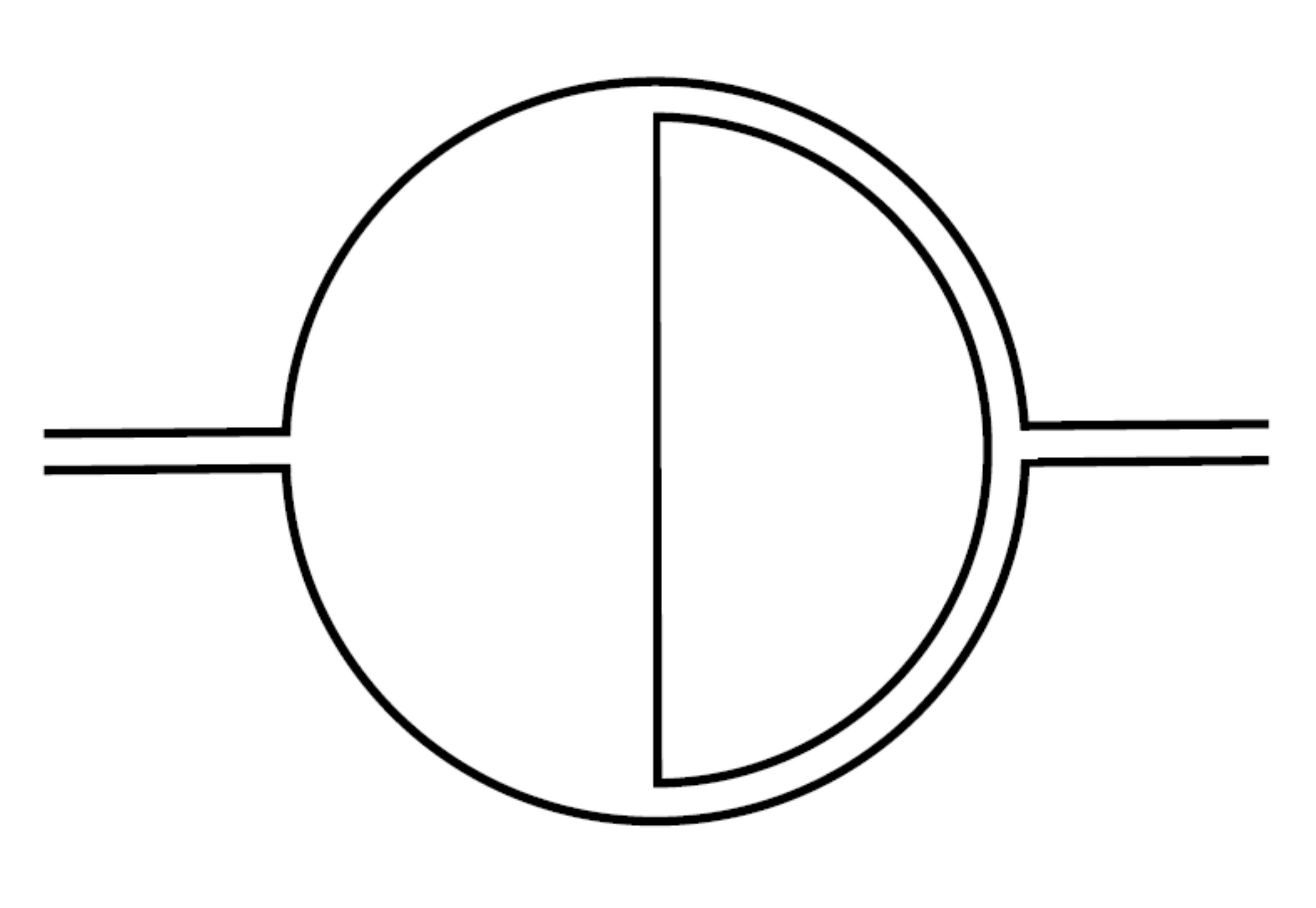} &&
\includegraphics[width=0.15\textwidth]{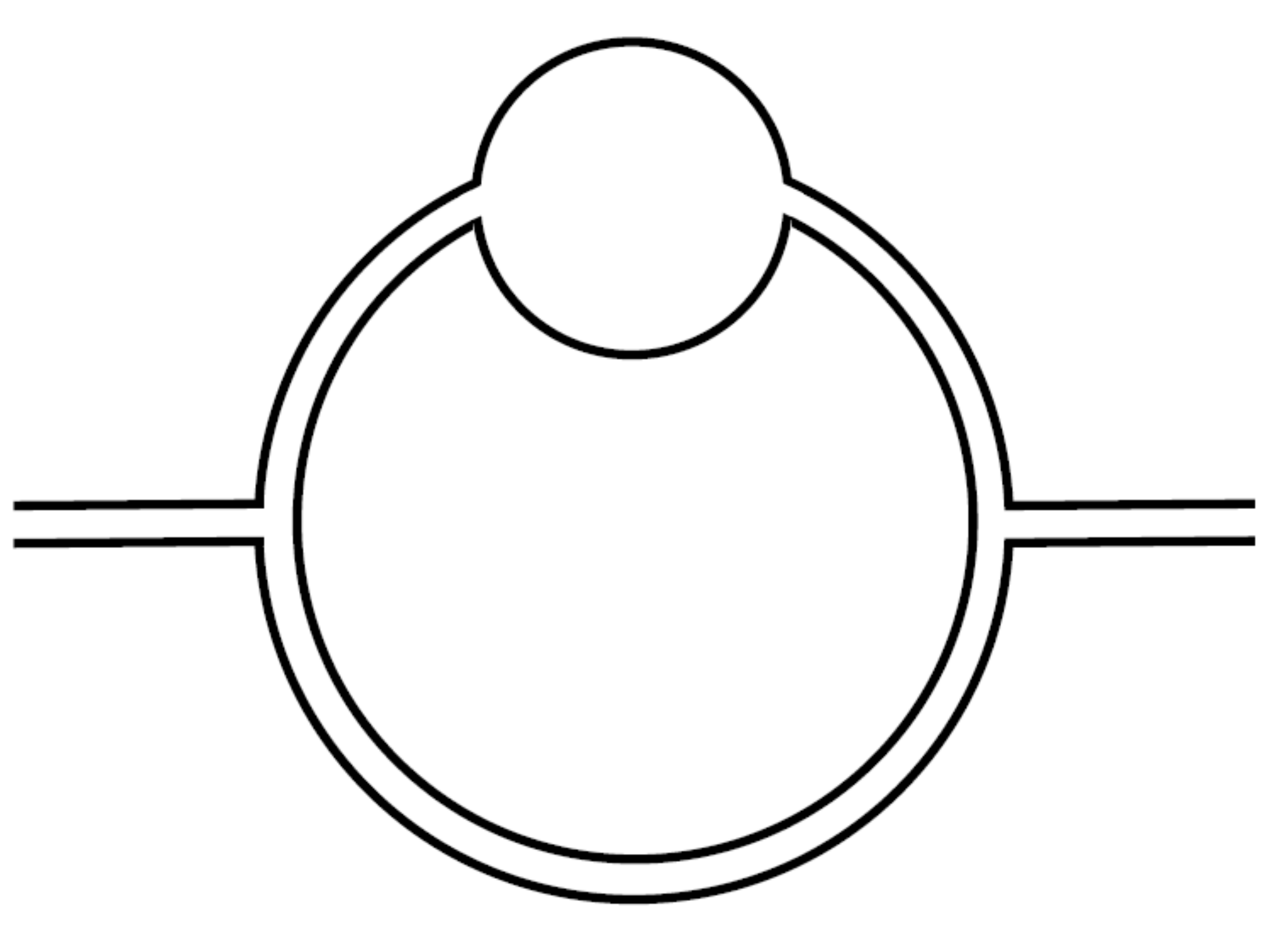} &&
\includegraphics[width=0.15\textwidth]{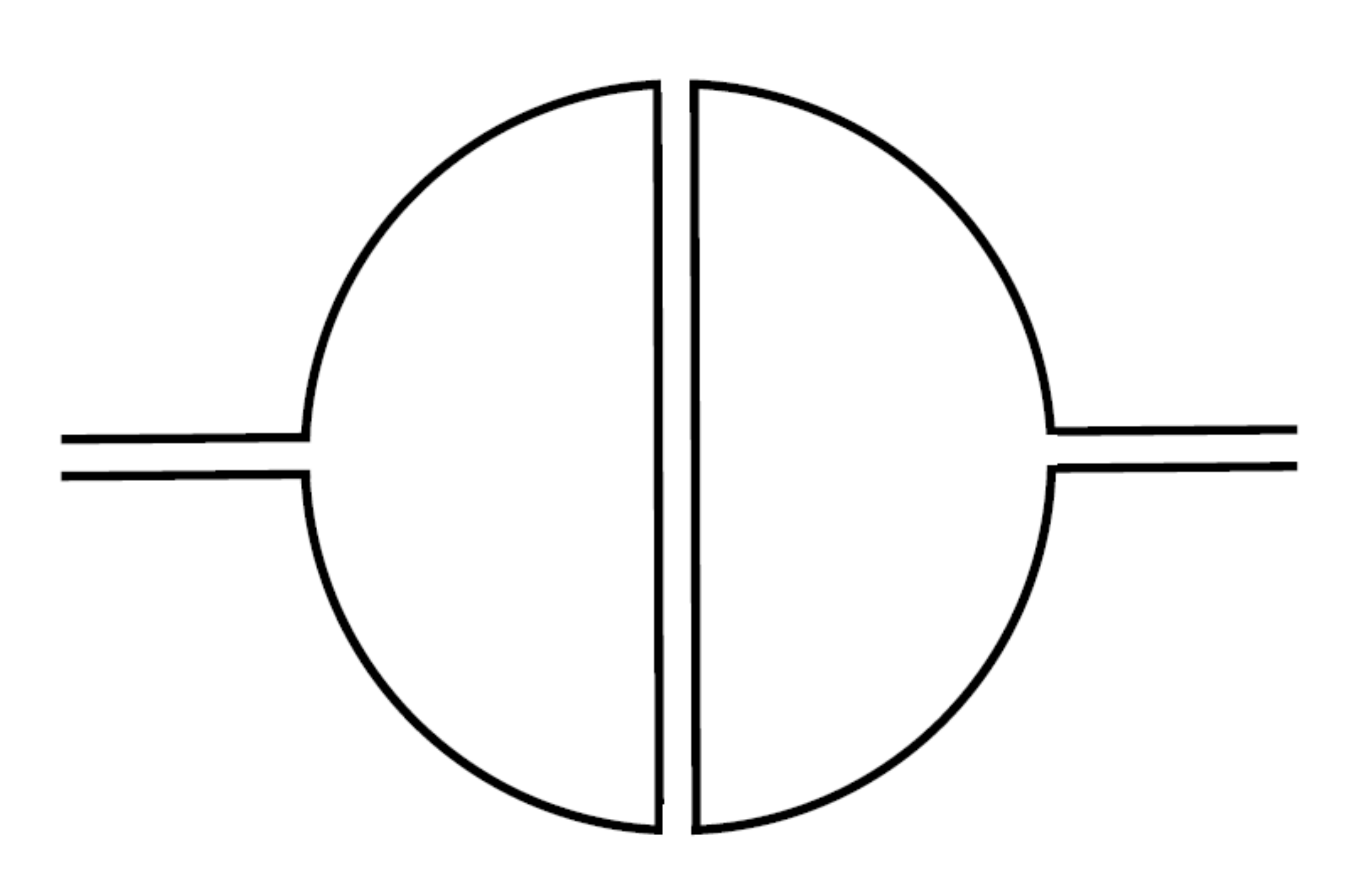}   \\
$\propto g_{YM}^4M N_f$ &&
$\propto g_{YM}^4M $ &&
$\propto g_{YM}^4M $ &&
$\propto g_{YM}^4M N_f$ &&
$\propto g_{YM}^4$
\end{tabular}
\caption{2-loop diagrams contributing to the gauge boson self-energy with their relative sizes. Diagrams a) - g) - purely gluonic contributions; h) - j) - contributions with fermionic degrees of freedom.}
\label{fig-four}
\end{figure}

\begin{figure}[h]
\centering
\begin{tabular}{ccccccccccc}
a)&& b)&& c) && d)  \\
\includegraphics[width=0.15\textwidth]{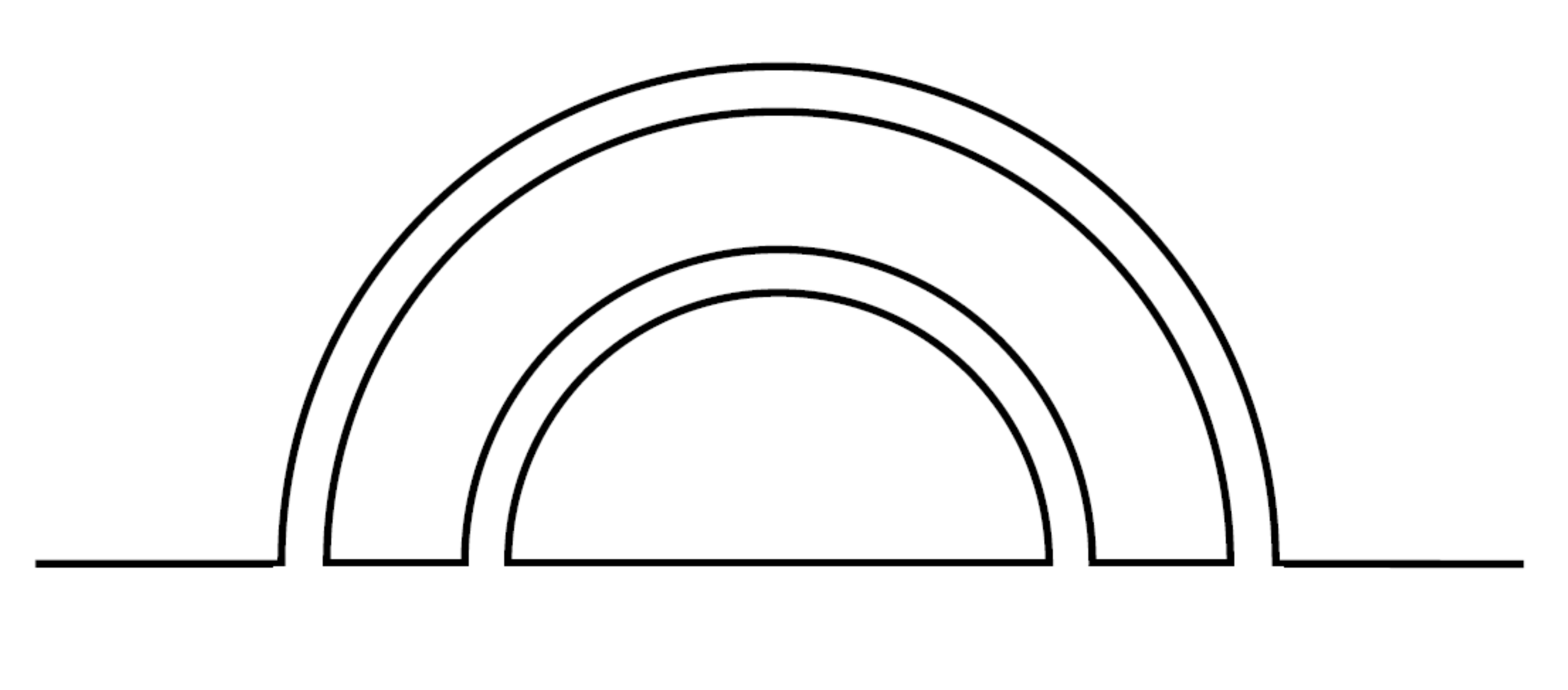} &&
\includegraphics[width=0.15\textwidth]{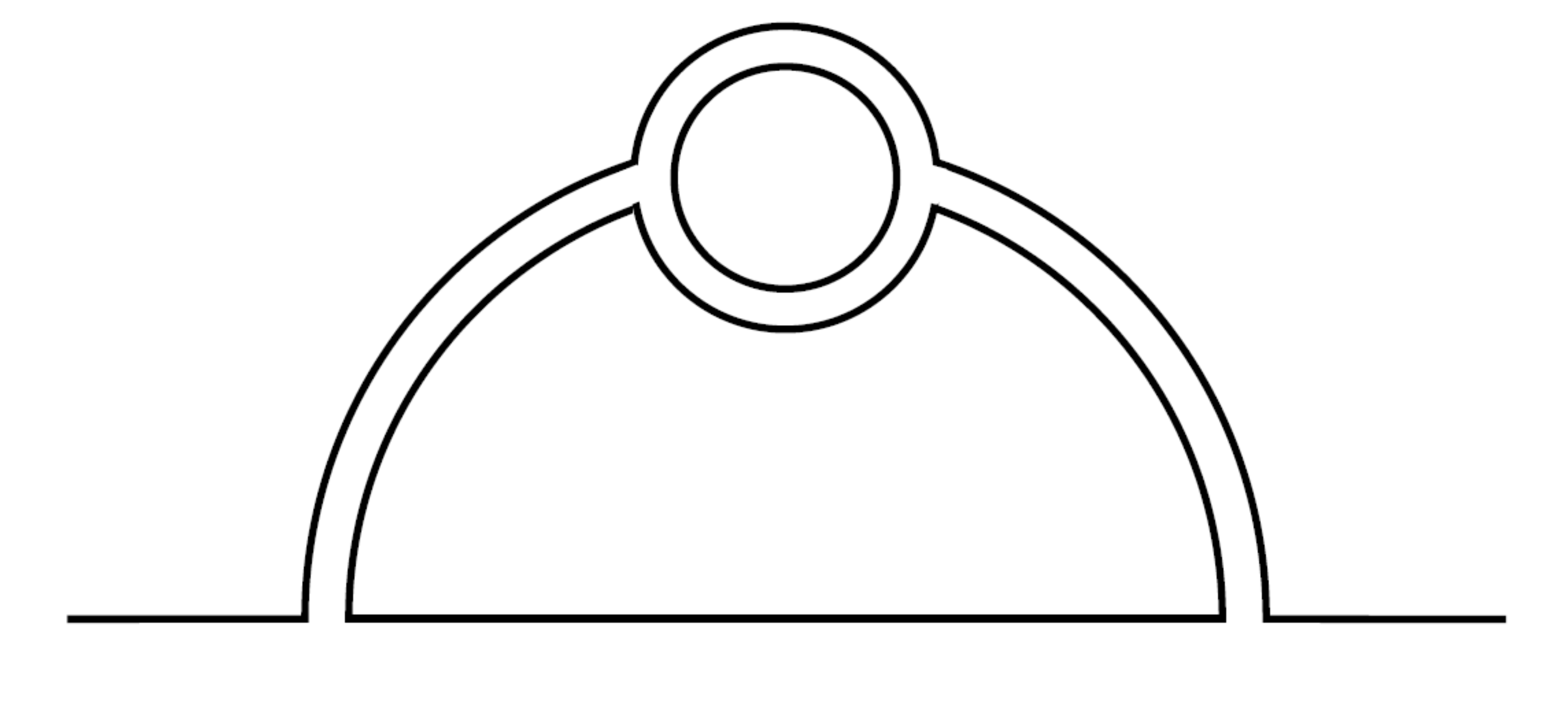}  &&
\includegraphics[width=0.15\textwidth]{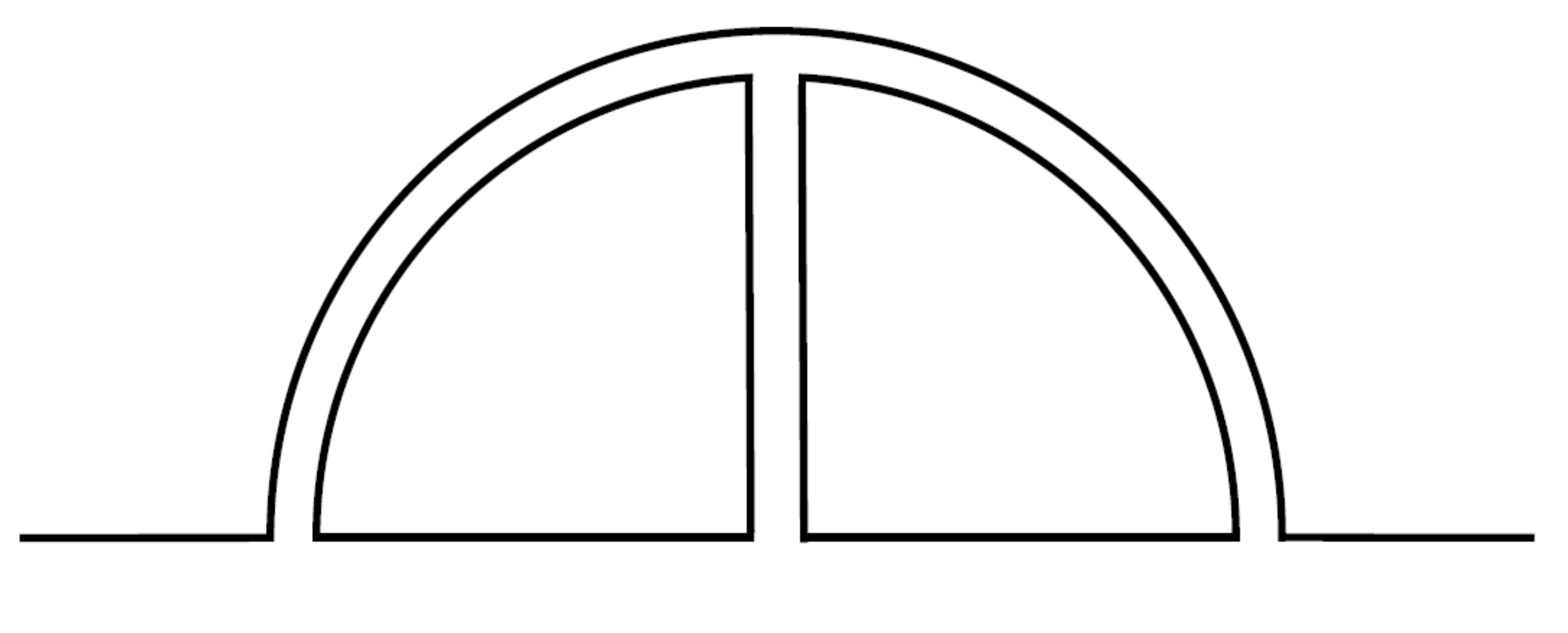} &&
\includegraphics[width=0.15\textwidth]{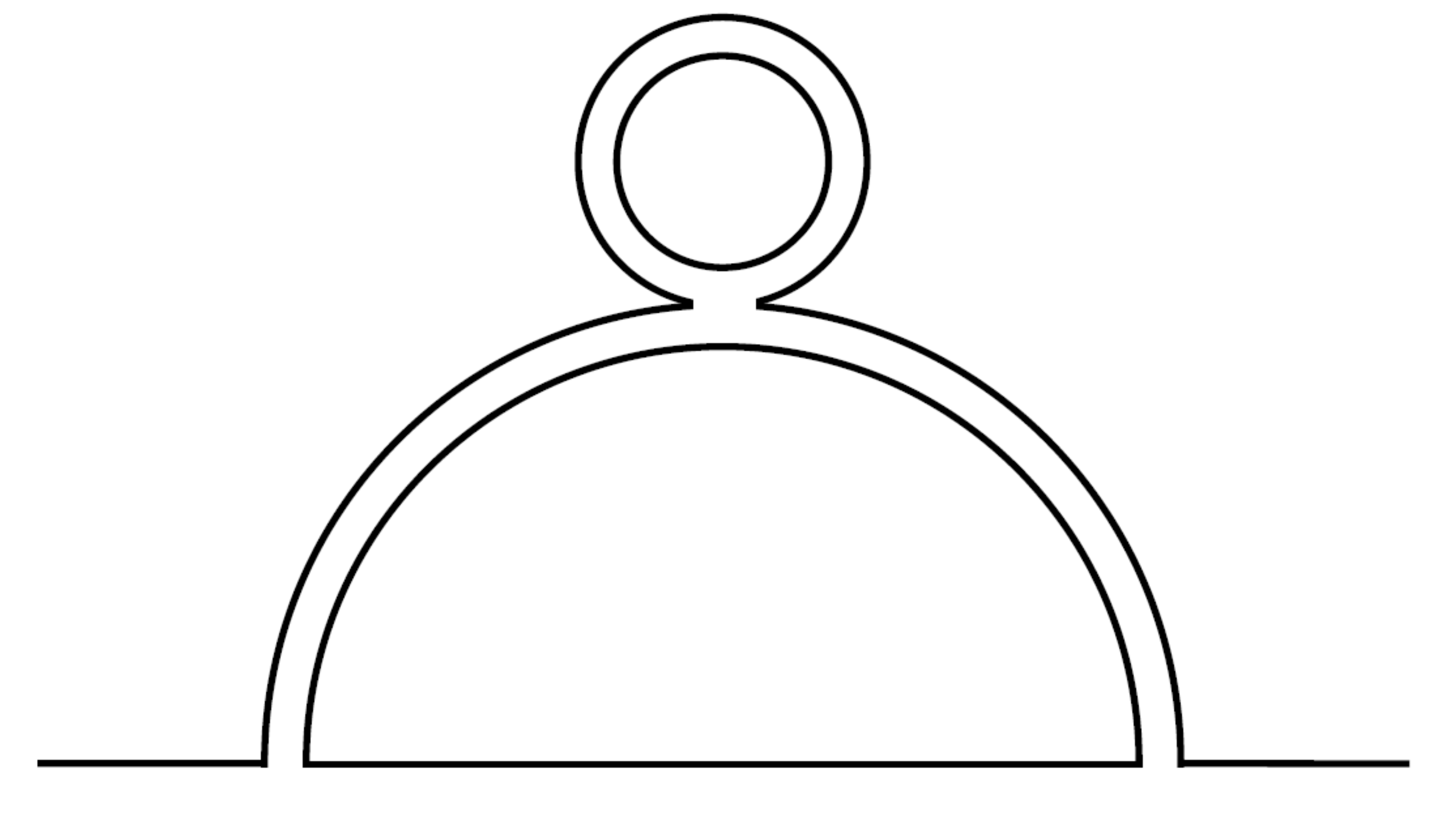} \\
$\propto g_{YM}^4M^2$ &&
$\propto g_{YM}^4M^2$ &&
$\propto g_{YM}^4M^2$ &&
$\propto g_{YM}^4M^2$  \\
&&&&&&\\
&&e) && f)&& \\
&&
\includegraphics[width=0.15\textwidth]{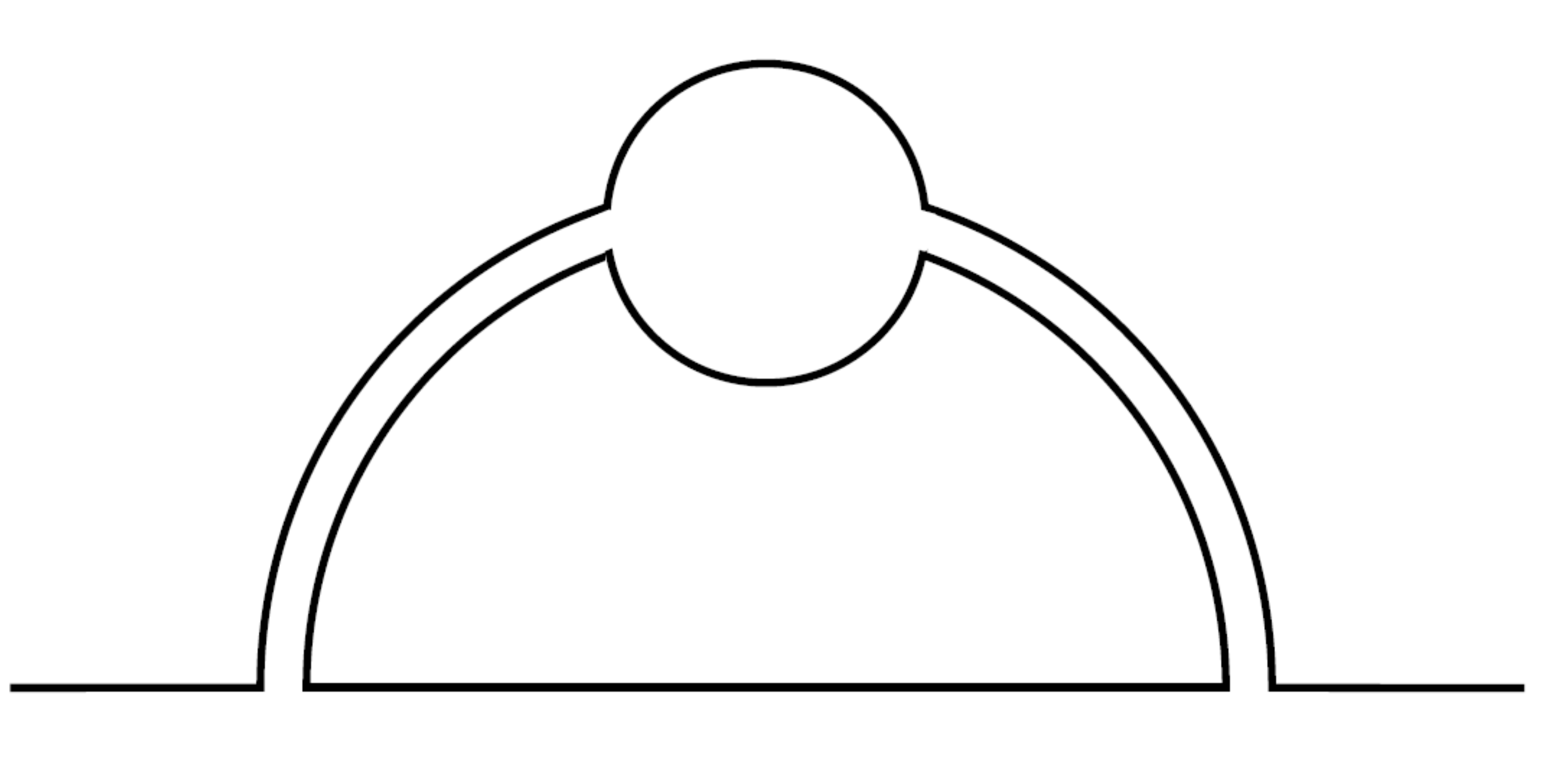} &&
\includegraphics[width=0.15\textwidth]{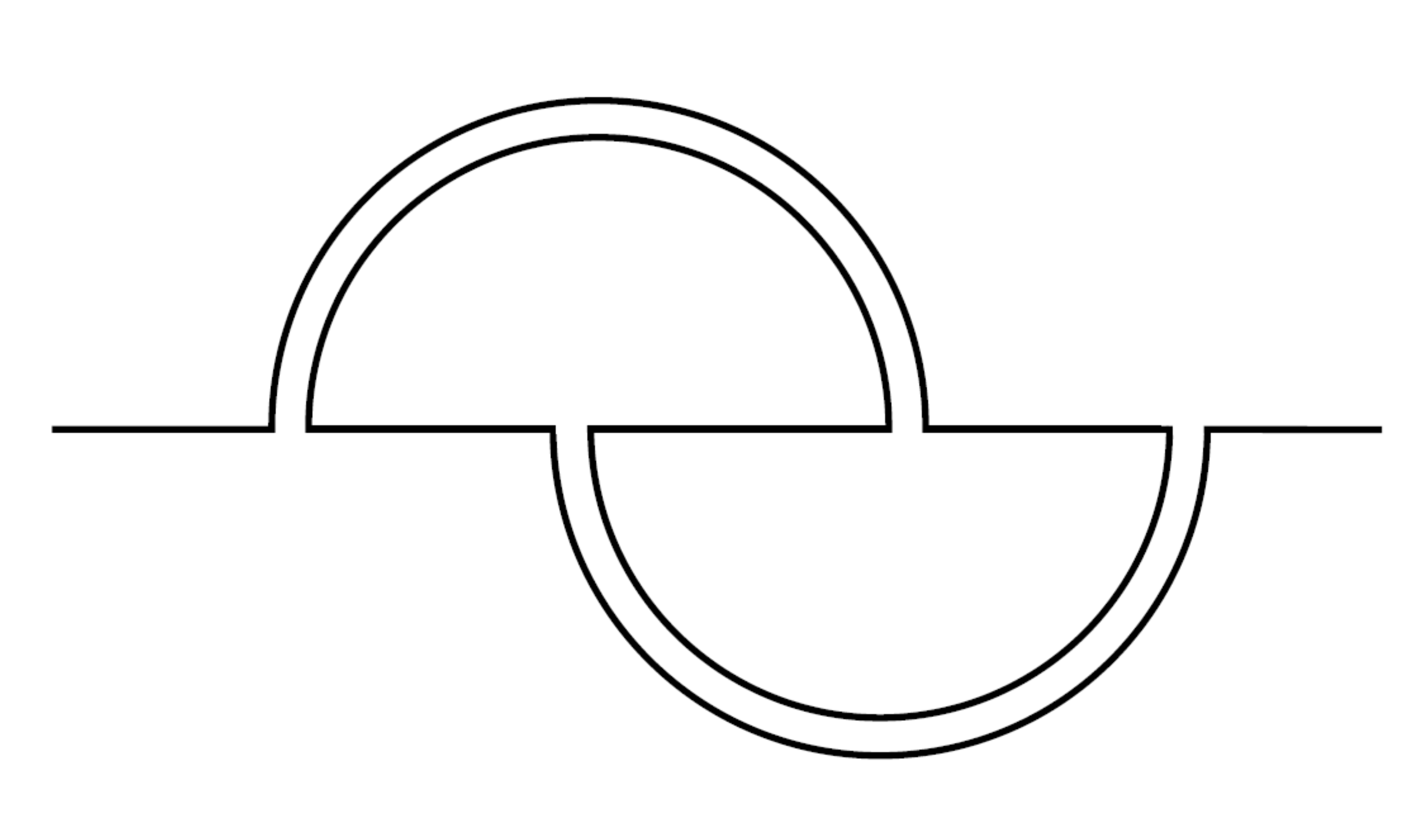}
&& \\
&&
$\propto g_{YM}^4 M N_f$ &&
$\propto g_{YM}^4$ &&
\end{tabular}
\caption{2-loop diagrams contributing to the quark self-energy with their relative sizes.}
\label{fig-five}
\end{figure}

The imaginary part of the one-loop self-energy vanishes when bare propagators are used due to kinematic constraints. It does not vanish when the resummed propagators are used, but that is equivalent to the two-loop self-energy which we discuss next. The relevant two-loop self-energy diagrams and their sizes for both gluons and quarks are shown in Fig.~\ref{fig-four} and \ref{fig-five}, respectively. It is then apparent again that gluon contributions dominate over the fermion ones by a factor of $M$ for both the gluon and quark self-energies. The size of the imaginary parts of the self energies is:
\ba
\label{im-self}
{\rm Im}\Pi \propto {\rm Im}\Sigma \propto \lambda^2
T^2,
\ea
which leads to the thermal widths being of the same size, $\Gamma^g \sim \Gamma^f \propto \lambda^2 T$ in the leading order. This is already enough to justify that quark loops do not have to be considered any more since the gluon contribution to a given quantity is always $M$ times bigger than the quark one, up to the factor of $N_f$, which is fixed constant and much smaller than $M$. This justifies omitting all quark contributions in the forthcoming analysis. One can then observe that the typical size of the propagator part of the correlation function is:
\ba
\label{gl1}
\int dp^0
G_{ra}(p)G_{ar}(p) &\sim& \frac{M^2}{(E_p^g)^2 \Gamma_p^g},
\sim \frac{M^2}{\lambda^2 T^3}
\ea
up to the logarithm.

The parametric estimate of the self-energy is of significant importance since it controls power counting of the scattering processes establishing the bulk viscosity coefficient. Consequently, as we discuss in the next subsection, it is the form of the self-energy at the one and two-loop order that controls the form of the collision kernel of the Boltzmann equation. In particular, by studying the self-energy one is able to find which processes contribute to the collision kernel and what is their sensitivity to different scales. In contrast to the shear viscosity, where the hard scale dominates its typical size, the bulk viscosity is sensitive to the soft scale as well. Since the soft scale is dictated by the size of the thermal mass, whenever we refer to it we mean momenta $p \sim {\rm Re}\Pi \sim \sqrt{\lambda}T$.

\subsection{Diagrammatic justification of the processes contributing to the Boltzmann equation \label{sec-just}}

Although the parametric estimate of bulk viscosity can be found by considering the one-loop diagram of the correlation function, in the thermal medium infinite number of multiple processes have to be included in the leading order. Equivalently, the infinite number of relevant loops need to be resummed. In the kinetic theory formulation this procedure is supposed to be captured by the collision term of the Boltzmann equation. The equivalence between the two approaches can be established by showing that at a given order of the coupling constant microscopic processes obtained in the diagrammatic representation have their counterparts in the collision kernel of the Boltzmann equation. Here we discuss this issue.

There is a twofold source of the need for resummation of infinite number of diagrams. Each of them is related to the presence of a different type of singularity. The first case has already been discussed and it is the pinching pole singularity, regulated by the thermal width $\Gamma^g$, where no other singularity occurs. The diagrams reflecting this type of singularity correspond to the number conserving, $2 \to 2$, processes. Then, within the one loop already discussed, any number of gluon exchanges between the side rails is possible. Any possible insertion of the permissible gluon exchanges, meant as rungs, is of the order of $\lambda^2$ and it is compensated by an accompanying $1/\Gamma^g$ factor coming from pinching poles of the pair of retarded and advanced propagators. Infinitely many such combinations are possible. The other type of singularity, characteristic for gauge theories, is the collinear singularity associated with the small angle between the scattering and scattered particles. This type of singularity governs the number changing processes, $1+N \to 2+N$, where $N$ is a number of hard particles taking part in a splitting of one hard gluon into two hard gluons. The collinear processes contribute to the bulk viscosity computation at the same order of the coupling constant as the number conserving processes. The splitting process occurs when a hard gluon traversing the medium interacts with another hard gluon via a soft momentum exchange and then emits an additional hard gluon so that both the emitted and the emitting gluons move almost collinearly within an angle $\theta \sim O(\sqrt{\lambda})$. The collinear region of propagating particles is always associated with the corresponding product of retarded and advanced propagators, which, if not dressed, has a singular behavior. The collinear singularities, similarly to the pinch singularities, are regulated by the thermal width, but in this case, the soft scale fixed by the thermal mass plays an essential role as well. As discussed in detail later, all hard gluons taking part in the process can interact infinitely many times via the soft exchanges with the thermal background and they have to be coherently resummed.

Since each type of singularity involves a resummation of the corresponding set of infinitely many diagrams, there are two integral equations that need to be solved, each of them associated with the corresponding type of singularity. We first focus on the physics of number conserving processes which can be represented by a set of diagrams involving the pinch singularities only. The case of collinear processes is discussed later.

\subsubsection{The case without collinear singularities \label{alina4}}

\begin{figure}[h]
\centering
\begin{tabular}{c}
\includegraphics[width=0.8\textwidth]{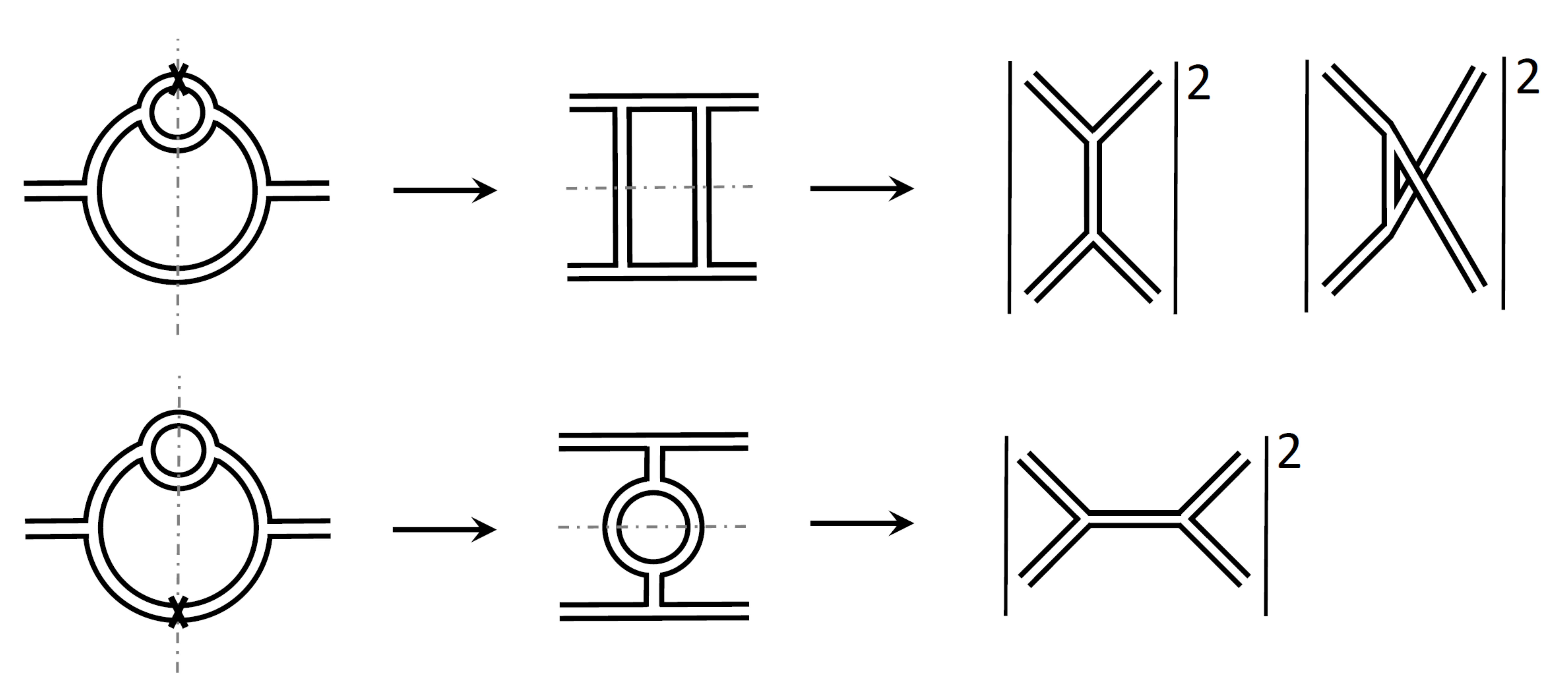}
\end{tabular}
\caption{The procedure of opening one line of the two-loop self-energy to reproduce rungs. The dashed line is the cutting line and the cross denotes the open line. The last column shows the scattering processes corresponding to the rungs shown.}
\label{cut}
\end{figure}

$2 \to 2$ processes are represented by rungs, which have to be inserted in the one-loop spectral function and then resummed. Finding the structures of rungs is not trivial and to do so one needs to rely on a few constraints: Ward identities, power counting and kinematic boundaries. The most essential constraint is imposed by the Ward identities which provide relations between the effective vertex and  dressed propagators and also dictate the way to maintain gauge invariance. Thus, the Ward identities should be used to obtain relations between the full on-shell imaginary part of the self-energy and possible rung insertions. This is discussed for QED transport coefficients in \cite{Gagnon:2006hi,Gagnon:2007qt} and for any SU($M$) one should expect similar relations. Accordingly, one can reproduce corresponding rungs by cutting the two-loop self energy diagrams in all possible ways and then opening one line in every diagram in all permissible ways. In Fig. \ref{cut} we show one schematic example. The two-loop diagram in Fig. \ref{fig-four} a) is cut through the two loops and the cross denotes lines which are open. In this way one gets two possible topologies of a rung. The lines which are cut, but not open, represent particles put on shell. Similarly, the external lines represent thermal on shell excitations. By opening the cut lines one reproduces $2\to2$ scattering processes shown in the right column in Fig. \ref{cut}. Therefore, the first row shows how to obtain $t$ and $u$ channels of scattering events, they are obtained from the same rung but with a different momentum flow along the rung, which is not shown explicitly. The second row presents how one gets the $s$ channel.

\begin{figure}[h]
\centering
\begin{tabular}{c}
\includegraphics[width=0.95\textwidth]{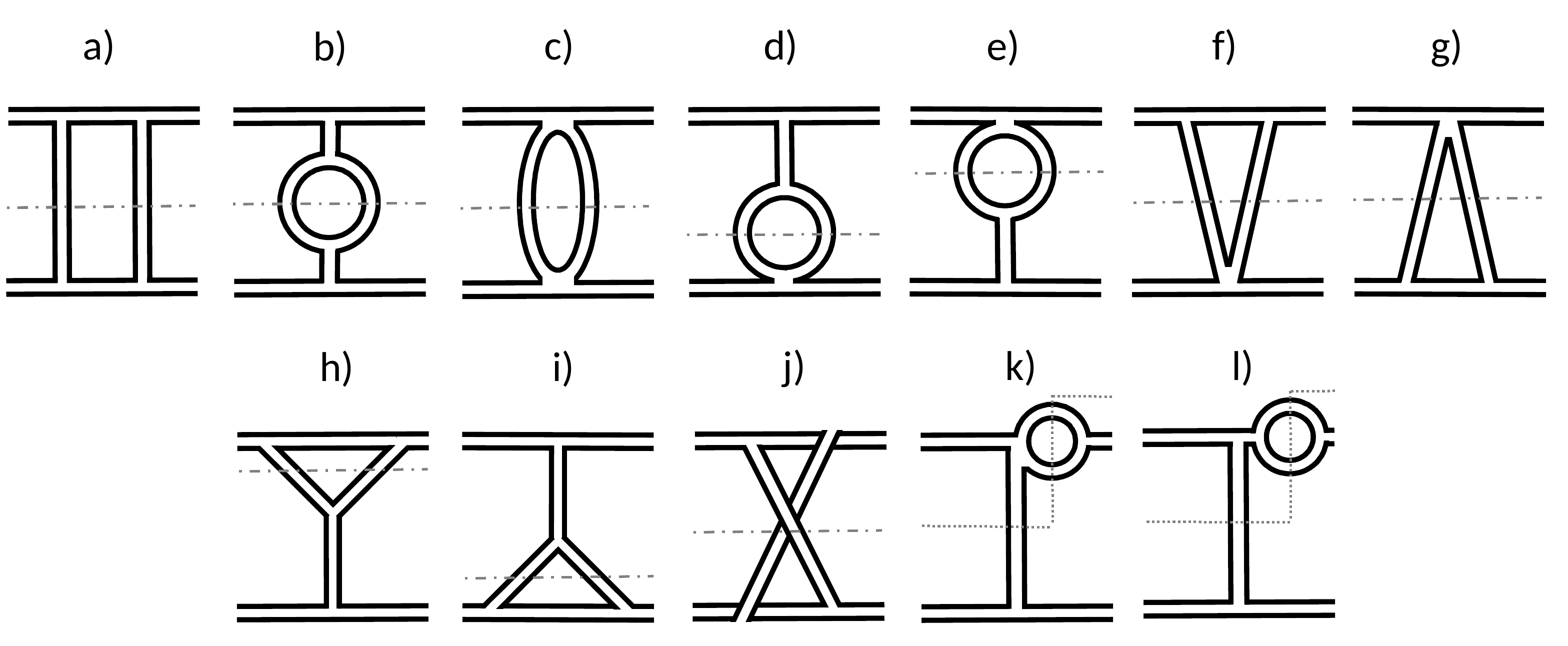}
\end{tabular}
\caption{The topological structures of rungs obtained by cutting the two-loop self energy. The dashed and dotted lines represent the allowed and forbidden cuts, respectively, through the diagrams.}
\label{topol}
\end{figure}

The topological structures obtained by using this procedure are depicted in Fig.~\ref{topol}. First, only the diagrams a)-e) in Fig. \ref{fig-four} of the two loop gluon self energy have to be examined when looking for the topological structures of rungs. The diagrams g) and h) of Fig.~\ref{fig-four} are tadpole diagrams with one-loop corrections and they contribute to the real part of self energy. The diagram f) is the one-loop diagram with a tadpole correction and it also provides a contribution to the resummed propagator. Therefore, in general, all diagrams containing tadpoles do not have to be investigated any more for this qualitative analysis. For the power counting analysis one should include all rungs which have $g_{YM}^4$ factor coming from the interaction vertices and which have one closed loop contributing a factor of $M$. The other factor of $M$, expected for the proper 't~Hooft coupling order is obtained when the external lines of rungs on the right-hand side are joined with other rungs or with each other. What is more, in Fig. \ref{topol} we present all topological structures arising only from the use of the Ward identity. The final relevant contributions to the kernel of the integral equation can be, however, found by using kinematic constraints and power counting arguments. The kinematic constraints are schematically represented by the dashed and the dotted lines. The dashed lines represent the cuts through rungs which are allowed by kinematics. The dotted lines, although coming from the Ward identity analysis, reflect forbidden processes since one on-shell massless particle cannot decay into two on-shell masslesss particles. Therefore, the structures k) and l) in Fig. \ref{topol} do not contribute to the kernel. Accordingly, only diagrams a)-j) constitute the kernel of the integral equation determined by the pinching singularity and they all contribute at the order $O(\lambda^2)$. It is also easy to observe that all the contributing rungs with the associated cuts may be converted to reproduce matrix elements in the scattering amplitude defining the collision term of the Boltzmann equation. The rungs a) and b), shown also in Fig.~\ref{cut}, represent a contribution to the scattering amplitude squared given by $t$, $u$, and $s$ channels. The diagram c) leads to the respective contribution from the contact interaction. The diagrams d)-j) reflect possible interference terms.

At hard scale all allowed rungs contribute at the order $O(\lambda^2)$, where it is enough to count the number of interaction vertices and the number of color loops. To see the relevant $M$-dependence it is more convenient to count closed loops of the spectral function shown in Fig.~\ref{fig-one} a) with the rungs inserted.

\begin{figure}[h]
\centering
\begin{tabular}{c}
\includegraphics[width=0.9\textwidth]{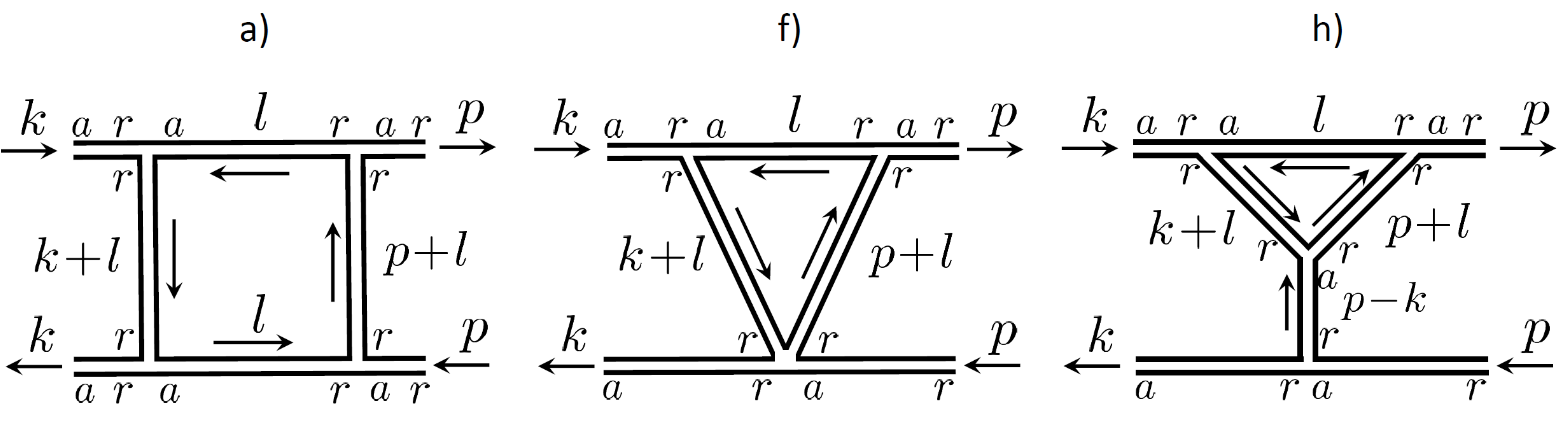}
\end{tabular}
\caption{Rungs a), f) and h) from Fig. 7
shown with momentum convention and one out of many $r$ and $a$ assignment.}
\label{asq}
\end{figure}

When the soft scale starts to play a role power counting of the diagrams presented in Fig.~\ref{topol} can change and not all diagrams are of the same size. The rung c) is not affected by the soft physics since all lines must be hard and on-shell. To do power counting of other diagrams with momenta of the order $O(\sqrt{\lambda}T)$, we use the $(r,a)$ basis. It is important to notice that there can be many diagrams of the same topology but with different $a$ and $r$ assignment and of different kinematic constraint; we discuss only a few exemplary cases. In Fig.~\ref{asq} we show diagrams from Fig.~\ref{topol}, where the $a$ and $r$ positions and the momentum convention are shown explicitely.

The expression corresponding to the rung a) is:
\ba
\mathcal{K}_{a} \sim \lambda^2 \int \frac{d^4l}{(2\pi)^4} G_{ra}(l)G_{ar}(l)G_{rr}(k+l)G_{rr}(p+l).
\ea
The size of the rung is estimated as follows. All incoming and outgoing momenta are hard, $k \sim p \sim O(T)$, and on-shell, while the loop momentum $l$ is soft $l\sim O(\sqrt{\lambda}T)$ and off-shell. In this case both $G_{ra}(l)$ and $G_{ar}(l)$ propagators are of the order $O(\lambda^{-1} T^{-2})$. Additionally, since both $G_{rr}(k+l)$ and $G_{rr}(p+l)$ are on-shell they contain delta functions to maintain energy-momentum conservation. When the loop momentum integration is performed the phase space $d^4l$ combined with the delta functions reduces to $d^2l$, which is $O(\lambda T^2) $ when $l$ is soft. Combining all these factors one gets $\lambda^2$ from the explicit interaction vertices, $\lambda$ from a phase space suppression and $\lambda^{-2}$ from two soft propagators which make this rung to be of the order $O(\lambda)$. The rung has therefore a different size at the soft scale than at the hard one, which is due to the Coulomb divergence characteristic for these scattering processes. This is, however, only a superficial difference since there is an additional mechanism which makes this rung contribute to the integral equation at the expected $O(\lambda^2)$ order. The best way to see it is to refer to the $2 \to 2$ collision kernel of the Boltzmann equation~\cite{Arnold:2003zc}, which is:
\ba
C^{gg\to gg}({\bf k})&=&\frac{1}{32} \int \frac{d^3p}{E_p(2\pi)^3} \frac{d^3k'}{E_{k'}(2\pi)^3}
\frac{d^3p'}{E_{p'}(2\pi)^3} |\mathcal{M}(k,p;k',p')|^2 (2\pi)^4 \delta^4(k+p-k'-p') \nonumber\\
&& \times n_(k)n_(p)(1+n_(k'))(1+n_(p')) [\chi({\bf k})+\chi({\bf p})-\chi({\bf k}')-\chi({\bf p}')],
\ea
where the functions $\chi$ represent a small nonequilibrium deviations from the Bose-Einstein distribution function. When $2 \to 2$ scattering processes represented by rungs a) and b) in Fig. \ref{topol} occur in the medium via the soft momentum exchange, which is when ${\bf k} - {\bf k}'= {\bf l}$, with ${\bf l} \sim \sqrt{\lambda}T$, then one encounters the following cancellation between the $\chi$ functions:
\ba
\label{chi}
\chi({\bf k}) - \chi({\bf k}') = -{\bf l} \cdot \nabla \chi({\bf k}) +O(l^2).
\ea
The prescription dictated by the Kubo formula has similar structure to the Boltzmann equation \cite{Jeon:1994if}, where the term $[\chi({\bf k})+\chi({\bf p})-\chi({\bf k}')-\chi({\bf p}')]$ needs to be squared to compute any transport coefficients from the Boltzmann equation. This introduces additional power of $\lambda$ from the soft momentum and softens the contribution of rung a) so that its final size is $O(\lambda^2)$. An analogous mechanism applies to the diagram b) where the two vertical soft lines cause $\lambda^{-2}$ enhancement, the explicit vertices and the phase space introduce $\lambda^3$ and the Boltzmann equation structure (\ref{chi}) - the factor $\lambda$, which altogether give the size $O(\lambda^2)$.

We also need to evaluate the interference terms, that is, the rungs d) - j). They are all of $O(\lambda^2)$ order when the off-shell exchange momentum is soft. To see this we first consider the rungs f) and g) (for the notation of the rung f) see Fig. \ref{asq}). They both contain one propagator with a soft momentum $l$ whose contribution is $O(\lambda^{-1})$, but this is canceled by an additional phase space suppression. Due to combination of two delta functions in the propagators $G_{rr}(k+l)$ and $G_{rr}(p+l)$ with the phase space $d^4l$, the latter one is reduced to $d^2l$ which leads to $d^2l \sim O(\lambda)$. When assessing the size of the rungs h), i) and j) the same arguments hold as before. The rung h) is shown in Fig. \ref{asq}) and the corresponding expression is:
\ba
\int \frac{d^4p}{(2\pi)^4} \mathcal{K}_{h} \sim \lambda^2 \int \frac{d^4p}{(2\pi)^4} \int \frac{d^4l}{(2\pi)^4} G_{ra}(l)G_{rr}(k+l)G_{rr}(p+l)G_{ra}(p-k).
\ea
The soft propagator $G_{ra}(l)$ introduces $O(\lambda^{-1})$ and the number of integrals over the loop momentum is reduced as previously so that we are left with $d^2l \sim O(\lambda)$. These two factors cancel each other leaving the rungs $O(\lambda^2)$. In rungs d) and e) the vertical line represents the soft propagator which is $\lambda^{-1}$. Including further the phase space suppression and the couplings from the explicit interaction vertices, one gets this rung of the expected $O(\lambda^2)$ size.

\subsubsection{The case with collinear singularities \label{alina5}}

Number changing processes contribute at the same order as $2 \leftrightarrow 2$ processes (up to logarithm). They are entangled in the same topological structures as number conserving processes, shown in Fig. 6, but emerge under different kinematic conditions. The mechanism responsible for their occurrence is also more complicated than the one discussed above and it is fully controlled by soft physics. Here we briefly and qualitatively discuss how they emerge and evaluate their sizes.

\begin{figure}[h]
\centering
\begin{tabular}{c}
\includegraphics[width=0.9\textwidth]{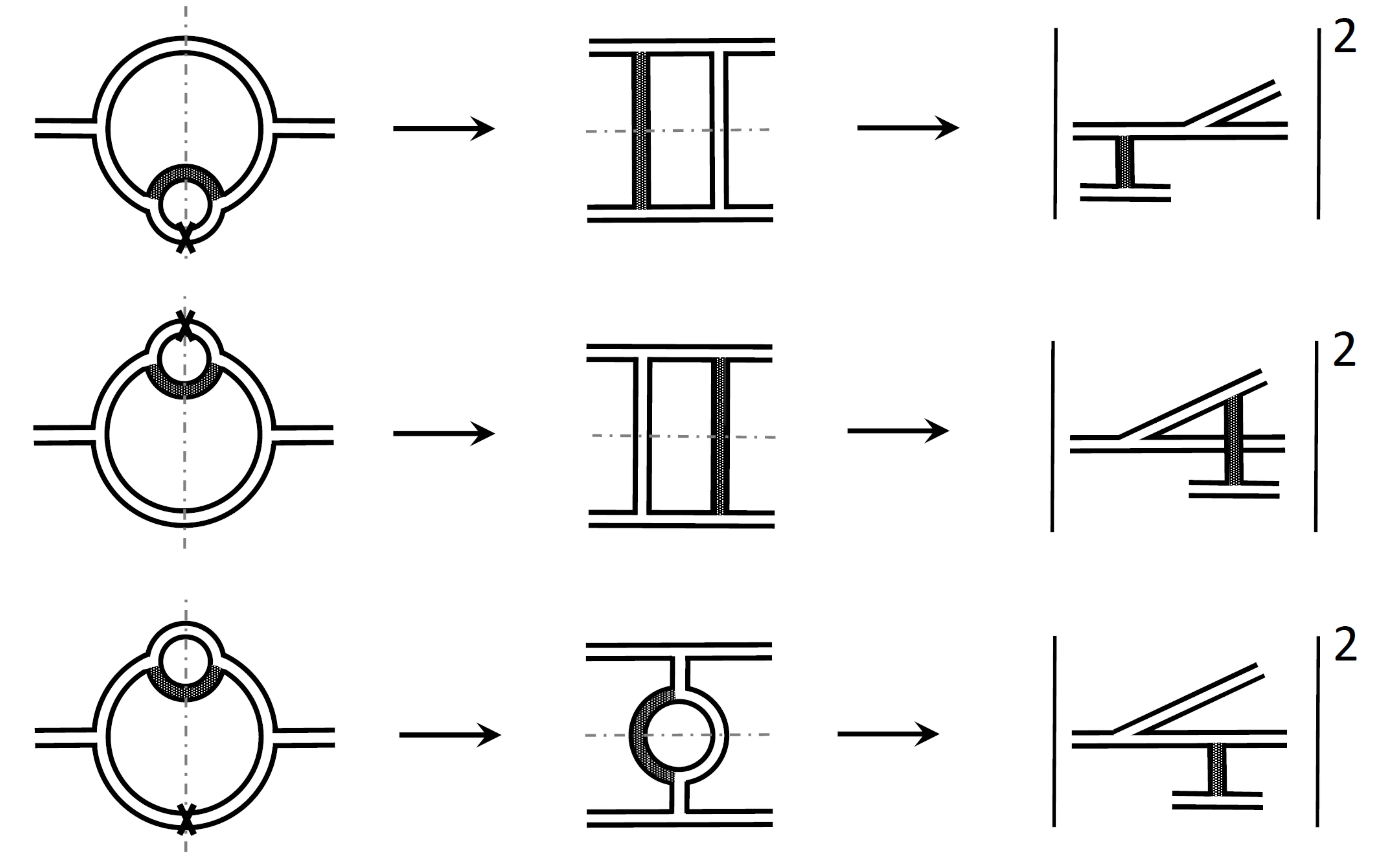}
\end{tabular}
\caption{The procedure showing how the number changing processes with a soft gluon exchange are reproduced from the two-loop self-energy. The dashed line is the cutting line, the black cross denotes the line opened to reproduce rungs, and the shaded line denotes a propagator with a soft momentum, which contains the hard thermal loop correction.}
\label{soft1}
\end{figure}

Collinear processes occur when one hard particle splits into two hard particles with an accompaniment of a soft gluon exchange with the thermal medium \cite{Arnold:2001ba,Arnold:2001ms,Arnold:2002ja}. The topological structures corresponding to these processes can be obtained in the procedure shown in Fig. \ref{soft1}. As presented, the rungs representing collinear processes are reproduced by opening one outer line of the two-loop self-energy. The line which is open is denoted by the black cross in the figure. The internal (shaded) lines of the self-energy represent propagators with soft momenta. They contain the hard thermal loop corrections, which is not shown explicitly in the two first columns of Fig. \ref{soft1}. Thus, whenever the cut is through the soft line it means that the hard thermal loop is cut. Consequently, all the cut lines and the external lines are hard and nearly on-shell. Specifically, in contrast to the number conserving processes, the thermal masses in the respective propagators must be included.

\begin{figure}[h]
\centering
\begin{tabular}{c}
\includegraphics[width=0.4\textwidth]{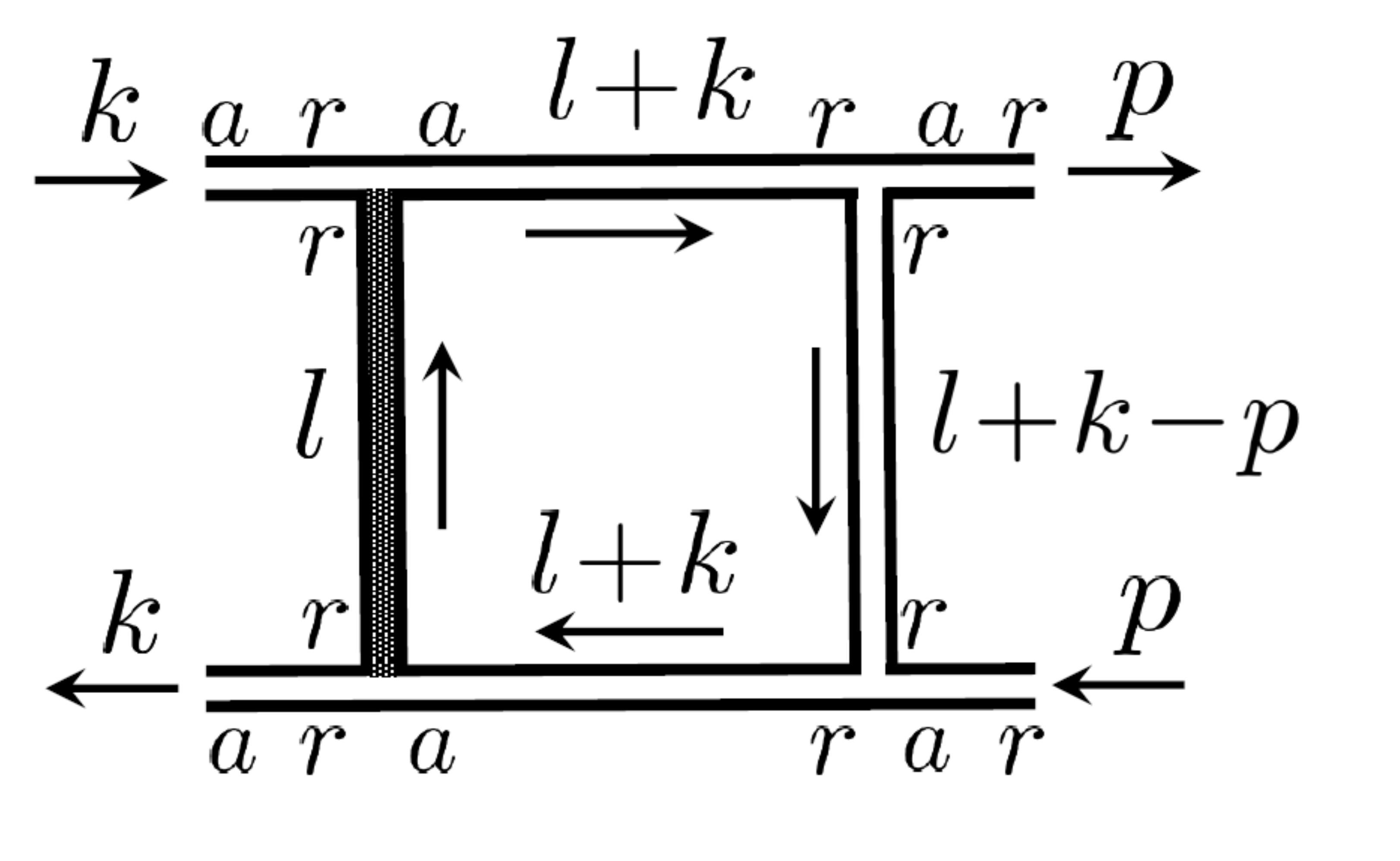}
\end{tabular}
\caption{The rung representing gluon splitting shown with the momentum convention and $a$ and $r$ assignment.}
\label{sqr1}
\end{figure}

To evaluate the size of the processes in Fig. \ref{soft1} we consider in detail the rung shown in Fig. \ref{sqr1}, reproduced with $a$ and $r$ positions and momentum convention. As before there is more than one layout of the $a$ and $r$ assignment and a complete analysis of the kernel of the spectral function has to include all possibilities. The size of this rung can be evaluated similarly to the case where collinear singularities are absent, but the power counting is more subtle. First, it is important to point out that whenever a soft line appears in the rung, it must be $G_{rr}$ propagator since it carries the distribution function to account for the interaction with the medium. $G_{rr}$ propagator, by contrast to other propagators, introduces $1/\sqrt{\lambda}$ enhancement in the soft momentum region. Moreover, the process under consideration is in the collinear regime when there is a pair of the adjacent retarded and advanced propagators with respect to a given momentum. If these propagators were bare their product would produce a singular behavior as their poles would nearly pinch the real axis in the contour integration. This is, however, cured by the inclusion of the self-energies, which leads to a finite expression. As in case of pinching pole approximation, diagrams containing the products $G_{ra}G_{ra}$ or $G_{ar}G_{ar}$ instead of $G_{ra}G_{ar}$ for the same momentum give much smaller contribution to the whole expression and can be neglected in the leading order analysis.

The expression corresponding to the rung shown in Fig. \ref{sqr1} is:
\begin{eqnarray}
&&
\int \frac{d^4k}{(2\pi)^4} (\dots)\int \frac{d^4p}{(2\pi)^4} \mathcal{K}(k,l,p)(\dots)  \\
&&
\sim
\lambda^2 \int \frac{d^4k}{(2\pi)^4} (\dots) \int \frac{d^4p}{(2\pi)^4}
\int \frac{d^4l}{(2\pi)^4} G_{rr}(l)G_{ar}(l+k)G_{ra}(l+k)G_{rr}(l+k-p) (\dots),\nonumber
\end{eqnarray}
where $(\dots)$ means the contribution from the external propagators, which is not needed to be shown explicitly. As mentioned, the external momenta are hard and nearly on-shell, $k \sim p \sim T$ and $k^2 \sim p^2 \sim O(\lambda T^2)$, while the loop momentum is soft $l \sim \sqrt{\lambda} T$. In this kinematic region the integral over the loop momentum is dominated by $dl_0 \sim O(\lambda T)$ in the frequency region, and $d^3{\bf l} \sim O(\lambda^{3/2}T^3)$. What is more, $G_{ra}(l+k)$ and $G_{ar}(l+k)$ propagators are both $O(\lambda^{-1})$ since they are dressed with the self-energies to cure pinch singularities. Additionally, since $G_{rr}(l+k-p)$ is in the collinear regime with $G_{ar}(l+k)$, it is also dressed and is of the order $O(\lambda)$. The properties of propagators impose that $(l+k)^2$ and $(l+k-p)^2$ are $O(\lambda T^2)$ and the same holds for $(l)^2$, which is soft and dressed with the HTL correction. These conditions are, in turn, equivalent to the fact that the angles between all participating particles are parametrically small so that they all propagate collinearly. The small angles are therefore $\theta_{kl} \sim \theta_{pl} \sim O(\sqrt{\lambda})$. The constraints on the angles impose constraints on the phase spaces, which is, $d^3p \sim |{\bf p}|^2 d|{\bf p}|\sin\theta_{pl} d\theta_{pl} d\phi \sim O(\lambda T^3) $ and $d^3k \sim |{\bf k}|^2 d|{\bf k}|\sin\theta_{kl} d\theta_{kl} d\phi \sim O(\lambda T^3) $. The loop momentum $l$ must be spacelike and since it is soft there is an additional Bose-Einstein enhancement making $G_{rr}$ be of the order $O(\lambda^{-3/2})$. Combining all these powers of $\lambda$ and the couplings coming from the explicit interaction vertices combined with the closed color loops one finds this rung to be $O(\lambda^2)$.

The presence of the self-energy in $G_{ar}(l+k)$ and $G_{ra}(l+k)$ propagators signals further interactions, which have not yet been explicitly shown nor discussed. In fact one can attach other lines to the side rails of the rung to reproduce processes involving a larger number of participating excitations. For example, one could add a hard line so that to obtain a double gluon emission. Such a process is however subleading \cite{Arnold:2003zc}. One could also add many soft lines to reflect the process of a hard excitation interacting many times with the medium via a soft exchange and then ending up with splitting into two hard particles. Attaching any number of soft lines to the side rails is possible and they all contribute $O(1)$ corrections. These processes, however, do not need to be explicitly included inside the diagram in Fig \ref{sqr1} since they are resummed within the integral equation for the bulk viscosity. Also, apart from the pair of propagators with pinching poles, there is also a pair of $G_{ar}(l+k)$ and $G_{ra}(l+k-p)$ propagators, which contain nearly pinching poles, where other insertions are possible. We do not examine them here since they are a part of the forthcoming collinear analysis, which investigates the emergence of an effective vertex in the collinear regime.

\begin{figure}[h]
\centering
\begin{tabular}{c}
\includegraphics[width=0.6\textwidth]{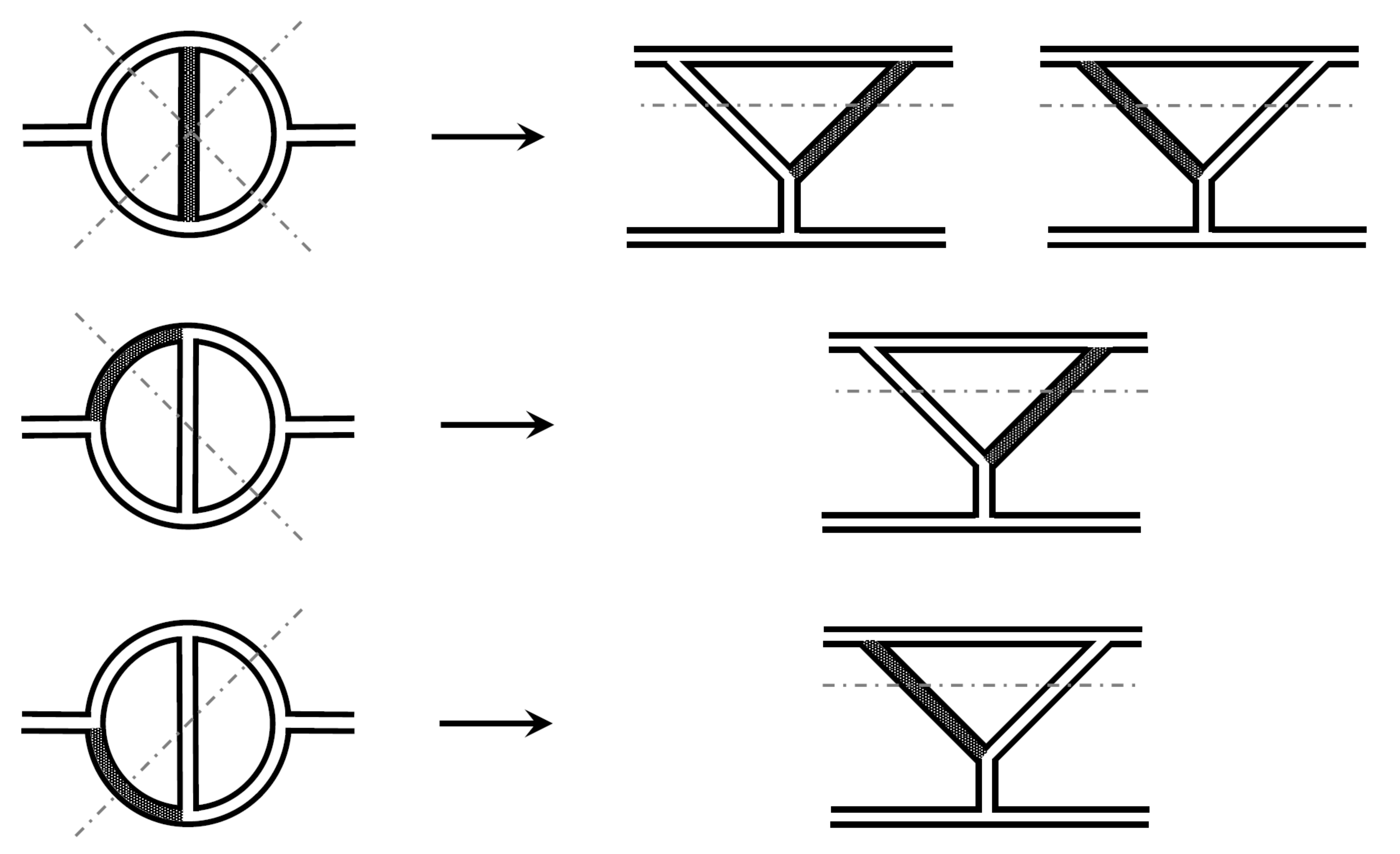}
\end{tabular}
\caption{The procedure showing how to obtain topological structures of rungs representing the interference terms between the number changing processes in the collinear regime. Only few representative structures are shown.}
\label{inf00}
\end{figure}

In Fig. \ref{soft1} we depicted how one can reproduce the collinear processes when one soft line appears in the two-loop self-energy. The rightmost column in Fig. \ref{soft1} presents the squares of amplitudes of these processes. For the entire analysis to be completed one also needs the interference terms. The leading order interference terms, which contain number changing processes, are those with only 3-gluon vertices; diagrams with 4-gluon vertices are suppressed. The diagrams in question are shown in Fig. \ref{inf00}. The figure presents the procedure of opening one cut line of the self-energy to reproduce the topological structures representing interference terms between collinear processes. To reproduce number changing processes the cutting line has to go through 3 lines of the 2-loop self-energy including the soft (shaded) line (as shown in Fig. \ref{inf00}). The line which can be open is the cut line which is a part of the side rails. Opening the internal (vertical) line would mean opening all color loops and it would lead to emergence of nonplanar diagrams, which are suppressed in large $M$ limit. One can also cut the self-energy so that the soft line remains uncut. Diagrams obtained in this way could only represent number conserving processes. In Fig. \ref{inf00} we show only a few typical topologies with respect to the position of the soft line. One can also realize that the same structures, but inverted upside down, are also possible. The latter ones would be the complex conjugates of those shown in Fig. \ref{inf00}. The interference terms are essentially the sums of the rungs shown in Fig. \ref{inf00} and their complex conjugates. Additionally, all structures shown can have different momentum and $a$ and $r$ assignment and the full computation of the imaginary part of the spectral function requires summation over all possibilities.

\begin{figure}[h]
\centering
\begin{tabular}{c}
\includegraphics[width=0.4\textwidth]{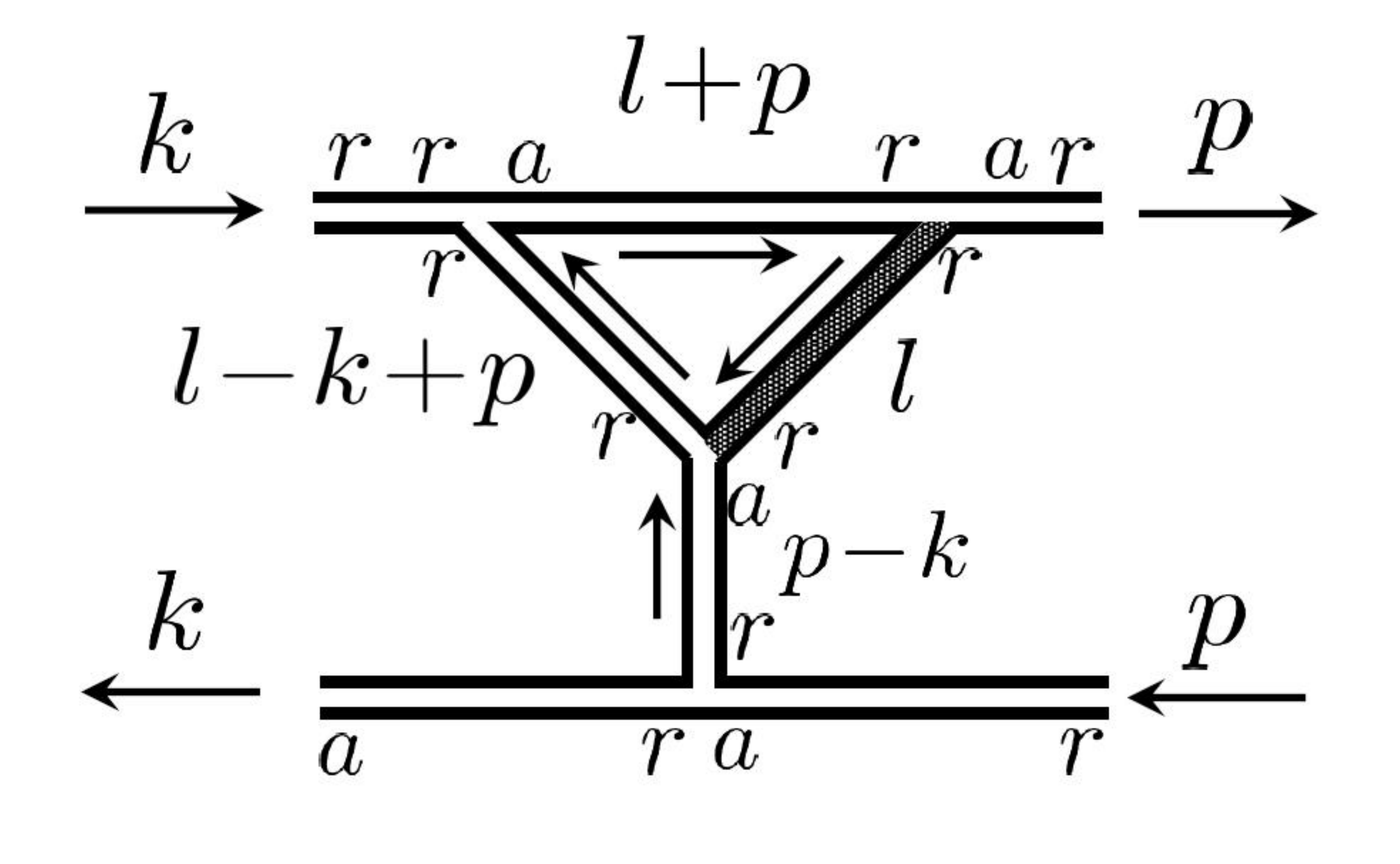}
\end{tabular}
\caption{A typical interference rung containing nearly collinear singularities.}
\label{inf11}
\end{figure}

To show that collinear splittings occur at the same order as $2 \to 2$ processes we examine one representative rung depicted in $r,a$ basis in Fig. \ref{inf11}. Other rungs with collinear singularities can be considered analogously. The expression corresponding to this rung is
\ba
&&\int \frac{d^4k}{(2\pi)^4} (\dots)\int \frac{d^4p}{(2\pi)^4} \mathcal{K}(\dots)  \\
&&\sim
\lambda^2 \int \frac{d^4k}{(2\pi)^4} (\dots) \int \frac{d^4p}{(2\pi)^4}
\int \frac{d^4l}{(2\pi)^4} G_{rr}(l)G_{rr}(l-k+p)G_{ar}(l+p)G_{ra}(p-k) (\dots),\nonumber
\ea
where $(\dots)$ stands for the insertion of propagators corresponding to the incoming and outgoing states and we also included integrals over all momenta since pairs of propagators with respect to all momenta can have nearly pinching poles in the collinear regime. In this particular case there are three such pairs: $G_{rr}(l-k+p)G_{ar}(l+p) \sim G_{ra}(l-k+p)G_{ar}(l+p)$, $G_{ra}(k)G_{ra}(p-k)$ and $G_{ar}(p)G_{ra}(p-k)$, which have singularities with respect to momentum $l$, $k$, and $p$, respectively. Notice that in case of $k$ momentum integration the propagators which have pinching poles are both denoted as $G_{ra}$ due to the notation and assignment of $r$ and $a$ with respect to $p$ in Fig. \ref{inf11}. However, taking into account that $G_{ra}(p-k)=G_{ar}(k-p)$ one obtains the expected $G_{ra}(k)G_{ar}(k-p)$ responsible for the emergence of the singularity. Due to all these constraints $G_{rr}(l)$, $G_{rr}(l-k+p)$, and $G_{ra}(p-k)$ have to be dressed and therefore they all are of $O(\lambda^{-1})$. In the leading order analysis $l$ has to spacelike, and thus $G_{rr}(l)$ is $O(\lambda^{-3/2})$. All these properties of propagators have their equivalence in the kinematic constraints, which reflect collinearity conditions, namely, small scattering angles $\theta_{kp}\sim \theta_{kl} \sim \theta_{pl}\sim O(\sqrt{\lambda})$. These, in turn, limit the respective phase spaces to $d^3k \sim d^3p \sim O(\lambda)$ and $d^3l \sim O(\lambda^{3/2})$. Additionally, the integral over $dl_0$ is dominated by the narrow frequency width, $\sim O(\lambda)$. Collecting all powers of coupling constant one finds this rung to be $O(\lambda^2)$.

\begin{figure}[h]
\centering
\begin{tabular}{c}
\includegraphics[width=0.8\textwidth]{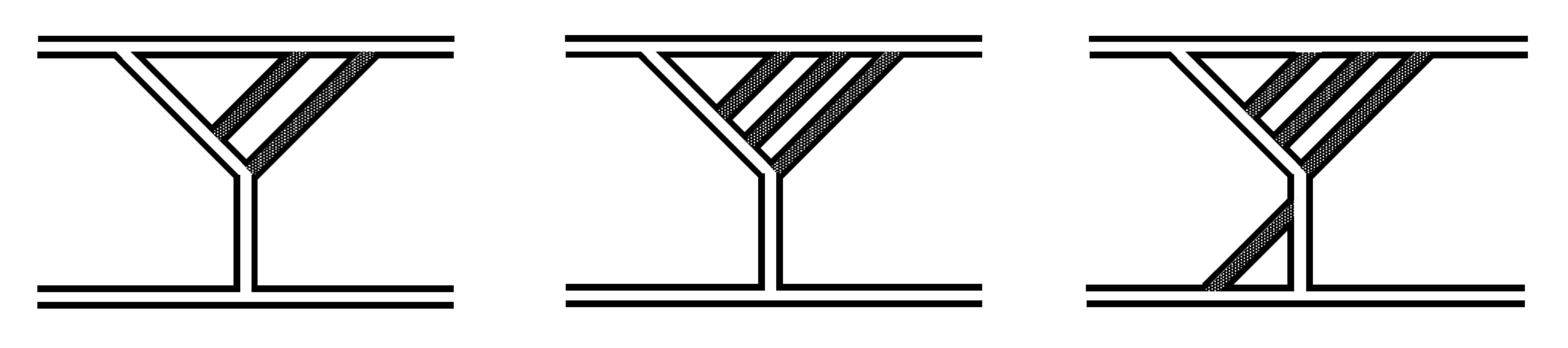}
\end{tabular}
\caption{Different possible ways of adding a soft line between the propagators with nearly pinching poles, which do not change the size of the rung. The momentum and $a$ and $r$ convention is the same as in
Fig. 12.}
\label{inf22}
\end{figure}

When assessing the size of the rung in Fig. \ref{inf11} one can realize that the enhancement in the rung's size coming from the collinear singularities is always balanced by the suppression coming from the phase space caused by the small scattering angle. Given that, there are more effects that need to be included in the leading order evaluation. One can attach infinitely many soft lines to a given pair of propagators with nearly pinching poles and still get the rung at the same order. This is schematically shown in Fig. \ref{inf22}, where a few exemplary insertions of a soft line are shown (the $r,a$ positions and the momentum convention is the same as in Fig. \ref{inf11}). The insertion of soft lines in the leading order is governed by a few rules. The lines have to be $G_{rr}$ propagators and they cannot cross each other since they must be ordered in time and coherent. Moreover, their insertion must follow the standard $a$ and $r$ assignments so that one has to have an odd number of $a$ in a given vertex. Also, a pair of propagators with the nearly pinching poles must appear, otherwise the rung is suppressed by some powers of $\lambda$. If all these rules are kept then attaching $G_{rr}$ soft line to the pair of lines with the nearly pinching poles always introduces $\lambda^{-3/2}$ from the very size of the propagator, $\lambda$ from the two explicit interaction vertices and a closed color loop, a pair of new propagators with the nearly pinching poles with the contribution $O(\lambda^{-2})$ and a phase space suppression $d^4l' \sim \lambda^{5/2}$. Altogether the insertion is $O(1)$ and thus infinitely many soft lines can be added in this way without changing the size of the rung. All possibilities have to be resummed and such a procedure reflects the diagrammatic representation of the LPM effect.

\begin{figure}[h]
\centering
\begin{tabular}{c}
\includegraphics[width=0.6\textwidth]{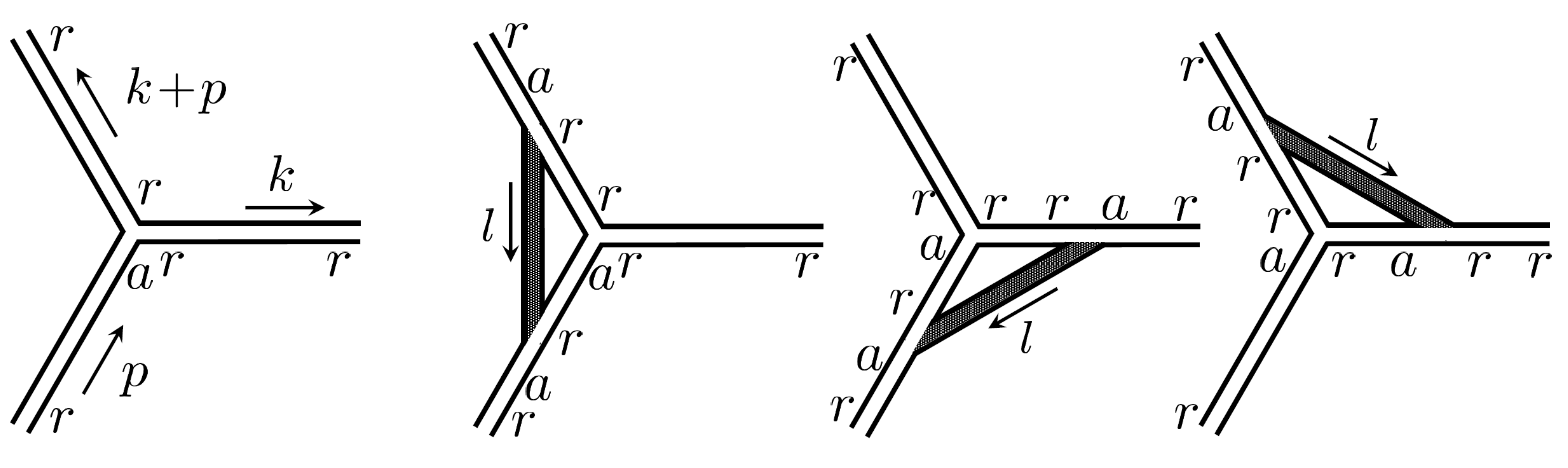}
\end{tabular}
\caption{One out of many vertices in $(r,a)$ basis which has to be considered in the collinear splitting analysis. Shown are also different possibilities of adding a soft line between a pair of propagators with nearly pinching poles.}
\label{inf33}
\end{figure}

The resummation of all possibilities of adding a soft line to a given rung is most efficiently done by finding an effective vertex. The vertex involves three hard and nearly on-shell lines, where all possible insertions of a soft line are included. One exemplary vertex in $(r,a)$ basis is shown in Fig. \ref{inf33}, where one soft line can be added in 3 possible ways. There are more such combinations since $r$ and $a$ can have a different layout. One has to include all of them to perform a full analysis and sum over all possible insertions of the soft line.

An inclusion of all possible combinations reflects the need for an integral equation, which needs to be solved to find a form of the effective vertex. The solution should be then inserted in the kernel of the integral equation established by the pinch singularities. This approach is, however, demanding within quantum field theory approach. The only essential point to notice is that insertion of any number of soft lines does not change the size of the vertex nor, consequently, the size of a rung where the effective vertex appears.

\begin{figure}[h]
\centering
\begin{tabular}{c}
\includegraphics[width=0.6\textwidth]{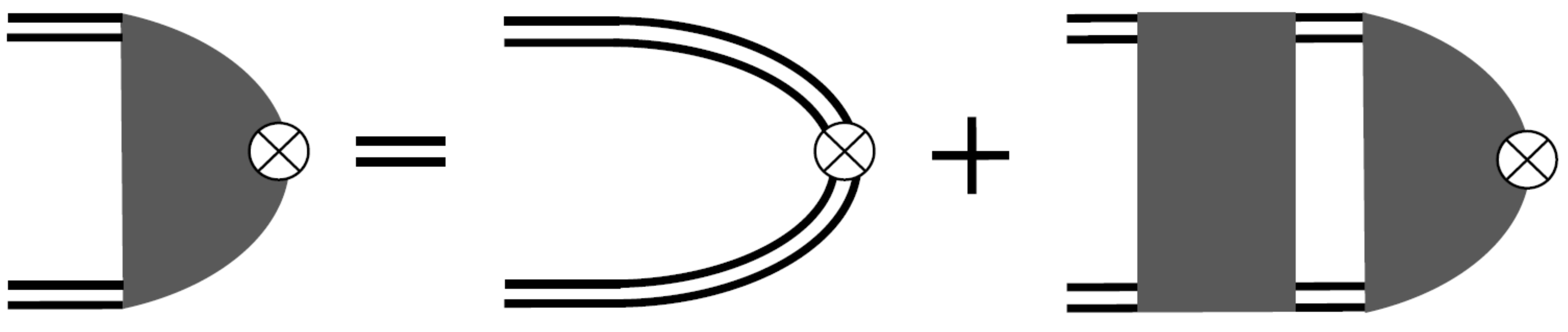}
\end{tabular}
\caption{The integral equation for bulk viscosity in the SU($M$) theory.}
\label{series}
\end{figure}

\begin{figure}[h]
\centering
\begin{tabular}{c}
\includegraphics[width=0.8\textwidth]{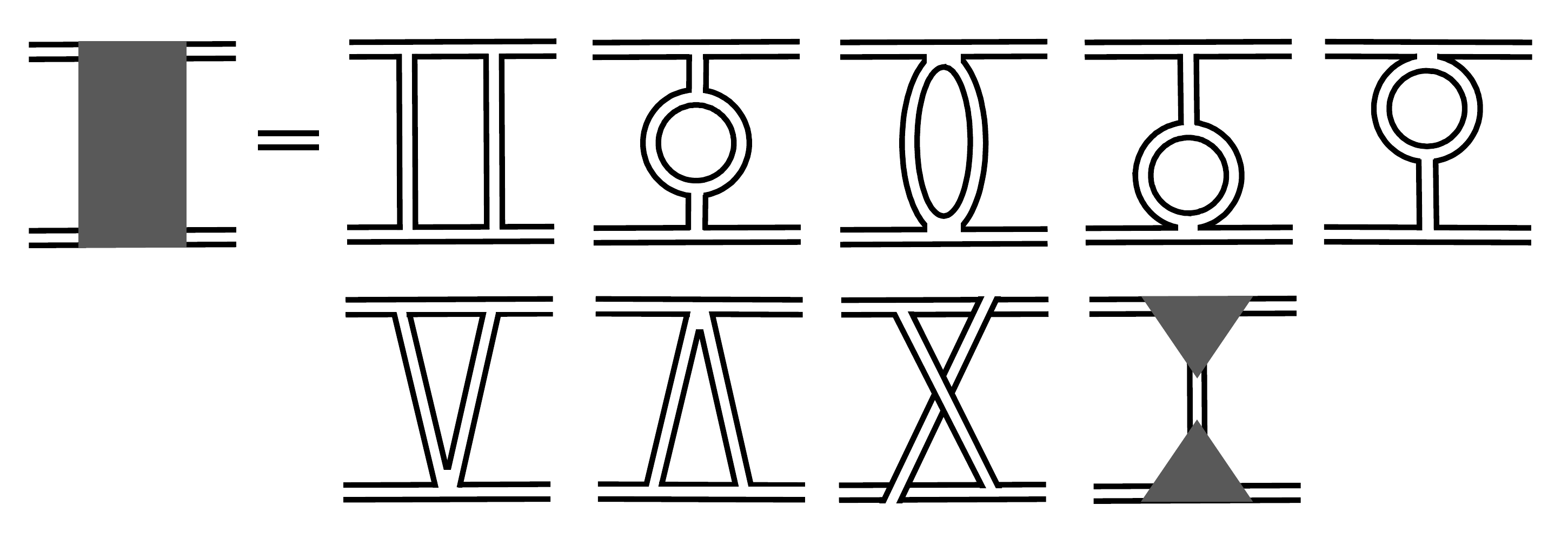}
\end{tabular}
\caption{The kernel of the integral equation with possible topological rung insertions. The rungs are shown schematically and the distinction between number conserving and number changing processes (or soft and hard momenta) is not denoted explicitly here since both these classes come from the same topological structures.}
\label{series-k}
\end{figure}

\begin{figure}[h]
\centering
\begin{tabular}{c}
\includegraphics[width=0.9\textwidth]{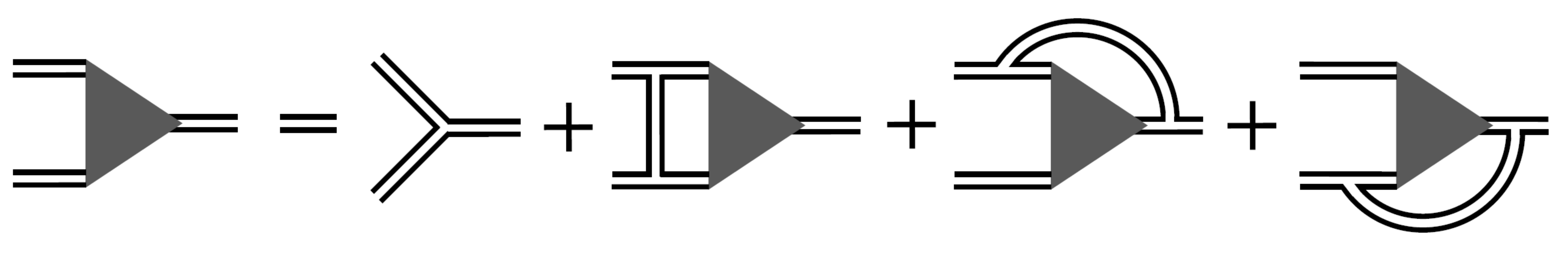}
\end{tabular}
\caption{Integral equation for the effective vertex characteristic of the collinear splittings. The hard and soft momenta are not distinguished.}
\label{series-v}
\end{figure}

\subsection{Integral equations \label{alina6}}

The bulk viscosity is controlled by the elementary scattering processes entangled in rungs discussed above and both $2 \to 2$ and effective $1 \to 2$ processes between gluons contribute at the same order in the 't~Hooft coupling $\lambda$. For a quantitative computation of the bulk viscosity coefficient all diagrams representing scattering events have to be resummed, which leads to the relevant integral equations. For the prescription given by the Kubo formula the integral equation is shown schematically in Fig. \ref{series}. The kernel of the equation is presented in Fig.~\ref{series-k}. It includes $2 \to 2$ processes and effective $1 \to 2$ processes as well. For number changing processes another integral equation needs to be solved. It is the equation for the effective vertex and is shown schematically in Fig. \ref{series-v}. The shaded regions denote resummed parts. For this schematic representation of the integral equations we do not distinguish between the propagators with hard and soft, HTL resummed, momenta, but it can be easily done taking into account the discussion in Sec.~\ref{sec-just}. In general, the leading order analysis requires all propagators and vertices to be dressed with the self-energies. In the last diagram in Fig. \ref{series-k} effective vertices are shown explicitely. They must be used in the leading order analysis as arbitrary many coherent interactions with the medium through the soft momentum exchange occur. The rung is thus  responsible for the interference terms as well as it reflects the LPM effect for the effective $1 \to 2$ processes.

The analytical computation of the bulk viscosity spectral function in terms of quantum field theory tools is very challenging and only qualitative picture can be sketched. The same physics is, however, embodied in the Boltzmann equation as long as the same elementary processes govern its collision kernel. As has been examined in this section, both $2 \to 2$ and $1 \to 2$ processes can be reproduced from the planar diagrams of the spectral function. Both classes of processes occur at the same order and so contribute to the kernel of the Boltzmann equation, discussed in detail in Refs. \cite{Arnold:2000dr,Arnold:2002zm,Arnold:2003zc}. It therefore justifies that the collision kernel of the Boltzmann equation captures the same physics as the kernel of the spectral function, shown in Fig. \ref{series-k}, and serves as a convenient way to compute transport coefficients. In particular, the analysis justifies the employment of the Boltzmann equation to calculate the bulk viscosity coefficient of the SU$(M)$ theory, as carried out in Ref. \cite{Arnold:2006fz} and summarized in Sec. \ref{kinetic} of this manuscript.

\subsection{Alternative diagrammatic approaches \label{alina7}}

In the previous subsections we discussed how to justify the Boltzmann equation using standard diagrammatic techniques. There are, however, other alternatives such as Kadanoff-Baym and the $n$PI ($n$-particle irreducible) formalisms, which deal with the Green function techniques to study real-time dynamics. Kadanoff-Baym equations \cite{Baym:1961zz} is a set of coupled equations of motion for two-point real-time Green functions, where terms with self-energies determine the system's dynamics. They, in general, hold for nonequilibrium situations and capture quantum effects, for a review see~\cite{Blaizot:2001nr}. Using a number of approximations one is able to derive a set of kinetic equations, see for example~\cite{Danielewicz:1982kk}. To compute any physical quantity one needs to specify the form of self-energy. As long as the coupling is weak, the analysis of the self-energy of a theory under consideration is uniform for different approaches at a given coupling order and holds for $n$PI formalism as well. Let us, therefore, focus on the latter approach in more detail.

The $n$PI formalism allows one to study systems close to thermal equilibrium, which involves perturbative calculations, as well as it provides an efficient way to analyze far-from-equilibrium dynamics and to include nonperturbative effects, see \cite{Cornwall:1974vz,Calzetta:1986cq} and for a review \cite{Berges:2004yj}. In general, the formalism relies on the effective action $\Gamma$, which is the generating functional for all correlation functions of the quantum theory. In practical calculations, one does not have to know the most general form of $\Gamma$ to reproduce all Green functions for arbitrarily large $n$, but, due to the equivalence hierarchy between different $n$PI effective actions, the expansion series can be truncated at a given loop order. For example, a self-consistent two-loop order approximation of a given quantity is completely captured by the 2PI effective action and all $(n>2)$PI effective actions are equivalent to it, analogously, three-loop order approximation is well described by the 3PI formalism. The method has been, in particular, successfully employed to compute transport coefficients of different systems. The shear and bulk viscosities of the real scalar $\lambda \phi^4$ theory were computed in Ref.~ \cite{Calzetta:1999ps} from the closed-time-path 2PI effective acion truncated at the four-loop order. In~\cite{Aarts:2005vc} the shear viscosity and the electrical conductivity of gauge theories were obtained in large $N_f$ limit to next-to-leading order in the $1/N_f$ expansion. Later, the electrical conductivity of QED was obatined in Refs.~\cite{Carrington:2007fp} and  \cite{Carrington:2007ea} using 2PI and 3PI effective action, respectively. The shear viscosity of QCD was calculated in Ref.~\cite{Carrington:2009xf}, where the analysis to derive the matrix elements of the collision kernel of the Boltzmann equation is equivalent to the one discussed in this manuscript in the leading order of 't Hooft coupling. Below we briefly summarize the basics of the $n$PI methods which lead to calculation of the shear viscosity coefficient.

To analyze processes which determine the collision kernel of the Boltzmann equation in the full leading order one needs to start with the 3PI effective action. As discussed above, the quark contribution is subleading in the 't Hooft coupling expansion, so we do not include it here as well, and the 3PI effective action (in the Feynman gauge) can be written as \cite{Carrington:2009xf}:
\begin{eqnarray}
\label{eff-action}
&&\Gamma[A,\eta,\bar\eta,G,\Delta,U,Y] = S_{\text{cl}}[A,\eta,\bar\eta]
+\frac{i}{2} \text{Tr}\,\ln G^{-1}_{12} + \frac{i}{2}\text{Tr} [(G^0_{12})^{-1}(G_{21}-G^0_{21})]
 \nonumber \\
& & \qquad\qquad-i \text{Tr}\, \ln \Delta_{12}^{-1}-i \text{Tr} [(\Delta^0_{12})^{-1} (\Delta_{21} - \Delta^0_{21})]
-i \Phi[A,G,\Delta,U,Y],
\end{eqnarray}
where $A$ and $\eta,\bar\eta$ are the gluon and ghost fields, respectively. The notation is compactified so that the Lorentz and color indices are not shown explicitely and the indices 1 and 2 refer to the time positions on the contour of the closed-time-path formalism. $S_{\text{cl}}$ is the classical action defined by the Lagrangian density (\ref{L-SUN}) adjusted to the Feynman gauge, that is, when the gauge fixing term and the ghost term are included. $G^0$ and $\Delta^0$ are the free gluon and ghost propagators and $G$ and $\Delta$ are the full ones. $U$ and $Y$ are the self-consistent 3-gluon and gluon-ghost vertices. $\Phi$ is the sum of all 3-particle irreducible diagrams, which contains the most essential information on the dynamics.

The free propagators as well as the free vertices are given by:
\begin{eqnarray}
\label{free-prop}
&(G^0_{12})^{-1} = -i\frac{\delta^2 S_{\text{cl}}}{\delta A_2 \delta A_1}, \qquad\qquad
(\Delta^0_{12})^{-1} = -i\frac{\delta^2 S_{\text{cl}}}{\delta \eta_2 \delta \bar \eta_1}, \\
\label{free-vert}
&\Omega^0_{132}=-\frac{\delta (G^0_{12})^{-1}}{\delta A_3}, \qquad
\Theta^0_{132}=-\frac{\delta (\Delta^0_{12})^{-1}}{\delta A_3}, \qquad
M^0_{1234} = -\frac{\delta^2(G^0_{12})^{-1}}{\delta A_4 \delta A_3},
\end{eqnarray}
where $\Omega^0_{132},\Theta^0_{132}$ and $M^0_{1234}$ are the 3-gluon, gluon-ghost, and 4-gluon vertices. The free 3-gluon and gluon-ghost vertices belong to respective chains of resummed vertices: $\Omega_{132}$ and $\Theta_{132}$. To obtain building blocks of the integral equations needed to resumm all diagrams at the leading order one needs to consider the equations of motion of the mean fields, propagators, and vertices. They are obtained from the stationarity of the action:
\begin{eqnarray}
\label{stationary}
&&
\frac{\delta \Gamma}{\delta A} = \frac{\delta \Gamma}{\delta \eta} = \frac{\delta \Gamma}{\delta \bar \eta} = 0, \\
\label{stationary2}
&&
\frac{\delta \Gamma}{\delta G} = \frac{\delta \Gamma}{\delta \Delta} = 0,\\
\label{stationary3}
&&
\frac{\delta \Gamma}{\delta U} = \frac{\delta \Gamma}{\delta Y} = 0.
\end{eqnarray}
These condtions are, however, not satisfactory to maintain the Ward identity for vertex functions associated with the gauge symmetry of the theory. The problem appears as a result of the truncations of the effective actions at a given order and the use of inconsistent approximation scheme, see, for example~\cite{Arrizabalaga:2002hn,Kobes:1990xf}. The solution to obtain gauge independent quantities and preserve the Ward identities is to rely on the resummed effective action \cite{VanHees:2001pf,vanHees:2002bv,Berges:2005hc,Reinosa:2006cm,Reinosa:2007vi}, which provides properly resummed vertices. The resummed effective action $\Gamma[A,\bar\eta,\eta]$ depends only on the expectation values of fields meant as:
\begin{equation}
\label{resum-act}
\Gamma[A,\bar\eta,\eta]  = \Gamma[A,\bar\eta,\eta,\tilde G[A,\bar\eta,\eta],\tilde \Delta[A,\bar\eta,\eta], \tilde U[A,\bar\eta,\eta],\tilde Y[A,\bar\eta,\eta]],
\end{equation}
where $\tilde G,\tilde \Delta, \tilde U, \tilde Y$ are the self-consistnet solutions which are obtained simultaneously when performing Eqs.~(\ref{stationary})-(\ref{stationary3}). The equivalence of (\ref{resum-act}) with Eq.~(\ref{eff-action}) holds at the exact level.

Having given the 3PI formalism one proceeds to obtain integral equations and prescription to reproduce matrix elements of the collision kernel. The whole procedure is presented in Ref.~\cite{Carrington:2009xf} and its main ingredients are as follows. As mentioned, there are two types of sets of vertices. The first one contains $\Omega$ and $\Theta$, which are defined by functional derivatives of the self-consistent solutions with respect to field expectation values, that is, by analogy to their free counterparts, see Eqs.~(\ref{free-vert}). Using next the equations of motion (\ref{stationary2}) and the effective action (\ref{resum-act}) one obtains the integral equations for $\Omega$ and $\Theta$ vertices. The kernels of these integral equations contain already topological rungs shown in Fig.~\ref{topol} (and also respective rungs with ghosts). These effective vertices carry therefore information on both the pinching and collinear singularities. The latter ones are indeed taken into account due to very structure of the integral equation for the effective vertices. The other set of verices includes self-consistent vertex functions $U$ and $Y$. These are defined by functional derivatives of the resummed effective action with respect to field expectation values. The corresponding integral equations are obtained directly from the equations of motion given by (\ref{stationary3}). The role of $U$ and $Y$ veritces is mainly to supply the consistency of the procedure so that some diagrams produced in $\Omega$ and $\Theta$ configurations are cancelled and thus double counting is avoided. In the end, the procedure of dealing with the set of these coupled equations for vertex functions shall be equivalent to the analysis discussed in the previous parts of this section.

\section{Bulk viscosity at intermediate coupling \label{int}}

In the previous section, we have discussed the behavior of the bulk viscosity in the SU$(M)$ gauge theory in the weak coupling limit. In the next sections, we will discuss the strong coupling behavior. In these two limits, we have well defined calculational tools, perturbation theory in the case of the weakly coupled limit and the AdS/QCD correspondence in the strongly coupled limit. When the coupling is neither weak nor strong the only reliable QCD results are from Euclidean Lattice QCD (LQCD) calculations. Unfortunately, direct extraction of viscosities from LQCD is very nontrivial since viscosities have to do with dissipation in real time while LQCD calculations are inherently static. Of course, if one can calculate full Euclidean correlation functions, they can be analytically continued to real time correlation functions. But since only discrete and finite number of data points are available from LQCD, this procedure in practice introduces large uncertainties.

In literature, efforts were made to extract information on the bulk viscosity from LQCD results using sum rules \cite{Kharzeev:2007wb,Karsch:2007jc,Moore:2008ws,Romatschke:2009ng,Meyer:2007dy,Meyer:2010ii,Astrakhantsev:2018oue}. In this section, we summarize the main points and point out why it is difficult to get any information on the bulk viscosity from the sum rules in particular and from LQCD results in general.

Extraction of the bulk viscosity from the sum rule relies on the Kubo formula for the bulk viscosity $\zeta$:
\ba
\lim_{\omega\to 0} {\rho_{PP}(\omega, \mathbf{0})\over \omega} = {\zeta \over \pi},
\label{eq:KKT_Kubo_PP}
\ea
where $\rho_{PP}$ is the spectral density for the pressure-pressure correlator. Hence one may expect that the bulk viscosity can be extracted from a sum rule involving $\rho_{PP}(\omega,\mathbf{0})/\omega$. Equivalently, it may be able to extract $\zeta$ from a sum rule involving the correlation function of the trace operator $\hat\Theta = \hat T^{\mu}_{\mu} = \hat T^{00} - 3\hat P$ because the energy-momentum conservation laws dictate that the zero wavenumber limit of $\hat T^{00}$ correlators vanish. Note that different forms of the trace operator can be used and they were briefly discussed in the previous section.

In Ref.~\cite{Karsch:2007jc}, low energy sum rules for the stress-energy tensor trace, $\Theta = T^{\mu}_{\mu}$, were derived and the following result was established:
\ba
\int_0^\infty {d\omega\over \pi }\, {\rho_{\Theta\Theta}(\omega, \mathbf{0})\over\omega}
=
\left( T{\partial\over \partial T} - 4\right)\langle \Theta_G \rangle_T
+ (\hbox{quark contribution}),
\label{eq:KKT}
\ea
where $\Theta_G$ is the gluon contribution to the stress-energy tensor trace and $\rho_{\Theta\Theta}$ is the spectral density for the $\Theta\Theta$ correlation function.\footnote{%
	In literature, the definition of the spectral density varies.
	In this paper, we use the definition
	$\rho_{\Theta\Theta}(x) = \langle [\hat\Theta(x), \hat\Theta(0)] \rangle$.
	The spectral density in Ref.~\cite{Karsch:2007jc} ($\rho_{\rm KT}$)
	and in Ref.~\cite{Romatschke:2009ng} ($\rho_{RS}$) are related by
	$\rho_{\Theta\Theta} = 2\rho_{\rm RS} = 2\pi\rho_{\rm KKT}$.}
As we have argued in the previous section, the quark contribution is negligible in the large $M$ limit and we will not consider it here, either. Consequently we will drop the subscript $G$. The trace average is:
\ba
\langle \Theta \rangle_T = \left(\epsilon - 3P\right) + \langle \Theta \rangle_0
\ea
where $\langle \Theta \rangle_0$ is the vacuum contribution.

In Ref.~\cite{Romatschke:2009ng}, re-derivation of the results with the direct subtraction of the vacuum spectral density led instead to:
\ba
\int_0^\infty {d\omega\over \pi} {\delta\rho_{\Theta\Theta}(\omega, \mathbf{0})\over \omega}
=
\left( 3s {\partial\over \partial s} - 4\right)\left(\epsilon - 3P\right),
\label{eq:SR}
\ea
where $\delta\rho_{\Theta\Theta}(\omega, {\bf k}) = \rho_T(\omega, {\bf k}) - \rho_0(\omega, {\bf k})$ is the deviation from the vacuum spectral density at finite temperature $T$. The difference between the two sum rules was attributed to the non-commutability of the limits $\lim_{\omega\to 0}$ and $\lim_{{\bf k} \to 0}$, see \cite{Romatschke:2009ng}. In Ref.~\cite{Meyer:2010ii}, the sum rule Eq. (\ref{eq:SR}) is re-cast as:
\ba
\int_{0}^\infty {d\omega\over \pi}
{\delta\rho_{*}(\omega,\mathbf{0}) \over\omega}
=
3(1 - 3c_s^2)(\epsilon+P) - 4(\epsilon - 3P)
\label{eq:Meyer}
\ea
where $\rho_*$ is the spectral density for the operator $\hat\Theta_* =  \hat T^{\mu}_{\mu} - (1 - 3c_s^2)\hat T^{00}$. The spectral density of the operator $\hat\Theta_*$ satisfies the same Kubo formula but has an added benefit that the limits $\lim_{\omega\to 0}$ and $\lim_{{\bf k}\to 0}$ commute.

The right hand side of the sum rule (\ref{eq:Meyer}) can be evaluated using the LQCD results. If one can then show that the left hand side is a well defined function of the bulk viscosity then these sum rules may be used to determine the bulk viscosity in the region of temperature where LQCD calculations can be performed.

A first attempt at relating the sum rule integral (the left hand side of Eq.~(\ref{eq:Meyer}) to the bulk viscosity was carried out in \cite{Kharzeev:2007wb,Karsch:2007jc}. In Ref.~\cite{Karsch:2007jc} the following ansatz was introduced:
\ba
{\delta\rho_{*}(\omega, \mathbf{0})\over \omega}
=
{9\zeta\over \pi} {\omega_0^2\over \omega_0^2 + \omega^2},
\label{eq:KKT_ansatz}
\ea
which does satisfy the Kubo formula and makes the sum rule integral in the left hand side of Eq. (\ref{eq:Meyer}) proportional to $\zeta\omega_0$. However, this form lacks contribution from frequencies higher than the unknown parameter $\omega_0$, see Ref.~\cite{CaronHuot:2009ns}. It turned out that the high frequency contribution is actually {\em negative} that largely cancels the low frequency contribution.

The fact that this ansatz is not adequate has been shown by \cite{Moore:2008ws,Romatschke:2009ng,Meyer:2010ii} both perturbatively and non-perturbatively. The biggest problem is that the right hand side of Eq. (\ref{eq:Meyer}) is {\em negative} while Eq.(\ref{eq:KKT_ansatz}) makes the left hand side strictly positive. This difference can be attributed to the the presence of the glueball \cite{Meyer:2010ii}. Hence, the sum rule Eq. (\ref{eq:Meyer}) is not particularly useful since it cannot be definitely established that the dominant contribution to the sum rule integral comes from the low frequencies.

If one cannot rely on the sum rule, then one needs to obtain the spectral density directly from LQCD calculations at least in the ${\bf k}= 0$ and $|\omega| \ll T$ limit. This is not an easy task as it involves analytic continuation when only a finite number of data points in the Euclidean space is known. First attempts in this direction were made in \cite{Meyer:2007dy,Meyer:2010ii}, which does show that $\delta\rho_*(\omega,\mathbf{0})/\omega$ has a peak at $\omega = 0$. Unfortunately, actual values and behavior of $\zeta/s$ obtained in this way contains too much uncertainty at this point. All one can conclude from the LQCD studies right now is that $\zeta/s = O(10^{-2}) - O(10^{-1})$ within $T_c < T < 1.65\,T_c$.

\section{Bulk viscosity at weak string and strong 't~Hooft couplings with zero flavors \label{nf0}}

\begin{figure}[h]
\centering
\begin{tabular}{c}
\includegraphics[width=\textwidth]{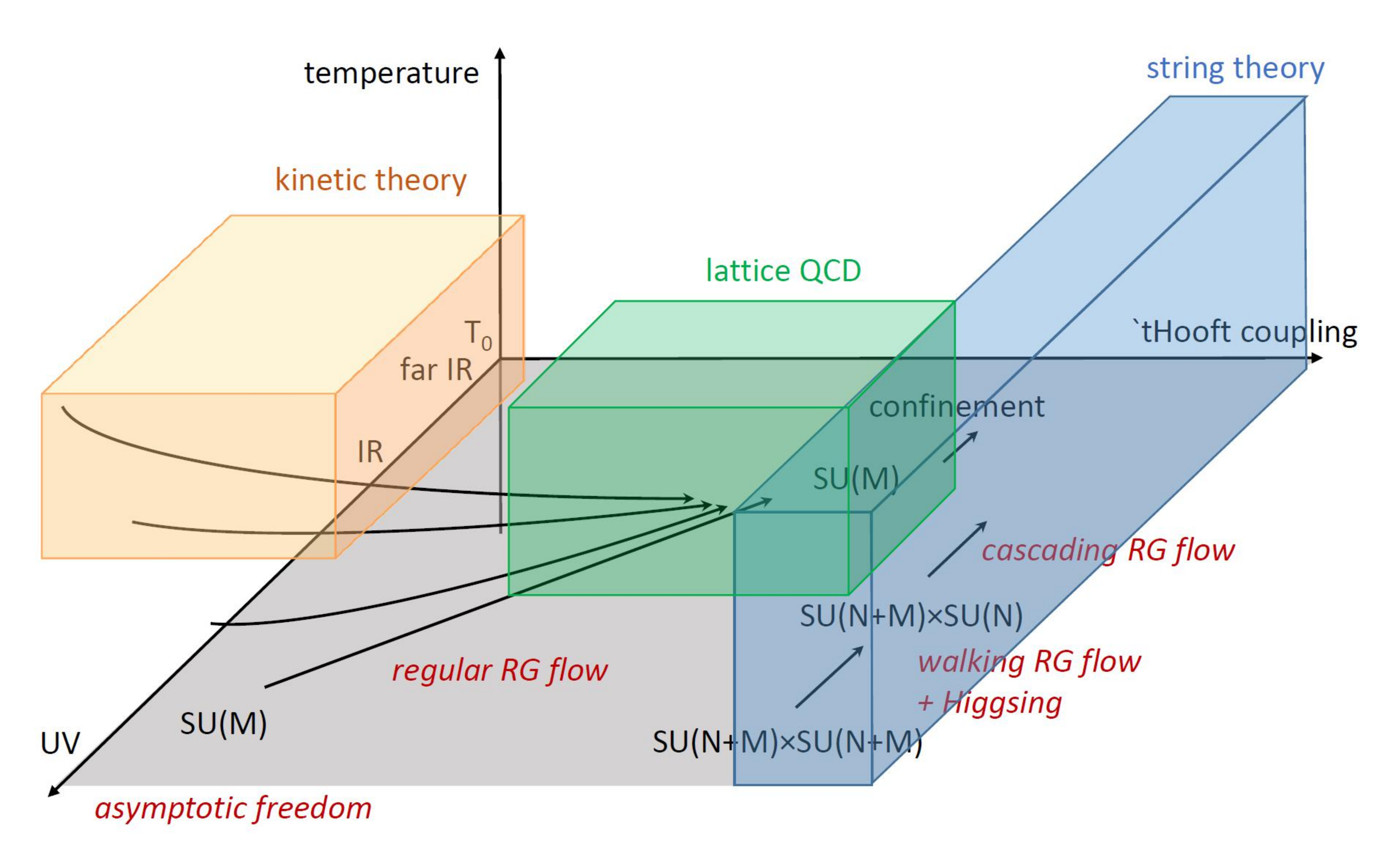}
\end{tabular}
\caption{The three different regimes of interest used here to study bulk viscosity. The yellow box denotes the regime of kinetic theory, the green box denotes the regime of LQCD, and the blue box denotes the regime of string theory. All these regimes are analyzed at high temperatures, i.e above the deconfinement temperatures, and the regular RG flows connecting the yellow and the green boxes are denoted by black curves. On the other hand, the cascading RG flows, that specifically arise from string theory, are shown here in the blue box.  All the three different RG flows lead to a consistent picture at low energies connecting the weak, intermediate and strong 't~Hooft couplings. However the UV pictures are very different, for example in the string side, i.e in the blue box, the UV gauge group is a product gauge group and is at strong 't~Hooft coupling.}
\label{allregions}
\end{figure}

In the previous section we studied the bulk viscosity at weak 't~Hooft coupling, and argued how the ratio of the bulk to shear viscosities should be interpreted at both weak and
strong couplings. As mentioned therein, the strong coupling result depends on the existence of a gravity dual of the resulting framework. Our aim here is to analyze SU$(M)$ gauge theory at various values of the 't~Hooft couplings and at high temperatures as depicted in Fig. \ref{allregions}. The three regimes of interest are shown in
Fig. \ref{allregions}: the yellow box denotes weak,  the green box denotes intermediate and the blue box denotes strong 't~Hooft couplings, all at high temperatures. The theories governing each of the three dynamics are also varied as we discussed above. The weak and the intermediate couplings are studied using kinetic and LQCD, whereas the strong coupling will be studied using string theory. The latter however is more elaborate because of its UV properties, and in fact differs quite a bit from what we expect from kinetic theory and LQCD.

Let us start with the kinetic theory which was discussed in details earlier. The regular RG flow of such a theory is governed by the black lines in Fig. \ref{allregions}. At low energies the YM coupling becomes very large and the theory confines. However if we increase the temperature, the coupling can be made smaller. In fact at high temperature the 't~Hooft coupling $\lambda \equiv g^2_{\rm YM} M$ can become very small even for large $M$. This is of course the regime where kinetic theory can be studied (see section \ref{weak} for more details), and is denoted by the yellow box in Fig. \ref{allregions}. One can similarly go to the intermediate coupling regime, whose dynamics is governed by LQCD.

What we now require is to understand the regime where the 't~Hooft coupling $\lambda$ can be very large for {\it both} weak and strong YM couplings. This is the regime where neither kinetic theory nor LQCD can help us, and therefore the only way we can have any analytic control is to use techniques of string theory. Of course when $M$, the number of colors, is small even string theory cannot provide a controlled laboratory, so it is the large $M$ limit that can be tackled using stringy techniques. This is then the regime of gauge/gravity dualities, i.e the dynamics at strong 't~Hooft  in the gauge theory side may now be done using a gravity dual description.

Clearly since string theory provides a UV complete picture, it is natural to ask what UV completion would mean in the present set-up. However before we go about exploring this side of the story, let us first elucidate the IR dynamics of the theory directly from the gravity dual description. The gravity dual description was originally provided in
\cite{metrics} (see also \cite{MQGP} for the mirror set-up which will be useful soon). In simple terms, the gravity dual is given in terms of a resolved warped-deformed conifold with fluxes with an additional black hole that provides the high temperature physics in the gauge theory side, i.e the physics above the deconfinement temperature.
In the absence of a black hole we expect minimal four-dimensional supersymmetry (that may be broken too), whose simplest description appear, on one side from wrapped D5-branes on a non-K\"ahler resolved conifold \cite{DEM}, and on the other side from fluxes on a resolved warped-deformed conifold alluded to above \cite{UVcom}. The ``resolution" parameter in the resolved warped-deformed conifold is responsible for the UV completion, that we will discuss soon (see also \cite{UVcom2} for a slightly different realization of the same story). In the following we want to discuss the background as well as the issue of supersymmetry, mostly for the IR part of the gauge theory. For simplicity we will concentrate on the Baryonic branch of the gauge theory where is the issue of supersymmetry is most prominently displayed. Later on, in section \ref{2bbg}, we will  concentrate on a more specific point in the moduli space of the corresponding gauge theory.

In the Baryonic branch, generated by $M$ wrapped D5-branes on a non-K\"ahler resolved conifold \cite{DEM}, the gravity dual for the IR physics may be given by the following type IIB background with three- and five-form
fluxes \cite{UVcom}\footnote{From here onwards we shall be using ($-, +, +, +$) convention to express the metric. This differs from the ($+, - , -, -$) convention used in sections \ref{weak} and \ref{int}.} :
 \bg\label{heidi}
&&ds^2 = {1\over \sqrt{h}} ~ds^2_{0123} + \sqrt{h}~ds^2_6 , \nonumber\\
&& {\cal F}_3 = {\rm cosh}~\beta e^{-2\phi} \ast_6 d\left(e^{2\phi} J\right), ~~~~~~ {\cal H}_3 = -{\rm sinh}~\beta ~d\left(e^{2\phi} J\right) , \nonumber\\
&&{\widetilde{\cal F}}_5 = -{\rm sinh}~\beta~{\rm cosh}~\beta\left(1 + \ast_{10}\right) {\cal C}_5(r)~d\psi \wedge \prod_{i=1}^2~\sin~\theta_i~ d\theta_i \wedge d\phi_i, \nd
where ($\theta_i, \phi_i, \psi$) are the angular coordinates, $\beta$ is the parameter of the Baryonic branch, $h$ is the warp-factor and $J$ is the fundamental (1, 1) form that is not closed. We have also denoted the dilaton by $\phi$,  and the five-form by ${\cal C}_5$. The internal metric $ds^2_6$ can be expressed as:
{\footnotesize
\bg\label{goeskalu}
ds^2_6 & = &  {\bf H}_1~ dr^2 + {\bf H}_2 (d\psi + {\rm cos}~\theta_1 d\phi_1 + {\rm cos}~\theta_2 d\phi_2)^2  + \sum_{i = 1}^2 {\bf H}_{2+i}
(d\theta_i^2 + {\rm sin}^2\theta_i d\phi_i^2) \\
& + & {\bf H}_5 ~\cos~\psi\left(d\theta_1d\theta_2 - \sin~\theta_1 \sin~\theta_2 d\phi_1 d\phi_2\right)  +
{\bf H}_5~  \sin~\psi\left(\sin~\theta_1d\phi_1 d\theta_2  + \sin~\theta_2d\phi_2 d\theta_1\right), \nonumber
\nd}
with ${\bf H}_i(r)$ being the additional warp-factors. Note that the two-spheres, denoted by ($\theta_1, \phi_1$) and
($\theta_2, \phi_2$) have different curvatures governed by ${\bf H}_3$ and ${\bf H}_4$ respectively, and their inequality will be responsible for UV completion.  The complexified three-form flux ${\cal G}_3$ then takes the following form \cite{UVcom}:

{\footnotesize
\bg\label{ashanti}
{{\cal G}_3 \over {\rm cosh}~\beta}  =  {ie^{-2\phi} {\bf M}_1 \over 2} ~E_1 \wedge \left(E_3 \wedge {\overline E}_3
- E_2 \wedge {\overline E}_2\right)
+  {ie^{-2\phi} {\bf M}_2 \over 2} ~E_1 \wedge \left( \overline{E}_2 \wedge {E}_3
- E_2 \wedge \overline{E}_3 \right),  \nd}
where ${\bf M}_i$ are certain functions expressed in terms of the vielbeins whose form may be ascertained from
eq (2.113) of \cite{UVcom}. The $E_i$ are defined with the following choice of the {\it almost} complex structure:
\bg\label{annihilation}
 (-i e^\phi \tanh~\beta, i, i), \nd
 which is integrable\footnote{The $\beta = 0$ limit has to be studied separately as discussed in \cite{DEM, UVcom}.}
  for a constant dilaton, otherwise the three-form flux ${\cal G}_3$ is defined as an ISD (Imaginary Self-Dual) form with respect to the almost complex structure \eqref{annihilation}. Note that \eqref{ashanti} is a (2, 1) form as one would expect from a supersymmetry-preserving background. Additionally, the choice of the Baryonic branch tells us that the gauge group is SU$(2M) \times SU(M)$, which is in fact one cascading step away from the confining SU$(M)$ gauge group that we seek! In the blue box of Fig. \ref{allregions} this may be seen as the second-last stage of the cascading RG flow before permanent confinement sets in.

 One can also give a physical meaning to the Baryonic branch directly from the wrapped five-brane picture. The
 $SU(2M) \times SU(M)$ gauge group implies that, along with $M$ wrapped D5-branes, we have $M$ D3-branes too.
The five-branes wrap the two-sphere parametrized by ($\theta_2, \phi_2$). For vanishing size of the two-sphere, the $M$ additional D3-branes preserve the same supersymmetries as the $M$ wrapped D5-branes. However if the two-sphere is of finite size, supersymmetry is completely broken, and the only way to preserve supersymmetry in this case would be to {\it dissolve} the D3-branes on the D5-branes.

\begin{figure}[h]
\centering
\begin{tabular}{c}
\includegraphics[width=\textwidth]{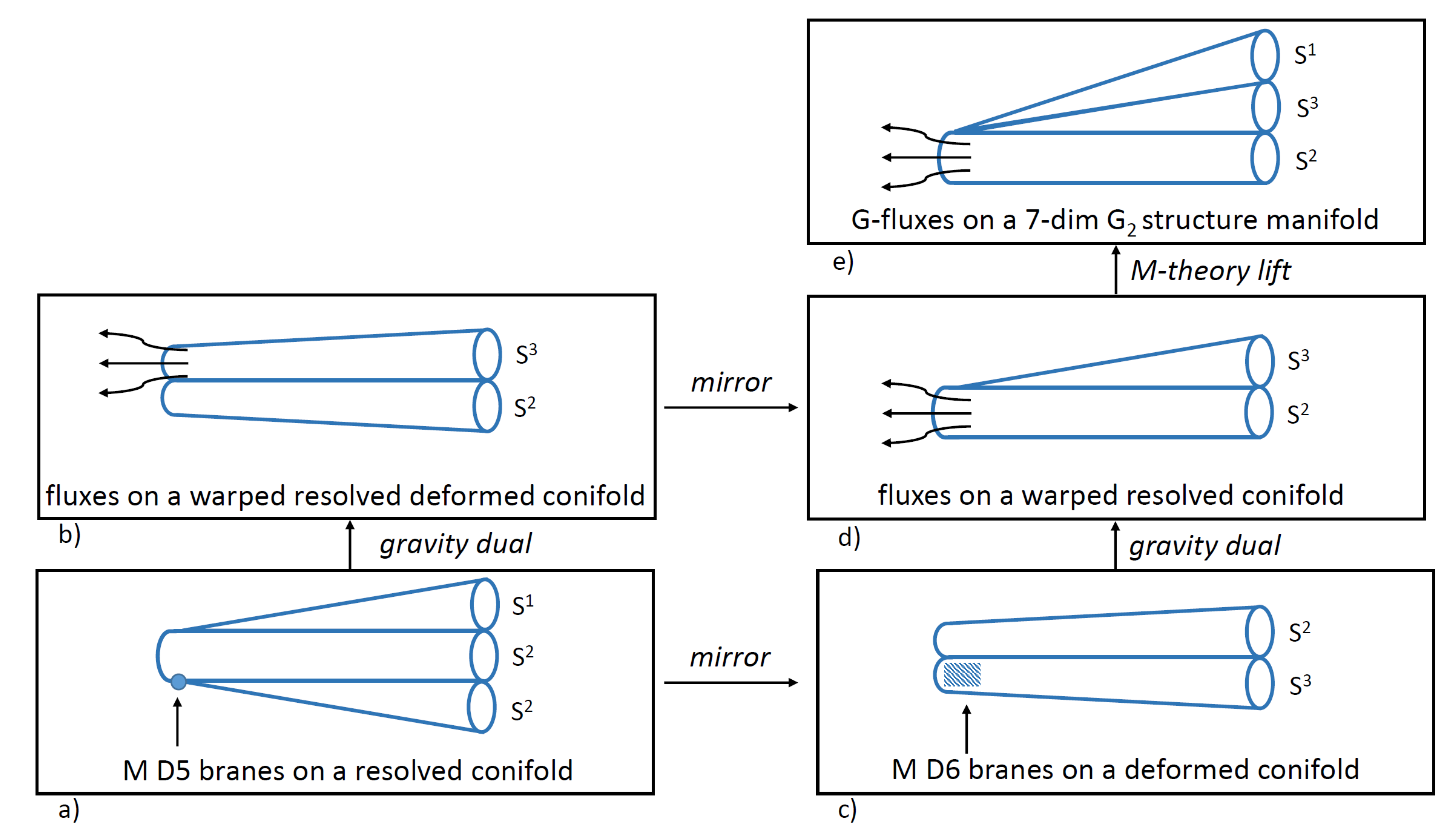}
\end{tabular}
\caption{The two configurations, one in type IIB and the other in the M-theory uplift of type IIA, on which all the computations of sections 4 and 5  respectively will be based upon. On the left is the type IIB picture with the gravity dual given by a resolved warped-deformed conifold with fluxes. On the right is the M-theory uplift of the type IIA gravity dual. The IIA gravity dual involes a non-K\"ahler resolved conifold with fluxes, whereas the M-theory uplift is a seven dimensional manifold with a $G_2$ structure. The type IIB computations will be done at high temperatures, i.e above the deconfinement temperatures, but with zero flavors. The type IIA, and also the M-theory uplift, will take into account both high temperatures as well as non-zero flavors. }
\label{2a2b}
\end{figure}

Being on the Baryonic branch does not give a well-defined UV picture. We will still need to find the UV completion of the model. This will be discussed a bit later,  but note that being in the Baryonic branch does tell us that if we make one Seiberg duality we will land in the confining SU$(M)$ gauge theory description. At non-zero temperatures, we will require black holes in the gravity side of our story. Since
this is the premise on which our calculations in this paper will be based on, let us elaborate the story a bit more. At zero temperature, the duality sequence that we shall use is laid out in Fig. \ref{2a2b}. On the bottom left corner, i.e box (a), is the gauge theory configuration discussed in \cite{metrics, DEM} with $M$ D5-branes wrapped on the two-sphere parametrized by ($\theta_2, \phi_2$). This is a non-K\"ahler resolved conifold because at $ r = 0$ there is a resolved two-sphere parametrized by ($\theta_1, \phi_1$). The usefulness of such a configuration will be spelled out a little later. The wrapped D5-branes on the non-K\"ahler resolved conifold give rise to the gravity dual background which is a
non-K\"ahler deformed conifold with three-form fluxes, much in the lines of \eqref{heidi}, \eqref{goeskalu} and
\eqref{ashanti} and is given by box (b) in Fig. \ref{2a2b}.
The computations performed in section \ref{nf0} will be based on this configuration, albeit with a black hole that will signify non-zero temperature, but with no flavors.

A mirror transformation, {\it a la} Strominger-Yau-Zaslow \cite{syz}, on both the type IIB boxes of Fig. \ref{2a2b} will produce the IR type IIA background whose gravity dual configuration involves a non-K\"ahler resolved conifold with fluxes, as shown in box (d) in the figure.
The M-theory uplift of this is given in the top right-hand box of Fig. \ref{2a2b}, i.e box (e), which is a seven-dimensional $G_2$ structure manifold with G-fluxes \cite{MQGP}. Our computations in section \ref{nfn0} will be based on this specific M-theory manifold albeit, again, with non-zero temperatures but now including non-zero flavors. Interestingly for the spectral function computation of section \ref{spectral}, we shall resort back to the type IIA picture.

Let us now come to the UV completion of these models that we alluded to earlier. In the type IIB side, this was first discussed in \cite{metrics}, but a full elaborations on the actual ingredients that constitute the UV degrees of freedom were given in \cite{3reg} and \cite{chenchen} and were named Regions 3 and 2 in
\cite{3reg}.  We expect the UV theory to be a strongly coupled conformal field theory, as this would be the closest to being asymptotically free. The reason for choosing a CFT $-$ and not an asymptotically free theory $-$ as the UV theory is because we require strong 't~Hooft coupling to allow for a gravity dual. In fact, a gravity dual description only exists if the corresponding gauge theory is strongly coupled at all scales, i.e strongly coupled from UV to IR. For large but finite number of colors, this means
that the requirement for asymptotic freedom is  {\it not} quite compatible with the existence of a gravity dual.
Therefore the closest we can come to asymptotic freedom is to allow for a CFT in the UV. In the limit of infinite number of  colors, the 't~Hooft coupling can be very large, yet the YM coupling can be made arbitrarily small.

One specific choice of a UV group that could lead to a CFT is SU$(N+M) \times SU(N + M)$, where we have introduced an extra parameter of $N$. In the present context, the choice of $N$ has a special meaning. In the type IIB theory, $N$ signifies the number of D3-branes whereas $M$ is the usual wrapped D5-branes. The two-cycle on which the D5-branes wrap, i.e the two-cycle parametrized by ($\theta_2, \phi_2$), should now be of vanishing size to preserve supersymmetry.
In the blue box of Fig. \ref{allregions}, we have denoted the UV group ${\rm SU}(N+M) \times {\rm SU}(N+M)$ that is shown to get Higgsed to a smaller group ${\rm SU}(N+M) \times {\rm SU}(N)$ at a certain IR scale. This is followed by a series of
cascading RG flows that eventually takes us to the confining gauge group SU$(M)$ at the far IR.

The complete RG flow that is depicted in the blue box of Fig. \ref{allregions} can be described rather succinctly from both type IIB as well as the type IIA theories. This will also answer all the questions that we put aside earlier. From the type IIB side, the UV CFT may be easily described by allowing additional $M$ anti-D5-branes distributed on the northern hemisphere of the resolved sphere parametrized by ($\theta_1, \phi_1$). These anti-D5-branes are stabilized against collapse by using fluxes, details of which have appeared in \cite{3reg, rgmaxim}. The string connecting the branes and the anti-branes are heavy, and they are integrated out at low energies. Thus at low energies we only see the cascading ${\rm SU}(N) \times {\rm SU}(N+M)$ theory. At high energies, the anti-brane degrees of freedom are integrated in and the $M$ D5-brane and the $M$ $\overline{\rm D5}$-branes combine to give $M$ D3-brane degrees of freedom. Together with $N$ D3-branes localized at the south pole of the resolved sphere, this leads to the UV CFT described above.  Therefore the three stages of operation namely, (1) emergence of CFT, (2)  Higgsing and (3) the cascading behavior, are all described neatly from the type IIB configuration of $N$ D3, $M$ D5 and and $M$
$\overline{\rm D5}$-branes on a non-K\"ahler resolved conifold with fluxes.

\begin{figure}[h]
\centering
\begin{tabular}{c}
\includegraphics[width=\textwidth]{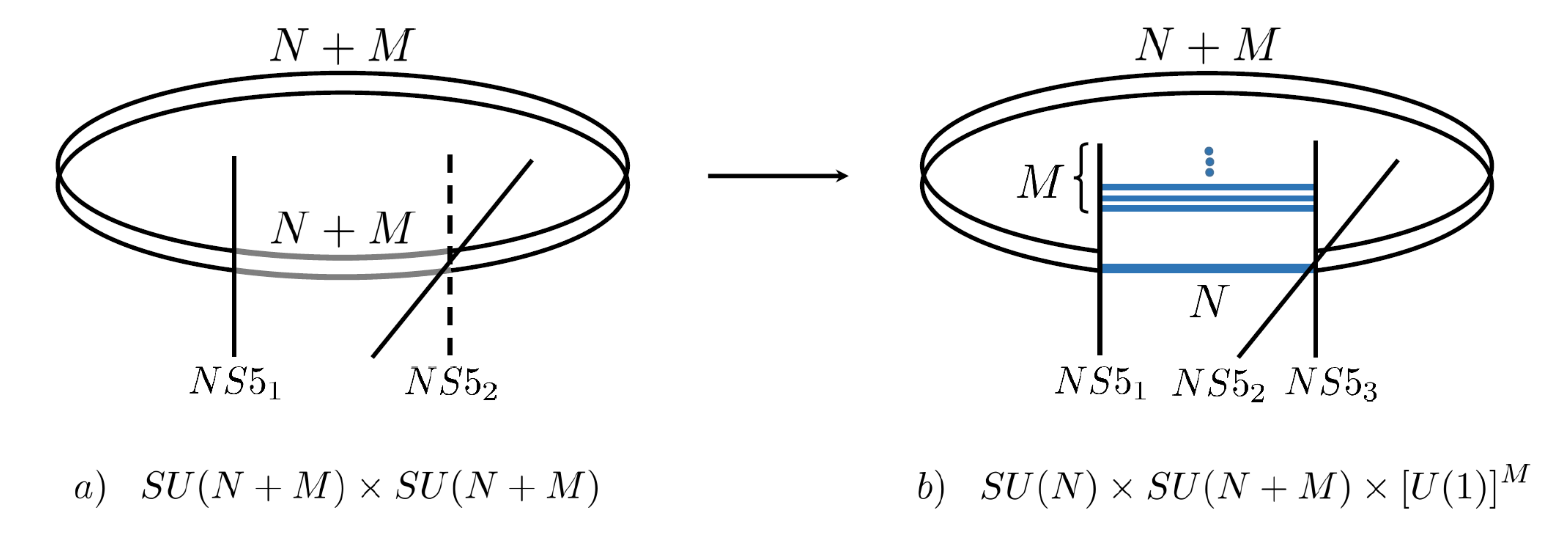}
\end{tabular}
\caption{The type IIA dual description of the UV completion as well as the Higgsing effect. Figure (a) represents a CFT whereas figure (b) shows how cascading theory can be realized.
The absence of a Coulomb branch in figure (a) indicates that one cannot move the straddling D4 branes around to Higgs the underlying
gauge theory. Also the NS5-branes are not bent, so there is no RG flow. In figure (b) and third NS5-brane is added without breaking
any supersymmetry. This is almost like the remnant of the ${\cal N}= 2$ Coulomb branch. Higgsing can now be done resulting in the bendings of the NS5-branes and triggering the RG flows.
The far IR will be a confining SU$(M)$ gauge theory.}
\label{2abranes}
\end{figure}

The correctness of our construction may also be ascertained from a T-dual type IIA configuration as shown in
Fig. \ref{2abranes}. This T-duality is a {\it single} T-duality along the $\psi$ direction and therefore should not be confused with the three T-dualities that we performed earlier to determine the mirror configuration. A single T-duality  of a conifold along $\psi$ direction, in the type IIB theory, leads to a configuration of two orthogonal NS5-branes in the type IIA side. In the presence of
$N+ M$ extra D3-branes in the type IIB side, the T-dual configuration is  shown on the left of Fig. \ref{2abranes}. The $M$ D3-branes have five-brane origins as discussed above, and so the configuration on the left of Fig. \ref{2abranes} gives us a CFT with a gauge group ${\rm SU}(N+M) \times {\rm SU}(N+M)$. The {reason} why this is a CFT comes from the fact that the NS5-branes are not {\it bent}. Clearly, any bendings of the NS5-branes would have lead to running couplings of the gauge theories on the D4-branes \cite{wittenMM}. These bendings can be achieved by having an unequal number of D4-branes on both sides of the NS5-branes. Such a feature may be achieved independently but does not seem to come naturally from the configuration on the left of Fig. \ref{2abranes}: a consequence due to the absence of Coulomb branches in  ${\cal N} = 1$ gauge theories.

However, all is not lost as these theories do have other branches, namely Baryonic, Mesonic and possible remnants of the ${\cal N} = 2$ Coulomb branches. Without going into too much details, which the readers may find in \cite{UVcom, UVcom2}, one may easily see that a branch in the moduli space arises by putting an extra NS5-brane along the dotted line in the left configuration of Fig. \ref{2abranes}. Happily, this does not break any extra supersymmetries but creates the necessary Higgsing effect that we require to jump-start the cascading process! On the right of
Fig. \ref{2abranes} we have shown how one may go from UV conformal to IR cascading behavior.  As should be obvious from Fig. \ref{2abranes}, moving the  $M$ D4-branes along parallel NS5-branes bends the NS5-branes, thus creating RG flows on the remaining D4-branes. The far IR physics is then exactly a confining SU$(M)$ gauge theory with decoupled
U$(1)$'s that we seek here. Switching on a non-zero temperature we can study the various transport coefficients.

In the gravity dual, the IR story is clear: this is given as in \eqref{heidi} and \eqref{goeskalu}. The UV degrees of freedom start appearing from Region 2 onwards as shown in \cite{3reg}, and as we go to large $r$ we are effectively in Region 3 where the three-form fluxes vanish and the background asymptotes to an $AdS_5$ space.
In this section we will use a slightly simplified form of this background and mainly concentrate on Region 1 $-$ to be at low energies $ - $ to study the
bulk viscosity at strong 't~Hooft but weak string coupling
in the absence of fundamental flavors. In the next section, we will put in the flavors and study the bulk viscosity as well as the
ratio of the bulk to shear viscosites at both strong 't~Hooft and strong string couplings, again concentrating on Region 1. For an earlier work on bulk viscosity with bottom-up approach, using two different AdS spaces at UV and IR and for a wide class of models, see \cite{benihana}. Note however that the study of bulk viscosity in \cite{benihana} differs from our study here in at least two respects. First, the model considered in \cite{benihana} has two fixed points: one at UV and the other at IR respectively. This differs from the IR confining model that we consider here. Secondly, the study of bulk viscosity in \cite{benihana} finds violation of the Buchel bound \cite{Buchel-bound}. Although this is possible in our set-up, by choosing a different lower bound for $d_1$ in \eqref{mathadulai} and \eqref{duldul}, we do not analyze such cases here.

\subsection{The type IIB dual background for large $N$ thermal QCD \label{2bbg}}

In \cite{rgmaxim} we made some preliminary study of bulk viscosity using the UV complete large $N$ thermal QCD model of \cite{metrics} with $N_f = 0$.
The metric that we took in \cite{rgmaxim} is of the form:
\bg\label{met}
ds_{10}^2 &=& e^{2A} \left[-e^{2B}dt^2+ dx^2 + dy^2 + dz^2 + e^{-4A-2B}\left({r^2 + 6a^2\over r^2 + 9a^2}\right)dr^2\right] \nonumber\\
& + & \frac{r^2 e^{-2A}}{6}(d\theta_1^2 +\sin^2~\theta_1~d\phi_1^2) + e^{-2A}\left({r^2 + 6a^2\over 6}\right)\left(d\theta_2^2 + \sin^2\theta_2 ~d\phi_2^2\right)\nonumber\\
&+&  \frac{r^2e^{-2A}}{9}\left({r^2 + 9a^2\over r^2 + 6a^2}\right)(d\psi +\cos~\theta_1~d\phi_1 +\cos~\theta_2~d\phi_2)^2.
\nd
Note that the internal space is a warped resolved conifold and not a resolved warped-deformed conifold as one would have expected from \eqref{goeskalu}. This is a simplifying assumption which helped us
to study bulk viscosity without worrying about the far IR regime of the gauge theory. Recall that the far IR regime of the gauge theory is governed by the blown-up
three-cycle of the resolved warped-deformed conifold. However since the small $r$ regime of the geometry is covered by the horizon radius $r_h$, our choice of the metric
\eqref{met} is not too far from the correct answer. The resolution parameter $a^2(r)$ is {\it not} the resolution parameter used in the brane side to control the UV behavior of the
theory. In the brane side, i.e in the gauge theory description, the $M$ D5-branes wrap the vanishing two-cycle of the resolved conifold parametrized by ($\theta_2, \phi_2$)
in a way that the D5-branes (and the
$N$ D3-branes) are at the south pole of the resolved 2-cycle, parametrized by ($\theta_1, \phi_1$), and the anti-D5-branes are distributed over the upper hemisphere of the 2-cycle.

In the language of the metric \eqref{met}, this means putting the $M$ $D5$-branes on the ($\theta_2, \phi_2$) 2-cycle has caused an asymmetry
quantified by the resolution parameter $a^2$.  From the discussion above this would mean that $a(r)^2=\mathcal{O}(\epsilon)$
and should have no terms that are zeroth order in $\epsilon$.  This can be confirmed by plugging the metric into the equations of motion.The Einstein's equations are:
\bg\label{einsto}
&& R_{\mu\nu} = -g_{\mu\nu}\left[\frac{{\cal G}_3\cdot\bar{{\cal G}}_3}{48{\rm Im}~\tau}+\frac{{\cal F}_5^2}{8\cdot5!}\right]+\frac{{\cal F}_{\mu abcd}{\cal F}_{\nu}^{abcd}}{4\cdot4!}
\nonumber\\
&& R_{mn} = -g_{mn}\left[\frac{{\cal G}_3\cdot\bar{{\cal G}}_3}{48{\rm Im}~\tau}+\frac{{\cal F}_5^2}{8\cdot5!}\right]+\frac{{\cal F}_{mabcd}{\cal F}_{n}^{abcd}}{4\cdot4!}
+\frac{{\cal G}_m^{bc}\bar{{\cal G}}_{nbc}}{4{\rm Im}~\tau}+\frac{\partial_m\tau\partial_n\bar{\tau}}{2|{\rm Im}~\tau|^2}.
\nd
here ${\cal G}_3, {\cal F}_5$ and $\tau$ are the complexified three-form flux, five-form flux and the axio-dilaton respectively as defined in \eqref{heidi} and \eqref{ashanti}.
To figure out how the wrapped D5-branes, inserted in the non-extremal system, affect the warp-factor, we can express the change as:
\bg\label{waha}
e^{-4A} = e^{-4A_0}\Bigl(1 + \epsilon P(r)\Bigr),
\nd
where $A_0 \equiv -{1\over 4} {\rm log}{L^4\over r^4}$ is the
conformal value and $\epsilon = {3g_s M^2\over 2\pi N}$ is our expansion parameter. The resolution parameter $a^2$ now may be expressed in the
following way:
\bg\label{comcon}
a(r)^2  \equiv  0 + \epsilon Q(r)
\nd
where, as we emphasized above, to zeroth order in $\epsilon$, the D5-branes wrap vanishing two-cycle. We start seeing non-zero resolution only from the first order in
$\epsilon$. The two functions,
$P(r)$ and $Q(r)$, are related via the following set of equations:
\bg\label{chukkaM}
P(r) &=& \int^r x^3 dx \left[D_1 -\int^x dy\left({15\over y^5} {d^2Q\over dy^2} - {51\over y^6} {dQ\over dy} + {72 Q(y)\over y^7} + {4\over y^5}\right)\right] + D_2 \\
&=& \frac{1}{4}\int^r dx \left[-\frac{15 x^3}{r_h^4}\left(1-\frac{r_h^4}{x^4}\right)\left(\frac{d^2Q}{dx^2}-\frac{dQ}{dx}\right)+\frac{144 Q(x)}{x^3} + \frac{2}{x}\right]dx
+ \widetilde{D}_1,\nonumber
\nd
where $D_1, D_2$ and $\widetilde{D}_1$ are constants that may be fixed from the boundary conditions. This has been discussed in details in \cite{rgmaxim}, and after the
dust settles, the functional form for $P(r)$ and $Q(r)$ can be explicitly represented in the following way:
\bg\label{sphinx3d}
&& Q(r) = \frac{r^2}{30}\left[-\ln\left(1 - {r_h^4\over r^4}\right) + \frac{r_h^2}{r^2}\ln\left(\frac{r^2-r_h^2}{r^2+r_h^2}\right)
+\frac{1}{2}\operatorname{dilog}\left(1 - {r_h^4\over r^4}\right)\right]\\
&&P(r) = \ln{r}+\frac{1}{5}\left[\ln\left(1-{r_h^4\over r^4}\right) - \frac{r_h^2}{r^2}\ln\left(\frac{r^2-r_h^2}{r^2+r_h^2}\right)-\frac{1}{8}\operatorname{dilog}\left(1-{r_h^4\over r^4}\right)\right], \nonumber
\nd
where both behave well in the limit $r \to r_h$ as one would have expected. In fact knowing the functional form for $Q(r)$ immediately tells us what the black-hole factor, $e^{2B}$,
in the metric \eqref{met} should be. This may be expressed by the following integral form:
\bg\label{dhua}
e^{2B(r, \epsilon)} = 1 + 4r_h^4 \int^r {dx\over x^3\left(x^2 + 9\epsilon Q(x)\right)} =  1 - {r_h^4\over r^4} - 36 \epsilon r_h^4 \int^r {Q dx \over x^7}, \nd
which reproduces the conformal result for vanishing $\epsilon$. Finally,
plugging \eqref{sphinx3d} to \eqref{waha} and \eqref{comcon} gives us  the ${\cal O}(\epsilon)$
corrections to the conformal values for the resolution and the warp-factors:
\begin{eqnarray}\label{fatur}
&& a(r)^2= 0 + \frac{\epsilon r^2}{30}\left[-\ln\left(1 - {r_h^4\over r^4}\right) + \frac{r_h^2}{r^2}\ln\left(\frac{r^2-r_h^2}{r^2+r_h^2}\right)
+\frac{1}{2}\operatorname{dilog}\left(1 - {r_h^4\over r^4}\right)\right] \\
&& e^{-4A} = \frac{L^4}{r^4}\left\{1 + \epsilon \left[\ln{r}+\frac{1}{5}\left(\ln\left(1 - {r_h^4\over r^4}\right) - \frac{r_h^2}{r^2}\ln\left(\frac{r^2-r_h^2}{r^2+r_h^2}\right)
-\frac{1}{8}\operatorname{dilog}\left(1 - {r_h^4\over r^4}\right)\right)\right]\right\}. \nonumber
\end{eqnarray}
The functional forms for $a^2, e^{2B}$ and $e^{-4A}$ are consistent with the general picture developed in \cite{metrics}, \cite{3reg} and \cite{boidyo}. In particular knowing
the ${\cal O}(\epsilon)$ correction to the black-hole factor is consistent with the ${\cal O}(\epsilon)$ corrections to the two black-hole factors $g_1$ and $g_2$ in \cite{metrics}. There are
also ${\cal O}(g_sN_f)$ corrections, from the $N_f$ flavors,
 that we do not consider here. This is relegated to section \ref{nfn0}.

\subsection{Details on the bulk viscosity computations from gravity dual \label{bvcomp}}

Bulk viscosity appears, in a system with a $SO(3)$ spatial symmetry, from the correlation of $T_{xx}$ at two different points in four-dimensional space-time with one point fixed at the
origin.  This means, as discussed earlier, in the gravity dual bulk viscosity may be computed from the fluctuations of the vielbeins $e_k$ with $k = 0, x$ and $r$. These fluctuations
may be divided into positive and negative frequencies, and are expressed as:
\bg\label{eileen}
\delta e^{\pm}_k = \int_{-\infty}^{+\infty} d\omega e_k(r) {\rm exp}(i\omega t)\left[\left(1 \pm i\epsilon \sum_{n = 0}^\infty p_{nk} \omega^{2n-1}\right) \Gamma_{0k}\left(r, \vert \omega\vert\right) + \epsilon \Gamma_{1k}\left(r, \vert\omega\vert\right) + {\cal O}\left(\epsilon^2\right)\right], \nonumber\\ \nd
where $\epsilon$ is the same non-conformality factor as before and $\omega$ is the frequency. The other parameters
appearing in \eqref{eileen} may be defined in the following way. The coefficients $p_{nk}$ are in general functions of $r$ as well as $\vert\omega\vert$ but not constants. With
constant $p_{nk}$, the bulk viscosity would vanish despite the existence of a complex piece in \eqref{eileen}. Note however that $\delta e_k \equiv \delta e_k(r, t)$
are all real functions of $r$ and $t$.

The coefficients $\Gamma_{0k}\left(r, \vert\omega\vert\right)$ and $\Gamma_{1k}\left(r, \vert\omega\vert\right)$ capture the essence of the bulk viscosity computations here.  In a
system with $SO(3)$ symmetry, $\Gamma_{0x}$ takes the following form:
\bg\label{_rose}
\Gamma_{0x}\left(r, \vert\omega\vert\right) = {\rm exp}\left[2 B(r, 0)\left(1 + {\vert\omega\vert^2\over 8\pi^2 T^2}\right)\right], \nd
where $B(r, 0)$ is given in \eqref{dhua} and $T$, the temperature, is proportional to $r_h$, the horizon radius.  We also expect $\Gamma_{0y} = \Gamma_{0z}$ to be
equal to $\Gamma_{0x}$. On the other hand, $\Gamma_{00}$ and $\Gamma_{0r}$ take the following form:
\bg\label{illumina}
\Gamma_{00} = \left[1 - {4B(r, 0)\over {\rm log}~\Gamma_{0x}}\right]\Gamma_{0x}, ~~~
 \Gamma_{0r} = {2[e^{-2B(r, 0)} - 1]\left[{\rm log}~\Gamma_{0x} - B(r, 0)\right] \Gamma_{0x} \over B(r, 0)}. \nd
Although \eqref{_rose} and \eqref{illumina} are related to conformal theory\footnote{In fact, as one would expect,
$\Gamma_{0x}$ and $\Gamma_{00}$ do form the non-normalizable modes at the AdS
boundary and hence couple to the required components of the energy momentum tensor of the corresponding gauge theory.},
we will use them to analyze the non-conformal regime of our model as the imaginary parts of the fluctuations
in \eqref{eileen} depend on $\Gamma_{0k}$ as well as $p_{nk}$. The latter are associated with extra sources coming from the distribution of anti-D5 branes in Regions 2 and 3. These regions have been described earlier and one may also see them in the blue box of Fig. \ref{allregions} when we approach the high energy regime. We can
quantify these sources in the following way\footnote{For details about the sources, the readers may refer
to section 4.4 of \cite{rgmaxim}.}:
\bg\label{unsaid}
\Delta_k(r, \omega) = 0 + \epsilon \left(\Delta_{1k}(r, \vert\omega\vert) + i \sum_{n = 0}^\infty \Delta^{(n)}_{2k}(r, \vert\omega\vert)\omega^{2n-1}\right) + {\cal O}(\epsilon^2), \nd
where we see that the imaginary part involves three infinite series of modes specified by the sources $\Delta^{(n)}_{20}, \Delta^{(n)}_{2x}$ and $\Delta^{(n)}_{2r}$. These modes
can also be expressed in terms of $p_{nk}$ appearing in the fluctuation \eqref{eileen}. For example $\Delta^{(n)}_{20}$ has the following expression:
\bg\label{thali}
\Delta^{(n)}_{20} & = &  p''_{n0} \Gamma_{00} + p'_{n0}\left[\left({5\over r} + 2B'_0\right)\Gamma_{00} + 2\Gamma'_{00} - A'_0 \Gamma_{00}\right]
- 3p'_{nx} A'_0\Gamma_{0x} \nonumber\\
&+& p_{n0} \left[\Gamma''_{00} + \Gamma'_{00}\left({5\over r} + 2 B_0' - A_0'\right)\right] - 3 p_{nx} \Gamma'_{0x} A'_0 - p_{nr} \Gamma'_{0r}\left(A_0' + B_0'\right)\nonumber\\
&-&  p'_{nr} \Gamma_{0r}\left(A'_0 + B'_0\right) + e^{-4(A_0 + B_0)}\big[3p_{(n-1)x} \Gamma_{0x} + p_{(n-1)r} \Gamma_{0r}\big], \nd
where $A_0$ is defined in \eqref{waha} and $B_0 \equiv B(r, 0)$ is given in \eqref{dhua}. Note that \eqref{thali}
involves five fluctuation modes, $p_{n0}, p_{nx}, p_{nr}, p_{(n-1)x}$ and $p_{(n-1)r}$; as well as the three $\Gamma_{0k}$'s defined in \eqref{_rose} and \eqref{illumina}. This means
knowing $\Delta^{(n)}_{2k}$, we will need at least five equations to solve for the fluctuations $p_{nk}$.  One may also construct the following recursion relations from \eqref{thali}:
\bg\label{illunoty}
&& \Delta^{(0)}_{20} =   p''_{00} \Gamma_{00} + p'_{00}\left[\left({5\over r} + 2B'_0 - A'_0\right)\Gamma_{00} + 2\Gamma'_{00}\right]
- 3p'_{0x} A'_0\Gamma_{0x} - p'_{0r} \Gamma_{0r}\left(A'_0 + B'_0\right)\nonumber\\
&&~~~~~~~~~ + p_{00} \left[\Gamma''_{00} + \Gamma'_{00}\left({5\over r} + 2 B_0' - A_0'\right)\right] - 3 p_{0x} \Gamma'_{0x} A'_0 - p_{0r} \Gamma'_{0r}\left(A_0' + B_0'\right)\nonumber\\
&& ~~~~~~~~~~~~~~~ ~~~ + e^{-4(A_0 + B_0)}\big[3p_{(-1)x} \Gamma_{0x} + p_{(-1)r} \Gamma_{0r}\big]\nonumber\\
&& \Delta^{(1)}_{20}  =   p''_{10} \Gamma_{00} + p'_{10}\left[\left({5\over r} + 2B'_0 - A'_0\right)\Gamma_{00}+ 2\Gamma'_{00} \right]
- 3p'_{1x} A'_0\Gamma_{0x} - p'_{1r} \Gamma_{0r}\left(A'_0 + B'_0\right) \nonumber\\
&&~~~~~~~~~ + p_{10} \left[\Gamma''_{00} + \Gamma'_{00}\left({5\over r} + 2 B_0' - A_0'\right)\right] - 3 p_{1x} \Gamma'_{0x} A'_0 - p_{1r} \Gamma'_{0r}\left(A_0' + B_0'\right)\nonumber\\
&& ~~~~~~~~~~~~~~~ ~~~ + e^{-4(A_0 + B_0)}\big[3p_{0x} \Gamma_{0x} + p_{0r} \Gamma_{0r}\big], \nd
and so on. Note that there are two types of non-derivative
terms in the first equation of \eqref{illunoty}: (a) the terms proportional to $p_{00}, p_{0x}$ and $p_{0r}$, and (b) terms proportional to $p_{(-1)x}$ and $p_{(-1)r}$. The latter have no dynamics, so maybe we could use them
to {\it cancel} the former terms in the following way:
\bg\label{cloverfield}
&& 3p_{(-1)x} \Gamma_{0x}  + p_{(-1)r} \Gamma_{0r} \equiv
-p_{00} \left[\Gamma''_{00} + \Gamma'_{00}\left({5\over r} + 2 B_0' - A_0'\right)\right] e^{4(A_0 + B_0)}    \nonumber\\
&& ~~~~~~~~~~~~~~~~ + \left[3 p_{1x} \Gamma'_{0x} A'_0 + p_{1r} \Gamma'_{0r}\left(A_0' + B_0'\right)\right]e^{4(A_0 + B_0)}. \nd
This would mean that by
knowing $p_{00}, p_{0x}$ and $p_{0r}$, one may not only build the next set of fluctuation modes from \eqref{illunoty} but also determine the functional forms for the
non-dynamical modes $p_{(-1)k}$. Unfortunately such an identification would either over-constrain the dynamics or lead to some apparent contradictions. To avoid this, we will
set:
\bg\label{adrenaline}
p_{(-1)0}  = p_{(-1)x} = p_{(-1)r} \equiv 0. \nd
In any case, identification like \eqref{cloverfield}  can never be used to cancel the non-derivative terms $p_{1k}$ with
$p_{0k}$ as both set of fluctuations are dynamical now. Thus generically we should assume the existence of $p_{(n-1)k}$ modes along with the $p_{nk}$ modes.

The next series of sources appear from $\Delta^{(n)}_{2x}$ and would follow similar strategy as above. These sources may be expressed in terms of the fluctuation modes $p_{nk}$
in the following way:
\bg\label{nor25}
\Delta^{(n)}_{2x} &= & p''_{nx} \Gamma_{0x} + p'_{nx}\left[\left({5\over r} -3A'_0 - B'_0\right)\Gamma_{0x} + 2 \Gamma'_{0x} \right] -p_{nr}\Gamma'_{0r} A'_0  \nonumber\\
&+& p_{nx} \left[\Gamma''_{0x} + \Gamma'_{0x} \left({5\over r} - 3A'_0 - B'_0\right)\right] - p_{n0} \Gamma'_{00} \left(A'_0 + B'_0\right) \nonumber\\
&-&  p'_{n0}\Gamma_{00}\left(A'_0 + B'_0\right)
- p'_{nr} A'_0 \Gamma_{0r} + e^{-4(A_0 + B_0)} p_{(n-1)x}\Gamma_{0x}, \nd
where this time four, instead of five, modes $p_{nx}, p_{n0}, p_{nr}$ and $p_{(n-1)x}$ are needed. As before, the zeroth and the first order recursion relations may be written as:
\bg\label{indigo}
\Delta^{(0)}_{2x} &=&  p''_{0x} \Gamma_{0x} + p'_{0x}\left[\left({5\over r} -3A'_0 - B'_0\right)\Gamma_{0x} + 2 \Gamma'_{0x} - \right]
- p'_{00}\Gamma_{00}\left(A'_0 + B'_0\right)
- p'_{0r} A'_0 \Gamma_{0r}\nonumber\\
&+& p_{0x} \left[\Gamma''_{0x} + \Gamma'_{0x} \left({5\over r} - 3A'_0 - B'_0\right)\right] - p_{00} \Gamma'_{00} \left(A'_0 + B'_0\right) -p_{0r}\Gamma'_{0r} A'_0  \nonumber\\
\Delta^{(1)}_{2x} &=&  p''_{1x} \Gamma_{0x} + p'_{1x}\left[\left({5\over r} -3A'_0 - B'_0\right)\Gamma_{0x} + 2 \Gamma'_{0x}\right]
- p'_{10}\Gamma_{00}\left(A'_0 + B'_0\right) - p'_{1r} A'_0 \Gamma_{0r} \nonumber\\
&+& p_{1x} \left[\Gamma''_{0x} + \Gamma'_{0x} \left({5\over r} - 3A'_0 - B'_0\right)\right] - p_{10} \Gamma'_{00} \left(A'_0 + B'_0\right) -p_{1r}\Gamma'_{0r} A'_0  \nonumber\\
&& ~~~~~~~~~~~~~~~~+ e^{-4(A_0 + B_0)} p_{0x}\Gamma_{0x}, \nd
where this time an input of $p_{0x}$ is needed to build the first order fluctuation equation. In a similar vein, we now construct the third series of sources associated with
$\Delta^{(n)}_{2r}$ in the following way:
\bg\label{sphinx}
\Delta^{(n)}_{2r} & = & p''_{n0} \Gamma_{00} + 3 p''_{nx}\Gamma_{0x} +  p'_{n0}\left[2\Gamma'_{00} + \Gamma_{00} \left(A'_0 + 2B'_0\right)\right]
+ p'_{nx} \left(6\Gamma'_{0x} + 3 A'_0\Gamma_{0x}\right) \nonumber\\
&+& p_{n0}\left[\Gamma''_{00} + \Gamma'_{00}\left(A'_0 + 2 B'_0\right)\right] + 3 p_{nx}\left(\Gamma''_{0x} + A'_0 \Gamma'_{0x}\right)
- p_{nr} \Gamma'_{0r}\left({5\over r} + B'_0 - A'_0\right)\nonumber\\
&-& p'_{nr}\left[\left({5\over r} + 2B'_0\right)\Gamma_{0r} - \Gamma_{0r}\left(A'_0 + B'_0\right)\right]
+ e^{-4(A_0 + B_0)} p_{(n-1)r} \Gamma_{0r}, \nd
with the input of four fluctuation modes $p_{nx}, p_{n0}, p_{nr}$ and $p_{(n-1)r}$ governing the dynamics. The recursion relations for the zeroth and the first order can be easily
expressed in terms of the fluctuation modes as:
\bg\label{teripolo}
\Delta^{(0)}_{2r} & = & p''_{00} \Gamma_{00} + 3 p''_{0x}\Gamma_{0x} +  p'_{00}\left[2\Gamma'_{00} + \Gamma_{00} \left(A'_0 + 2B'_0\right)\right]
+ p'_{0x} \left(6\Gamma'_{0x} + 3 A'_0 \Gamma_{0x}\right) \nonumber\\
&+& p_{00}\left[\Gamma''_{00} + \Gamma'_{00}\left(A'_0 + 2 B'_0\right)\right] + 3 p_{0x}\left(\Gamma''_{0x} + A'_0 \Gamma'_{0x}\right)
- p_{0r} \Gamma'_{0r}\left({5\over r} + B'_0 - A'_0\right)\nonumber\\
&-& p'_{0r}\left[\left({5\over r} + 2B'_0\right)\Gamma_{0r} - \Gamma_{0r}\left(A'_0 + B'_0\right)\right] \nonumber\\
\Delta^{(1)}_{2r} & = & p''_{10} \Gamma_{00} + 3 p''_{1x}\Gamma_{0x} +  p'_{10}\left[2\Gamma'_{00} + \Gamma_{00} \left(A'_0 + 2B'_0\right)\right]
+ p'_{1x} \left(6\Gamma'_{0x} + 3 A'_0 \Gamma_{0x}\right) \nonumber\\
&+& p_{10}\left[\Gamma''_{00} + \Gamma'_{00}\left(A'_0 + 2 B'_0\right)\right] + 3 p_{1x}\left(\Gamma''_{0x} + A'_0 \Gamma'_{0x}\right)
- p_{1r} \Gamma'_{0r}\left({5\over r} + B'_0 - A'_0\right)\nonumber\\
&-& p'_{1r}\left[\left({5\over r} + 2B'_0\right)\Gamma_{0r} - \Gamma_{0r}\left(A'_0 + B'_0\right)\right]
+ e^{-4(A_0 + B_0)} p_{0r} \Gamma_{0r}. \nd
At this stage let us ask whether the above three set of equations, \eqref{thali}, \eqref{nor25} and \eqref{sphinx},
are enough to determine the five unknown functions\footnote{Note that $p_{(n-1)0}$ do not appear in any of the equations.},
$p_{n0}, p_{nx}, p_{nr}, p_{(n-1)x}$ and $p_{(n-1)r}$. It would seem we need at least two more equations. However a careful look tells us that
the first equations in each of the three recursion
series, \eqref{illunoty}, \eqref{indigo} and \eqref{teripolo}, are enough to determine the three functions $p_{00}, p_{0x}$ and $p_{0r}$ provided the sources $\Delta^{(0)}_{2k}$
and the boundary
conditions are adequately specified. Similar arguments apply for the next three functions, $p_{10}, p_{1x}$ and $p_{1r}$: once we specify the sources $\Delta^{(1)}_{2k}$
and the boundary conditions, this would in principle fix the functional forms
for all $p_{1k}$. Thus it seems that the above three equations
\eqref{thali}, \eqref{nor25} and \eqref{sphinx} should suffice.

For the present case we will work out the equation satisfied by $p_{0x}$ as this is the {\it only} component relevant for bulk viscosity. This will be explained soon (see also
\cite{rgmaxim}). In fact what is required is not $p_{0x}$, rather $p'_{0x}$, and therefore we will work out the equation for $Y_x(r, \vert\omega\vert) \equiv p'_{0x}$.
In the process we will also see how to write the equations for $p_{00}$ and $p_{0r}$.
To start, let us define
a few variables $f_i, g_i$ and $h_i$ using which the zeroth order equations in \eqref{illunoty}, \eqref{indigo} and \eqref{teripolo} may be re-expressed in the following way:
\bg\label{TUTa}
&&p''_{00} f_1  + p'_{00} f_2 + p'_{0x} f_3 + p'_{0r} f_4 = \Delta^{(0)}_{20} + \sum_{k = 0}^2 p_{0k} f_{5+k} \nonumber\\
&&p'_{00} g_1  + p''_{0x} g_2 + p'_{0x} g_3 + p'_{0r} g_4 = \Delta^{(0)}_{2x} + \sum_{k = 0}^2 p_{0k} g_{5+k} \nonumber\\
&&p''_{00} h_1  + p'_{00} h_2 + p''_{0x} h_3 + p'_{0x} h_4 + p'_{0r} h_5 = \Delta^{(0)}_{2r} + \sum_{k = 0}^2 p_{0k} h_{6+k} \nd
where we have identified $p_{01} \equiv p_{0x}$ and $p_{02} \equiv p_{0r}$ to avoid clutter.
One may define similar equations for the first order fluctuation equations, namely for the $f_{1k}$ using the recursion relations. The various coefficients appearing in
\eqref{TUTa} may be written as:
\bg\label{pikkola}
&&f_1 =  \Gamma_{00}, ~~ f_2 =  \left({5\over r} + 2B'_0 - A'_0\right) \Gamma_{00} + 2\Gamma'_{00},  ~~ f_4 =  - \Gamma_{0r}\left(A'_0 + B'_0\right)\nonumber\\
&& h_4 =  6\Gamma'_{0x} + 3 A'_0 \Gamma_{0x}, ~~~h_5 = - \left({5\over r} + B'_0 - A'_0\right)\Gamma_{0r}, ~~~ f_3 = - 3A'_0\Gamma_{0x}\\
&& g_1 = - \Gamma_{00}\left(A'_0 + B'_0\right), ~~~
g_2 =  \Gamma_{0x}, ~~~ g_3 = \left({5\over r} - B'_0 -3A'_0\right) \Gamma_{0x} + 2 \Gamma'_{0x} \nonumber\\
&& g_4 = -  A'_0 \Gamma_{0r}, ~~~ h_1 =
\Gamma_{00}, ~~~ h_3 = 3 \Gamma_{0x}, ~~~ h_2 = 2\Gamma'_{00} + \Gamma_{00} \left(A'_0 + 2B'_0\right), \nonumber\\
&& g_5 = \Gamma'_{00} \left(A'_0 + B'_0\right), ~~~~ g_6 = -\Gamma''_{0x} - \Gamma'_{0x} \left({5\over r} - 3A'_0 - B'_0\right), ~~~~g_7 = A'_0 \Gamma'_{0r}\nonumber\\
&& f_5 = -\Gamma''_{00} - \Gamma'_{00} \left({5\over r} + 2B'_0 - A'_0\right), ~~~~ f_6 = 3 A'_0 \Gamma'_{0x}, ~~~~ f_7 = \left(A'_0 + B'_0\right) \Gamma'_{0r}\nonumber\\
&& h_6 = -\Gamma''_{00} - \Gamma'_{00} \left(2B'_0 + A'_0\right), ~~ h_7 = -3\left(\Gamma''_{0x} + A'_0 \Gamma'_{0x}\right), ~~
h_8 = \Gamma'_{0r} \left({5\over r} + B'_0 - A'_0\right), \nonumber\nd
where $\Gamma_{0k}$ have been defined in \eqref{_rose} and \eqref{illumina}; and $A_0$ and $B_0$ are defined in \eqref{waha} and \eqref{dhua} respectively. It is
interesting to note that the LHS of the equations in \eqref{TUTa}, i.e the coefficients of $p''_{0k}$ and $p'_{0k}$, are mostly functions of $\Gamma_{0k}$ whereas the RHS of
\eqref{TUTa}, i.e the coefficients of $p_{0k}$, are {\it all} functions of the derivatives of $\Gamma_{0k}$.

The set of equations \eqref{TUTa} are highly non-linear and solving them will in general be a non-trivial exercise. Therefore it might be instructive to first solve a slightly
simpler system than \eqref{TUTa} to gain some familiarity with the solutions and then proceed to address the full set of equations.  In the following subsection we analyze a
simpler case, and in the next subsection we will study the full system.

\subsubsection{A toy example in full details \label{amandas}}

To study a toy example from \eqref{TUTa}, the first question is:
how can we simplify the set of equations in
\eqref{TUTa}? This is where the observation that we made above could become useful, namely, we can assume that the derivatives of $\Gamma_{0k}$ are much smaller than
$\Gamma_{0k}$ at some $r > > r_h$. This would immediately imply:
\bg\label{tvader}
f_{5+k} =  g_{5+k} = h_{6 + k} ~ \approx ~ 0, \nd
making the RHS of all the equations in \eqref{TUTa} to only depend on the sources $\Delta^{(0)}_{2k}$. Note that \eqref{tvader} {\it does not} imply absorbing the $p_{0k}$
terms in the definition of the sources because the sources are independent of the bulk fluctuations. Nor does this imply  invoking relations like \eqref{cloverfield}, since such a
procedure is generically prone to errors. Thus \eqref{tvader} would be the only way to simplify \eqref{TUTa}.

With this in mind the next set of procedures may be elaborated in the following way.
Using
\eqref{pikkola}, let us define another set of functions as:
\bg\label{reileen}
&&F_1(r) = {\rm exp}\left[\int^r \left({f_2 g_4 - g_1 f_4 \over f_1 g_4}\right) dx\right], ~~~~~~ F_2(r) = {\rm exp}\left[\int^r \left({f_3 g_4 - g_3 f_4 \over f_4 g_2}\right) dx\right]\nonumber\\
&&F_3(r) = {\rm exp}\left[\int^r \left({h_2 g_4 - g_1 h_5 \over h_1 g_4}\right) dx\right], ~~ F_4(r) = {\rm exp}\left[-\int^r \left({h_4 g_4 - g_3 h_5 \over h_3 g_4 - g_2 h_5}\right) dx\right],
\nonumber\\ \nd
which will help us to avoid cluttering of formulae later when we write the equations for the fluctuations $p_{nk}$. Note that these functions are all expressed in terms of
certain definite integrals (the lower bounds of these integrals could be $r_h$ or $r = 0$, but these details will be irrelevant). There are also four other functions that are not expressed
in terms of integrals. They may be expressed as:
\bg\label{edenlac}
&&G_1(r) = {g_2 f_4\over f_1 g_4}, ~~~~~~~ G_3(r) = {h_3 g_4 - g_2 h_5\over h_1 g_4} \nonumber\\
&& G_2(r, \vert\omega\vert) = {\Delta^{(0)}_{20} g_4 - \Delta^{(0)}_{2x} f_4 \over f_1 g_4}, ~~~~
G_4(r, \vert\omega\vert) = {\Delta^{(0)}_{2r} g_4 - \Delta^{(0)}_{2x} h_5 \over h_1 g_4}, \nd
where $\Delta^{(0)}_{2k}$ are the zeroth order sources that appear in \eqref{TUTa}. Note that $G_2$ and $G_4$ are functions of $r$ as well as $\vert\omega\vert$ because they
depend on the sources $\Delta^{(0)}_{2k}(r, \vert\omega\vert)$. Therefore with \eqref{pikkola}, \eqref{reileen} and \eqref{edenlac}, we are ready to write the equation governing
the fluctuation $Y_x(r, \vert\omega\vert) \equiv p'_{0x}$ as:
\bg\label{skawaii}
a_{11} {d^2Y_x\over dr^2} + a_{21} {dY_x\over dr} + {a}_{31} Y_x = a_{41}, \nd
 which is a second order differential equation and therefore would require boundary conditions, both at the cut-off $r = r_c$ as well as at the horizon radius $r = r_h$, to determine
 the functional behavior precisely. The coefficients $a_{I1}$ appearing in \eqref{skawaii} are non-trivial functions of $F_i$ and $G_i$ variables, defined in \eqref{reileen} and
 \eqref{edenlac}, and can be written as:
 \bg\label{mado}
&& a_{11} \equiv {\left(G_3 - G_1\right)F_3 \over \left(F_3/F_1\right)'}, ~~~~ a_{41} \equiv {d\over dr}\left[{(G_4 - G_2)F_3 \over \left(F_3/F_1\right)'}\right] -G_2 F_1, ~~~~
k \equiv \left({F_3\over F_1}\right)'\nonumber\\
&&a_{21} \equiv  {1\over k}\left[ 2{d\over dr}\left(G_3F_3 - G_1F_3\right) + {\left(G_3 - G_1\right)F_3} {d^2\over dr^2}\left({F_3\over F_1}\right)\right]
- {1\over k F_4} {d\over dr} \left(G_3 F_3 F_4\right) \nonumber\\
&& ~~~~~~~~ + {F_3\over k F_1 F_2}
{d\over dr}\left(G_1 F_1 F_2\right), \nonumber\\
&& a_{31} \equiv {d\over dr}\left[{1\over k}{d\over dr}\left(G_3 F_3  -G_1 F_3\right) - {1\over k F_4} {d\over dr}\left(G_3 F_3 F_4\right)
+ {F_3\over k F_1 F_2} {d\over dr}\left(G_1 F_1 F_2\right)\right] \nonumber\\
&& ~~~~~~~~~ + {1\over F_2} {d\over dr} \left(G_1 F_1 F_2\right), \nd
where $a_{41} = a_{41}(r, \vert\omega\vert)$ and all other $a_{I1}$ are functions of $r$. This implies $Y_x = Y_x(r, \vert\omega\vert)$ as expected. Solving \eqref{skawaii} would
provide the fluctuation mode $p_{0x}$. Once $p_{0x}$ is known, we can use it to determine the next fluctuation mode, $p_{00}$. Let us now define
$p_{00} \equiv Y_0(r, \vert\omega\vert)$, instead of $p'_{00}$, and write the equation for $Y_0$ in the following way:
\bg\label{lateshow}
{dY_0\over dr} & = & {1\over F_1(r)}\int^r dx G_1(x) F_2(x) F_1(x) {d\over dx}\left[{Y_x(x, \vert\omega\vert)\over F_2(x)}\right]
+ {1\over F_1(r)}\int^r dx F_1(x) G_2(x, \vert\omega\vert) \nonumber\\
& = & {1\over F_3(r)}\int^r dx G_3(x) F_4(x) F_3(x) {d\over dx}\left[{Y_x(x, \vert\omega\vert)\over F_4(x)}\right]
+ {1\over F_3(r)}\int^r dx F_3(x) G_4(x, \vert\omega\vert),  \nonumber\\ \nd
 where one may use either of the two set of expressions on the RHS of \eqref{lateshow} to solve for $Y_0$. The equality between the two expressions can be argued easily
 from \eqref{TUTa}. Finally, knowing $Y_x$ and $Y_0$, one may use any of the three equations in \eqref{TUTa} to solve for $Y_r(r, \vert\omega\vert) \equiv p'_{0r}$.

 Let us now work out the first order fluctuations for our case invoking \eqref{tvader}.
 Again we expect  three set of fluctuations of the form $p_{10}, p_{1x}$ and $p_{1r}$, similar to the three set  of
 fluctuations $p_{00}, p_{0x}$ and $p_{0r}$ respectively for the zeroth order case. The equations satisfied by the first order fluctuations are a slight variations of \eqref{TUTa}, namely:
 \bg\label{TUTb}
&&p'_{10} g_1  + p''_{1x} g_2 + p'_{1x} g_3 + p'_{1r} g_4 = \Delta^{(1)}_{2x}  - e^{-4(A_0 + B_0)}p_{0x} \Gamma_{0x} \nonumber\\
&&p''_{10} h_1  + p'_{10} h_2 + p''_{1x} h_3 + p'_{1x} h_4 + p'_{1r} h_5 = \Delta^{(1)}_{2r} - e^{-4(A_0 + B_0)}p_{0r} \Gamma_{0r} \nonumber\\
&&p''_{10} f_1  + p'_{10} f_2 + p'_{1x} f_3 + p'_{1r} f_4 = \Delta^{(1)}_{20} - e^{-4(A_0 + B_0)}\left(3p_{0x} \Gamma_{0x} + p_{0r} \Gamma_{0r}\right),  \nd
 where $f_i, g_i$ and $h_i$ are exactly the ones appearing in \eqref{pikkola}; $\Gamma_{0k}$ are as in \eqref{_rose} and \eqref{illumina}; and $A_0$ and $B_0$ are the zeroth
 order values in \eqref{waha} and \eqref{dhua} respectively. However not everything remain the same: the RHS of the equations \eqref{TUTb} have two kind of sources, (a) the
 first order sources $\Delta^{(1)}_{2k}$, and (b) sources appearing from the zeroth order in fluctuations, $p_{0x}$ and $p_{0r}$. These changes in sources imply that
  $G_4(r, \vert\omega\vert)$ and $G_2(r, \vert\omega\vert)$ in \eqref{edenlac} may be replaced by:
\bg\label{dunkirk}
{\widetilde G}_4(r, \vert\omega\vert) & = & {\widetilde{\Delta}^{(1)}_{2r} g_4 - \widetilde{\Delta}^{(1)}_{2x} h_5 \over h_1 g_4} \nonumber\\
&\equiv &
 {\left(\Delta^{(1)}_{2r} - e^{-4(A_0 + B_0)}p_{0r} \Gamma_{0r}\right)g_4 - \left(\Delta^{(1)}_{2x}  - e^{-4(A_0 + B_0)}p_{0x} \Gamma_{0x}\right)h_5 \over h_1 g_4}\\
 {\widetilde G}_2(r, \vert\omega\vert) & = & {\widetilde{\Delta}^{(1)}_{20} g_4 - \widetilde{\Delta}^{(1)}_{2x} f_4 \over f_1 g_4} \nonumber\\
  &\equiv & {\left(\Delta^{(1)}_{20} - e^{-4(A_0 + B_0)}\left(3p_{0x} \Gamma_{0x} + p_{0r} \Gamma_{0r}\right)\right)g_4
 - \left(\Delta^{(1)}_{2x}  - e^{-4(A_0 + B_0)}p_{0x} \Gamma_{0x}\right) f_4 \over f_1 g_4}, \nonumber \nd
 respectively and not naively by replacing $\Delta^{(0)}_{2k}$ with $\Delta^{(1)}_{2k}$ in $G_2$ and $G_4$. Note that  there are
 no additional changes to $G_1(r)$ and $G_3(r)$ in \eqref{edenlac}. The above observation immediately tells us that the equation satisfied by
 $p'_{1x} \equiv Y_{1x}(r, \vert\omega\vert)$ should be:
 \bg\label{colonia}
 a_{11} {d^2Y_{1x}\over dr^2} + a_{21} {dY_{1x}\over dr} + a_{31} Y_{1x} = \widetilde{a}_{41}, \nd
where we see that the coefficients appearing in the LHS of \eqref{colonia} are the same as the ones appearing in \eqref{skawaii}
 with $a_{11}, a_{21}$ and $a_{31}$ as given in \eqref{mado}. The only difference from \eqref{skawaii} is the replacement of $a_{41}$ by $\widetilde{a}_{41}$, where:
 \bg\label{emmaw}
  \widetilde{a}_{41} \equiv {d\over dr}\left[{(\widetilde{G}_4 - \widetilde{G}_2)F_3 \over \left(F_3/F_1\right)'}\right] - \widetilde{G}_2 F_1. \nd
  Similarly the equation for $p'_{10} \equiv Y_{10}(r, \vert\omega\vert)$ will be similar to \eqref{lateshow} with the replacement of $Y_x$ by $Y_{1x}$ and $G_2$ and $G_4$ by
  $\widetilde{G}_2$ and $\widetilde{G}_4$ respectively. Once we know $Y_{1x}$ and $Y_{10}$, we can use \eqref{TUTb} to determine the equation for $Y_{1r}$. This way the
  first order fluctuations may be completely determined.

 The picture is now clear for the generic order fluctuations. If we want to study the $n$-th order fluctuations $Y_{nx}, Y_{n0}$ and $Y_{nr}$, all we need is to rewrite the sources,
 $\Delta^{(n)}_{2x}, \Delta^{(n)}_{20}$ and $\Delta^{(n)}_{2r}$ by adding the fluctuations $Y_{(n-1)x}$ and $Y_{(n-1)r}$  exactly in a way elaborated in \eqref{dunkirk}, i.e:
 \bg\label{watson}
 && \widetilde{\Delta}^{(n)}_{2r} \equiv \Delta^{(n)}_{2r} - e^{-4(A_0 + B_0)} \Gamma_{0r}\int^r Y_{(n-1)r}(y, \vert\omega\vert) dy \\
 && \widetilde{\Delta}^{(n)}_{2x} \equiv \Delta^{(n)}_{2x} - e^{-4(A_0 + B_0)} \Gamma_{0x}\int^r Y_{(n-1)x}(y, \vert\omega\vert) dy\nonumber\\
  && \widetilde{\Delta}^{(n)}_{20} \equiv  \Delta^{(n)}_{20} - e^{-4(A_0 + B_0)}\int^r\left[3Y_{(n-1)x}(y, \vert\omega\vert) \Gamma_{0x}(r)
  + Y_{(n-1)r}(y, \vert\omega\vert) \Gamma_{0r}(r)\right] dy. \nonumber \nd
  Once these sources are specified  we can construct $\widetilde{G}_4$ using
  $\widetilde{\Delta}^{(n)}_{2r}$ and $\widetilde{\Delta}^{(n)}_{2x}$; and $\widetilde{G}_2$ using $\widetilde{\Delta}^{(n)}_{20}$ and
  $\widetilde{\Delta}^{(n)}_{2r}$ using the definitions in \eqref{dunkirk}; and finally $\widetilde{a}_4$ using \eqref{emmaw}. The equations for  $Y_{nx}, Y_{n0}$ and $Y_{nr}$
  would then follow the steps outlined above.

\subsubsection{Towards exact solutions for the fluctuations \label{galgad}}

  To study exact solutions for the system of equations in \eqref{TUTa}, one way would be to eliminate the $p_{0k}$ pieces on the RHS by rearranging the set of equations there.
  However a slightly simpler approach is to keep the RHS only as a function of $p_{0x}$ and eliminate the others. This leads to the following set of equations:
  \bg\label{atlantis}
&& k_{[1} l_{6]}  p''_{00} + k_{[2} l_{6]} p'_{00} + k_{[3} l_{6]} p''_{0x} + k_{[4} l_{6]} p'_{0x} + k_{[5} l_{6]} p'_{0r} = {\bf \Delta}_1\left(r, \vert\omega\vert; p_{0x}\right) \nonumber\\
&& l_{[1} m_{6]}  p''_{00} + l_{[2} m_{6]} p'_{00} + l_{[3} m_{6]} p''_{0x} + l_{[4} m_{6]} p'_{0x} + l_{[5} m_{6]} p'_{0r} = {\bf \Delta}_3\left(r, \vert\omega\vert; p_{0x}\right) \\
&& k_{[1} m_{6]}  p''_{00} + k_{[2} m_{6]} p'_{00} + k_{[3} m_{6]} p''_{0x} + k_{[4} m_{6]} p'_{0x} + k_{[5} m_{6]} p'_{0r} = {\bf \Delta}_2\left(r, \vert\omega\vert; p_{0x}\right), \nonumber \nd
 where, as mentioned above, we kept the RHS as functions of $p_{0x}$ only. The set of equations \eqref{atlantis} are in some sense more symmetrical than the earlier
 set of equations \eqref{TUTa}. The coefficients are expressed in terms of brackets which may be defined as:
 \bg\label{desert}
 k_{[a}l_{b]} \equiv k_a l_b - k_b l_a. \nd
 This formalism has some distinct advantages that will be clear soon. Note also that, in \eqref{atlantis}, there are no second derivatives of $p_{0r}$ which in turn will help us to
 rearrange the set of equations further. But before we do so, let us define the coefficients appearing in \eqref{atlantis}. The $k_i$ are defined in the following way:
 \bg\label{hearts}
  && k_1 = f_1 g_7, ~~~~ k_2 = f_2 g_7 - f_7 g_1, ~~~~ k_3 = - f_7 g_2\nonumber\\
 && k_4 = f_{[3} g_{7]}, ~~~~ k_5 = f_{[4} g_{7]}, ~~~~ k_6 = f_{[5} g_{7]}, ~~~~ k_7 = f_{[6} g_{7]},  \nd
 where $k_6$ and $k_7$ will be used to describe the sources ${\bf\Delta}_1$ and ${\bf\Delta}_2$
 in \eqref{atlantis} below.  All the $k_i$ are in turn constructed out of the ($f_i, g_k$) coefficients
 defined earlier in \eqref{pikkola}. In a similar vein, the $l_i$ coefficients are defined as:
 \bg\label{redh}
 && l_1 = h_1 g_7, ~~~~ l_2 = h_2 g_7 - g_1 h_8, ~~~ l_3 = h_3 g_7 - g_2 h_8 \\
 && l_4 = h_4 g_7 - g_3 h_8, ~~~ l_5 = h_5 g_7 - g_4 h_8, ~~~ l_6 = h_6 g_7 - g_5 h_8, ~~~ l_7 = h_7 g_7 - g_6 h_8, \nonumber \nd
  where the ($h_i, g_k$) coefficients, used here to define $l_i$, are given in \eqref{pikkola}. As before, the ($l_6, l_7$) coefficients will be used below to describe the sources
  ${\bf\Delta}_1$ and ${\bf\Delta}_3$. Finally the $m_i$ coefficients may be defined in the following way:
  \bg\label{pumpp}
  && m_1 = h_8 f_1 - h_1 f_7, ~~~~ m_2 = f_2 h_8 - h_2 f_7, ~~~ m_3 = - h_3 f_7 \\
 && m_4 = h_8 f_3 - f_7 h_4, ~~~ m_5 = h_8 f_4 - f_7 h_5, ~~~ m_6 = h_8 f_5 - f_7 h_6, ~~~ m_7 = h_8 f_6 - f_7 h_7, \nonumber \nd
where again the ($h_i, f_k$) coefficients are given in \eqref{pikkola}, and $m_6$ and $m_7$ will be used to describe the sources ${\bf\Delta}_2$ and ${\bf\Delta}_3$. The sources
${\bf\Delta}_k$   may now be expressed as:
\bg\label{julias}
&& {\bf\Delta}_1(r, \vert\omega\vert; p_{0x}) \equiv \left(\Delta^{(0)}_{20} g_7 - \Delta^{(0)}_{2x} f_7\right) l_6 - \left(\Delta^{(0)}_{2r} g_7 - \Delta^{(0)}_{2x} h_8\right) k_6
+ p_{0x}k_{[7}l_{6]}\\
&& {\bf\Delta}_2(r, \vert\omega\vert; p_{0x}) \equiv \left(\Delta^{(0)}_{20} g_7 - \Delta^{(0)}_{2x} f_7\right) m_6 - \left(\Delta^{(0)}_{20} h_8 - \Delta^{(0)}_{2r} f_7\right) k_6
+ p_{0x}k_{[7}m_{6]}\nonumber\\
&&{\bf\Delta}_3(r, \vert\omega\vert; p_{0x}) \equiv \left(\Delta^{(0)}_{2r} g_7 - \Delta^{(0)}_{2x} h_8\right) m_6 - \left(\Delta^{(0)}_{20} h_8 - \Delta^{(0)}_{2r} f_7\right) l_6 + p_{0x}l_{[7}m_{6]}.
\nonumber \nd
The new sources are combinations of the original sources $\Delta^{(0)}_{2k}$, the coefficients defined in \eqref{hearts}, \eqref{redh}, \eqref{pumpp} and \eqref{pikkola}; and $p_{0x}$. These
equation explicitly take us away from the simplifying assumption \eqref{tvader}, and so are only valid when no approximations are made\footnote{In fact both sides of all the equations
in \eqref{atlantis} would vanish in the limit \eqref{tvader}.}.  Additionally, the dependence of all the sources only on $p_{0x}$ means that any further rearrangements of the sources
will not have new dependences on other fluctuation modes. This means one may eliminate $p'_{0r}$ pieces from \eqref{atlantis} to simplify them further in the following way:
\bg\label{ekkuk}
&&\beta_1 p''_{00} + \beta_2 p'_{00} + \beta_3 p''_{0x} + \beta_4 p'_{0x} = {\bf \Delta}_{[2, 3]}\nonumber\\
&&\alpha_1 p''_{00} + \alpha_2 p'_{00} + \alpha_3 p''_{0x} + \alpha_4 p'_{0x} = {\bf \Delta}_{[1, 2]}, \nd
which mix the sources ${\bf\Delta}_1$ and ${\bf\Delta}_2$ as well as ${\bf\Delta}_2$ and ${\bf\Delta}_3$. We could also write another equation, parametrized by $\gamma_i$ coefficients,
that mix the sources ${\bf\Delta}_1$ and ${\bf\Delta}_3$, but that won't be necessary for us. The new sources may be expressed in the following way:
\bg\label{casinv}
{\bf \Delta}_{[1, 2]} \equiv {\bf \Delta}_1 k_{[5} m_{6]} - {\bf \Delta}_2 k_{[5} l_{6]}, ~~~~
{\bf \Delta}_{[2, 3]} \equiv {\bf \Delta}_2 l_{[5} m_{6]} - {\bf \Delta}_3 k_{[5} m_{6]}, \nd
which explicitly show that they are not only linear with respect to the fluctuation mode $p_{0x}$ but also that no other modes show up in the definition \eqref{casinv}. The precise coefficients
of $p_{0x}$ appearing in the sources above are respectively:
\bg\label{kakabari}
k_{[5}m_{6]} k_{[7}l_{6]} - k_{[5}l_{6]} k_{[7}m_{6]}, ~~~~~ l_{[5}m_{6]} k_{[7}m_{6]} - k_{[5}m_{6]} l_{[7}m_{6]}, \nd
which do not vanish generically, although special cases with vanishing coefficients could appear. Of course in the limit \eqref{tvader} everything vanishes, but since we are no longer
considering the simplifying condition \eqref{tvader},  we will assume non-zero coefficients. This consideration also allows us to express the other coefficients in \eqref{ekkuk},
namely $\alpha_i$ and $\beta_i$, in the following suggestive way:
\bg\label{shaver}
&& \alpha_1 \equiv k_{[5}m_{6]} k_{[1}l_{6]} - k_{[5}l_{6]} k_{[1}m_{6]}, ~~~~ \beta_1 \equiv k_{[1}m_{6]} l_{[5}m_{6]} -l_{[1}m_{6]} k_{[5}m_{6]}\nonumber\\
&& \alpha_2 \equiv k_{[2}l_{6]} k_{[5}m_{6]} - k_{[2}m_{6]} k_{[5}l_{6]}, ~~~~ \beta_2 \equiv k_{[2}m_{6]} l_{[5}m_{6]} - l_{[2}m_{6]} k_{[5}m_{6]} \nonumber\\
&& \alpha_3 \equiv k_{[3}l_{6]} k_{[5}m_{6]} - k_{[5}l_{6]} k_{[3}m_{6]}, ~~~~ \beta_3 \equiv k_{[3}m_{6]} l_{[5}m_{6]} -l_{[2}m_{6]} k_{[5}m_{6]} \nonumber\\
&& \alpha_4 \equiv k_{[4}l_{6]} k_{[3}m_{6]} - k_{[4}m_{6]} k_{[5}l_{6]}, ~~~~ \beta_4 \equiv k_{[4}m_{6]} l_{[5}m_{6]} - l_{[4}m_{6]} k_{[5}m_{6]}, \nd
which again do not generically vanish. At this stage the {\it signs} of the various $\alpha_i$ and $\beta_i$ coefficients are not important, but they could be worked out by carefully studying the
relative terms. The relative terms depend on the ($k_i, l_i, m_i$) coefficients defined in \eqref{hearts}, \eqref{redh} and \eqref{pumpp} respectively which in turn are
expressed in terms of coefficients given in \eqref{pikkola}. We also expect $\alpha_i \ne \beta_i$ as well as ${\alpha_i\over \alpha_j} \ne {\beta_i\over \beta_j}$ for all $i \ne j$, which may be
inferred from \eqref{shaver}.

Something interesting happens here. Eliminating $p_{00}$ from \eqref{ekkuk} lands us directly to an equation for $p'_{0x} \equiv Y_x$ whose form is similar to what we had earlier when we
analyzed a toy example. This means, as in \eqref{reileen} therein, we can define the following functions:
\bg\label{mcnally}
&& J_3(r) = {\rm exp}\left(\int^r {\beta_2(y) \over \beta_1(y)} dy\right), ~~~~~~~ J_4(r) = {\rm exp}\left(\int^r {\beta_4(y) \over \beta_3(y)} dy\right)\nonumber\\
&&J_1(r) = {\rm exp}\left(\int^r {\alpha_2(y) \over \alpha_1(y)} dy\right), ~~~~~~~ J_2(r) = {\rm exp}\left(\int^r {\alpha_4(y) \over \alpha_3(y)} dy\right), \nd
using the integrals of the functions defined in \eqref{shaver}, assuming neither $\alpha_i$ nor $\beta_j$ vanish. If any of the $\alpha_i$ or $\beta_j$ vanish, the analysis has to be changed
completely to get the requisite equation for $Y_x$.

We can also define another set of functions using $\alpha_i, \beta_j$ and the sources ${\bf \Delta}_{[a, b]}$ that do not involve integrals, much like the ones in \eqref{edenlac}. They are:
\bg\label{winnipeg}
&& P_3(r, \vert\omega\vert) = {{\bf\Delta}_{[1, 2]}\over \alpha_1}, ~~~~~~~ P_4(r, \vert\omega\vert) = {{\bf\Delta}_{[2, 3]}\over \beta_1} \nonumber\\
&&P_1(r) = {\alpha_3(r)\over \alpha_1(r)}, ~~~~~ P_2(r) = {\beta_3(r)\over \beta_1(r)}, ~~~~~ k \equiv {d\over dr} \left({J_1\over J_3}\right), \nd
where as before we have $-$ similar to $G_2$ and $G_4$ in \eqref{edenlac} $-$ $P_1$ and $P_2$ that are functions of both $r$ and $\vert\omega\vert$ because of their dependences on the
sources ${\bf\Delta}_{[1, 2]}$ and ${\bf\Delta}_{[2, 3]}$ respectively. Thus using \eqref{mcnally} and \eqref{winnipeg}, we can write the equation for $Y_x$ in the following way:
\bg\label{clayoven}
a_{12} {d^2Y_x\over dr^2} + a_{22} {dY_x \over dr} + a_{32} Y_x = a_{42}, \nd
similar to \eqref{skawaii}. The coefficients $a_{I2}$ are defined in somewhat similar form to \eqref{mado} in the following way:
\bg\label{bhishonM}
&&a_{12} = {(P_1 - P_2)J_1\over k}, ~~~~ a_{42} = -P_4 J_3 + {d\over dr}\left({P_3 J_1 - P_4 J_1\over k}\right)\\
&& a_{32} = J_4 {d\over dr}\left({P_2 J_3\over J_4}\right) + {d\over dr} \Bigg[{1\over k} {d\over dr}\left(P_1J_1 - P_2 J_1\right) + {J_1 J_4\over k J_3}{d\over dr}\left({P_2 J_3\over J_4}\right)
-{J_2\over k} {d\over dr} \left({P_1 J_1\over J_2}\right)\Bigg] \nonumber\\
&& a_{22} = {d\over dr}\left({P_1 J_1 - P_2 J_1\over k}\right) + {1\over k}\Bigg[{d\over dr}\left(P_1 J_1 - P_2 J_1\right) +  {J_1 J_4\over J_3} {d\over dr}\left({P_2J_3\over J_4}\right)
-J_2 {d\over dr} \left({P_1 J_1\over J_2}\right)\Bigg]. \nonumber \nd
Note that, although the analysis is similar to what we had for \eqref{skawaii}, there is an important difference now. The RHS of the equation \eqref{clayoven}, defined using
$a_4$ is constructed with $P_3$ and $P_4$ which are in turn defined in \eqref{winnipeg}. Both $P_3$ and $P_4$ are linear in $p_{0x}$ as may be seen from \eqref{casinv} and
\eqref{julias}. Thus $a_4$ in \eqref{bhishonM} differs from $a_4$ in \eqref{mado} by the presence of $p_{0x}$, implying \eqref{clayoven} to be a third order equation in $p_{0x}$.

We can use the above set of equations to formulate the equation for $p_{00}$, instead of $p'_{00}$, as we had in \eqref{lateshow}. Needless to say,
the equation for $Y_0(r, \vert\omega\vert) \equiv p_{00}$ follows similar route as before, and we can write the equation for $Y_0$ in the following way:
\bg\label{bjjgaram}
{dY_0\over dr} & = & {1\over J_1(r)}\int^r  dy J_1(y) P_3(r, \vert\omega\vert)  - {1\over J_1(r)} \int^r dy {P_1(y) J_1(y)\over J_2(y)} {d\over dy}\left[Y_x(y, \vert\omega\vert) J_2(y)\right] \nonumber\\
& = & {1\over J_3(r)}\int^r  dy J_3(y) P_4(r, \vert\omega\vert)  - {1\over J_3(r)} \int^r dy {P_2(y) J_3(y)\over J_4(y)} {d\over dy}\left[Y_x(y, \vert\omega\vert) J_4(y)\right], \nonumber\\ \nd
where the equality between the two sides is the consequence of \eqref{clayoven}. The way we have constructed the sources $P_3$ and $P_4$ in \eqref{winnipeg}, $Y_0$ do not appear
on the RHS of \eqref{bjjgaram} and therefore knowing $Y_x$ we would not only know:
\bg\label{paris4}
p_{0x}(r, \vert\omega\vert) \equiv \int^r dy Y_x(y, \vert\omega\vert), \nd
but also $p_{00}$. We may then use any one of the three equations in \eqref{atlantis} to determine $p_{0r}$. This way all the zeroth order fluctuation modes may be easily determined. For the
first order, and consequently the higher order fluctuation modes, one will have to rely on the recursion relations \eqref{thali}, \eqref{nor25} and \eqref{sphinx} for $\Delta^{(n)}_{20}$,
$\Delta^{(n)}_{2x}$ and $\Delta^{(n)}_{2r}$ respectively. These may be worked out with some effort, but we will not do so here as these fluctuation modes are not important for computing
the bulk-viscosity to the order that we want to analyze here.

The story however does not end here as there are additional constraints on the $p_{nk}$ modes that appear from the flux EOMs, namely the five-form, the three-forms and the axio-dilaton
EOMs. We can also get another equation from the cross-term in the  metric, namely the $rt$ component of the metric. All these should further constrain the fluctuation modes, and there
is a worry that these additional EOMs may over-constrain the system rendering them inconsistent. The scenario is subtle, so let us proceed carefully. First, and to ${\cal O}(\epsilon)$, we may ignore the three-form EOMs as
they start changing the equations only to ${\cal O}(\epsilon^2)$. Similarly, once we switch off the $g_sN_f$ corrections we are also effectively switching off the contributions from the
axio-dilaton EOMs. On the other hand we {\it cannot} ignore the five-form and the $rt$ EOMs. They will constrain the $p_{nk}$ modes, and it is easy to see how the $rt$ component of the metric
EOM does this:
\bg\label{bonda}
3p'_{nx} \Gamma_{0x} + 3p_{nx}\left(\Gamma'_{0x} - B_0' \Gamma_{0x}\right) - p_{nr} \Gamma_{0r}\left({5\over r} - 2A'_0\right) =
\sum_{k = 0}^2 \left(c_{nk} \Delta^{(n)}_{2k} + c_{[q]nkm}\partial^m_r p_{qk}\right), \nonumber\\  \nd
where the summation convention for $k$ follows the same as in \eqref{TUTa}. The other coefficients appearing in \eqref{bonda} are defined in the following way:
$c_{nk}$ are constants
that one may determine from the way the sources arrange themselves in the
$rt$ EOM, whereas $c_{[q]nkm}$ are functions of $r$ such that:
\bg\label{beblon}
c_{[q]nkm} \equiv 0~~ {\rm for} ~~ m \ge 3, \nd
which should be obvious from the construction itself. Also should be obvious are the two possible categories for $c_{[q]nkm}$, namely $c_{[n-1]nkm}$  and $c_{[n]nkm}$, for which
we could define the RHS of \eqref{bonda}. In fact once we choose the mode $p_{nk}$, there are {\it nine} possible choices of $c_{[n]nkm}$ for the allowed values of $k$ and $m$
in \eqref{bonda} and \eqref{beblon} respectively.
In fact this is where the above mentioned constraint show up: one can determine the functional forms of the coefficients $c_{[q]nkm}$ and the constants $c_{nk}$ by comparing with the
LHS of \eqref{bonda}.  One may also get these coefficients directly from the $rt$ EOM. We expect these two ways of getting these coefficients  to match because
in the absence of the sources i.e for the conformal case, the extra $rt$ equation {\it did not} over-constrain the system \cite{rgmaxim}.

Motivated by the above discussions, one may now give similar arguments for the five-form EOM, where the constraint  equation takes the following form:
\bg\label{bell}
&&\sum_{k = 0}^1 (2k+1) \Bigg\{p''_{nk} \Gamma_{0k} + p'_{nk} \left[2\Gamma'_{0k} + \Gamma_{0k}\left({5\over r} - 4A'_0\right)\right] +
p_{nk} \left[2\Gamma''_{0k} + \Gamma'_{0k}\left({5\over r} - 4A'_0\right)\right]\Bigg\}\nonumber\\
&& ~~~~~~~~~~~ - 4A'_0\left(p'_{nr} \Gamma_{0r} + p_{nr} \Gamma'_{0r}\right) = \sum_{k = 0}^2 \left(d_{nk} \Delta^{(n)}_{2k} + d_{[q]nkm}\partial^m_r p_{qk}\right), \nd
for every choice of $n$, and
with $d_{nk}$ and $d_{[q]nkm}$ being the coefficients similar to
$c_{nk}$ and $c_{[q]nkm}$ respectively
in \eqref{bonda} with $ d_{[q]nkm}$ vanishing for $m \ge 3$ as \eqref{beblon}. As before,
the RHS of  \eqref{bell} may be expressed in terms of the modes $p_{nk}$ and their derivatives
which may be compared with the LHS of \eqref{bell}. The system will be consistent when all the coefficients on both sides match.

There is also a simpler way to see why the coefficients on both sides of the equations in \eqref{bonda} and \eqref{bell} would match, once the RHS of these equations have been
specified in terms of the sources  and the modes. This is because, all the three equations in \eqref{thali}, \eqref{nor25} and \eqref{sphinx} may be expressed as:
\bg\label{cindesi}
\Delta^{(n)}_{2k} ~\equiv~ \sum_{l = 0}^2 f^{(k)}_{[q]nlm} \partial^m_r p_{ql}, \nd
with $f^{(k)}_{[q]nlm}$ being constrained in the same way as in \eqref{beblon}, implying that the RHS of either of the two equations \eqref{bonda} and \eqref{bell} take the following form:
\bg\label{slbang}
 \sum_{k = 0}^2 \left(\sum_{l = 0}^2 b_{nk}f^{(k)}_{[q]nlm} \partial^m_r p_{ql}+ b_{[q]nkm}\partial^m_r p_{qk}\right), \nd
 where $b$ can be either $c$ or $d$ for \eqref{bonda} and \eqref{bell} respectively. In this form \eqref{slbang} may easily be made to match with the LHS of the respective equations.

 Finally, let us give a reason why the RHS of the two equations \eqref{bonda} and \eqref{bell} are expressed in terms of the sources $\Delta^{(n)}_{2k}$ and the modes $p_{nk}$ and
 $p_{(n-1)k}$. For \eqref{bonda} it is easy to justify since it is the Einstein equation for the $rt$ component and therefore should depend on the sources and the fluctuation modes. To
 ${\cal O}(\epsilon)$ we expect only a linear combination of the form given as the RHS of \eqref{bonda}. On the other hand, in the five-form EOM \eqref{bell}, the fluxes used to balance the
 system against any collapse \cite{rgmaxim} would in turn induce three-brane sources on the anti-D5 branes.  The fluctuation modes should also affect these sources, and therefore
 the RHS of \eqref{bell} is expressed as a linear combination of the sources and the fluctuation modes to ${\cal O}(\epsilon)$, justifying the above analysis.

\subsection{The speed of sound in the strongly coupled plasma \label{sound}}

We are now ready to do the two set of computations  related to bulk viscosity: the speed of sound and the bound on the ratio of bulk viscosity to shear viscosity. The latter is again related
to the speed of sound \cite{Buchel-bound},
so it will suffice to compute the speed of sound in the strongly coupled plasma. However before we go about computing the sound speed, let us
present the generic formula for the ratio of the bulk viscosity $\zeta$ to the entropy density $s$ (a specific case was studied earlier in \cite{rgmaxim}) for an appropriate choice of the quadrant:
\bg\label{bulky}
{\zeta\over s} = {3\epsilon Y_x(r_h, 0) r_h\over 64}\left[3 + {13r_c\over r_h}\left({r_c^4\over r_h^4} - {16\over 13}\right){Y_x(r_c, 0)\over Y_x(r_h, 0)}\right], \nd
where $Y_x(r, \omega) \equiv p'_{0x}(r, \omega)$ satisfies the differential equation given in \eqref{clayoven}. The result for the ratio of bulk viscosity to entropy
density in a different quadrant can also be written down, and even their equivalence may be shown  as in \cite{rgmaxim}, but we will not do so here. Instead we will analyze the
sound speed in the medium using all the ingredients we have collected so far.

One of the ingredients that we shall use extensively to compute the sound speed is the entropy density $s$. This has already appeared in \eqref{bulky} above, but
the $s$ appearing above is only the conformal result as the ratio \eqref{bulky} is already proportional to $\epsilon \equiv {3g_sM^2\over 2\pi N}$. What we now need
is the non-conformal correction to $s$. This may be written as:
\bg\label{ento2b}
s = {\pi r_h^3 \sin~\theta_1 \sin~\theta_2 \sqrt{g_s N} \over 2\sqrt{27} \kappa_{10}^2} \left[ 1 - {6a^2\over r^2} + {\cal O}(a^4)\right], \nd
where the non-conformal corrections to $s$ enters through the resolution parameter $a^2$ given in \eqref{fatur}. We have chosen zero bare resolution parameter for
simplicity and therefore, as evident from \eqref{fatur}, a non-zero resolution already implies non-conformality in this set-up. One may worry that a zero bare resolution
parameter may fail to capture the essential ingredients for a UV completion \cite{metrics, rgmaxim}. However that is not much of a concern here as we are not exploring the
UV physics. Thus a cut-off $r_c$ will prominently feature in our results, as evident from \eqref{bulky} already. 
However the end results of the bulk and shear viscosities as well as their ratio 
will be cut-off {\it independent}, clarifying their IR nature, as will be demonstrated soon.   
However since we are using Wilsonian method, we 
will continue with this construction with an explicit cut-off $r_c$. In section \ref{nfn0} and beyond we will use the full set of quantum corrections, where the cut-off is taken to infinity, and therefore observables will appear with the QCD scale $r_d$. 

There is however one issue that we do want to emphasize at this point and it has to do with the sign of the first expression in \eqref{fatur}.
Of course we naively expect
$a^2$ to be positive, but the expression \eqref{fatur} involves various functions of log and dilog, so it will be instructive to check the sign of \eqref{fatur}. Let us
therefore start
by defining $x \equiv {r_h^2\over r_c^2} << 1$, using which we can express \eqref{fatur} by:
\bg\label{pperri}
a^2 =  -{\epsilon r_c^2\over 60}  \sum_{n = 1}^\infty {x^{2n}\over (2n-1)n^2}, \nd
which is negative definite. This may trigger an alarm because $a$ now becomes imaginary. Note that this problem does not arise if there is a bare resolution
parameter $a_0$, however small (as one may tune $\epsilon$ to be smaller than the smallest $a_0$). The way out of this conundrum is to notice that {\it all} expressions
of fluxes etc involve $a^2$ and not $a$. Further, $a^2$ appears in the metric \eqref{met} as a combination $r^2 + 6a^2$, and since we are only exploring the region
$r \ge r_h$, the sign of $a^2$ does not create any problem here too. On the other hand, when there is no black-hole, $r_h$ vanishes, and so does $a^2$ \eqref{pperri}.
All this has also appeared in \cite{boidyo} $-$ see discussions around figure 3 therein $-$
 for a more generic choice of $a^2$ given as eq. (2.63) in \cite{boidyo}.  We can of course resort to a more conservative approach by writing an expression for $|a|$
 instead, and we shall do so in \eqref{a} in the next section wherein a non-zero bare resolution parameter will also be taken into account.

Coming back,
the entropy density computed above in \eqref{ento2b} is proportional to powers of $r_h$, so it vanishes when $r_h \to 0$. Additionally when $r \to r_h$, the entropy
density receives corrections that take us away from the conformal value. These corrections may be easily quantified as powers of $\epsilon$ but we won't analyze it
here\footnote{For example, to first order in $\epsilon$ and for $r \to r_h$ we can sum up the series \eqref{pperri}, or use \eqref{fatur}, to show that the entropy
density may be expressed as: $$s = s_0\left[1+ {\epsilon\over 5}\left(\log~4 - {\pi^2\over 12}\right)\right]$$
\noindent where $s_0$ is the conformal value for the entropy density that can be read up from \eqref{ento2b}. To this order we can see that there is no $r_h$ dependence at the horizon.}.
Instead,
at this point it may be instructive to point out the steps that went in the
computation of the entropy density $s$. This would in turn effect the computation of the sound speed $c_s$, since it depends upon $s$ via:
\bg\label{desjardin}
c_s^2 =  {d\log~T\over d\log~s} = {s\over T} \left({dT/dr_h\over ds/dr_h}\right). \nd
The entropy density may be determined directly from supergravity by first computing the energy-momentum tensors and then dividing the result by the temperature $T$.
The energy-momentum tensor, on the other hand, arises from the variation of the action of the the form given by eq (3.120) of \cite{metrics}. One may add a Gibbon-Hawkings
term to it to control the boundary behavior, as evident from equations (3.121) and (3.123) of \cite{metrics}, but that does not alter the required linear term for our case. One may
also add counter-terms to holographically renormalize the subsequent  action, but since we are using a finite cut-off $r_c$, it is not necessary to add them at this stage. This aspect has already been alluded to earlier, and here we see a more concrete realization of this. Putting everything together, the sound speed for $r > r_h$ will be given by:
\bg\label{speedu}
c_s^2 & = & {1\over 3} + {2\epsilon\over 45}\left[x ~\log\left({1-x\over 1+x}\right) - \log\left(1-x^2\right)\right]\nonumber\\
&=&  {1\over 3} + {g_sM^2\over 15\pi N}\left[{r_h^2\over r_c^2}~\log\left({r_c^2-r_h^2\over r_c^2 + r_h^2}\right)
-  \log\left(1 - {r^4_h\over r^4_c}\right)\right],  \nd
where $x$ is the same parameter used in \eqref{pperri} before. Expectedly, the sound speed reduces to
$c_s = {1\over \sqrt{3}}$ in the conformal limit, and is {\it smaller}
than ${1\over \sqrt{3}}$ when non-conformal corrections are included. One may justify this by looking at
either of the two expressions in \eqref{speedu}: the two terms, that account for the non-conformal corrections, are
negative definite\footnote{This may be easily seen from the first expression in \eqref{speedu} written in terms of the
variable $x$ in the following way:
$$c^2_s = {1\over 3} - {2\epsilon\over 45} \sum_{n= 1}^\infty {x^{2n}\over n(2n-1)}$$
\noindent which is by construction smaller than $c_s = {1\over \sqrt{3}}$. Note that, for vanishing $\epsilon$ we get back the conformal result for the
sound speed as one would expect.} when $x < 1$ (or $r_h < r_c$). In the
limit $r_h << r_c$, the sound speed \eqref{speedu} may be approximated by:
\bg\label{katsyn}
c_s^2 = {1\over 3} - {g_sM^2\over 15\pi N} \left({r_h\over r_c}\right)^4. \nd
The above limit is not without its merit as we expect $r_c$ to be much bigger than $r_h$, even if we restrict the dynamics completely to Region 1 of \cite{metrics}. We can
now use \eqref{bulky} and \eqref{katsyn}, to express the ratio of the bulk viscosity to shear viscosity in the following suggestive way:
\bg\label{bishop}
{\zeta\over \eta} = {135\pi\over 32 x^2}\left(3 - {16\alpha_x\over \sqrt{x}}\right)\left({1\over 3} - c_s^2\right) + {39\pi \alpha_x\epsilon\over 16 x^{5/2}}, \nd
where $\eta = {1\over 4\pi}$ is taken at its conformal value to this order in $\epsilon$, $x = {r^2_h\over r^2_c}$ as before, and $\alpha_x \equiv  {Y_x(r_c, 0)\over Y_x(r_h, 0)}$ is the ratio if the two fluctuations.
We have also defined, without loss of
generality, $Y_x(r_h, 0) \equiv {1\over r_h}$ for $x << 1$. Note that the expression \eqref{bishop} is valid
in the limit:
\bg\label{kbloom}
{3g_sM^2\over 2\pi N} = \epsilon \to 0, ~~~~~~ x \to \epsilon^{1/9}, \nd
and therefore we can see how the last term in \eqref{bishop} vanishes. This would be a useful way to
separate the scales in the problem. We could have also gone for more generic analysis where we do not impose \eqref{kbloom}. In this case we can rewrite
\eqref{bishop}, using the expression \eqref{katsyn} for the speed of sound $c_s^2$, in the following way:
\bg\label{europe}
{\zeta\over \eta}  = {135\pi\over 32 x^2}\left(3 - {16\alpha_x \over \sqrt{x}} + {13 \alpha_x \over x^{5/2}}\right)\left({1\over 3} - c_s^2\right).\nd
However such an expression hides the scale dependences of each terms and is also a bit cumbersome to analyze. For simplicity therefore we will resort to an expression like \eqref{bishop} with a suppression factor going like the last term in \eqref{bishop}. In fact this works as long as the second term in \eqref{bishop} appears with a plus sign. If there is a relative minus sign, an expression like \eqref{europe} is not only useful but also necessary. 

We expect the ratio \eqref{bishop} to be positive definite, as \eqref{bulky} is positive definite. The second term is already positive, and the first term can become positive
if $\alpha_x$ is constrained in the following way:
\bg\label{sherbrooke}
\alpha_x < {3\sqrt{x}\over 16} ~~~~ \Longrightarrow ~~~~ {Y_x(r_c, 0)\over Y_x(r_h, 0)} < {3r_h\over 16 r_c}. \nd
There is something puzzling about \eqref{bishop} that we should clarify right now. The way we have expressed \eqref{bishop} would seem to put an additional
constraint on the ratio $\alpha_x$ of the fluctuations as evident from \eqref{sherbrooke}. However such a constraint does not seem to follow from
\eqref{bulky}. In fact  as long as $x^2 < {13\over 16}$ both \eqref{bulky} {\it and}
\eqref{bishop} should be positive definite. Since the expression \eqref{bishop} is basically a rewriting of \eqref{bulky} using the expression
\eqref{katsyn}, it implies that
\eqref{bishop} should not introduce any additional constraint of the form \eqref{sherbrooke} on the ratio $\alpha_x$. Then why is there a new constraint?
One way to argue for this would be to observe that
the expression \eqref{bishop} is generic  in the sense that it may be re-expressed as:
\bg\label{candyj}
{\zeta\over \eta} = a(x) \left({1\over 3} - c_s^2\right) +  b(x), \nd
where $a(x)$ and $b(x)$ are variations of the coefficients appearing in \eqref{bishop}. If $b(x)$ is not proportional to $\epsilon$, this generalization will suffer\footnote{In addition to being inconsistent at the conformal limit.}
 from the
appearance of explicit cut-off dependences of the
respective variables that cannot be expressed in an implicit scale-separated way as in \eqref{europe}. Once however an expression like \eqref{europe} is realized for \eqref{candyj}, there will be no additional constraints and the expression will be similar to \eqref{bulky}. We will exploit this angle to our advantage when we go for the case where $c_s^2$ itself generalizes.

An example of such generalizations appears with ${\cal O}(\epsilon)$ corrections to \eqref{speedu} and \eqref{katsyn} that may change the coefficients of \eqref{bishop}. These corrections appear
from ${\cal O}(\epsilon)$ corrections to the temperature $T$, which we had identified to the horizon radius $r_h$. To see this first let us take the cut-off temperature $T_c$
used in \cite{metrics}, which may be expressed as:
\bg\label{catseye}
T_c  \equiv  {B'(r_h, \epsilon)\over 2\pi} {\rm exp}\left[2B(r_h, \epsilon) - B(r_c, \epsilon) + 2A(r_h, \epsilon)\right], \nd
where $e^{2B}$ and $e^{2A}$ are defined in \eqref{dhua} and \eqref{fatur} respectively. To ${\cal O}(\epsilon)$ the functional forms for the various parameters,  i.e
$g(r) \equiv e^{2B(r, \epsilon)}$ and $h(r) \equiv e^{-4A(r, \epsilon)}$,  appearing in \eqref{catseye}
may
be determined exactly as:
\bg\label{monee}
&&g'(r_h) = {4\over r_h} + {6\epsilon\over 5r_h}\left(\log~4 - {\pi^2\over 12}\right) \nonumber\\
&&g(r_c) = 1 - x^2 - {3\epsilon\over 20}\sum_{n = 1}^\infty {x^{2(n+1)}\over n^2(n+1)(2n-1)} \nonumber\\
&& h(r_h) = {L^4\over r_h^4}\left[1 + \epsilon ~\log ~x + {\epsilon\over 5}\left(\log~4 - {\pi^2\over 48}\right)\right], \nd
where $L^4 \equiv {27 g_sN\over 4}$, and note the appearance of higher powers of $x$ in the black-hole factor $g(r)$ defined at the cut-off $r_c$. This series has
similarity with the series \eqref{pperri} defined for the resolution parameter $a^2$. The connection is of course spelled out earlier in \eqref{dhua}, and once we
plug  \eqref{monee} in \eqref{catseye}, the temperature may be expressed as:
\bg\label{silverb}
T_c & = & {{r_h\over \pi L^2}\left[1 + {3\epsilon\over 10}\left(\log~4 - {\pi^2\over 12}\right)\right]\over \Big[1 + {\epsilon\over 5}\left(\log~4x^5 - {\pi^2\over 48}\right)\Big]^{1/2}
\Big[1 - x^2 - {3\epsilon\over 20}\sum_{n = 1}^\infty {x^{2(n+1}\over n^2(n+1)(2n-1)}\Big]^{1/2}} \\
&=& {r_h\over \pi L^2} \left(1 + {x^2\over 2}\right)\left[1 + {\epsilon\over 5}\left(\log~4 - {11\pi^2\over 96}\right) - {\epsilon\over 4}~\log~x + {9\epsilon\over 2}
\left({x^4 \over 2+x^2}\right) + {\cal O}(x^6)\right], \nonumber \nd
where the second line is in the limit $x << 1$.  The $\epsilon$ corrections are exactly the ones that one would expect from switching on non-conformalities in the
system. However note that even in the limit $\epsilon \to 0$, our expression for $T_c$
seems to have an additional factor of the form:
\bg\label{nolight}
T_c = {r_h\over \pi L^2 \sqrt{1 - x^2}}, \nd
which implies the cut-off dependence of the temperature. Clearly when we make $r_c \to \infty$ we recover the conformal result, but the appearance of $x$ in \eqref{nolight}
as well as in \eqref{silverb} means that UV completion is necessary to argue for the physical value of $T_c$ here. Naively taking $r_c \to \infty$
for non-zero $\epsilon$ will {\it not} give us the correct
answer here, which of course resonates well with the UV completion discussed in \cite{metrics}.

Thus there is a way to holographically renormalize the system, following the procedure given in \cite{metrics}, that would take care of the log pieces in the metric
and other variables in the problem. Once this is accomplished one may, in some restrictive sense, take $r_c \to \infty$. This is a specific UV completion wherein
the UV cap gives rise to an asymptotically conformal theory. For such a case the temperature does take a physical value which may be expressed as:
\bg\label{kalopedo}
T_c =  {r_h\over \pi L^2}\left[1 + {\epsilon\over 5}\left(\log~4 - {11\pi^2\over 96} - {5\over 4} \log~{r_h\over \Lambda}\right)\right] + {\cal O}(\epsilon^2), \nd
where $\Lambda$ is related to the QCD scale for this model.
The above is the so-called boundary temperature of \cite{metrics} that we define at far UV. We will however need to define the temperature at any given scale, not just
the UV, to avoid issues  like \eqref{nolight} in the absence of any non-conformalities.
Let us therefore take the following definition of the temperature:
\bg\label{flowers}
T = r_h\left(a_1 + \epsilon a_2\right), \nd
where $a_1$ and $a_2$, which can be functions of $x = {r_h^2\over r_c^2}$,  will be determined below.  Note that $T$ and $T_c$ are similar when $a_i$ take
specific values  extracted from \eqref{kalopedo}. In general however, $T$ should be the temperature that would occur naturally in this framework. This means
we need to change slightly the formula for entropy in \eqref{ento2b} by replacing $r_h^3$ in \eqref{ento2b} by $r_h^4/T$, with $T$ given by \eqref{flowers}. The sound
speed will also change from \eqref{speedu} to the following:
\bg\label{attic}
c_s^2 & = & {1\over 3} -{2\epsilon\over 45}\sum_{n = 1}^\infty {x^{2n}\over n(2n-1)}\left(1 + {2x\over a_1} {da_1\over dx}\right)\\
&+& {4x^2\over 9a_1^2}\left[\left({da_1\over dx}\right)^2\left(1 - {2\epsilon a_2\over a_1}\right) + {2\epsilon} {da_1\over dx} {da_2\over dx}\right]
+ {8x\over 9a_1}\left[{da_1\over dx}\left(1 - {\epsilon a_2\over a_1}\right) + \epsilon {da_2\over dx}\right], \nonumber \nd
where the expected cut-off dependence appears from $x$ as before.  Clearly when $a_1 = {1\over \pi L^2}$ and $a_2 = 0$, we recover the sound speed computed
in \eqref{speedu}. However now, when both $a_1$ and $a_2$ are functions of $x$, the $\epsilon = 0$ limit gives us:
\bg\label{veroMPP}
c_s^2 = {1\over 3}  + {4x\over 9a_1}\left({da_1\over dx}\right)\left[2 + {x\over a_1}\left({da_1\over dx}\right)\right], \nd
which takes us away from the conformal value of $c_s^2 = {1\over 3}$ in the conformal limit. This is not what we expect here, so we can use \eqref{veroMPP} to
determine the functional form for $a_1(x)$. There are clearly two possible solutions for $a_1(x)$, namely:
\bg\label{petals}
a_1(x) = b, ~~~~~~~~ a_1(x) = {b\over x^2}, \nd
where $b$ is yet an undetermined constant. The second choice is not acceptable in a theory that is holographically renormalizable, as it blows up when the
cut-off is taken to infinity. This implies that $T$ in \eqref{flowers} can only be:
\bg\label{thorns}
T = r_h\left(b + \epsilon a_2(x)\right), \nd
with constant $b$. What value can $a_2(x)$ take? To determine this we will need to study the full holographically renormalized temperature. This is in general a
tedious exercise, but we can get a hint from the renormalized boundary temperature $T_c$ that we determined earlier in \eqref{kalopedo}. To the first order in
$\epsilon$, the renormalized boundary temperature depends on $\log~r_h$. This tells us that we can make the following ansatze for $a_2(x)$:
\bg\label{court}
a_2(x) = c_1(x) + c_2(x)\log ~x, \nd
where $c_1(x)$ and $c_2(x)$  are polynomials in $x$ that do not have either $\log~x$ or $x^{-n}$ pieces. The two functions $c_1(x)$ and $c_2(x)$ contribute
to the full sound speed in the following way:
\bg\label{seedsof}
c_s^2 &=& {1\over 3}  - {2\epsilon\over 45}\sum_{n = 1}^\infty {x^{2n}\over n(2n-1)} + {8\epsilon x\over 9a_1}\left({da_2\over dx}\right) \nonumber\\
&=& {1\over 3}  + {8\epsilon c_2(x)\over 9b} - {2\epsilon\over 45}\sum_{n = 1}^\infty {x^{2n}\over n(2n-1)}
+ {8\epsilon x\over 9b}\left({dc_1\over dx} + \log~x ~{dc_2\over dx}\right), \nd
where we see that the result is not so different from our earlier value for sound speed \eqref{speedu}.  The difference lies in the additional term proportional to
$da_2/dx$, which in turn would depend on how $c_1(x)$ and $ c_2(x)$ depend on $x$.  If $c_n(x) = -\vert c_n\vert$ with constant $c_n$, and $x << 1$, the
sound speed is simple and is given by\footnote{Both the expressions of generalized sound-speed, 
\eqref{seedsof} and \eqref{notielle}, appear with explicit cut-off dependent terms that vanish in the limit when the cut-off is taken to infinity, i.e when $x \to 0$. This is of course the consequence of using Wilsonian method as we emphasized repeatedly. We can alternatively write cut-off independent sound-speed as in 
\eqref{strawberry100} which works for {\it any} UV completions. Thus all observables will be cut-off independent as we demonstrate soon. One may also compare this with the sound-speed computed in 
\eqref{cs}, which uses full quantum corrections where the cut-off is taken to infinity, therefore as a consequence only the QCD scale $r_d$ appears. See also footnote \ref{lunamey} for more details.} :
\bg\label{notielle}
c_s^2 = {1\over 3}  -{8\epsilon \vert c_2\vert\over 9b} - {2\epsilon\over 45}\left({r_h^4\over r_c^4}\right) + {\cal O}(\epsilon^2), \nd
where $b > 0$, and the signs are dictated by the fact that the beta function is negative and so the sound speed is {\it smaller} than $1/\sqrt{3}$.  The additional term
in the sound speed \eqref{notielle} means that the ratio of bulk to shear viscosities, i.e \eqref{bishop}, changes to:
\bg\label{dahlia}
{\zeta\over \eta} = {135\pi\over 32 x^2}\left(3 - {16\alpha_x\over \sqrt{x}}\right)\left({1\over 3} - c_s^2\right)
+ {\pi \alpha_x\epsilon\over x^{5/2}}\left({39\over 16} + {60 \vert c_2\vert \over b}\right) - {45\pi \epsilon \vert c_2\vert\over 4b x^2}, \nd
where expectedly when $\vert c_2\vert = 0$ we recover \eqref{bishop} with scales separated as in \eqref{kbloom}. The relative minus sign for the last term appearing above however creates an issue now, so it is more advisable to resort to the expression of the form \eqref{europe} where the scale separation is more 
implicit. In other words, we want to express \eqref{dahlia} alternatively as:

{\footnotesize
\bg\label{dadissue}
{\zeta\over \eta} = {135\pi\over 32 x^2}\left[3 - {16\alpha_x\over \sqrt{x}} + \left(\left(26 + {640\vert c_2\vert\over b}\right){\alpha_x\over \sqrt{x}} - {120\vert c_2\vert\over b}\right) {b\over 2bx^2 + 40\vert c_2\vert}\right] \left({1\over 3} 
-c_s^2\right). \nd}
The RHS now crucially depends on $\alpha_x$ i.e on
the ratio of fluctuations $Y_x(r_c, 0)$ and $Y_x(r_h, 0)$ satisfying \eqref{clayoven}. The equation \eqref{clayoven} is difficult to solve, partly because of our ignorance of the
precise sources $a_4$ defined in \eqref{winnipeg} using ${\bf \Delta}_{[1, 2]}$ and ${\bf \Delta}_{[2, 3]}$ via \eqref{casinv} (which in-turn are defined in terms of $P_i$ and $J_k$ functions in \eqref{bhishonM}, extracted from 
\eqref{mcnally} and \eqref{winnipeg}). Nevertheless, using the constraint
\eqref{sherbrooke}, allows us to make the following ansatze for $\alpha_x$:
\bg\label{MPPm}
\alpha_x = {3 \sqrt{x}\over 16} \mathbb{F}(x), \nd
where $\mathbb{F}(x)$ is another function of $x$. What constraints do $\mathbb{F}(x)$ satisfy? From 
\eqref{sherbrooke}, $\mathbb{F}(x) < 1$ for all values of $x < 1$.  We also require $x^2 < {13\over 16}$ as 
mentioned before, which we can constrain further by inserting a constant $d_1 > 1$ allowing $x^2 < {13\over 16 d_1}$
without loss of generalities.  Thus $\mathbb{F}(x)$ may be expressed as a series in $x^2$. This series 
could be summed up\footnote{Let us denote the functional form for $\mathbb{F}(x)$, expressed as series in 
powers of $x^2$, in the following way: $$\mathbb{F}(x) = \sum_{n= 1}^\infty b_n x^{2n} \equiv 
{ax^2 \over b - cx^2},$$
whereas the RHS is an ansatze for the series sum. The way we have constructed this, for small $x^2$, 
$\mathbb{F}(x)$ is proportional to $x^2$. The positivity of the function gives $x^2 < {b\over c}$ which could be equated to the constraint $x^2 < {13\over 16}$. The function $\mathbb{F}(x)$ itself should be less that 1 for all $x^2 < 1$. This provides additional constraint of the form $x^2 < {b\over a + c}$. This could now be equated to $x^2 < {13\over 16 d_1}$ with $d_1$ defined above. Putting everything together essentially reproduces \eqref{mostlytan} up to an irrelevant overall constant that we can absorb in the definition of 
$d_1$.} providing the following ansatze for 
$\mathbb{F}(x)$: 
\bg\label{mostlytan}
\mathbb{F}(x) = {16(d_1 - 1) x^2\over 13 - 16 x^2}, \nd
where $d_1$ is the constant used earlier, and in fact because of this constant \eqref{mostlytan} is 
always positive definite, and expectedly smaller than 1. The way the series is constructed 
does not include terms proportional to $\sqrt{x}$ or $x^{3/2}$ in \eqref{mostlytan}. Combining 
\eqref{mostlytan} with \eqref{MPPm} and plugging this in \eqref{dadissue} gives us the following  
suggestive form for the 
ratio of bulk to shear viscosities:
\bg\label{lizshort}
y \equiv {\zeta\over \eta} = {405\pi b d_1\over 32\left(20 \vert c_2\vert + bx^2\right)} \left({1\over 3} - c_s^2\right), \nd
where the cut-off dependence, compared to \eqref{dadissue} or \eqref{dahlia}, appears
explicitly through the denominator of \eqref{lizshort} and {\it implicitly} through \eqref{notielle}.  Although
in the absence of the precise knowledge of $b, d_1$ and $c_2$, it appears that this is the best we can do at this stage, taking a $x$ derivative of $y$ in \eqref{lizshort} gives us:
\bg\label{oxelmey}
{\partial y\over \partial x} = - {405 \pi b d_1 \over 16\left(20 \vert c_2\vert + bx^2\right)} 
\left[{bx\over 20 \vert c_2\vert + bx^2}\left({1\over 3} - c_s^2\right) - c_s {\partial c_s\over \partial x}\right] = 0,
\nd
showing that the ratio of the bulk to shear viscosities is cut-off 
independent! As a corollary to above, since the cut-off independence of shear viscosity is already demonstrated in \cite{metrics}, it follows from \eqref{oxelmey} that the bulk viscosity is also cut-off independent.
The result appears to be almost miraculous, so question is what happened. First note that if we naively take the cut-off to infinity, i.e $x \to 0$, 
the sound speed and the ratio of the bulk to shear viscosities become:
\bg\label{strawberry100}
c_s^2 = {1\over 3} - {8\vert c_2\vert \epsilon \over 9b}, ~~~~~ 
{\zeta\over \eta} = {9\pi d_1\epsilon \over 16}. \nd
From this we can see how the miracle happens. On one hand, the ratio of bulk to shear viscosities is always proportional to the value quoted in \eqref{strawberry100} times a function that depends on the cut-off $x^2$. 
On the other hand, $\left({1\over 3} - c_s^2\right)$ is always proportional to the value quoted in 
\eqref{strawberry100}  times a function in $x^2$ that is exactly the {\it inverse} of the function that appears with the ratio of the bulk to shear viscosities. Thus they cancel proving not only the IR nature of the ratio, but also the IR natures of  both the bulk and the shear viscosities!

This is almost what we wanted, although one concern remains related to the choices \eqref{MPPm} and 
\eqref{mostlytan}.
How are we justified in the selective choices of the coefficients in the above equations? How do we even know that
such choices will solve the EOMs?
The answer to both the questions lies in the specific
UV completion, or more appropriately on the distribution of the anti D5-branes in Regions 2 and 3 (see the blue box in Fig. \ref{allregions}).
Once we plug  \eqref{MPPm} and \eqref{mostlytan} in \eqref{clayoven}, we can in
principle determine the form of the sources ${\bf \Delta}_{[1, 2]}$ and ${\bf \Delta}_{[2, 3]}$ in $a_4$, given via \eqref{winnipeg}. One can then re-arrange the
anti D5-distributions to match with the functional forms of ${\bf \Delta}_{[1, 2]}$ and ${\bf \Delta}_{[2, 3]}$,  justifying the ansatze \eqref{MPPm}. Notice that the ansatze \eqref{MPPm} combined with 
\eqref{mostlytan} implies that
$Y_x(r_c, 0) = {\alpha_x\over r_h}$. Plugging this in \eqref{clayoven} gives us:
\bg\label{katyb}
{4a_{12} x^3} \alpha''_x + \left({6 a_{12} x^2} - {2r_h a_{22} x^{3/2}}\right)
\alpha'_x + r_h^2 a_{32} \alpha_x = r_h^3 a_{42}, 
 \nd
where the $a_{mn}$ coefficients are defined in \eqref{bhishonM} in terms of the $P_n$ and $J_m$ sources. It is not too hard to see that \eqref{katyb} may be easily satisfied by choosing appropriate sources in
Regions 2 and 3. For example, taking $r_h = 1$ in appropriate units, one particular set of solution would be the following:
\bg\label{boyerased}
a_{12} = - {a_2\over x}, ~~~ a_{22} = {a_2\over \sqrt{x}}, ~~~ a_{32} = 3 a_2, ~~~ 
a_{42} =  - {21632 a_2 \alpha_x \over (13 - 16 x^2)^2}, \nd
where $a_2$ is a constant and $\alpha_x$ is defined in \eqref{MPPm}. The overall signs for $a_{12}$ and $a_{42}$ can be inferred from the definitions of the coefficients given in \eqref{bhishonM}. Thus if:
\bg\label{minaharker}
P_2 > P_1, ~~~~~ P_4 > P_3, \nd
with $P_i$ defined in \eqref{winnipeg}, and assuming that the derivatives in \eqref{bhishonM} do not offset the sign assignments, \eqref{boyerased} gives us a distribution of anti-branes in Regions 2 and 3 that would allow for IR independent observables. Interestingly, \eqref{boyerased} is not the only allowed solution. In fact our claim is that as long as the theory is holographically renormalizable, there would always exist distributions of the form \eqref{boyerased} that would effectively make the bulk and shear viscosities as well as their ratio to be cut-off independent. 

Once this is settled, note that the negative definite last-term in \eqref{dadissue} cannot be very large as all the three constants appearing there, namely $d_1, \vert c_2\vert$ and
$b$, are finite numbers. In fact for $x < 1$, it is easy to establish the following range of the function 
$\mathbb{F}(x)$:
\bg\label{pennyd}
{60\vert c_2\vert \over 13 b + 320 \vert c_2\vert} < \mathbb{F}(x) < 1,
 \nd
assuming $x$ is always away from zero. (When $x = 0$, $\mathbb{F}(x)$ vanishes so one will have to look for 
appropriate UV completion that allows for a non-zero $\mathbb{F}(x)$ at the boundary.) This also means that the cut-off dependent  terms in \eqref{dadissue} will dominate over the two negative definite
terms.  Expectedly, this is consistent with the overall positivity of the ratio \eqref{lizshort}. However we now need the lower bound on $d_1$. To determine this, let us first assume that the sound-speed $c_s^2$ quoted in \eqref{strawberry100} is the renormalized sound-speed. This immediately reproduces the 
following 
ratio of the bulk to shear viscosities: 
\bg\label{mathadulai}
{\zeta\over \eta}  =  {81 \pi b d_1\over 128 \vert c_2\vert}\left({1\over 3} - c_s^2\right), \nd
which is cut-off independent and may be justified from the positivity arguments that we presented earlier. We expect this bound to {\it not} violate the original Buchel-bound \cite{Buchel-bound}, which was
presented for weak string and strong 't~Hooft couplings\footnote{Although note that there does exist a possibility, by choosing an appropriate $d_1$, to violate the bound \cite{Buchel-bound} as alluded to earlier. This should lead to interesting physics whose implications, if any, will not be investigated here.}. Combining this with \eqref{pennyd} then gives the following lower bound for $d_1$ in \eqref{mostlytan}:
\bg\label{duldul}
 d_1 \ge {256 \vert c_2\vert  \over 81\pi b }  > 1, \nd
which would eventually control the behavior of the fluctuation modes studied in section \ref{bvcomp}. In the next section, we will study the sound speed and viscosity bound
with non-zero fundamental flavors and with string coupling of order 1. We will re-derive some of the above results, but in a different regime of the parameter space. Such an analysis will hopefully shed light on the underlying universality of the results derived here\footnote{In {\bf {Appendix} \ref{UBC}} we will compare our 
results, derived here and in section \ref{nfn0}, with some of the other related works like \cite{EO}.}.



\section{Bulk viscosity at strong string and strong 't~Hooft couplings with non-zero flavors \label{nfn0}}

In the previous section we saw how one may study bulk viscosity, sound speed and the bound on the ratio of the bulk  to shear viscosities at strong 't~Hooft coupling using a gravity dual
in type IIB theory. At this stage one may attempt few improvements in the present scenario by including both the flavor degrees of freedom as well as the UV regions. One may even ask
the questions in the regime where the string coupling itself is of order 1, which of course still maintains strong 't~Hooft coupling in the gauge theory side.
The latter is however harder to study because it is the regime where even S-duality does not help. The question then is whether we can say something concrete in this regime of parameter space.

One simple answer to the enigma may be to T-dualize the system to type IIA, by including the flavor branes, and then lift the configuration to M-theory.  This should in principle
accomplish the task, except that the T-dual scenario leads to a configuration of intersecting NS5-branes with the intersection region being blown up to a {\it diamond} \cite{diamond, DOT}.
This is
not necessarily bad, and in fact in the past useful results have been drawn out of this configuration \cite{meson}, but the requirement of keeping track of the NS5 degrees of freedom
may thwart a simple analysis of the system. What we are looking for is a configuration with manifold and fluxes that we could use to succinctly address similar set of questions as in the
previous section, avoiding the unnecessary requirement of including extra degrees of freedom.
This is exactly where the {\it mirror} dual of the type IIB framework becomes handy. In fact, lattice-compatible results pertaining to glueball spectroscopy were obtained in \cite{glueball Sil+Yadav+Misra} and P(article)D(ata)G(roup)-compatible results pertaining to meson spectroscopy were obtained in \cite{meson Yadav+Misra+Sil}, by working  with the mirror dual.

\subsection{The mirror type IIA model and its M-theory uplift \label{rannT}}

As discussed above, and also alluded to in Fig. \ref{2a2b}, the model that we want to use here is the M-theory uplift of the type IIB scenario that we studied earlier. This is the MQGP model of
\cite{MQGP, transport-coefficients} where at weak string coupling we have a type IIA description. One of important procedure that goes in the construction of \cite{MQGP, transport-coefficients} is the so-called {\it delocalized} mirror symmetry via the Strominger-Yau-Zaslow (SYZ) prescription \cite{syz}. This prescription involves a two-step procedure: one, by viewing the Calabi-Yau manifold as a special Lagrangian ${\bf T}^3$ fibered over a base that is taken very large, and two, by performing three T-dualities over the ${\bf T}^3$ fiber. In this sub-section, we will provide some discussions on the details of the procedure.

The first requirement of a large base is important. This has to do with nullifying the contributions from open-string disc instantons with boundaries that appear as non-contractible 1-cycles in the special Lagrangian (sLag) ${\bf T}^3$ fibered over the base. To see this more clearly, let us define three delocalized T-dual coordinates ($x, y, z$)
which are basically proportional to ($\phi_1, \phi_2, \psi$) coordinates respectively that we encountered earlier. These coordinates are valued in the fiber torus ${\bf T}^3$ via \cite{MQGP}:
\begin{equation}
\label{xyzdefs}
x = s_1 \phi_1,~~~ y = s_2  \phi_2,~~~  z= s_3 \psi,
\end{equation}
where $s_i$ are constants whose values may be derived from \cite{anke1}.  Interestingly, the choice of the coordinates ($x, y, z$) allows us to study the local geometry of the underlying manifold. Furthermore,
using the results of \cite{M.Ionel and M.Min-OO (2008)} the following conditions, as shown in
\cite{transport-coefficients, EPJC-2}, are satisfied:
\begin{eqnarray}
\label{sLag-conditions}
& & i^* J \approx 0,~~~~{\rm Im} \left( i^*\Omega\right) \approx 0,\nonumber\\
& & {\rm Re} \left(i^*\Omega\right)\sim{\rm volume \ form}\left({\bf T}^3(x,y,z)\right),
\end{eqnarray}
for the  underlying ${\bf T}^2$-invariant special Lagrangian manifold of \cite{M.Ionel and M.Min-OO (2008)} for resolved and deformed conifold. This immediately implies that, if the underlying resolved warped-deformed conifold is predominantly either complelely resolved or deformed, the local sLag ${\bf T}^3$ of \eqref{xyzdefs} is then the  required sLag to allow for the SYZ mirror construction via local T-dualities.

{Let us analyze thus further by taking the type IIB background given in \eqref{met} but now with $e^B = 1$. The latter requirement is to just simplify the ensuing discussion.
As we saw above, to enable use of SYZ-mirror duality via three T dualities, one is required to take a large base. This immediately means taking large complex structures of the aforementioned two two-tori of the sLag
${\bf T}^3(x,y,z)$ fibration. One may easily implement this via the following considerations \cite{chenGT}:
\bg\label{SYZl}
d\phi_{k}\rightarrow d\phi_{k} - f_{k}(\theta_{k})d\theta_{k}, ~~~
d\psi\rightarrow d\psi + \sum_{k = 1}^2  f_k(\theta_k)~\cos ~ \theta_k d\theta_k,
\nd
for appropriately chosen large values of $f_{k}(\theta_{k})$ with $k = 1, 2$.
This choice does not change the local NS three-form flux, as was shown in \cite{anke1, chenGT}. Globally the underlying manifold can be a non-K\"ahler manifold as we discussed earlier. This is the advantage of using the
($x, y, z$) coordinates.  On the other hand,
 the fact that one may be allowed  to choose large values of $f_{k}(\theta_{k})$, was justified later in \cite{MQGP}.  The main idea is basically the requirement that the metric obtained after SYZ-mirror transformation, applied to the
 non-K\"{a}hler  resolved warped-deformed conifold,  {\it should} resemble, at least locally, a non-K\"{a}hler warped resolved conifold. This means after incorporating \eqref{SYZl} to \eqref{met},
 the ($x, y, z$) coordinates
 discussed in \eqref{xyzdefs} will parametrize the local behavior succinctly. The global considerations will follow afterwards as shown in \cite{katzo, chenGT}\footnote{To justify the delocalization method while constructing the type IIA mirror {\it a la} SYZ triple-T-duality prescription \cite{syz} and its subsequent M-theory uplift one may argue the following.
Consider the example of the mirror of a D5-brane wrapping the resolved $S^2$ with fluxes as studied in the first reference of \cite{anke1}. The M-theory uplift can be made free of delocalization ensuring that one can construct a permissible $G_2$ structure manifold for the entire domain of validity of the delocalized coordinates. For example, in the delocalized large-complex structure limit and after a fixed $\psi$ coordinate rotation, one obtains the SYZ mirror to be D6-brane wrapping a non-K\"ahler deformed conifold. Now, as shown in section 6 of \cite{anke1}, one can define an appropriate set of vielbeins to construct an explicit $G_2$ structure in terms of which the M-theory uplift of the previously obtained type IIA mirror could be rewritten, and which is valid for all values of $\psi$.  In other words,
the mirror for $\psi=\psi_0$ coincides with the triple-T-dual-fixed-$\psi$ rotated type IIA mirror obtained assuming delocalization. This  essentially states that the type IIA mirror in equation (6.23) of the first reference in \cite{anke1} obtained by descending to type IIA from arbitrary-$\psi$ M-theory uplift will be the same as the fixed-$\psi_0$ type IIA mirror of equation (5.64) obtained using delocalization for $\psi=\psi_0$. Hence we could just replace $\psi_0$ by
$\psi$ in the type IIA mirror obtained assuming delocalization.This therefore implies that the type IIA mirror is effectively free of the delocalization restriction.}.

{In the local geometry, we can now perform three T-dualities\footnote{Now also switching on $e^B$ in \eqref{met}.},
first along coordinate $x$, then along coordinate $y$ and finally along coordinate $z$, to get the local mirror manifold.
The details of this construction, utilizing the results of \cite{anke1}, \cite{chenGT} was first worked out in
\cite{MQGP}. The local mirror captures all the right properties of the expected dual configuration in the type IIA side,
and then one may use the coordinates ($\phi_1, \phi_2, \psi$) to express the global metric as $ds^2_{\rm IIA}$ (see \cite{anke1, katzo, chenGT, MQGP} for details). An additional ingredient that appears naturally from the SYZ procedure from the type IIB three and five-form fluxes as well as the axio-dilaton, is the one-form type IIA potential
${\cal A}$. Such a one-form is useful to construct the M-theory uplift of the mirror type IIA as was shown in \cite{MQGP}\footnote{As is standard in such constructions, the one-form ${\cal A}$ may not be globally defined, although it's field strength will be. In the type IIB side such one-form will lead to either a  RR two-form field or the axion depending on the T-duality direction.}.
The global M-theory metric takes the following form:
\bg\label{Mtheorymet}
ds^2_{11} = e^{-\frac{2\varphi}{3}} \left[-g_{tt}dt^2 + g_{\mathbb{R}^3}\sum_{i = 1}^3 dx_i^2 +  g_{rr}dr^2  +   ds^2_{\rm IIA}(\phi_i, \theta_i, \psi)\right]
+ e^{\frac{4{\varphi}}{3}}\left(dx_{11} + {\cal A}\right)^2, \nonumber\\
\nd
where $\varphi$ is the type IIA dilaton that appears from the mirror transform of the type IIB dilaton. Once the dilaton is allowed to take a non-trivial value, both in type IIB as well as in the mirror type IIA side, one starts seeing the effects of the flavors. This is simply because, in the type IIB side, non-trivial axio-dilaton shows up only when we switch on $N_f$ seven-branes. Of course, not all the $N_f$ seven-branes are required to be local D7-branes, but having D7-branes make the mirror picture more transparent as these would eventually contribute to the dilaton $\varphi$ in the type IIA side. Once the dust settles,  the $g_{\mathbb{R}_3}$ and $g_{tt}$  components
appearing in \eqref{Mtheorymet} may be defined
in the following way\footnote{Note that, unless mentioned otherwise, we shall always assume $\log~r$, in expressions like \eqref{metdef}, is written as $\log~{r_{~}\over r_d}$ with $r_d$ being the scale associated with ${\rm D5}- \overline{\rm D5}$ separation  to keep the ratio dimensionless. Compared to the previous section, the analysis in this section and next uses full quantum corrections where the cut-off is taken to infinity so only the scale $r_d$ appears. To avoid clutter,
we will also take $r_d = 1$ so that $r$ remains dimensionless. \label{lunamey}}:
\bg\label{metdef}
& & g_{\mathbb{R}^3} =  {r^2\over \sqrt{4\pi g_sN}}\Bigg\{1  - {3g_s M^2\over 4\pi N} \left[1 + {3g_sN_f\over 2\pi}\left(\log~r + {1\over 2}\right)
+ {g_sN_f\over 4\pi}\log\left({\alpha_{\theta_1}\alpha_{\theta_2}\over 4\sqrt{N}}\right)\right] \log~ r \Bigg\} \nonumber\\
& & g_{tt} = {(r^4-r_h^4)\over r^2\sqrt{4\pi g_sN}} \Bigg\{1  +  {3g_s M^2\over 4\pi N} \left[1 + {3g_sN_f\over 2\pi}\left(\log~r + {1\over 2}\right)
+ {g_sN_f\over 4\pi}\log\left({\alpha_{\theta_1}\alpha_{\theta_2}\over 4\sqrt{N}}\right)\right] \log~ r \Bigg\}, \nonumber\\ \nd
where $r_h$ is the horizon radius, and both $g_sN_f$ as well as ${g_sM^2\over N}$ are expectedly
small\footnote{In section \ref{nf0} we took $g_s \to 0$ with $N, M$ very large and $N_f$ vanishing
such that ${g_sM^2\over N} << 1$ and $g_sN_f = 0$. Here we take $g_s < 1$ and $N_f = {\cal O}(1)$ with
$N, M$ still very large. Again ${g_sM^2\over N} << 1$, but $g_sN_f < 1$. The latter can be implemented, for example, by choosing $g_s \sim 0.4$ and $N_f \sim 2$. Such a choice will guarantee that
$\left(g_sN_f\right)^m \left({g_sM^2\over N}\right)^n << 1$ even for $n = 1$ and $m = \mathbb{Z}$. Note however that
$g_s\to 0$ does not always imply $g^2_{YM} \to 0$. We can have $g^2_{YM} = {\cal O}(1)$ when $N_f \ne 0$. This will be elaborated in section \ref{gsnf1}. \label{7khoon}}.
Note also that both the metric components are independent of the resolution parameter $a^2$. In fact the only metric component that depends on the resolution parameter would be the $g_{rr}$ component, whose explicit value is given by:

{\footnotesize
\bg\label{metdef22}
& & g_{rr}= {r^2\sqrt{4\pi g_sN}\over r^4-r_h^4}\left({6a^2 + r^2\over 9a^2 + r^2}\right)\Bigg\{1  - {3g_s M^2\over 4\pi N} \left[1 + {3g_sN_f\over 2\pi}\left(\log~r + {1\over 2}\right)
+ {g_sN_f\over 4\pi}\log\left({\alpha_{\theta_1}\alpha_{\theta_2}\over 4\sqrt{N}}\right)\right] \log~ r \Bigg\} \nonumber\\
\nd}
\noindent where the full structure for $a^2$ will be given later. The functional forms for the coefficients appearing in \eqref{metdef} and \eqref{metdef22} are determined by mapping the local metric to the warped resolved conifold
metric\footnote{Recall that globally we can only put a non-K\"ahler metric on the resolved conifold \cite{DEM}.} with a resolution parameter $a^2$. In addition to that,  and in the MQGP limit of \cite{MQGP}, the $\alpha_{\theta_k}$ factors for
$k  = 1, 2$ are angular coordinates such that
around:
\bg\label{harper}
 \theta_1\sim {\alpha_{\theta_1}\over N^{\frac{1}{5}}}, ~~~~~~~ \theta_2\sim {\alpha_{\theta_2}\over {N^{\frac{3}{10}}}},
 \nd
we can allow the decoupling of the
five-dimensional spacetime $M_5(t,x_{1,2,3},u)$ from the internal six-dimensional space $M_6(\theta_{1,2},\phi_{1,2},\psi,x_{10})$. This decoupling is affected by making the
Kaluza-Klein (KK) modes very heavy.

The above discussions more or less summarizes the mirror construction as well its M-theory uplift. However it would be instructive to compare this with the type IIA brane construction of Fig. \ref{2abranes}, which deals with both the UV and the IR brane configurations. The IR picture is of course the
Klebanov-Strassler construction which is got by making a
single T-duality along a direction orthogonal to the wrapped D5-brane world volume, i.e along $z$ of (\ref{xyzdefs}). This  yields the RHS of Fig. \ref{2abranes}, if we ignore the parallel NS5-brane. In other words, we get $M$
D4-branes straddling between a pair of orthogonal $NS5$-branes whose world-volume directions are parametrized by
 $(\theta_1,x)$ and $(\theta_2,y)$ \cite{uranga, mukhi}.  The mirror picture discussed here is then got by making two further T-dualities along $x$ and $y$ directions. Each of these T-dualities would yield Taub-NUT spaces from
 the corresponding NS5-branes \cite{Tong}. The $N_f$ flavor D7-branes would yield $N_f$ D6-branes that are then uplifted to M-theory as KK monopoles \cite{senKK}. These are also Taub-NUT spaces.  Combining everything together then leads to a seven-dimensional manifold with a $G_2$ structure and with G-fluxes. This configuration is precisely equivalent to the uplift of the wrapped D5-branes on a warped resolved conifold of \cite{chenGT, anke1, DEM}.

\subsection{Quasi-normal modes, attenuation constant and the sound speed \label{lostgls}}

Let us now discuss the main ingredient of our construction, namely the quasi-normal modes in the dual gravitational background. The procedure involves a few steps that we lay down
in the following.

Building up on the ideas developed in \cite{klebanov quasinormal} and \cite{EPJC-2}, and using gauge-invariant combinations of metric perturbations invariant under infinitesimal diffeomorphisms, in other
words:
\bg\label{nova}
 h_{\mu\nu} ~\rightarrow ~ h_{\mu\nu}+\nabla_{(\mu}\xi_{\nu)},
 \nd
 as discussed in \cite{klebanov quasinormal}, the gauge-invariant combination of scalar modes of (M-theory) metric\footnote{As discussed above, this corresponds to the local uplift of the delocalized Strominger-Yau-Zaslow \cite{syz} type IIA mirror of the holographic type IIB dual of \cite{metrics} of large-$N$ thermal QCD, having integrated out the six angular
 directions as in \cite{Herzog-vs}, up to NLO in $N$ in the MQGP limit of \cite{MQGP}.} perturbations was constructed in \cite{EPJC-2}. A discussion of the same appears in Appendix {\bf A}.

 Next, we work near the decoupling limit prescribed in \eqref{harper}, and choose the other three angular coordinate $(\psi, \phi_{1, 2})$ in the mirror metric \eqref{Mtheorymet} as
  $\psi=2n\pi$, with $n=0,1,2$ and small $\phi_{1,2}$. We also choose our radial variable henceforth as  $u\equiv\frac{r_h}{r}$. Using these we can define:
\bg\label{Bdef}
{{\cal B}(u)\over 2N} = {3g_s M^2\over 4\pi N} \left[1 + {3g_sN_f\over 2\pi}\left(\log~{r_h\over u} + {1\over 2}\right)
+ {g_sN_f\over 4\pi}\log\left({\alpha_{\theta_1}\alpha_{\theta_2}\over 4\sqrt{N}}\right)\right] \log ~{r_h\over u}, \nd
in the context of the gravitational dual of large-$N$ thermal QCD with $N_f \neq 0$,  where $N_f$ is the number of flavors. This functional form of ${\cal B}(u)$ appears in the construction of
the gauge-invariant $Z_s(u)$ in the following way:

{\footnotesize
\bg\label{gaugu}
Z_s(u)=H_{yy} \left(q^2 + {q^2 u^4\over \pi^2 T^2}-w^2-\frac{{\cal B}^\prime(u) q^2 u^5 N_f g_s^2 }{2 N}\right)+q^2 \left(u^4 - 1\right)
   H_{tt}+2 q w H_{tx}+w^2 H_{xx}, \nd}
\noindent where the $H_{ab}$ functions are given in appendix {\bf A}, and $T$ is the temperature whose form will be given below. Note that the
upshot of appendix {\bf A} is essentially the construction of the gauge-invariant $Z_s(u)$ that will satisfy certain EOM to be elaborated in the following\footnote{Our emphasis here would
be to determine the  EOM up to NLO in $N$.}.

In obtaining an EOM for $Z_s(u)$, we will make use of $q_3 = \frac{q}{\pi T}, w_3 = \frac{w}{\pi T}$ where $T$ is temperature that appears in \eqref{gaugu} above. We will express $T$
in terms of all the variables that appear in the metric. To proceed, and for later brevity, we start by defining the following quantity:
\bg\label{jkay85}
{\cal C}_{kj}(u) \equiv 1 + {g_sN_f \over 4\pi} \log \left({\alpha_{\theta_1} \alpha_{\theta_2}\over 4\sqrt{N}}\right) + {3g_s N_f\over 2\pi} \left(k ~\log~{r_h\over u} + {2j -1\over 2}\right), \nd
where ($k, j$) will be integers. Now
assuming the resolution to be larger than the deformation in the resolved warped-deformed conifold in the type IIB background of \cite{metrics} in the MQGP limit, and using the
decoupling limit \eqref{harper}, the temperature $T$ may be expressed in the following way (see also \cite{EPJC-2}):
\bg\label{T}
T & = &  {\partial_r\vert G_{00}\vert \over 4\pi\sqrt{\vert G_{00}\vert G_{rr}}} \nonumber\\
&=&  {}^{\rm lim}_{u \to 1} ~~ {r_h\over \pi \sqrt{4\pi g_s N}} \left(1 + {3a^2\over 2r_h^2}\right)\left[1 - {3g_sM^2\over 4\pi N} {\cal C}_{11}(u) \log~r_h\right],
\nd
where $G_{\mu\nu}$ is the M-theory metric \eqref{Mtheorymet}, and
${\cal C}_{11}(u)$ may be extracted from \eqref{jkay85}. We can also go to the limit where $\alpha_{\theta_i}$ are ${\cal O}(1)$ numbers. This way the temperature may be written
completely in terms of the resolution parameter $a^2$ and the horizon radius $r_h$. Interestingly when $a^2 >> r_h^2$, the temperature is expressed in terms of inverse $r_h$. Otherwise,
the temperature is proportional to $r_h$. In the limit of vanishing flavors, small {\it bare} resolution parameter,
and large cut-off, the expression for the temperature becomes identical to what we took
on the type IIB side (see \eqref{thorns} and \eqref{court} and discussion below). The bare resolution parameter in type IIB side, as given in \eqref{fatur}, was taken to be zero. A natural question then is to ask what happens if we take non-zero
bare resolution parameter\footnote{Note that allowing a bare resolution parameter in the type IIB side allows us to perform the SYZ mirror transformation more efficiently \cite{anke1}. Here however we will use the word {\it bare} to denote the part of the resolution parameter that is independent of $g_sN_f$ and ${g_sM^2\over N}$. Of course the $r_h$ independent piece of $a(u)$ vanishes.}.  A
particular choice of $a(u)$ can be:
\bg\label{a}
a(u)  = \left[b + \frac{g_sM^2}{N}\left(c_1 + c_2 \log~ r_h\right)\right]r_h, \nd
this way $b$ may serve as the bare resolution parameter in \eqref{T} and $c_1(u), c_2(u)$ are some slowly varying functions of the $u$ parameter (not to be confused
with $b$ and $c_1, c_2$ taken in \eqref{thorns} and \eqref{court}).
One may compare \eqref{a}
with the type IIB resolution parameter \eqref{fatur} in the limit $b \to 0$. The functional forms in the two cases are similar, but not identical. This is intentional because the choice \eqref{a}
allows us to perform computations in the mirror side more efficiently compared to the choice \eqref{fatur}. This in turn
will also effect some of our final results, so comparison with the type IIB side
will have to be done more carefully. In fact writing \eqref{court} as: 
\bg\label{mentalist}
a_2(x) = c_1 + c_2(x)\log~ x + c_3(x)\log^2 x, \nd
where we have added a ${\rm log}^2 x$ piece with a coefficient $c_3$ to the already existing result, 
 one can show that (using $c_{20} \equiv {3^{3/2}\over 256{\pi}^{3/2}L^2}$):
\begin{eqnarray}
\label{c1-c2-c3_defs}
& & b\vert_{\eqref{thorns}} = \frac{3^{\frac{3}{2}}(2+3b^2)}{8\sqrt{\pi}L^2},~~ c_1(x)\vert_{\eqref{court}} = \frac{4\cdot 3^{\frac{3}{2}}c_1(x)\pi}{8\sqrt{\pi}L^2},~~c_3(x) = -\frac{3^{\frac{5}{2}}(2 + 3 b^2)g_s N_f}{64\sqrt{\pi}L^2}\\
& & c_2(x)\vert_{\eqref{court}} = 
-4\pi c_{20}\left[4 - 16\pi b c_2(x) + 6 b^2\right] 
+ g_s N_f(2 + 3 b^2)c_{20}\left[3 - 2 \log\left(\frac{\alpha_{\theta_1}\alpha_{\theta_2}}{4\sqrt{N}}\right)\right].
\nonumber
\end{eqnarray}
With these definitions at hand, we are now ready to write down the equation of motion for $Z_s(u)$ appearing in \eqref{gaugu}. This may be expressed in the following way:
\begin{equation}\label{Z-EOM}
Z_s^{\prime\prime}(u) = m(u) Z_s^\prime(u) + l(u) Z_s(u),
\end{equation}
which is a second-order differential equation in $u$ whose solutions will tell us the precise gauge-invariant variables that we seek here. This equation depends on two non-trivial
functions of $u$, namely $m(u)$ and $l(u)$,  whose functional form will be important. Both these functions may be expressed in terms of ${\cal C}_{kj}(u)$ in \eqref{jkay85} as well as certain
other functions that we shall elaborate in the following. We start with $m(u)$ which may be written as:
\bg\label{novharpx}
m(u) \equiv && \frac{1}{4
   \left(u^4-1\right) \left[{q_3}^2 \left(u^4-3\right)+3 {w_3}^2\right]^2} \Bigg\{{{\cal A}_1(u)\over u} + {g_sM^2\over N} \bigg[8bu\left(c_1 + c_2 \log~r_h\right) {\cal A}_2(u) \nonumber\\
  && ~~~~ - {9b^2 u\over 2\pi} {\cal C}_{21}(u) {\cal A}_3(u) + {3\over 2\pi u} {\cal C}_{23}(u) {\cal A}_4(u)
- {9\over \pi u} {\cal C}_{21}(u){\cal A}_5(u)\bigg]\Bigg\}, \nd
where note the appearance of ${\cal C}_{21}(u)$ and ${\cal C}_{23}(u)$  defined in \eqref{jkay85} as well as ${\cal A}_i(u)$ that form the various coefficients above.  The function ${\cal A}_1(u)$ may be
written as:
\bg\label{A1}
{\cal A}_1(u) \equiv && 4 \left({q_3}^2 \left(u^4-3\right)+3 {w_3}^2\right) \Big[{q_3}^2 \left(b^2 \left(u^8+2 u^4-3\right) u^2+7 u^8-8 u^4+9\right) \nonumber\\
&& ~~~~~ -{w_3}^2\left(b^2 \left(5 u^4-3\right) u^2+3 \left(u^4+3\right)\right)\Big], \nd
which expectedly simplifies for vanishing $b$. On the other hand, at the boundary when $u$ vanishes,  ${\cal A}_1(0)$ is proportional to $\left(\omega_3^2 - q_3^2\right)^2$ which is
now expressed in terms of $T$ defined at $u \to 0$ instead. Similarly,  ${\cal A}_2(u)$ may be written in the following way:
\bg\label{A2}
{\cal A}_2(u) \equiv {q_3}^4\left(u^{12}-u^8-9 u^4+9\right)-2 {q_3}^2 \left(u^8-12 u^4+9\right) {w_3}^2+3 \left(3-5 u^4\right) {w_3}^4, \nonumber\\ \nd
which is again proportional to $\left(\omega_3^2 - q_3^2\right)^2$ at the boundary $u \to 0$. When $b$ vanishes, ${\cal A}_2$ is unaffected, but the term itself comes multiplied with $b$ in
\eqref{novharpx}, so decouples completely.

 The third term in \eqref{novharpx}  is proportional to $b^2$, so we expect it to decouple in the limit of vanishing $b$.  To see what happens at the boundary, i.e when $u \to 0$, we express
 ${\cal A}_3(u)$ as:

{\footnotesize
\bg\label{A3}
{\cal A}_3(u) \equiv {3 {q_3}^4 \left(u^8-4 u^4+3\right)^2+2 {q_3}^2 \left(10 u^{12}-43 u^8+60 u^4-27\right) {w_3}^2+\left(17
   u^8-48 u^4+27\right) {w_3}^4\over u^4-3}, \nonumber\\ \nd}
\noindent  which is expectedly proportional to $\left(\omega_3^2 - q_3^2\right)^2$, but the term itself decouples because it appears together with a factor of $u$ in \eqref{novharpx},
 much like the previous term in \eqref{novharpx}.

 The remaining two coefficients, ${\cal A}_4(u)$ and ${\cal A}_5(u)$, are in similar vein as \eqref{A1}, \eqref{A2} and \eqref{A3} and share much of the same properties as above. They
 take the following form:
 \bg\label{A4A5}
 &&{\cal A}_4(u) \equiv  {\left(u^4-1\right) \left({q_3}^2 \left(u^4-3\right)+3 {w_3}^2\right) \left({q_3}^2 \left(7
   u^4-3\right)+3 {w_3}^2\right)\over u^4 - 3}\\
&& {\cal A}_5(u) \equiv    {\left(u^4-1\right) \left({q_3}^4 \left(4 u^8-25 u^4+15\right)+5 {q_3}^2 \left(5 u^4-6\right) {w_3}^2+15
   {w_3}^4\right) \over u^4 - 3}, \nonumber \nd
 and become proportional to $(\omega_3^2 - q_3^2)^2$ when $u \to 0$ at the boundary, but do not decouple in a simple way as before. In this sense they share the property of the first term
 in the definition \eqref{novharpx}. The boundary behavior of $m(u)$ can then be given by the following limiting expression:
 \bg\label{mub}
 m(0) = {}^{\rm lim}_{u \to 0} ~{36 - \epsilon\left[{\cal C}_{23}(u) - 10 {\cal C}_{21}(u)\right]\over 12 u}, \nd
 where $\epsilon \equiv {3g_sM^2\over 2\pi N}$ is the same expansion parameter that we used in section \ref{nf0}. Both the ${\cal C}_{2j}$ factors behave as $\log ~u$, but are suppressed
 by $\epsilon$ as well as $g_sN_f$ \eqref{jkay85} (any constant factors get suppressed by $\epsilon$ from \eqref{novharpx}). Thus  $m(0)$ seems to blow up as ${1\over u}$ or
 ${\log~u \over u}$. This preliminary analysis however is naive
 because precisely in this regime the UV cap modifies the boundary behavior appropriately to avoid any such pitfalls. Therefore a more
 relevant question to ask is the behavior of $m(u)$ at the horizon, i.e when $u \to 1$. We will analyze this below, but before that let us discuss the behavior of the other function $l(u)$ appearing
 in \eqref{Z-EOM}.

The expression for $l(u)$ turns out to be very large so we shall suffice ourselves by demonstrating that the horizon $u=1$ is an irregular singular point whenever $N\neq0$.
In the following we give below the expansion of $l(u)$ about $u=1$ to see the same:

{\footnotesize
\bg\label{lu1}
l(u\to 1) = && {\omega_3^4 b^2 \left(6 + q_3^2\right)\over 128 q_3^4 (u -1)^3} +
{3b^2 \omega_3^4\over 1024 \pi q_3^2 (u - 1)^3}\left({g_sM^2\over N}\right) \left[{\cal C}_{21}(1) + {16\pi\over 3b} \left(6 + q_3^2\right)\left(c_1 + c_2 \log~r_h\right)\right]\nonumber\\
&& ~~~~~~~~~~~~~  + {\cal O}\left[{1\over (u -1)^2}\right], \nd}
\noindent where ${\cal C}_{21}(1)$ may be extracted from \eqref{jkay85} by putting $u = 1$ therein. For vanishing bare resolution parameter \eqref{lu1} vanishes, so a minimal resolution is
necessary to see the behavior at the horizon.

The above expression for $l(u)$ near the horizon is what we need, and we could also go to the $u \to 1$ limit for $m(u)$ in \eqref{novharpx} to determine its behavior at the horizon. However
the results are expressed in terms of both $q_3$ and $\omega_3$. To elaborate further, we need to first express $\omega_3$ in terms of $q_3$ and then identify the subsequent behavior of
$m(u)$ and $l(u)$ at the horizon. To this effect, we make the following ansatz:
\begin{eqnarray}
\label{dispersion}
\omega_3 = \left(\frac{1}{\sqrt{3}} + \alpha \frac{g_s M^2}{N}\right)q_3 + \left(-\frac{i}{6} + \beta\frac{g_s M^2}{N}\right)q_3^2,
\end{eqnarray}
and substitute into $m(u)$ and $l(u)$. Here $\alpha(u)$ and $\beta(u)$ are certain functions whose values will be determined near the horizon.
We then first perform a small $q_3$ expansion, followed by an expansion around $u=1$ and lastly a large $N$ expansion. The procedure is straightforward albeit a little tedious. After
the dust settles,
we come up with the following expansions for $m(u)$ by keeping terms up to ${\cal O}\left(q_3,\frac{g_sM^2}{N}\right)$ and the most singular term in $u$
near $u=1$, namely:
\bg\label{mu1}
&& m(u\to 1)  =  {b^2 - 6\over 6(u -1)} - {b^2\over 16\pi (u - 1)}\left({g_sM^2\over N}\right)\left[{\cal C}_{21}(1)+ {32\pi\alpha\over \sqrt{3}} + {16\pi \over 3b}\left(c_1 + c_2 \log~r_h\right)\right]
\nonumber\\
&&~~~ - {ib^2q_3\over 3\sqrt{3}(u - 1)} + {iq_3 \sqrt{3} b^2\over 12\pi (u - 1)} \left({g_sM^2\over N}\right)\left[ {\cal C}_{21}(1) - i8\pi \beta - {20\pi\alpha \over \sqrt{3}}
- {8\pi\over b}\left(c_1 + c_2 \log~r_h\right)\right], \nonumber\\ \nd
with ${\cal C}_{21}$ as in \eqref{jkay85}. For vanishing bare resolution parameter, there is a further simplification: $m(u \to 1)$ behaves simply as ${1\over 1 - u}$ as may be easily seen
from \eqref{mu1}.  On the other hand, the behavior of $l(u)$ at the horizon may be read more directly from \eqref{lu1} as:
\bg\label{lu11}
&&l(u\to 1)  =  {bq_3 \over 288(u-1)^3}\left({g_sM^2\over N}\right)\left[2\sqrt{3}\left(3b\beta - ic_1\right) - 9 i\alpha b - 2i\sqrt{3} c_2 \log~r_h\right] \nonumber\\
&& ~~ +  {b\over 96(u-1)^3}\left({g_sM^2\over N}\right)\left(2\sqrt{3} \alpha b + c_1 + c_2 \log~r_h\right) + {b^2\over 96(u -1)^3}\left({1\over 2} - {iq_3\over \sqrt{3}}\right),  \nonumber\\ \nd
which expectedly vanishes for vanishing bare resolution parameter, and has the required irregular singular point.

The functional forms for $m(u)$ and $l(u)$, expressed using the dispersion relation \eqref{dispersion}, and analyzed near the horizon $u \to 1$ is essentially the regime that we
want to concentrate here. We can also study the system at the boundary by attaching an appropriate UV cap controlling, in turn, the behavior of $m(u)$ and $l(u)$, but this will not be the
emphasis of this section.  Our aim would be to explore the near horizon behavior where
one sees  $u=1$ as an irregular singular point of (\ref{Z-EOM}). To proceed, let us make the following ansatz for $Z_s(u)$:
\bg\label{zadu}
Z_s(u) ~= ~  e^{S(u)},
\nd
where we shall assume
$\left[S^\prime(u\sim1)\right]^2 > |S^{\prime\prime}(u\sim1)|$. This derivative requirement essentially converts \eqref{Z-EOM} to a simple quadratic equation in $S'(u)$ with coefficients $m(u)$ and
$l(u)$. The solutions are:
\bg\label{quadu}
S'(u\to 1) = {}^{\rm lim}_{u\to 1} ~\frac{1}{2}\left(m(u)\pm\sqrt{m^2(u)+4l(u)}\right). \nd
At this stage it would be interesting to ask what happens when the derivative condition is not satisfied. Clearly in this case we will get a second order inhomogeneous differential equation
which becomes homogenous when the bare resolution parameter vanishes. Generically it is harder to deal with the inhomogeneous case, because of the complicated forms of
$m(u)$ and $l(u)$, and the homogenous form is not a suitable choice for the system undergoing SYZ transformations \cite{syz, anke1}. Thus the simplification and calculability attained
from the derivative requirement guarantee not only analytic control, but also solutions not far from the regime of interest. With this in mind,
the next set of steps are standard\footnote{Although we do not undertake here, a more generic analysis away from
$u = 1$ can be performed and from there the limiting form of \eqref{sup} can be ascertained. Needless to say, the results match.}.
Choosing the minus sign in \eqref{quadu}, one obtains the following:
\bg\label{sup}
S'(u\to 1) & =& {\left(b^2 - 6\right)\sqrt{3} -2ib^2 q_3\over 12 \sqrt{3} (u -1)}
-{\sqrt{{bg_sM^2\over N}\left({\cal B}_1 + {{\cal B}_2\over 3}\right) + b^2\left({1\over 2} - {iq_3\over \sqrt{3}}\right)} \over 4\sqrt{6} (u - 1)^{3/2}} \\
&-&{b^2\over 32\pi (u - 1)}\left({g_sM^2\over N}\right)\left[{\cal C}_{21}(1) -{32\pi\alpha\over \sqrt{3}} - {16\pi\over 3b}\left(c_1 + c_2 \log~r_h\right)\right]\nonumber\\
&+& {i\sqrt{3}b^2q_3\over 24(u-1)}\left({g_sM^2\over N}\right)\left[{\cal C}_{21}(1) - i8\pi\beta - {8\pi\over 3b}\left(c_1 + c_2 \log~r_h\right) - {20\pi\alpha\over \sqrt{3}}\right], \nonumber \nd
which has a simple pole structure of $-{1\over 2(u - 1)}$ in the limit of vanishing bare resolution parameter $b$.  The other parameters appearing in \eqref{sup} are ${\cal C}_{21}(u)$
defined in \eqref{jkay85}, and the two functions ${\cal B}_1$ and ${\cal B}_2$ defined in the following way:
\bg\label{kalikit}
&&{\cal B}_1 \equiv  2\sqrt{3} \alpha b + c_1 + c_2 \log~r_h \nonumber\\
&&{\cal B}_2 \equiv  3\sqrt{3} b\left(2\beta - i \sqrt{3} \alpha\right) - 2i\sqrt{3}\left(c_1 + c_2 \log~r_h\right). \nd
Let us now assume that
$q_3\rightarrow0$ as $N^{-1-\kappa}$ with $\kappa>0$. One might worry that imposing this one would obtain, near $u=1$ $-$ which is an irregular singular point $-$
a solution of the type $e^{S(u)} = (1 - u)^{\gamma}F(u)$ implying  $u=1$ to be a regular singular point. This does not happen, and therefore demanding the
vanishing of the residue of $S^\prime(u)$ at $u=1$ gives the following values for $\beta$, $b$, and 
$\alpha$:
\bg\label{nullres}
&& \beta = -{{3i}{\cal C}_{21}(1)\over 64} - {i\sqrt{6}\left(c_1 + c_2\log~r_h\right)\over 72}\nonumber\\
&&b \approx \sqrt{6}, ~~~\alpha = {\sqrt{3}{\cal C}_{21}(1)\over 32\pi}  - {c_1 + c_2\log~r_h\over 6\sqrt{2}},
 \nd
where ${\cal C}_{21}(u)$ is defined in \eqref{jkay85}. In fact this is all we needed to determine both the sound speed $c_s$ as well as the attenuation constant $\Gamma$ because the first
term in \eqref{dispersion}, i.e the term proportional to $q_3$, gives us the sound speed as\footnote{Recall that we can express the dispersion relation \eqref{dispersion} in terms
of sound speed $c_s$, shear viscosity $\eta$ and bulk viscosity $\zeta$ as: $$ \omega_3 = c_s q_3 -{i\pi\over 2s}\left(\zeta + {4\eta\over 3}\right)q_3^2 $$ \noindent where $s$ is the entropy
density. This means $\alpha(u)$ in \eqref{dispersion} is related to $c_s$ and $\beta(u)$ in \eqref{dispersion} is related to shear viscosity and bulk viscosity combination, or the attenuation
constant $\Gamma$. However since we are analyzing the system close to the horizon, i.e $u \to 1$, the relevant parameters for us will be $\alpha(1)$ and $\beta(1)$.}:
\bg\label{cs}
c_s \equiv {1\over \sqrt{3}} + {\sqrt{3}\over 32\pi}\left({g_sM^2\over N}\right) {\cal C}_{21}(1) - {g_sM^2\over 6\sqrt{2} N}\left(c_1 + c_2\log~r_h\right), \nd
where one can see that the result differs from the conformal answer of ${1\over \sqrt{3}}$ expectedly by the ${g_sM^2\over N}$ and $g_sN_f$ factors.  Even in the absence of the fundamental
flavors $N_f$, the sound speed deviates from the conformal answer. The form of the deviation is consistent with what we had earlier in \eqref{speedu}, or more generically in \eqref{seedsof}, although the precise factors
differ. This is understandable in the light of the different choices of the supergravity parameters in the type IIB and the M-theory pictures. Interestingly, if one drops the ${\cal O}\left(\frac{\epsilon r_h^4}{r_c^4}\right)$  terms (as $|\log ~r_h|\gg r_h^4$) in addition to the ${\cal O}(\epsilon^2)$ terms in \eqref{notielle} in the type IIB computation of $c_s$, the $c_s$ in \eqref{notielle} from a type IIB computation can then be shown to match up with the M theory result (for $N_f=0$, explicitly in ${\cal C}_{21}(1)$) given in equation \eqref{cs} for the following choice of the bare resolution parameter\footnote{There is of course one crucial difference between \eqref{cs} and \eqref{notielle} in the sense that the former is the renormalized answer, so only depends on the scale $r_d$, whereas the latter is the Wilsonian cut-off dependent answer, and therefore depends on the cut-off radius $r_c$.}: 
\bg\label{cuerpo}
b = \frac{64\sqrt{3}|c_2|}{8\sqrt{2}\pi(|\log ~r_h||c_2| - |c_1|) - 3^{\frac{3}{2}}}. \nd
To ensure that $b>0$ and $b={\cal O}(1)$, as has been assumed in this section, this would imply a further 
constraint on $|c_{1,2}|$, namely: 
\bg\label{100days}
|\log ~r_h||c_2| - |c_1| = {\cal O}(1) |c_2|\cap  {\cal O}(1)>\frac{3^{\frac{3}{2}}}{8\sqrt{2}\pi|c_2|}. \nd
However, it must be kept in mind that the M-theory-uplift calculation presented here is expected to yield a better result because of two reasons: (i) it is essentially a pure supergravity calculation (unlike the type IIB which also involves branes), and (ii) it permits considering finite $g_s$ thus encompassing the full quantum corrections.

The attenuation constant $\Gamma$ may now be easily extracted from \eqref{dispersion} by plugging in the value of $\beta$ from \eqref{nullres}. To NLO in $N$, $\Gamma$ may be
written as:
\bg\label{Gamma}
\Gamma  \equiv {1\over \pi T}\left[{1\over 6} + {3 g_sM^2\over 64\pi N}\left({\cal C}_{21}(1) + {8\sqrt{6}\pi\over 27}\left(c_1 + c_2\log~r_h\right)\right)\right], \nd
where $T$ is the temperature, and again we see that even in the absence of fundamental flavors, the attenuation constant differs from the conformal value of
${1\over 6\pi T}$.  The
parameter ${\cal C}_{21}(1)$, defined in \eqref{jkay85}, becomes identity when $N_f = 0$, so the deviation from the conformal value is solely governed by
${g_sM^2\over N}$.

\subsection{The case with a vanishing bare resolution parameter \label{zerbare}}

Let us now discuss what happens if one sets $b=0$ in (\ref{a}).  We briefly dwelt on this earlier, wherein we saw how \eqref{novharpx} and \eqref{lu1} behave when $b$ vanishes:
\eqref{lu1} completely decouples but some remnants of \eqref{novharpx}, as seen from \eqref{A1}, \eqref{A2}, \eqref{A3} and \eqref{A4A5}, survive.
Interestingly, this now makes $u=1$ a regular singular point of (\ref{Z-EOM}). To proceed, let us then rewrite $Z_s(u)$ using an analytic function $F(u)$ in the following way:
\bg\label{fudef}
Z_s(u) \equiv  (1- u)^{-{i\over 8}\sqrt{\epsilon {\cal C}_{23}(1)}} F(u), \nd
where ${\cal C}_{23}(1)$ can be evaluated from  \eqref{jkay85} by putting $u = 1$ in the required expression and we have defined $\epsilon \equiv {3g_sM^2\over 2\pi N}$ as the
non-conformality factor. With the definition \eqref{fudef},
the EOM (\ref{Z-EOM}) becomes:
\begin{equation}
\label{F-EOM}
1024 F^{\prime\prime} + \left(\frac{a_1 + a_2 \log~ u}{u}\right)F^\prime(u) + \frac{b_2}{u^4}F(u) = 0,
\end{equation}
which is a second order homogenous differential equation with coefficients defined by parameters $a_1, a_2$ and $b_2$.  The ${1\over u}$ and
${\log~u\over u}$ terms are remnants of the equivalent terms in \eqref{mub}. The $a_1$ and $a_2$ coefficients take the following form:
\bg\label{dolce}
a_1 = {384\over \pi} \left[-8\pi + {9g_sM^2\over N} \left({\cal C}_{21}(1) + 24 g_sN_f\right)\right], ~~ a_2 = - {10368\over \pi^2}\cdot g_s N_f \cdot {g_sM^2\over N}, \nd
where ${\cal C}_{21}(1)$ is extracted from ${\cal C}_{21}(u)$ in \eqref{jkay85}. It is interesting to note that the combined expression with $a_1$ and $a_2$ may be succinctly
expressed as:
\bg\label{kandy}
a_1 + a_2\log~u = {384\over \pi} \left[-8\pi + {9g_sM^2\over N} \left({\cal C}_{21}(u) + 24 g_sN_f\right)\right], \nd
which simply converts ${\cal C}_{21}(1)$  in \eqref{dolce}
to ${\cal C}_{21}(u)$.  This is expected from the way we represented the EOM for $Z_s$ in \eqref{Z-EOM}. On the other hand, the form for $b_2$
in \eqref{F-EOM} may be expressed as:
\bg\label{b2}
b_2 = {2q_3^2 {\cal C}_{23}(u)\over 9\pi}\left({g_sM^2\over N}\right)\left({\cal D} + 2\sqrt{3} + 6\right)^2\left({\cal D} + 2\sqrt{3} - 6\right)^2, \nd
where ${\cal C}_{23}(u)$ is given in \eqref{jkay85} (note the appearance of ${\cal C}_{23}(u)$  instead of ${\cal C}_{23}(1)$, much like what we have in \eqref{kandy}) and ${\cal D}$
is defined in the following way:
\bg\label{semil}
{\cal D} \equiv 6(\alpha + \beta q_3){g_sM^2\over N} - iq_3,\nd
where we see that there are terms in $b_2$ \eqref{b2} that are of ${\cal O}\left(\epsilon^2\right)$ which would help us to simplify the third term in the EOM \eqref{F-EOM}.
On the other hand, looking at \eqref{kandy} and \eqref{b2},
simplifications in both the second and the third terms in \eqref{F-EOM} can happen if we go from ${\cal C}_{21}(u)$ to ${\cal C}_{21}(1)$. Implementing this,
$F(u)$ takes the following  form:
\bg\label{FUsol}
F(u) & = & \left({1\over u}\right)^{{1\over 2}(a_1 + |a_1 - 1|-1)}\Bigg[d_1 e^{-{{\sqrt{-b_2}}\over u}}
{}_1{\bf F}_1\left({1\over 2}\left(|a_1-1|+1\right); |a_1-1|+1; {2\sqrt{-b_2}\over u}\right)\nonumber\\
&& ~~~~~~~~ + {2^{-{1\over 2}|a_1-1|}d_2\over \sqrt{\pi}}\left({\sqrt{-b_2}\over u}\right)^{-{1\over 2}|a_1-1|}
{\bf K}_{{1\over 2}|a_1-1|}\left({\sqrt{-b_2}\over u}\right)\Bigg], \nd
where $d_1$ and $d_2$ are constants. We can also
motivate the replacement ${\cal C}_{21}(u) \to {\cal C}_{21}(1)$ in both \eqref{kandy} and \eqref{b2} in the following way. In
\eqref{kandy}, this amounts to dropping the $\log~u$ term near $u = 0$ compared to the $\log~N$ term in the large $N$ limit i.e making $a_2 = 0$ in \eqref{F-EOM}. In \eqref{b2}, this
amounts to just keeping terms of ${\cal O}\left({g_sM^2\over N}\right)$ as the $\log~u$ term in ${\cal C}_{21}(u)$ is already suppressed by ${g_sN_f}$ (see \eqref{jkay85}).

The $u \to 0$ limit may seem a bit puzzling
because so far we have analyzed the system near the horizon i.e near $u \to 1$. However in \eqref{Z-EOM}, for vanishing bare resolution parameter, as we saw earlier,
$l(u)$ vanishes and the EOM is solely governed by $m(u)$ \eqref{novharpx}. This may be defined both at the boundary \eqref{mub} and at the horizon \eqref{mu1}. Thus extrapolating $F(u)$
to the boundary is still well defined, modulo the subtlety of including a UV cap.

Let us now go to the various choices of the parameters $d_1$ and $d_2$ in the solution \eqref{FUsol}.
If one sets $d_2 =0$ then the small-$u$ expansion of $F(u)$ will be given by the small-$u$ expansion of the first part of the solution \eqref{FUsol}, i.e
the $d_1$ part of the solution in \eqref{FUsol}, as:
\bg\label{fu0}
& & F(u) =  \left(\frac{1}{u}\right)^{\frac{1}{2} \left({a_1}+|a_1-1|-1\right)} d_1 e^{-\frac{\sqrt{-{b_2}}}{u}} \, _1{\bf F}_1\left(\frac{1}{2}   \left(|a_1-1|+1\right);|a_1-1|+1;\frac{2 \sqrt{-{b_2}}}{u}\right) \nonumber\\
   & & = \frac{1}{\Gamma \left(\frac{1}{2}\left(|a_1-1|+1\right)\right)}\biggl\{2^{-\frac{1}{2} |a_1-1|-\frac{1}{2}} d_1 \left(-\sqrt{-{b_2}}\right)^{-\frac{1}{2} |a_1-1|-\frac{1}{2}}\\
   & & \times  ~  u^{\frac{1-{a_1}}{2}+\frac{1}{2}} \Gamma \left(|a_1-1|+1\right) e^{-\frac{\sqrt{-{b_2}}}{u}} \left[(-{b_2})^{\frac{1}{4}
   |a_1-1|+\frac{1}{4}} e^{\frac{2 \sqrt{-{b_2}}}{u}}+\left(1+O\left(u^1\right)\right)\right]\biggr\}. \nonumber
\nd
To analyze the boundary conditions, first let us make ${\rm Im}~b_2 = 0$. There is already a problem at this stage, but let us still carry on.
At the boundary $u = 0$, if ${\rm Re}~b_2 <  0$ then \eqref{fu0} blows up as
${\rm exp}\left({\sqrt{|{\rm Re}~b_2|}\over u}\right)$. On the other hand, if ${\rm Re}~b_2 > 0$, then
\eqref{fu0} oscillates infinitely fast as  ${\rm exp}\left({i\sqrt{|{\rm Re}~b_2|}\over u}\right)$.
This behavior will persist even if we include the $a_2 \log~u$ piece in \eqref{F-EOM}.
Thus to be able to impose Dirichlet boundary condition on $F(u)$, i.e impose $F(u)=0$ at the boundary, one needs to set $b_2=0$. Now, substituting $b_2=0$ in (\ref{F-EOM}), one obtains:
\begin{equation}
\label{myas}
F(u) = {16 \sqrt{2 \pi} d_3 \over \sqrt{a_2}}~ {\rm exp}\left[{(a_1-1024)^2\over 2048 a_2}\right] {\bf erf}\left({a_1+a_2 \log~u-1024\over 32 \sqrt{2a_2}}\right) + d_4,
\end{equation}
where $d_3$ and $d_4$ are constants. We can fix $d_4$ in terms of $d_3$ by demanding Dirichlet boundary condition on $F(u)$. This immediately gives us:
\bg\label{slapb}
d_4 = \pm 16 {\sqrt{2\pi \over a_2}} {\rm exp}\left[{(a_1 - 1024)^2\over 2048 a_2}\right] d_3. \nd
Plugging \eqref{slapb} in \eqref{myas} now determines $F(u)$ up to an overall constant. This form of $F(u)$ may be used in \eqref{fudef} to determine the gauge invariant combination
$Z_s(u)$.
This is almost what we need, except for an important caveat. Putting $b_2 = 0$ (or ${\rm Im}~b_2 = 0$) in \eqref{b2} gives us:
\bg\label{marpac}
{\cal D} + 2\sqrt{3} \pm 6 = 0, \nd
where ${\cal D}$ is defined in terms of $\alpha, \beta$, $q_3$ as well as ${g_sM^2\over N}$
in \eqref{semil}. Since the RHS of \eqref{marpac} is a c-number, and $\beta$ defined in \eqref{dispersion} is a pure imaginary number (at least at the horizon),
the equation \eqref{marpac} can only be solved if:
\bg\label{chukka}
\alpha = \left(1-{1\over \sqrt{3}}\right) \left({g_sM^2\over N}\right)^{-1}, ~~~~~\beta \equiv {i\over 6} \left({g_sM^2\over N}\right)^{-1}. \nd
The above forms of $\alpha$ and $\beta$ are unfortunately not acceptable as they will not only lead to the wrong sound speed and
attenuation constant, but also take us away from the perturbative regime where
all our computations were focussed. One might think that this could be rectified if  we had started off with a non-zero ${\rm Im}~b_2$,
but unfortunately the conclusions don't  change much as $F(u)$ would
still oscillate infinitely fast or blow up.

The above failure is a near miss, but it teaches us an important lesson about the choice of the function $F(u)$: the boundary conditions are subtle and important, but one still needs to
choose the function carefully, as any arbitrary choice may take us away from the perturbative regime of interest. Therefore
at this stage there are two ways to fix the function $F(u)$. One, we may not
impose a Dirichlet boundary condition, and allow a non-normalizable functional form for $F(u)$. Two, we again allow for a Dirchlet boundary condition, but choose the functional form for
$F(u)$ a little differently from the previous choice \eqref{fu0}. The latter case is easier to implement, so we start by setting $d_1 = 0$ in \eqref{FUsol}. This gives us:
\begin{eqnarray}
\label{K-u=0}
 F(u) &=&  \left(\frac{1}{u}\right)^{\frac{1}{2} \left({a_1}+|a_1-1|-1\right)}
 {2^{-{1\over 2}|a_1-1|}d_2\over \sqrt{\pi}}\left({\sqrt{-b_2}\over u}\right)^{-{1\over 2}|a_1-1|}
{\bf K}_{{1\over 2}|a_1-1|}\left({\sqrt{-b_2}\over u}\right) \\
   &=& \left(\frac{1}{u}\right)^{\frac{1}{2} \left({a_1}+|a_1-1|\right)} d_2 e^{-\frac{\sqrt{-{b_2}}}{u}}
   \left(\frac{\sqrt{-{b_2}}}{u}\right)^{-\frac{1}{2} |a_1-1|} \left(\frac{2^{-\frac{1}{2} |a_1-1|-\frac{1}{2}}
   u}{\sqrt[4]{-{b_2}}}+O\left(u^2\right)\right).\nonumber
\end{eqnarray}
We see that if ${\rm Im}~b_2 = b_2 = 0$ then we would encounter similar problem as in \eqref{chukka}. On the other hand, if we allow ${\rm Re}~b_2 < 0$,
then we can control the amplitude of oscillation from the ${\rm Im}~b_2$ piece, provided:
\bg\label{tagram}
\vert {\rm Re}~b_2\vert   >>  \left({\rm Im}~b_2\right)^2. \nd
The above set of conditions does help us to solve for $F(u)$ as before allowing the required Dirichlet boundary condition at $u = 0$, although the procedure for
getting the exact functional form for $F(u)$ is not as straightforward as in \eqref{myas}. However the condition ${\rm Re}~b_2 < 0$ now leads to the following
condition on $\alpha$ and $\beta \equiv i\gamma$:
\bg\label{chulkut}
2\pi\epsilon\left(4\sqrt{3} \gamma - 12\alpha\right)q_3 - 12\sqrt{3}q_3 >  4\pi\epsilon\left(\gamma q_3^2 + 12\sqrt{3} \alpha\right) -3( q_3^2 + 24), \nd
where $\epsilon = {3g_sM^2\over 2\pi N}$ is the non-conformal factor. Although the above condition gets further refined by \eqref{tagram},
getting $\alpha$ and $\gamma$ satisfying \eqref{chulkut} can at least indicate the behavior of $\alpha$ and $\beta$ with respect to ${g_sM^2\over N}$.

A careful look at \eqref{chulkut} tells us that if both $\alpha$ and $\gamma$ are proportional to $\epsilon$, then $q_3$ gets constrained. This cannot be right, so
it seems the only way to satisfy \eqref{chulkut} would be to take $\alpha$ and $\gamma$ to be {\it inversely} proportional to $\epsilon$, much like what we
had in \eqref{chukka} before. Such a choice will again take us away from the perturbative regime of interest. Thus it seems the only way to analyze the behavior of
$\alpha$ and $\beta$ from the boundary $u = 0$ point of view
is to allow for a non-normalizable $F(u)$. This resonates well with the analysis of fluctuation modes of the metric
in section  \ref{nf0} where $p_{nk}$ and $\Gamma_{0k}$ functions were both non-normalizable functions. Note that
we did not encounter these issues while studying the $b \ne 0$ case because the analysis was performed at the horizon $u = 1$ where
these subtleties were not visible.

\subsection{Shear viscosity, entropy and the bulk viscosity bound \label{suhaag}}

After this detour, it is time to go back to our analysis of bulk viscosity and the bound on the ratio of the bulk to shear viscosities. To proceed, we will first
quantify the functional forms of $f_1(\theta_1)$ and $f_2(\theta_2)$ in \eqref{SYZl} in the following way \cite{NPB}:
\bg\label{mozart}
f_1(\theta_1) = {{\rm cot}~\theta_1\over \alpha_N}, ~~~~~ f_2(\theta_2) = -\alpha_N~ {\rm cot}~\theta_2, \nd
where $\alpha_N$ and the choice \eqref{harper}
ensure large base for implementing the SYZ \cite{syz} mirror transformation. Recall the necessity of a large base in our set-up to nullify certain
disc instanton contributions.  The choice \eqref{mozart} is essential to compute transport coefficients and entropy in the M-theory uplift of the mirror set-up. We can now
combine this with the value of the bare resolution parameter $b = \sqrt{6}$  that we got in \eqref{nullres}, and
using results of \cite{MQGP}, \cite{EPJC-2}, we can show that the shear viscosity near the horizon takes the following form:

{\footnotesize
 \bg\label{etavalue}
 \eta = {N^{9/10} r_h^3 \Upsilon \sqrt{g_s\pi}\over \alpha_N \alpha^2_{\theta_1} g_s^2}\left\{{\pi\over 20} + {3g_sM^2\over 80 N}\left[\log~r_h \left(\hat{{\cal C}}_{21}(1)
 -{g_sN_f\over 8\pi} \log~N\right)
 -{2\sqrt{6}\pi\over 5}\left(c_1 + c_2\log~r_h\right)\right]\right\}, \nonumber\\  \nd}
 \noindent where  $\hat{{\cal C}}_{21}(1)$ is defined as ${{\cal C}}_{21}(u)$ with $u = \alpha_{\theta_i} = 1$ in the definition \eqref{jkay85}.
  Note also the appearance of $\alpha_{\theta_1}$
  and not $\alpha_{\theta_2}$ in \eqref{etavalue}. This is because $\theta_1$ and $\theta_2$ defined in \eqref{harper} approach zero at
  different rates so the former got selected in the
  computation\footnote{Although for $N \sim 100 - 1000$, one may notice that  $\theta_1$  and $\theta_2$ as given in \eqref{harper}
  are not too different, so it does not really matter that much which one is chosen in \eqref{etavalue}.}.  We have also introduced a coefficient $\Upsilon$ in
  the formula  \eqref{etavalue} for $\eta$, whose value will be fixed soon.

It is now time to compute the entropy density $s$. The procedure for computing $s$ remains similar to what we did in section \ref{nf0}, although the choice of the
mirror variables differ from the type IIB case. This implies that the entropy density at the horizon may now be expressed as:
\bg\label{entoden}
s = {64 \pi^{3/2} N^{3/4} r_h^3\over \alpha_N \alpha^5_{\theta_1} g_s^{9/4}}\left\{1 + {3g_sM^2\over 4\pi N}\left[{\pi\sqrt{6}\over 2}\left(c_1 + c_2\log~r_h\right) +
\hat{\cal{C}}_{23}(1) \log~r_h  + \hat{\cal{C}}_{01}(1)\right]\right\}, \nonumber\\ \nd
where $\hat{\cal{C}}_{kj}(1)$ is defined for ${\cal C}_{kj}(u)$ with $u = \alpha_{\theta_i} = 1$ in \eqref{jkay85}.
One may now compare \eqref{entoden}
with \eqref{ento2b} as well as the entropy computed in \cite{metrics} where we see similar suppressions with respect to ${g_sM^2\over N}$ and
$g_sN_f$. The  precise coefficients understandably differ because of the different choices of
variables alluded to above. One may get away from this by
choosing a uniform definition of the variables in all the models. However this suffers from a reduction in the efficiency of computations of physical quantities in some models
and increase in others\footnote{As an example, if we choose zero bare resolution parameter in both type IIB and the M-theory uplift of the
mirror type IIA models, the efficiency
of computing the sound speed and the attenuation constant in M-theory reduces considerably, whereas in type IIB it becomes much enhanced.}.

The above discussion however does not spell out a {\it failure} to compare the physical quantities in different models; rather one should interpret the
validity of different results to be at different range of parameter values. For the choice of parameters in the mirror set-up, and using \eqref{etavalue} and \eqref{entoden}, we
can now express the ratio between shear viscosity and entropy density as:

{\footnotesize
\bg\label{etaos}
{\eta\over s} = {1\over 4\pi} + {3g_sM^2\over 128\pi^3 N}\left[g_sN_f
\log\left({16N\over r_h^{24 + \log~N}}\right) - (8\pi + 6g_sN_f) - {36\sqrt{6}\pi\over 5}\left(c_1 + c_2\log~r_h\right)\right], \nd}
where note the absence of the parameter $\alpha_N$ from \eqref{mozart}. We also
wrote the first term as ${1\over 4\pi}$. In the absence of the ${g_sM^2\over N}$ correction, this should be the conformal result \cite{KSS}, and therefore
we have used this to fix the parameter $\Upsilon$ in \eqref{etavalue} as:
\bg\label{covenant}
\Upsilon \equiv {320\over \pi g_s^{3/4} N^{3/20} \alpha^3_{\theta_1}}. \nd
There are a few issues regarding the ratio \eqref{etaos} that we should take into account now. First, observe the appearance of an inherent scale in \eqref{etaos}. This
appears through the $\log~r_h$ term above as $\log~{r_h\over r_d}$, where $r_d$ is the scale defined earlier.  

The $r_d$ dependence in \eqref{etavalue} for example should remind us of a similar $r_c$ dependence of shear viscosity in the type IIB side as given in
eq (3.198) of \cite{metrics}. The introduction of UV cap to the geometry contributed an additional piece as eq (3.200) in \cite{metrics}.  This eventually led to the
ratio of the shear viscosity to the entropy density being given by eq (3.222) therein that depended upon the UV degrees of freedom ${\cal N}_{uv}$ as $e^{-{\cal N}_{uv}}$.
The result for infinite UV degrees of freedom was exactly ${1\over 4\pi}$, so we should expect similar result for our case too.
However the analysis of ${\eta\over s}$ in
\eqref{etaos} is done at the horizon with a scale $r_d$, and one may easily see that the scale dependence is $\log~r_d$ which is an expected answer for a
QCD like model. This means that, even with a QCD scale inserted in, we expect ${\eta\over s}$ to be at least bigger than ${1\over 4\pi}$ so that the KSS bound \cite{KSS} is not
violated. In \eqref{etaos} it is easy to see that the $r_h$ dependent  terms  are positive definite because $\log~{r_h\over r_d} < 0$.
Thus if we define $c_1$, which is as yet an unfixed function, as:
\bg\label{dansor}
c_1 \equiv -\vert\sigma \vert - {5\sqrt{6}\over 9}\left({1\over 3} + {g_sN_f\over 4\pi}\right), \nd
with $\sigma$ as another undetermined function, then ${\eta\over s} > {1\over 4\pi}$. Addition of the UV cap can then change the result accordingly, but we will not elaborate on this
here anymore. At this stage, it will simply suffice to see that the KSS bound is not violated.

All the ingredients are at hand now to compute both the bulk viscosity $\zeta$
as well as the ratio of the bulk to shear viscosities i.e ${\zeta\over \eta}$. As we saw earlier, the shear and the
bulk viscosities are connected by the following relation:
\bg\label{circle}
{1\over 2sT}\left(\zeta + {4\eta\over 3}\right) = \Gamma, \nd
where $s$ is the entropy density \eqref{entoden}, $T$ is the temperature \eqref{T} and $\Gamma$ is the attenuation constant \eqref{Gamma}. One can therefore use
\eqref{circle} to express the ratio ${\zeta\over s}$ in terms of $\Gamma$ and ${\eta\over s}$ as:

{\footnotesize
\bg\label{paxton}
{\zeta\over s} = {g_sM^2\over 32\pi^2 N}\left[3{\cal C}_{21}(1)  + 8{\hat{\cal C}}_{01}(1)
+ \kappa_0\left(c_1 + c_2\log~r_h\right)- 8\log~r_h \left({\hat{\cal C}}_{21}(1) - {\hat{\cal C}}_{23}(1)
+ {g_sN_f \over 8\pi} \log~N\right)\right], \nonumber\\ \nd}
where $\kappa_0 \equiv {364\pi \sqrt{6}\over 45}$, ${\cal C}_{jk}(1)$ is given by \eqref{jkay85} for $u = 1$, and ${\hat{\cal C}}_{jk}(1)$ is given by
\eqref{jkay85} with $u = \alpha_{\theta_i} = 1$. The overall factor of ${g_sM^2\over N}$ is interesting and crucial:
it tells us that the ratio  \eqref{paxton} is {\it zero} for conformal theories. This is
of course consistent with what we had in section \ref{nf0}, and we note that the bulk viscosity may be easily derived, to this order in ${g_sM^2\over N}$, by simply
multiplying \eqref{paxton} by the conformal entropy density. Any non-conformal corrections to $s$ will change the bulk viscosity only to higher orders in
${g_sM^2\over N}$. Note also that, in the limit of vanishing fundamental flavors i.e $N_f = 0$, the ratio \eqref{paxton} takes the following form:
\bg\label{vanishing}
{\zeta\over s} = {g_sM^2\over 32\pi^2 N}\left[11 + {364\pi \sqrt{6}\over 45}\left(c_1 + c_2\log~{r_h\over r_d}\right)\right], \nd
where we have inserted back the scale $r_d$ (which was taken to be 1 so far). Looking at \eqref{vanishing} one might be tempted to compare it with the bulk viscosity
that we got in \eqref{bulky}, which was expressed using the fluctuation mode $Y_x$ satisfying \eqref{clayoven}. In fact \eqref{bulky} had $r_h/r_c$ dependence whereas \eqref{paxton} has $r_h/r_d$ dependence; and their exact factors differ.
This has already been alluded to earlier because of the different choices of the parameters in the two theories. Additionally, as we discussed in
section \ref{zerbare}, the ratio \eqref{bulky} is derived for vanishing bare resolution parameter whereas \eqref{paxton} is derived with non-zero bare resolution parameter.  This of course is not the only difference. The zero bare resolution case, according to section \ref{zerbare}, involves study of quasi-normal frequencies whereas the result
\eqref{bulky} is derived from the study of fluctuation modes $Y_x$. The point of 
comparison\footnote{Despite the fact that the former uses Wilsonian method with a cut-off $r_c$ whereas the latter uses full quantum corrections with a scale $r_d$.} between the two results maybe that both involve certain non-normalizable functions at the boundary $u = 0$. Plugging in the non-normalizable function $F(u)$ in \eqref{fudef} will help us find $\alpha$ and $\beta$ in \eqref{dispersion}, which
in turn may be compared to \eqref{paxton}.

Finally, the ratio of bulk to shear viscosities may now be determined from \eqref{paxton}, to first order in ${g_sM^2\over N}$,  by taking the conformal limit of \eqref{etaos}. The
result is similar to what we have in \eqref{paxton} up to a factor of $4\pi$:

 {\footnotesize
\bg\label{paxton2}
{\zeta\over \eta} = {g_sM^2\over 8\pi N}\left[3{\cal C}_{21}(1)  + 8{\hat{\cal C}}_{01}(1)
+ \kappa_0\left(c_1 + c_2\log~{r_h\over r_d}\right)- 8\log~{r_h\over r_d} \left({\hat{\cal C}}_{21}(1) - {\hat{\cal C}}_{23}(1)
+ {g_sN_f \over 8\pi} \log~N\right)\right], \nonumber\\ \nd}
where $\kappa_0$ is defined earlier and we have inserted back $r_d$, the QCD scale.  To see whether \eqref{paxton2} does not violate the Buchel bound
 \cite{Buchel-bound} we will have to determine $c_1$ and $c_2$ in \eqref{paxton2}. In \eqref{dansor} we expressed $c_1$ in terms of a negative definite
function $-\vert\sigma\vert$ at the horizon $u = 1$, assuming $c_2 $ to be a positive definite quantity there.
However underneath this choice was the assumption that both the bare resolution parameter $b$ and the full resolution parameter $a$ in \eqref{a} remain positive definite. As long as $b > 0$, this could still be made true with $c_2 > 0$. However $b$ can be zero, as we saw in sections \ref{nf0} and \ref{zerbare},
and in this case $c_2 > 0$ will make $a < 0$ in \eqref{a} with the choice of $c_1$ in \eqref{dansor}, rendering the whole construction meaningless. One might think that
$c_1$ could be changed, but then the KSS bound \cite{KSS} will be affected.  Therefore it seems the only way to avoid any contradictions is to take $c_2 = -\vert c_2\vert$
with:
\bg\label{em18}
\vert c_2 \vert  ~ \le ~ {5\sqrt{6}g_sN_f\over 216\pi} \left(24 + \log~N\right), \nd
at the horizon $u = 1$. It should be clearly noted that \eqref{dansor} and \eqref{em18} arise by demanding the KSS bound \cite{KSS} is satisfied, and that the resolution parameter \eqref{a} is positive definite\footnote{To see this, define
$N$, without any loss of generality, as
$N \equiv {\rm exp}\left({w\over g_sN_f}\right)$ near the horizon $u = 1$ with some appropriately chosen function $w$ that takes very large value.}.

With this at hand, it is now time to see if the ratio of bulk to shear viscosities \eqref{paxton2} preserve the Buchel bound \cite{Buchel-bound}.
We will start with the simplest case
of vanishing flavor i.e $N_f = 0$.  Referring back to sound speed \eqref{cs} and the ratio \eqref{paxton2}, we get:
\bg\label{cmukha}
&&{\zeta\over \eta} = {g_sM^2\over 8\pi N} \left[11 + {364\sqrt{6}\pi\over 45} \left(c_1 + c_2\log~{r_h\over r_d}\right)\right]\nonumber\\
&& {1\over 3} - c_s^2 = {g_sM^2\over 16\pi N} \left[-1 + {40\sqrt{6}\pi\over 45}\left(c_1 + c_2 \log~{r_h\over r_d}\right)\right], \nd
where $c_1$ and $c_2$ now satisfy \eqref{dansor} and \eqref{em18} respectively. Since $\log~{r_h\over r_d} < 0$, all terms in ${\zeta\over \eta}$ in \eqref{cmukha}
are positive definite. In the limit where $r_d  > r_h$, the ratio of bulk to shear viscosities may be
related to the sound speed as:
\bg\label{830am}
{\zeta\over \eta} = {91\over 5}\left({1\over 3} - c_s^2\right) + {201\over 80\pi}\left({g_sM^2\over N}\right), \nd
which clearly satisfies the Buchel-bound \cite{Buchel-bound}. Interestingly, for
$r_d >> r_h {\rm exp}\left({11 + \vert c_1\vert\over \vert c_2 \vert}\right)$, one may ignore the second piece in \eqref{830am} and the ratio \eqref{paxton2} may solely
be expressed in terms of ${1\over 3} - c_s^2$. In either case, one may easily infer from \eqref{cmukha} that the
Buchel-bound is {\it always} satisfied, at least for vanishing fundamental flavors $N_f$.

What happens when $N_f \ne 0$, i.e when we switch on fundamental flavors? Both, the ratio of bulk over shear viscosities and sound speed, have been computed
above in \eqref{paxton2} and \eqref{cs} respectively. So it's time to combine them to see whether the specified combination of them satisfy the Buchel-bound. It is
easy to see that the bulk to shear ratio \eqref{paxton2} may now be expressed as:

{\footnotesize
\bg\label{mor446}
{\zeta\over \eta} = {91\over 5}\left({1\over 3} - c_s^2\right) + {g_sM^2\over 16\pi N} \left[16 {\hat{\cal C}}_{01} + {121\over 5} {\cal C}_{21} +
16 \log~{r_d\over r_h}
\left({\hat{\cal C}}_{21} - {\hat{\cal C}}_{23} + {g_sN_f\over 8\pi} \log~N\right)\right], \nd}
where the ${\cal C}_{ij}$ and ${\hat{\cal C}}_{ij}$  terms may be extracted from \eqref{jkay85} using the limits $u = 1$ and $u = \alpha_{\theta_i} = 1$ respectively.
By switching off the $g_sN_f$ terms one gets \eqref{830am} from \eqref{mor446}, so
the question now is whether the ${g_sM^2\over N}$ terms
in \eqref{mor446} can again be positive definite.

It turns out, with some algebraic manipulations, one may rewrite the ${g_sM^2\over N}$ terms appearing on the RHS of \eqref{mor446} in the following suggestive
way:
\bg\label{vedpra}
{\zeta\over \eta} & = &  {91\over 5}\left({1\over 3} - c_s^2\right) +  {g_sM^2\over 16 \pi N}\Bigg[{201\over 5}
+ {121\over 20\pi}\left(\log\left(\alpha_{\theta_1}\alpha_{\theta_2}\right) + {603\over 121}\right)g_sN_f \nonumber\\
& +& {2g_sN_f\over \pi} \left(\log~N - {603\over 10}\right) \left(\log~{r_d\over r_h} - {201\over 80}\right) - \sigma_0 g_sN_f\Bigg], \nd
where $\sigma_0 \equiv {201\over 20\pi}\left(\log~4 + {603\over 20}\right) \approx 100.86$ is a positive coefficient. Since $N$ is very large and $r_d >> r_h$, every
term in \eqref{vedpra} can be shown to be positive definite,
and the negative piece $\sigma_0g_sN_f$ does not change anything as long as $\log~N~\log~{r_d\over r_h} >> 160$. The latter is not a
constraint  as we saw above\footnote{To see that the terms on the RHS of \eqref{vedpra} can be positive definite, choose $N \equiv {\rm exp}\left(n_0 + 60.3\right)$ with
$n_0$ being a very large number approaching infinity,
and $r_d \equiv r_h {\rm exp}\left(n_1 + 2.5125\right)$ with $n_1$ being another large number, not necessarily infinite.
The condition for positivity of the
RHS of \eqref{vedpra} is $n_0 n_1 > 160$. This is easily achieved because going to the gravity dual description forces us to choose both $n_0$ and $n_1$ very large.}.
Thus generically our model satisfies the Buchel-bound \cite{Buchel-bound}, and comparing
\eqref{830am} and \eqref{vedpra}, we see that there is in fact a new bound on the ratio of bulk to shear viscosities given by:
\bg\label{newbound}
{\zeta\over \eta} ~> ~ {91\over 5}\left({1\over 3} - c_s^2\right). \nd

\section{Type IIA spectral function and the viscosity bound at strong coupling with non-zero flavors \label{spectral}}

In the above section we found an interesting bound \eqref{newbound} for the ratio of bulk to shear viscosities at strong string and strong 't~Hooft couplings. In fact the {\it form} of the bound seems consistent over the whole strong 't~Hooft coupling regime as is obvious from the weak string coupling bound that we got earlier in \eqref{mathadulai}: both
\eqref{newbound} and \eqref{mathadulai} are proportional to ${1\over 3} - c_s^2$, although their coefficients differ.
On the other hand, the weak 't~Hooft coupling bound differs by being proportional to the square of the strong coupling bound as shown in \eqref{sellroth}. The reason for the different results at the two ends of the couplings has been motivated in section \ref{kinetic}. Loosely, it is the ratio of the shear viscosity over entropy density that creates the difference at the two ends. At weak  'tHooft coupling the ratio is not a constant and is given by
\eqref{shear}, whereas at strong 't~Hooft couping we expect the ratio to be a constant \cite{KSS}. This is a reasonably strong argument for justifying the difference between the two bounds, despite the fact that we have no control on the dynamics at the intermediate coupling regime as argued in section \ref{int}. The very weak coupling results have been justified in great details, and in sections \ref{nf0} and \ref{nfn0}, we provided some justifications for the strong coupling results. However one might be interested in deriving the bound at strong coupling directly from the spectral function, as such an analysis will be in line with the discussions of section \ref{int}.
Further, we make the following observations:
\vskip.1in

\noindent $\bullet$ The ratio of the bulk viscosity $\zeta$ to entropy density $s$ is of ${\cal O}\left({g_sM^2\over N}\right)$
as we saw in \eqref{paxton}, and the ratio of the shear viscosity $\eta$ to the entropy density is dominated by the conformal result plus an ${\cal O}\left({g_sM^2\over N}\right)$  correction term from \eqref{etaos}. This means up to
${\cal O}\left({1\over N}\right)$  the ratio $\zeta/\eta$ would mimic $\zeta/s$.

\vskip.1in

\noindent $\bullet$ The  gauge and the metric perturbations may be required to be considered simultaneously
$-$ see subsection 4.2 of \cite{D.Tong_Cracow_13} and references therein.

\vskip.1in

\noindent $\bullet$ The correlation of gauge fluctuations,
$\langle {\cal A}_{x^i}{\cal A}_{x^i}\rangle$ for $i=1, 2, 3$, along the same direction could hence mimic the spirit
behind the correlation of the metric perturbations,
$\langle h_{x^ix^i}h_{x^ix^i}\rangle$,  along the $x^i$ axis relevant to the evaluation of bulk viscosity as, for example in \cite{Gubser:2008yx} or in section \ref{nf0}.

\vskip.1in

\noindent The above three observations provide the necessary motivation for this section. Therefore we would like to evaluate the aforementioned gauge-field correlation function (in the hydrodynamical limit using the prescription of \cite{Minkowskian-correlators}) and see if one obtains a linear bound seen in \eqref{mathadulai} and \eqref{newbound}. Even if this may not be explicitly tied to $\zeta/\eta$, we feel the result obtained in this section, in itself, is sufficiently interesting.

\subsection{Background gauge fluxes and perturbations on the flavor branes \label{spec2}}

Our starting point is configuration of $N_f$ D6-branes in the type IIA mirror set-up. For our purpose, we will however isolate one D6-brane and study world-volume dynamics on it. Alternatively, one can view this as D6-brane
probing a non-K\"ahler warped-resolved conifold with $N_f$ flavor D6-branes.
The DBI action for a single $D6$ brane is given as:
\begin{equation}\label{D6DBI}
S_{D6}=-T_{D6}\int d^{7}\xi~ e^{-\varphi}\sqrt{\det{\left(g+B+F\right)}},
\end{equation}
with $2\pi\alpha^{\prime}=1$. Here the worldvolume directions of the $D6$ brane are denoted by the coordinates:
$\left(t, x_1,x_2,x_3,Z,\theta_2,\varphi_2\right)$, with $\left( t, x_1, x_2, x_3\right)$ as the usual Minkowski coordinate,
$Z$ as the newly defined radial direction and two angular coordinate ($\theta_2$, $\varphi_2$); $Z$ is related to the usual radial coordinate $r$ as $r = r_{h}e^{Z}$ and $\varphi_2$ is the local value for the angle $\phi_2$ (for more details, see sections {\bf 3} and {\bf 4} of \cite{meson Yadav+Misra+Sil}).

In the above, and as before, $\varphi$ denotes the type IIA dilaton which is the triple T-dual version of type IIB dilaton. The pullback metric and the pullback of the NS-NS $B$ field on the worldvolume of the $D6$ brane are denoted as $g$ and $B$ in (\ref{D6DBI}). $F$ is the field strength for a U$(1)$ gauge field $A_{\mu}$, where the only nonzero component of the same is the temporal component $A_{t}$. In the gauge $A_{Z}=0$, the only nonzero component of $F$ is $F_{Zt}=-F_{tZ}$. Combining together the symmetric $g$ field and the anti-symmetric $B$ field as
$G \equiv g+B$ and expanding the DBI action up to quadratic order in $A$, we get:
\begin{equation}\label{D6DBI2}
S_{D6}=\frac{T_{D6}}{4}\int d^{7}\xi~ e^{-\varphi}\sqrt{-G}\left(G^{\mu\alpha}G^{\beta\gamma}F_{\alpha\beta}F_{\gamma\mu}-\frac{1}{2}G^{\mu\nu}G^{\alpha\beta}F_{\mu\nu}F_{\alpha\beta}\right).
\end{equation}
The second term in (\ref{D6DBI2}), is coming because of the anti-symmetric $B$ field in $G$. As none of the fields in the DBI action depends on the angular coordinates $\varphi_2$, the integrand in equation (\ref{D6DBI2}) is independent of the same. Also we choose to work around the same small values of both $\theta_2$ and $\theta_1$ given by \eqref{harper} earlier.
Hence, the upshot is that the integration over $\theta_2$ and $\varphi_2$ is trivial and we denote $\Omega_{2}$ as the factor one gets after the integration over $\theta_2$ and $\varphi_2$, such that:
\begin{equation}\label{D6DBI3}
S_{D6}=\frac{T_{D6}\Omega_2}{4}\int d^{4}x~dZ~ e^{-\varphi}\sqrt{-G}\left(G^{\mu\alpha}G^{\beta\gamma}F_{\alpha\beta}F_{\gamma\mu}-\frac{1}{2}G^{\mu\nu}G^{\alpha\beta}F_{\mu\nu}F_{\alpha\beta}\right).
\end{equation}
The equation of motion for the temporal gauge field $A_{t}(Z)$ as obtained from the above lagrangian in (\ref{D6DBI3}) is given as:
\begin{equation}\label{EOMAt}
\partial_{Z}\left(e^{-\varphi}\sqrt{-G}~G^{tt}G^{ZZ}\partial_{Z}A_{t}(Z)\right)= 0.
\end{equation}
At this point we can use the precise functional forms for the background data, namely $G_{tt}$, $G_{ZZ}$  as well as the dilaton $\varphi_2$, to rewrite the above equation in the following form:

{\footnotesize
\bg\label{AtEOM-1}
{{\bf C}\over A'_t(Z)}&= &3a^2\left[g_sN_f ~\log\left({N^2\over e^{3+6Z}}\right) + 8\pi\right] \alpha^4_{\theta_1} -
3a^2 \left[g_sN_f ~\log\left({N\over e^{3Z}}\right) + 4\pi\right] \alpha^4_{\theta_2}\\
&+& 2 \left[g_sN_f ~\log\left({N\over e^{3Z}}\right) + 4\pi\right] \alpha^4_{\theta_1} r_h^2 e^{2Z} -
3g_sN_f~\log~r_h\left[2 \alpha^4_{\theta_1} r_h^2 e^{2Z} + 3a^2\left(2\alpha^4_{\theta_1}
- \alpha^4_{\theta_2}\right)\right] , \nonumber \nd}
where $a^2$ is the resolution parameter,  ($\alpha_{\theta_1}, \alpha_{\theta_2}$) are the two angular values
in \eqref{harper} and ${\bf C}$ is the integration constant. In
 the large $Z$ and small $a^2$ limit, (\ref{AtEOM-1}) yields:
\bg\label{AtEOM-2}
 & &   2 \alpha_{\theta_1}^4 {r_h}^2 e^{2 Z} {A_t}'(Z) \left[g_sN_f~\log\left({N\over r_h^3 e^{3Z}}\right) + 4\pi\right]
 = {\bf C}, \nd
 which appears from the fact that the second line in \eqref{AtEOM-1} dominates over the first line. The large $Z$ limit is  also the large $r$ limit where one might be concerned about UV issues appearing from AdS cap. This is however not much of a worry at this stage because as long as $r_h e^Z >> a$, \eqref{AtEOM-2} continues to hold.
  With this in mind,  the solution to (\ref{AtEOM-2}) is:
   \bg\label{solution-At-EOM}
  \langle A_t(Z)\rangle & = &  {\bf C}_1- {{\bf C} e^{-\frac{2}{3} \left(\frac{4 \pi }{g_s  {N_f}}+\log ~N\ \right)}
  {\bf Ei}\left[\frac{2}{3}~\log\left({N \over r_h^3 e^{3Z}}\right) + {8\pi \over 3 g_s N_f}\right] \over
 6 \alpha_{\theta_1}^4 g_s  {N_f}} \nonumber\\
   & =  & \frac{{\bf C} e^{-2 Z}}{12 \alpha_{\theta_1}^4 g_s  {N_f} {r_h}^2 Z}+ {\bf C}_1 + {\cal O}\left(\frac{1}{Z^2}\right),
   \nd
 where we have used $\langle A_t\rangle$ to express the background value to avoid confusion. The other parameters
 appearing in \eqref{solution-At-EOM} are
 ${\bf C}_1$, which is yet another constant and ${\bf Ei}$, which is the exponential integral\footnote{${\bf Ei}(x)
  \equiv -\int_x^\infty {e^{-t}\over t}~dt$.  This definition can be used for positive values of $x$, but the integral has to be understood in terms of the Cauchy principal value due to the singularity of the integrand at zero. }.
 In the second line of \eqref{solution-At-EOM} we have shown the first dominant piece in the large $Z$ limit.  Higher powers of ${1\over Z}$ can then be ignored. This background value also prepares us to study the fluctuation of the gauge field components.
 For example we can express the gauge field appearing in \eqref{D6DBI3} as:
\bg\label{oohswt}
{A}_{\mu}(x, Z)=\delta^{t}_{\mu}\langle A_{t}(Z) \rangle +{\cal A}_{\mu}(x,Z),
\nd
where the fluctuation ${\cal A}_{\mu}$ only exists along the directions $\mu=$ ($ t, x_1, x_2, x_3$) due to the particular gauge choice and depends only on the radial variable $Z$.
Including the perturbations in the lagrangian of the DBI action (\ref{D6DBI}), one gets:
\begin{equation}
\mathcal{L}=e^{-\varphi}\sqrt{\det\left(g+B+F+{\cal F}\right)},
\end{equation}
with ${\cal F}$ as the field strength for the gauge field fluctuations. Now defining ${\cal G} \equiv g+B+F$ and again expanding the above lagrangian up to quadratic order in the gauge field fluctuation one gets:
\begin{equation}\label{D6DBI4}
\mathcal{L}=-\frac{1}{4}e^{-\varphi}\sqrt{-{\cal G}}\left({\cal G}^{\mu\alpha}{\cal G}^{\beta\gamma}{\cal F}_{\alpha\beta}{\cal F}_{\gamma\mu}
-\frac{1}{2}{\cal G}^{\mu\nu}{\cal G}^{\alpha\beta}{\cal F}_{\mu\nu}{\cal F}_{\alpha\beta}\right).
\end{equation}
Writing the field strength ${\cal F}$ in terms of the gauge field fluctuation ${\cal A}_{\mu}$ and after doing some simplifications in terms of the interchange of indices, one can write the above lagrangian as:
\bg\label{lag}
\mathcal{L}&= & e^{-\varphi}\sqrt{-{\cal G}}\left({\cal G}^{\mu[\alpha}{\cal G}^{\beta]\gamma}
\partial_{[\gamma}{\cal A}_{\mu]}
-\frac{1}{2}{\cal G}^{[\alpha\beta]}{\cal G}^{\mu\nu}\partial_{[\mu}{\cal A}_{\nu]}
\right)\partial_{\alpha}{\cal A}_{\beta} \nonumber\\
 & = & \partial_{\alpha}\Biggl[e^{-\varphi}\sqrt{-{\cal G}}\left({\cal G}^{\mu[\alpha}{\cal G}^{\beta]\gamma}
\partial_{[\gamma}{\cal A}_{\mu]}
-\frac{1}{2}{\cal G}^{[\alpha\beta]}{\cal G}^{\mu\nu}\partial_{[\mu}{\cal A}_{\nu]}
\right){\cal A}_{\beta}\Biggr]\\
& - & \partial_{\alpha}\Biggl[e^{-\varphi}\sqrt{-{\cal G}}\left({\cal G}^{\mu[\alpha}{\cal G}^{\beta]\gamma}
\partial_{[\gamma}{\cal A}_{\mu]}
-\frac{1}{2}{\cal G}^{[\alpha\beta]}{\cal G}^{\mu\nu}\partial_{[\mu}{\cal A}_{\nu]}
\right)\Biggr] {\cal A}_{\beta}. \nonumber
\nd
The second line in equation (\ref{lag}) is a total derivative term and equating the last line to zero for any arbitrary
${\cal A}_{\beta}$ gives the equation of motion for the gauge field fluctuation:
\begin{equation}\label{EOMtildeA}
\partial_{\alpha}\Biggl[e^{-\varphi}\sqrt{-{\cal G}}\left({\cal G}^{\mu[\alpha}{\cal G}^{\beta]\gamma}
\partial_{[\gamma}{\cal A}_{\mu]}
-\frac{1}{2}{\cal G}^{[\alpha\beta]}{\cal G}^{\mu\nu}\partial_{[\mu}{\cal A}_{\nu]}
\right)\Bigg] = 0.
\end{equation}
The total derivative term in \eqref{lag} does not necessarily have to vanish at $Z \to \infty$, as there could be non-normalizable modes serving as sources for the dual gauge theory operators. Our EOM in \eqref{EOMtildeA} however is not affected by this, and in the following section we will discuss possible solutions of \eqref{EOMtildeA}.

\subsection{Equation of motion for gauge field fluctuations \label{spec3}}

To derive the equation of motion for the gauge field fluctuation, we first need to write down the fluctuating field as the following Fourier decomposed form:
\begin{equation}\label{doctor}
{\cal A}_{\mu}(t,x_1,Z) = \int \frac{d^4 k}{(2\pi)^4}e^{-i \omega t+i q x_1}{\cal A}_{\mu}(\omega, q, Z),
\end{equation}
where we assume the fluctuation to have momentum along $x_1$ direction only, with $k_0 = \omega$, $k_1$  proportional to $q$ and $k_2, k_3$ arbitrary.
Now, the equation (\ref{EOMtildeA}) has a free index $\beta$ and for $\beta=$ ($t, x_1, x_2, x_3, Z$), one gets a total of five equations of motion. For example for $\beta = Z$, plugging in \eqref{doctor} in \eqref{EOMtildeA} yields:
\bg\label{laura1}
\omega{\cal G}^{tt}(\partial_{Z}{\cal A}_{t})-q{\cal G}^{x_1x_1}(\partial_{Z}{\cal A}_{x_1})=0, \nd
where the RHS vanishes because of the antisymmetry of ${\cal G}_{[\alpha\beta]}$. The dilaton does not appear because it is independent of the four-dimensional spacetime coordinates.
The above equation relates ${\cal A}'_t$ with
${\cal A}'_{x_1}$. On the other hand if we take $\beta = t$, we get the following EOM:
\bg\label{laura2}
\partial_{Z}\left(e^{-\varphi}\sqrt{-{\cal G}}~{\cal G}^{tt}{\cal G}^{ZZ}\partial_{Z}{\cal A}_{t}\right)
 = e^{-\varphi}\sqrt{-{\cal G}}~{\cal G}^{tt}{\cal G}^{x_1x_1}\left(\omega q {\cal A}_{x_1}+q^2{\cal A}_{t}\right), \nd
 which now relates ${\cal A}'_t$ with ${\cal A}_t$ and ${\cal A}_{x_1}$.  A somewhat similar equation
 appears when we choose $\beta = x_1$ in \eqref{EOMtildeA}, namely:
\bg\label{laura3}
 \partial_{Z}\left(e^{-\varphi}\sqrt{-{\cal G}}~{\cal G}^{x_1x_1}{\cal G}^{ZZ}\partial_{Z}{\cal A}_{x_1}\right)
= e^{-\varphi}\sqrt{-{\cal G}}~{\cal G}^{tt}{\cal G}^{x_1x_1}\left(\omega q {\cal A}_{t}+\omega^2{\cal A}_{x_1}\right). \nd
At this stage one can easily verify that pluging in \eqref{laura1} in \eqref{laura2}, we can get \eqref{laura3}. This shows that the above three equations \eqref{laura1}, \eqref{laura2} and \eqref{laura3} are self-consistent.  Finally, one may
find the equations for $\beta = x_2$ and $\beta = x_3$. We expect them to be equivalent, and  are given by:
\bg\label{laura4}
\partial_{Z}\left(e^{-\varphi}\sqrt{-{\cal G}}~{\cal G}^{\beta\beta}{\cal G}^{ZZ}\partial_{Z}{\cal A}_{\beta}\right)
= e^{-\varphi}\sqrt{-{\cal G}}~{\cal G}^{\beta\beta}\left(q^2{\cal G}^{x_1x_1}+\omega^2{\cal G}^{tt}\right){\cal A}_{\beta}.
\nd
To proceed further, we will have to define gauge invariant variables. For our case, there would be two such variables
$E_{x_1}$ and $E_\beta$ with $\beta = x_2$ or $x_3$, expressed in the following way:
\bg\label{hostiles}
E_{x_1} \equiv q {\cal A}_t + \omega {\cal A}_{x_1}, ~~~~ E_\beta \equiv E_T = \omega {\cal A}_\beta. \nd
With these new variable the three equations in \eqref{laura1}, \eqref{laura2} and \eqref{laura3}
can be cast into a {\it single} second order equation involving $E_{x_1}$.
Even more obviously, the fourth one for $\beta=x_2$ or $x_3$, can be rewritten in terms of the new
variable $E_T$. Moreover, in the zero momentum limit, i.e in the limit $q\rightarrow 0$, it can be shown that the equation involving $E_{x_1}$ is the same as the one involving $E_{T}$, given as:
\begin{equation}\label{EOMET}
\partial_{Z}\left(e^{-\varphi}\sqrt{-{\cal G}}~{\cal G}^{x_2x_2}{\cal G}^{ZZ}\partial_{Z}E_{T}\right)
= e^{-\varphi}\sqrt{-{\cal G}}~{\cal G}^{x_2x_2}\left(\omega^2{\cal G}^{tt}\right)E_{T},
\end{equation}
implying that in the zero momentum limit all we need is to solve one second order differential equation. This is of course a huge simplification, and one can even rewrite (\ref{EOMET}) in the following suggestive way:
\begin{equation}\label{bale}
\partial_{Z}\biggl[P(Z)\partial_{Z}E_{T}(Z)\biggr] + \omega^2 Q(Z) E_{T}(Z)=0,
\end{equation}
where all functions appearing above are only functions of the $Z$ variable. In fact
$P(Z)$ and $Q(Z)$ may be easily seen from \eqref{EOMET} to take the following form:
\bg\label{rosamund}
P(Z)\equiv e^{-\varphi}\sqrt{-{\cal G}}{\cal G}^{ZZ}{\cal G}^{x_2x_2}, ~~~~~
Q(Z) \equiv -e^{-\varphi}\sqrt{-{\cal G}}{\cal G}^{x_2x_2}{\cal G}^{tt}. \nd
The {\it suggestive} way alluded to above is that the above equation \eqref{bale} can be recast in a
Schr\"odinger like form by certain redefinition of the variables involved in the following way:
\begin{equation}
\label{Schrodie}
\left(\partial^{2}_{Z}+ \mathbb{V}_{E_{T}}\right)\mathbb{E}_{T} = 0,
\end{equation}
with $\mathbb{E}_T$ defined as $\mathbb{E}_{T}(Z) \equiv \sqrt{P(Z)}E_{T}(Z)$, and $\mathbb{V}_{E_T}$ is the potential term that is expressed as:
\bg\label{whotoy}
\mathbb{V}_{E_{T}} \equiv  \frac{1}{4P^2}\left(\frac{\partial P}{\partial Z}\right)^2
-\frac{1}{2P}\left(\frac{\partial^2 P}{\partial Z^2}\right) + \frac{\omega^2 Q}{P}.
\nd
The Schr\"odinger like equation is a valid description in the zero momentum limit. Once we go away from that limit, we will have more equations for the fluctuations with different choices of potentials. This is a more complicated scenario and fortunately our present analysis does not call for that. Nevertheless, the potential \eqref{whotoy} is still highly
non-trivial, as both $P(Z)$ and $Q(Z)$ take non-trivial values when expressed in terms of the background metric and dilaton in \eqref{rosamund}. For example, the
 function $P(Z)$ that may be written as:
\bg\label{runimara}
 P(Z) \equiv && -{g_sN_f ~e^{-3Z}\left(e^{4Z} -1\right) \mathbb{P}_1\mathbb{P}_2\mathbb{G}_3\over g_s^{3/2}}
 \left[r_h^2 e^{2Z} - {{\bf C}^2 e^{-4Z}\over 4 \alpha^8_{\theta_1} \left(g_sN_f\right)^2 r_h^4 \mathbb{P}_2^2}\right]^{1/2} \\
 && +  \Biggl[{e^{-4Z} \mathbb{P}_3 \mathbb{G}_2 \over g_s^{3/2}\left(g_sN_f\right)^2 \mathbb{P}_2^2} +
 {r_h^2 e^{2Z} \left(\alpha^2_{\theta_1} \mathbb{P}_2 -\alpha^2_{\theta_2}\mathbb{P}_3\right)\over g_s^{3/2}}\Biggr]
{g_sN_f ~e^{-5Z} \left(e^{4Z} - 1\right) \mathbb{P}_1 \mathbb{G}_1 \over
 \left[ r_h^2 e^{2Z} - {{\bf C}^2 e^{-4Z}\over 4 \alpha^8_{\theta_1} \left(g_sN_f\right)^2 r_h^4 \mathbb{P}_2^2}\right]^{1/2}}, \nonumber \nd
where ${\bf C}$ is a constant that appeared first time in \eqref{AtEOM-1}. The other parameters that appear above are
$\mathbb{G}_i$ and $\mathbb{P}_i$. All the $\mathbb{G}_i$'s depend only on the fixed parameters of the theory, and
are defined by:
\bg\label{clmehta}
\mathbb{G}_1 \equiv {{}^6\sqrt{3} a^2\over 32\sqrt{2} \pi^{3/2} N^{1/6} r_h \alpha^4_{\theta_1}\alpha^2_{\theta_2}}, ~~
\mathbb{G}_2 \equiv {\alpha^2_{\theta_2} {\bf C}^2 \over 4 r_h^4 \alpha^8_{\theta_1}}, ~~
\mathbb{G}_3 \equiv {r_h \over 16 \sqrt{3} 3^{5/6} \pi^{3/2}\alpha^4_{\theta_1} N^{1/10}}, \nd
where $a$ is the resolution parameter \eqref{a} and $\alpha_{\theta_i}$ are defined in \eqref{harper}. The other variables appearing in \eqref{clmehta} are the $\mathbb{P}_i$'s out of which only $\mathbb{P}_1$ is a constant. They are defined in the following way:
\bg\label{ginlynn}
&&\mathbb{P}_1 \equiv 9\sqrt{2} \alpha^3_{\theta_1} - 4\sqrt{3} N^{1/5} \alpha^2_{\theta_1} -2\sqrt{3} \alpha^2_{\theta_2}, ~~~ \mathbb{P}_2 \equiv \log\left({N\over r_h^3 e^{3Z}}\right) + {4\pi \over g_sN_f} \nonumber\\
&& \mathbb{P}_3 \equiv  g_sN_f \log\left({N^2\over r_h^6 e^{3+6Z}}\right) + 8\pi = \left(2\mathbb{P}_2 - 3\right) g_sN_f, \nd
where we have laid out clearly the $g_sN_f$ dependences of each of the coefficients. One may see that the $g_sN_f$ independent terms appear only from $\mathbb{P}_2$ and $\mathbb{P}_3$. In a similar vein, we can also work out the $Q(Z)$ piece in \eqref{rosamund}. This is given by:
\bg\label{wonder}
{Q(Z)} & \equiv & g_sN_f ~\mathbb{G}_4 \mathbb{P}_2 - {\sqrt{3} g_sN_f ~N^{17/30}\alpha_{\theta_1}^2
e^{-2Z} \mathbb{G}_1
\over r_h \sqrt{g_s}}\left(\alpha_{\theta_1}^2 \mathbb{P}_2 + 2 \alpha_{\theta_2}^2\right)\\
&-& g_sN_f \left(\mathbb{P}_1 + 4\sqrt{3} N^{1/3} \alpha_{\theta_1}^2\right) \left[ \mathbb{G}_5 \mathbb{P}_2
-{4\pi N^{16/15} \mathbb{G}_1 e^{-2Z}\over \sqrt{g_s}}\left(\alpha_{\theta_1}^2\mathbb{P}_2 +
3\alpha_{\theta_2}^2\right)\right], \nonumber \nd
where $\mathbb{P}_2$ and $\mathbb{G}_1$ have already been defined in \eqref{ginlynn} and \eqref{clmehta}
respectively, but $\mathbb{G}_4$ and $\mathbb{G}_5$ are new. They can be related to, say, $\mathbb{G}_3$ in the following way:
\bg\label{pmoran}
\mathbb{G}_4 \equiv {2\sqrt{2} \pi^{1/4}\over 3^{5/6}}\left({N^{23/20} \sqrt{\mathbb{G}_3}\over \sqrt{r_h}}\right), ~~~~
\mathbb{G}_5 \equiv 2 \sqrt{6\pi}\left({N\mathbb{G}_3 \over r_h \sqrt{g_s}}\right). \nd
With these set of definitions, the functional forms for $P(Z)$ and $Q(Z)$ are fully determined, although there is one issue that one may want to clarify at this point. This has to do with the presence of terms with relative {\it minus} sign inside the
square root in \eqref{runimara}.  To avoid \eqref{runimara} to develop complex values, we require:
\bg\label{tree}
r_h e^Z ~ > ~ \left({{\bf C} \over 8\pi \alpha^4_{\theta_1}}\right)^{1/3}, \nd
where we have used the fact that $g_sN_f \mathbb{P}_2 \ge 4\pi$ in the limit $g_s \to 0$ (see also \eqref{ginlynn}). This seems to constrain short distances, but since $r > r_h$ \eqref{tree} do not put strong constaints. In fact we can take small $Z$, large $N$ and vanishing momentum limits to re-express the potential \eqref{whotoy} in the following way:
\bg\label{vampirella11}
\mathbb{V}_{E_T} & = & \alpha(Z)\Bigg\{{\pi m^2_{0^{++}}\over m^2_0}\left({6b^2 + 1\over 9b^2+1}\right) +
{3\omega^2 \over 2}\left({g_sM^2\over N}\right)\left[{1\over 4}\log~r_h -
{\pi \beta b\left(1 + \log~r_h\right) \over \left(6b^2 + 1\right)\left(9b^2 + 1\right)}\right] \nonumber\\
& + & {\pi \omega^2 \over 4} + {3\omega^2~ \log~r_h\over 32\pi}\left(g_sN_f\right) \left({g_sM^2\over N}\right)\left[\log\left(\alpha_{\theta_1}
\alpha_{\theta_2}\right) + 6 ~\log\left({d_o r_h\over \sqrt{N}}\right)\right]\Bigg\}, \nd
The way we have expressed the above potential, one may clearly see how the various terms in the sum are increasingly suppressed by $g_sN_f$ and ${g_sM^2\over N}$.  The constant $\beta$ appearing above is
related to $c_i$ in \eqref{a} as $\beta = c_1 = c_2$  for simplicity\footnote{Alternative $a(u)$ in \eqref{a} can be expressed as $a(u) = \left[b + {\beta g_sM^2\over N}(1+\log~r_h)\right]r_h$ where $\beta$ appears as an overall coefficient of $g_sM^2/N$ piece. For the present case, where we study {\it weakly} coupled type IIA theory as opposed to the {\it strongly} coupled type IIA treatment in sec. \ref{nfn0}, $\beta$ appears as a constant only. Henceforth, unless mentioned otherwise, this will the case that we shall consider here. \label{internet}};
and $m_{0^{++}}$ is the mass of the lightest
glueball given via:
\bg\label{doclaura}
m_{0^{++}} = m_0\left( \frac{r_h}{\sqrt{4\pi g_s N}}\right), \nd
and parametrized by the scale $m_0$. This is computed using M-theory metric perturbation, much like the analysis we had in section \ref{nfn0}, and is further detailed in \cite{glueball Sil+Yadav+Misra}.
We have also used
\eqref{doclaura}
to define $d_o$ as:
\bg\label{shelbi}
d_o \equiv  {\sqrt{e} \over 4^{1/4}} \approx 1.1658 , \nd
which is a constant.  Note that in \eqref{doclaura} the first term in independent of $\omega^2$ and only depends on
$Z, b^2$ and the glueball mass.
The glueball mass also features
in the definition of $\alpha(Z)$, that appears in \eqref{vampirella11}, in the following way:
\bg\label{bluenails}
\alpha(Z) \equiv {1\over 4\pi Z^2} \left({6b^2 + 1\over 9b^2 + 1}\right) {m_0^2\over {m_{0^{++}}}}, \nd
where $b$ is the bare resolution parameter that is defined in \eqref{a}. Comparing the definition of the glueball mass in \eqref{doclaura}, we see that $\alpha(Z)$ is proportional to $g_sN$, but suppressed by ${1\over Z^2}$. This clearly indicates that the potential \eqref{vampirella11} goes as ${1\over Z^2}$ for small $Z$.

Note that $Z=0$ (horizon) is a regular singular point of \eqref{Schrodie} and the exponents of the indicial equation near $Z=0$ can  then be written as ${1\over 2} \pm i \mathbb{I}$, where $\mathbb{I}$ is defined as:
\bg\label{prelim}
\mathbb{I} \equiv {m_0 \omega \over 4m_{0^{++}}}\sqrt{6b^2 +1 \over 9b^2 + 1}\left[1 + {9 \over 8 \pi^2}
\left({g_sM^2\over N}\right) g_sN_f ~\log^2 r_h\right]. \nd
The functional form for $\mathbb{I}$ shows that it is suppressed by both $g_sN_f$ and ${g_sM^2\over N}$ so to zeroth order there is only a piece that depends on the bare resolution parameter $b$, the frequency $\omega$,  the horizon radius $r_h$ and the 't~Hooft  coupling $g_sN$. We can also express the solution for the Schr\"odinger type equation \eqref{Schrodie} using $\mathbb{I}$ as:
\bg\label{rama}
\mathbb{E}_T = Z^{{}^{{1\over 2} - i\mathbb{I}}}~\mathbb{F}_T(Z), \nd
with $\mathbb{F}_T(Z)$ being a function that is analytic at the horizon radius $r_h$ (or at $Z = 0$). The above equation tells us the precise behavior of the eigenfunction $\mathbb{E}_T(Z)$ at $Z = 0$. This is useful but not exactly relevant for the present case, as what we actually need is the form for $\mathbb{E}_T(Z)$ when $Z \gg 1$. The question then is how will the $Z = 0$ analysis be useful for the large $Z$ domain.

The answer lies in our choice of the ansatze \eqref{rama} that in fact serves as a good ansatze even when $Z \gg 1$. In other words, the exponent of the indicial equation $\mathbb{I}$ that we computed in \eqref{prelim} still remains a valid solution for large $Z$. What changes for large $Z$ is the functional form for $\mathbb{F}_T(Z)$.

Of course there is yet another change in \eqref{Schrodie}: it is the functional form for the potential
$\mathbb{V}_T(Z)$ that we computed earlier in \eqref{vampirella11}. Naturally since \eqref{vampirella11} was for small $Z$, this should change. The change is easy to work out, and may be written in the following way:
\bg\label{amberlin}
\mathbb{V}_{E_T}(Z) \stackrel{Z\gg1}{\longrightarrow} {\bf A} + {\bf B} e^{-2Z},
\nd
where we are ignoring higher powers of $e^{-2Z}$ that would appear from the relevant higher powers of $e^{-2Z}$ in $P(Z)$ and $Q(Z)$ in \eqref{runimara} and \eqref{wonder} respectively. The ${\bf A}$ and ${\bf B}$ appearing in
\eqref{amberlin}  are not constants, with ${\bf A}$ defined as:
\bg\label{ADdeaf}
{\bf A} \equiv - 1  + {3\over \mathbb{P}_2} + {9 \over 4 {\mathbb{P}}^2_2}, \nd
where $\mathbb{P}_2$ is given in \eqref{ginlynn}. The function $\mathbb{P}_2$ is defined with $N, r_h$ and $Z$, and one may take appropriate limits in terms of either of these parameters. Before we do this, let us write the expression for ${\bf B}$ in terms of the background parameters:
{\footnotesize
\bg\label{pfiffer}
{\bf B} & = & {m_0^2\omega^2 \over m^2_{0^{++}}} + {3(2\mathbb{P}_2 - 3)(\alpha^2_{\theta_1} - 2
\alpha^2_{\theta_2}) \over 2\beta_o^2\alpha^2_{\theta_2}\mathbb{P}_2^2}
\left({b\over \beta_o} + {g_sM^2\over N}\right)^2\left[\mathbb{P}_4 + {1\over 2} +
{\mathbb{P}_6\over \mathbb{P}_2} - \left(2 - {3\over \mathbb{P}_2}\right)\log~r_h\right]\\
& + & {3m_0^2 \omega^2 \mathbb{P}_5 (Z + \log~r_h)  \over 32 \pi^2 m^2_{0^{++}}}\left({g_sM^2\over N}\right)
\left[ 1+ {3\over 64\pi^2}\left({g_sM^2\over N}\right)(Z + \log~r_h) (\mathbb{P}_5 - 4~\log~2) g_sN_f\right]g_sN_f,
\nonumber \nd}
where the successive suppressions with respect to ${g_sM^2\over N}$ as well as $g_sN_f$ are shown. The term independent of all these is proportional to $b/\beta_o$ where $b$ is the resolution parameter and $\beta_o \equiv
\beta ~\log(er_h)$ with $\beta = c_1 = c_2$ in \eqref{a}.  The other parameters appearing in \eqref{pfiffer} are defined in the following way:
\bg\label{unbreakable}
&& \mathbb{P}_4 = 3~\log~r_h - 2\mathbb{P}_2 + \mathbb{P}_6, ~~~~
\mathbb{P}_6 = \mathbb{P}_2 - {3\alpha^2_{\theta_2} \over \alpha^2_{\theta_1} - 2 \alpha^2_{\theta_2}}\\
&&\mathbb{P}_5 = 12\left(1 - \log~2\right) +  4~\log(\alpha_{\theta_1}\alpha_{\theta_2}) - 2~\log~N + 24\left(Z + \log~r_h\right)
+ {16\over g_s N_f}.  \nonumber \nd
 One can now take the form of the potential, given
in \eqref{amberlin}, and the wave-function ansatze, given in \eqref{rama}, and plug them in the Schr\"odinger-type equation
\eqref{Schrodie} to obtain the following equation for $\mathbb{F}_T(Z)$:
\bg\label{diffeq-F}
\mathbb{F}"_T(Z)
+ \left({1 - 2 i \mathbb{I}\over Z}\right)\mathbb{F}'_T(Z) +
\left({1\over {\bf A} + {\bf B} e^{-2 Z}} -{1 + 4{\mathbb{I}}^2\over 4Z^2}\right) \mathbb{F}_T(Z)  = 0,  \nd
where $\mathbb{I}$ is still given by \eqref{prelim}. The above second order differential equation is rather hard to solve
because of the presence of the exponential term $e^{-2Z}$. However, since we seek the spectral function only in the limit of large $Z$ where $e^{-2Z}$ vanishes, we can easily remove the problematic term from our equation
\eqref{diffeq-F}. Doing this yields the following form for $\mathbb{E}_T(Z)$:
\bg\label{eerie10}
\mathbb{E}_T(Z) =  Z^{{}^{{1\over 2} - i \mathbb{I}}}
\left[ {\bf C}_+ {\rm exp}\left(-{iZ\over \sqrt{\bf A}}\right) + {\bf C}_- {\rm exp}\left({iZ\over \sqrt{\bf A}}\right)\right], \nd
where $C_+$ and $C_-$ are two integration constants whose values will be determined later. To extract the actual fluctuation $E_T(Z)$ from \eqref{eerie10}, we need the functional form for $P(Z)$ in the large $Z$ limit. This is easy to extract from \eqref{runimara} and may be written as:
\bg\label{titan}
P(Z) = g_sN_f~\mathbb{G}_6 \mathbb{P}_2 e^{2Z}, \nd
which is as expected proportional to $g_sN_f$, and $\mathbb{P}_2$ is defined in \eqref{ginlynn}. The other coefficient
$\mathbb{G}_6$ appearing above can be extracted from some combinations of $\mathbb{G}_i$ and $\mathbb{P}_i$
in \eqref{clmehta} and \eqref{ginlynn} respectively at large $N$. Here we write this simply as:
\bg\label{1000}
\mathbb{G}_6 \equiv  {N^{1/10} r_h^2 \over n_o \alpha^2_{\theta_1} g_s^{3/2}}, \nd
where $n_o$ is a numerical constant given by $n_o = 4\sqrt{2} 3^{1/3} \pi^{3/2} \approx 45.43$. Combining
\eqref{eerie10}, \eqref{titan} and \eqref{1000} together and looking at the definition of $E_T(Z)$ given just after
\eqref{Schrodie}, we can finally determine the form of the fluctuation at large $Z$ as:
\bg\label{kumbaya}
E_T(Z) =  {Z^{{}^{{1\over 2} - i \mathbb{I}}}
\left[ {\bf C}_+ {\rm exp}\left(-{iZ\over \sqrt{\bf A}}\right) + {\bf C}_- {\rm exp}\left({iZ\over \sqrt{\bf A}}\right)\right]
e^{-Z} \over
\sqrt{g_sN_f~\mathbb{G}_6 \mathbb{P}_2}}. \nd
Few comments are in order related to the form of \eqref{kumbaya}. First we see that the suppression factor is
$\left(g_sN_f\right)^{-1/2}$. From here it would  seem like this does not have a natural zero flavor i.e $N_f = 0$ limit. However when combined with $\mathbb{P}_2$, $g_sN_f~\mathbb{P}_2$ does have a zero flavor limit, and is given by:
\bg\label{jacryan}
{}^{\lim}_{N_f \to 0}~\big[g_sN_f~\mathbb{P}_2\big] = 4\pi, \nd
which one may also verify directly at the level of the Schrodinger equation \eqref{Schrodie}.  Secondly, the
integration constants $C_\pm$ appearing in \eqref{kumbaya}, can in principle be complex valued. So this will require us to investigate few possibilities associated with various choices of $C_\pm$ satisfying the boundary conditions. Let us start by investigating the form for ${\bf A}$ given in \eqref{ADdeaf}. First let us assume that $Z$ goes to infinity as:
\bg\label{mrglass}
Z ~ \approx  ~ {1\over 3} \left(\log~N + {4\pi\over g_sN_f}\right). \nd
The above would make sense because $N \to \infty$ and $g_sN_f \to 0$. In this limit $\mathbb{P}_2$ may be replaced by $- 3~\log~r_h$. In other words ${\bf A}$ in \eqref{ADdeaf} becomes:
\bg\label{mroyalmela}
{\bf A} =  -1 - {1\over \log~r_h} + {\cal O}\left({1\over \log^2 r_h}\right), \nd
for large $\log~r_h$ so that the inverse suppression in \eqref{mroyalmela} makes sense. Assuming this is possible,
plugging \eqref{mroyalmela} in
\eqref{kumbaya} would imply the following form for fluctuation $E_T(Z)$:
\bg\label{oculta}
E_T(Z) =  {Z^{{}^{{1\over 2} - i \mathbb{I}}}
\left[ {\bf C}_+ {\rm exp}\left(-{Z\over \sqrt{1 + {1\over \log~r_h}}}\right) + {\bf C}_
- {\rm exp}\left({Z\over \sqrt{1 + {1\over \log~r_h}}}\right)\right]
e^{-Z} \over \sqrt{g_sN_f~\mathbb{G}_6 \mathbb{P}_2}}, \nd
where we have suppressed inverse $\log^2r_h$ dependences. Note that the functional form for $E_T(Z)$ is {\it not} the only way to express $E_T(Z)$ from \eqref{kumbaya}. For example if the horizon radius goes as:
\bg\label{anastar}
r_h ~\approx~ {\rm exp}\left({4\pi\over 3g_sN_f} - Z_{\rm uv}\right), \nd
in the limit of very large $Z_{\rm uv}$ and vanishing $g_sN_f$, then one may rewrite $\mathbb{P}_2$ simply in terms of
$\log~N$ and not $\log~r_h$. This means ${\bf A}$ in \eqref{ADdeaf} in-turn will be expressed in terms of $\log~N$ and not $\log~r_h$, implying:
\bg\label{listener}
E_T(Z) =  {Z^{{}^{{1\over 2} - i \mathbb{I}}}
\left[ {\bf C}_+ {\rm exp}\left(-{Z\over \sqrt{1 - {3\over \log~N}}}\right) + {\bf C}_
- {\rm exp}\left({Z\over \sqrt{1 - {3\over \log~N}}}\right)\right]
e^{-Z} \over \sqrt{g_sN_f~\mathbb{G}_6 \mathbb{P}_2}}. \nd
From the multiple ways of expressing \eqref{kumbaya}, for example \eqref{oculta} and \eqref{listener}, one might worry that the final result would  be dependent on our approximation scheme. However we will show in section
\ref{onshala} that this will not be the case.

Finally, the functional form for $E_T(Z)$ in the zero momentum limit matches with the functional form for $E_{x_1}$ as may be seen from \eqref{hostiles}. This will help us to express the on-shell action completely in terms of known parameters appearing in \eqref{kumbaya}, allowing us to compute the spectral function more efficiently. This is the topic that we turn to next in the following section.

%

\subsection{On-shell action and the strong coupling spectral function \label{onshala}}

In the previous section we managed to find the functional form for the gauge field fluctuation $E_T(Z)$ in the large
$Z$ and in the zero momentum limit. What we now want is the four-dimensional on-shell action. This can be easily
extracted from the {\it boundary} piece of the Lagrangian \eqref{lag}. Earlier we had used \eqref{lag} to determine the EOM for $\mathbb{E}_T(Z)$ and subsequently for $E_T(Z)$. Plugging in the EOM in \eqref{lag} then leaves us only with the boundary term, that we shall label as the on-shell four-dimensional action $\mathbb{S}_4$.
This takes the following form:
\bg\label{lodgers}
\mathbb{S}_4 &= & \frac{\Omega_2 T_{D6}}{4}\int d^{4}x~dZ ~ \partial_\alpha\left[ e^{-\varphi}\sqrt{-{\cal G}}
\left({\cal G}^{\mu [\alpha}{\cal G}^{\beta]\gamma}\partial_{[\gamma}{\cal A}_{\mu]}
-\frac{1}{2}{\cal G}^{[\alpha\beta]}{\cal G}^{\mu\nu}\partial_{[\mu}{\cal A}_{\nu]}
\right){\cal A}_{\beta}\right]\nonumber\\
 &= & \frac{\Omega_2T_{D6}}{2}\int d^{4}x\Biggl[ e^{-\varphi}\sqrt{-{\cal G}}
\left(({\cal G}^{tZ})^2{\cal A}_{t}\partial_{Z}{\cal A}_{t}- \sum_{a = 0}^3 {\cal G}^{x_ax_a}
{\cal G}^{ZZ}{\cal A}_{x_a}\partial_{Z}{\cal A}_{x_a}
\right)\Biggr]^{Z_{\rm uv}}_{Z_{h}},  \nonumber\\
\nd
where $x_0 \equiv t$ and $\Omega_2$ is the same volume of the two-sphere that we had in \eqref{D6DBI3}.  Note that we took $Z_h$ to be the lower limit of $Z$ to be consistent with the lower bound \eqref{tree}\footnote{Using \eqref{tree} one may easily show that $Z_h \ge {1\over 3}~ \log\left({{\bf C} \over 8\pi \alpha^4_{\theta_1} r_h^3}\right)$.}.
However what we seek here is in fact the on-shell action at the boundary $Z = Z_{\rm uv}$, so the near-horizon
geometry is not too relevant for us. At the boundary
$F_{tZ}=-F_{Zt}=0$, so we must set ${\cal G}^{tZ}=0$ and replace $\sqrt{-{\cal G}}$ by $\sqrt{-G}$. Incorporating these changes, the boundary value of the on-shell action is now given as:
\bg\label{pawmil}
\mathbb{S}_4 =-\frac{\Omega_2T_{D6}}{2}\int d^{4}x \left[e^{-\varphi}\sqrt{-{G}}{G}^{ZZ}
\left(\sum_{a = 0}^3 {G}^{x_a x_a}{\cal A}_{x_a}(Z, -k)\partial_{Z}{\cal A}_{x_a}(Z, k)\right) \right]_{Z_{\rm uv}}
\nd
Using the gauge field EOM \eqref{laura1}, but now resorting to the metric $G_{\mu\nu}$ instead of
${\cal G}_{\mu\nu}$,  and the result in {\bf Appendix \ref{peridotv}}, the above action can be rewritten in terms of the gauge invariant variables $E_{x_1}$, $E_{x_2}$ and $E_{x_3}$ as:
\bg\label{duckling}
\mathbb{S}_4 =-\frac{\Omega_2T_{D6}}{2}\int d^{4}x \left[e^{-\varphi}\sqrt{-{G}}~{G}^{ZZ} \sum_{a = 1}^3{G}^{x_ax_a}
\left(\frac{E_{x_a}(Z, -k)\partial_{Z}E_{x_a}(Z, k)}{\omega^2- k_a^2}\right)\right]_{Z_{\rm uv}},
\nd
with $k_a^2$ given in \eqref{final2}, and
at this point we will be concerned about the $x_1$ part of the fluctuation. In other words, we only want to study the
behavior of $E_{x_1}$ at zero momentum. At zero momentum, according to \eqref{hostiles}, the fluctuations
$E_{x_1}$ and $E_T$ follow the same equation \eqref{EOMET}. Using such an identification, we can define:
\bg\label{stephanie}
E_{x_1}(Z, k) \equiv {E_{0}(k) E_T(Z) \over E_{x_1}(Z_{\rm uv}, k)}, \nd
where one may match the Lorentz indices using \eqref{hostiles}. Plugging
\eqref{stephanie} in \eqref{duckling} and using $E_0(k) E_0(-k) = 1$, it is easy to see that the zero momentum limit
yields the following action for the $x_1$ piece of the fluctuation:
\bg\label{asidana}
\mathbb{S}_4^{(1)}= - {\Omega_2 T_{D6} \over  2 \omega^2}\int d^4x \Biggl[e^{-\varphi}\sqrt{-G}~{G}^{ZZ}
{G}^{x_1x_1}\left(\frac{\partial_{Z}E_{T}(Z)}{E_{T}(Z)}\right)\Biggr]_{Z_{\rm uv}}.
\nd
Before moving ahead, let us make couple of observations. One, $E_T(Z)$ is exactly the fluctuation \eqref{kumbaya} that we derived earlier and is therefore subjected to take either of the two possible limits \eqref{oculta} and
\eqref{listener} that we mentioned above. Two, the coefficient of $E'_T(Z)/E_T(Z)$ looks very similar to
$P(Z)$ in \eqref{rosamund}, so one might think that it can take the functional form \eqref{runimara}. This is unfortunately not the case because $P(Z)$ in \eqref{rosamund} and \eqref{runimara} involves
${\rm det}~{\cal G}_{ab}$ whereas the coefficient of $E'_T(Z)/E_T(Z)$ in \eqref{asidana} involves
${\rm det}~G_{ab}$. The former differs from the latter by the presence of $F_{ab}$.

The above discussion more or less sets out the tone for the rest of the computations. There are two parts to the computation that we will indulge in the following. One is the coefficient of $E'_T(Z)/E_T(Z)$  in \eqref{asidana}
and the other is $E'_T(Z)/E_T(Z)$ itself. To condense some of the subsequent formulae, let us define:
\bg\label{chubrit}
\mathbb{P}_7 \equiv \mathbb{P}_2 + 3~\log~r_h, \nd
where $\mathbb{P}_2$ is given in \eqref{ginlynn}. The coefficient of $E'_T(Z)/E_T(Z)$ can then be represented in the following way:

{\footnotesize
\bg\label{tellnoone}
e^{-\varphi}\sqrt{-G}~{G}^{ZZ}{G}^{x_1x_1} \equiv {{\bf \Sigma}_{11} \over 16\sqrt{3}n_o N^{1\over 10}
\alpha^4_{\theta_1} \alpha^4_{\theta_2} g^{3/2}_s\left(6b^2 +  e^{2Z}\right)} +
{3 {\bf \Sigma}_{22} \over 2 n_o N^{-{1\over 10}} \alpha^2_{\theta_1} \alpha^4_{\theta_2}\left(6b^2 + e^{2Z}\right)^2},
\nonumber\\ \nd}
where we are suppressing higher orders $1/N$ terms, and $n_o$ is a numerical constant that appeared in
\eqref{1000}. Note that both the denominators are suppressed differently with respect to $N, \alpha_{\theta_1}$
and $e^{2Z}$. The numerators are non-trivial functions of $e^{2Z}$, and they will govern the behavior of the spectral function. Let us therefore study them carefully by first writing out the form for ${\bf \Sigma}_{11}$:
\bg\label{deadofwinter}
{\bf \Sigma}_{11} = -g_sN_f ~r_h^2 (1-e^{-4Z})(9b^2 + e^{2Z})(2\alpha^4_{\theta_1} e^{2Z}
- 3 b^2 \alpha^4_{\theta_2})(-9b^2 e^{-2Z} + 2 \mathbb{P}_2)\mathbb{P}_1, \nonumber\\ \nd
 where we see that it is proportional to $g_sN_f$. This makes sense because in the absence of the flavor D6-branes
 we won't see this contribution. The forms for $\mathbb{P}_1$ and $\mathbb{P}_2$ are given earlier in
 \eqref{ginlynn}, where $\mathbb{P}_2$ is a function of $Z$ and $r_h$ but $\mathbb{P}_1$ is independent of both of them. At this stage we can make \eqref{deadofwinter} vanish by choosing:
 \bg\label{tarazu}
 2\mathbb{P}_2 - 9 b^2 e^{-2Z} ~ = ~ 0. \nd
 Few questions immediately arise from \eqref{tarazu}. What is the logic behind the choice \eqref{tarazu}, instead of making the other bracketed terms in \eqref{deadofwinter}, to vanish?  What would happen if we make the other bracketed terms in \eqref{deadofwinter} vanish? The answers to both the questions lie on the following observation: since $b^2$ as well as $\alpha_{\theta_i}$ pieces cannot be large, the first three brackets in \eqref{deadofwinter} cannot vanish. Making them zero would lead to contradictions. Therefore from \eqref{mrglass} we see that $Z$ can be very large, and we can use this to fix the value of $r_h$ using \eqref{tarazu}. This gives us:
 \bg\label{insaafka}
 r_h = {\rm exp}\left[-{3b^2\over 2 N^{2\over 3} {\rm exp}\left({8\pi\over 3g_sN_f}\right)}\right] \equiv 1 - \epsilon^2, \nd
 where $\epsilon$ is a small number that can be derived from above. One may also see that \eqref{insaafka} cannot be related to \eqref{anastar}. This is because of our choice between \eqref{mrglass} and \eqref{anastar}: we are allowed either of them, but not both. Coincidentally, we can choose either \eqref{oculta} or \eqref{listener}, but not both. As one may easily verify that the choice \eqref{mrglass} only, and therefore \eqref{oculta}, can be consistent with \eqref{tarazu}.  The caveat however is that, since $\log~r_h$ is no longer a large number, the expansion in
 \eqref{mroyalmela} cannot be terminated and we shall require the exact form for ${\bf A}$ in \eqref{mroyalmela}. We will discuss a way out of this soon.

 After the dust settles, there will be no ${\bf \Sigma}_{11}$ term, and so we have to go to the next term given by
 ${\bf \Sigma}_{22}$. The next term incorporates both $g_sN_f$ as well as ${g_sM^2\over N}$, and takes the following form:
 \bg\label{zaman}
 {\bf \Sigma}_{22} &=&  g_sN_f \left({g_sM^2\over N}\right) b\beta r_h^2 e^{-6Z} (e^{4Z} - 1) (1 + \log~r_h)\nonumber\\
 &\times&  \Big[\alpha^4_{\theta_2}
 \left(6 e^{2Z}\alpha_b ~\log~r_h + \mathbb{L}(Z)\right) - \alpha^4_{\theta_1}\left(12 e^{6Z}~\log~r_h - 2\mathbb{K}(Z)\right)
  \Big], \nd
where we see that the term is dependent on $r_h^2$ as well as various other factors of $\log~r_h$.  There are also $e^Z$ and $N$ dependences that will take large values, so we will have to be careful taking the limits at large $Z$ and large $N$.  The various other quantities appearing above are $\alpha_b, \mathbb{K}(Z)$ and $\mathbb{L}(Z)$
that will be defined in the following. First let us start with $\alpha_b$:
\bg\label{chennai}
\alpha_b \equiv 54 b^4 + 18 b^2 e^{2Z}  + e^{4Z} ~ \to ~ e^{4Z}, \nd
where on the right we have shown its behavior at large $Z$: the resolution parameter $b^2$ being small, does not contribute anything to $\alpha_b$.  In the same vein, $\mathbb{K}(Z)$ is defined in the following way:
\bg\label{vananda}
\mathbb{K}(Z) \equiv  -162 b^4 e^{2Z} - 54 b^2 e^{4Z} + \left(2\mathbb{P}_7 - 3\right) e^{6Z}, \nd
 where we have defined $\mathbb{P}_7$ in \eqref{chubrit} above. Using this definition for $\mathbb{P}_7$, and
 \eqref{mrglass} for $Z$ that we took earlier, one can easily show that:
 \bg\label{juani}
 \mathbb{P}_7 ~  = ~ 0, \nd
 leading to some simplification in \eqref{vananda}. It also means that for large $Z$, $\mathbb{K}(Z)$ goes as
 $-3 e^{6Z}$. This is consistent with the other coefficient for $\alpha^4_{\theta_1}$ as evident from \eqref{zaman}.
 Finally, the last term $\mathbb{L}(Z)$ takes the following form:
 \bg\label{don2}
 \mathbb{L}(Z) \equiv 972 b^6 - 27 b^4 \left( 4\mathbb{P}_7 - 11\right)e^{2Z} - 18 b^2 \left( 2\mathbb{P}_7 - 1\right)
 e^{4Z} - 2 \mathbb{P}_7~ e^{6Z} ~ \to ~ 18 b^2 e^{4Z}, \nonumber\\ \nd
 where the large $Z$  behavior is solely governed by the vanishing of $\mathbb{P}_7$  in \eqref{juani}.  In fact plugging
 the limiting values of \eqref{chennai}, \eqref{vananda} and \eqref{don2} in \eqref{zaman} and then in
 \eqref{tellnoone}, leads us to the following
 behavior of \eqref{tellnoone} for large $Z$:

 {\footnotesize
 \bg\label{kohlapuri}
 e^{-\varphi}\sqrt{-G}~{G}^{ZZ}{G}^{x_1x_1}
 =  - 9 g_sN_f ~ \kappa_o r_h^2\left(1 + \log~r_h\right)\left[ {\left(1 + 2~\log~r_h\right)\alpha^2_{\theta_1} \over
 \alpha^4_{\theta_2}} - { \log~r_h \over \alpha^2_{\theta_1}}\right] {g_sM^2\over N} , \nd}
with $\kappa_o$ being a constant that depends on $N$ as $\kappa_o \equiv {b\beta N^{0.1}\over n_o}$, where $n_o$ remains the same numerical constant that appeared in \eqref{1000}.

Before moving ahead, let us pause briefly to examine the situation at hand.
The crucial outcome of \eqref{kohlapuri} is the dominance of ${\bf \Sigma}_{22}$ over ${\bf \Sigma}_{11}$ because of the imposed constraint \eqref{tarazu}. This further lead to the form of the horizon radius given in \eqref{insaafka}
that is of order 1. This in turn gives us the high temperature limit, and so one might ask if there is a way to analyze the spectral function for {\it small} $r_h$.  Otherwise an expression like ${\bf A}$ in \eqref{mroyalmela} does not have a good expansion in terms of inverse $\log~r_h$. The situation is subtle because we would still have to impose \eqref{tarazu} to eliminate
${\bf \Sigma}_{11}$ piece in \eqref{tellnoone}. How can we then avoid the outcome \eqref{insaafka} for the horizon radius?

A way out of this conundrum is to {\it not} impose \eqref{mrglass} that determines $Z$ from the start, and instead use
\eqref{tarazu} to fix $Z$. This means \eqref{mroyalmela} for ${\bf A}$ does not hold anymore although the form of
${\bf A}$ in \eqref{ADdeaf} continues to hold. $Z$ then satisfies:
\bg\label{carmilla}
Z + {3\over 2}~ b^2 e^{-2Z} = {1\over 3} ~\log~N - \log~r_h + {4\pi \over 3g_sN_f}, \nd
which is extracted from \eqref{tarazu}.  The RHS has $\log~ r_h$ and, as discussed above, we cannot use either
\eqref{insaafka} or \eqref{anastar} for $r_h$. Instead we will use a different way, as shown in {\bf Appendix} \ref{saldana}, to determine the horizon radius
by demanding the vanishing of the effective number of the three-brane charges in the original type IIB side.
Solving \eqref{carmilla} then gives us the following value for $Z$:
\bg\label{mircalla}
Z  = Z_{\rm uv}  \equiv {1\over 3}~\log~N - \log~r_h + {4\pi \over 3 g_sN_f} +
{1\over 2}  ~\mathbb{W}_n \left[ -{3b^2 r_h^2\over N^{2/3}} {\rm exp}\left(-{8\pi \over 3g_sN_f}\right)\right], \nd
where $\mathbb{W}_n$  is the analytic continuation of the product log function with integer $n$. By construction this is a large positive number because $N$ is large whereas $r_h$ is a very small number. Plugging \eqref{mircalla} in
\eqref{ADdeaf} then gives us the following value for ${\bf A}$:
\bg\label{kirnstein}
{\bf A} = -1 + {2N^{2/3}\over 3b^2 r_h^2} ~{\rm exp}\left({8\pi\over 3 g_sN_f}\right) +
{N^{4/3}\over 9 b^4 r_h^4} ~{\rm exp}\left({16 \pi\over 3 g_sN_f}\right) ~ \to ~
{N^{4/3}\over 9 b^4 r_h^4} ~{\rm exp}\left({16 \pi\over 3 g_sN_f}\right), \nonumber\\ \nd
 which is expectedly different from \eqref{mroyalmela}.  The form of ${\bf A}$ shows that it is in fact a very large number because in addition to it being inversely proportional to a small number, i.e $r_h << 1$ as mentioned above, it  is also exponentially dependent on a  large number as $g_sN_f \to 0$. This will be useful for us because large
 ${\bf A}$ can simplify the expression for $E_T(Z)$ in \eqref{kumbaya}. We will come back to this soon.

 Let us now compute the coefficient \eqref{tellnoone}, which in turn means computing ${\bf \Sigma}_{11}$ and
 ${\bf \Sigma}_{22}$. As mentioned earlier, ${\bf \Sigma}_{11}$ vanishes, so we only need to compute
 ${\bf \Sigma}_{22}$ at large $Z$. For this we will need the limiting values for $\alpha_b, \mathbb{K}(Z)$ and
 $\mathbb{L}(Z)$ in \eqref{chennai}, \eqref{vananda} and \eqref{don2} respectively. The limiting value for $\alpha_b$ remains $e^{6Z}$ as before, but the limiting values for $\mathbb{K}(Z)$ and $\mathbb{L}(Z)$ change because we can no longer apply \eqref{juani} anymore. They now take the following values:
 \bg\label{laracroft}
 \mathbb{K}(Z) = 2 \mathbb{P}_7 - 3, ~~~~~ \mathbb{L}(Z) = - 2 \mathbb{P}_7, \nd
 where $\mathbb{P}_7$ is given in \eqref{chubrit}. Plugging \eqref{laracroft} in \eqref{zaman} and using \eqref{tarazu}, gives us the following value for the coefficient in \eqref{tellnoone}:

{\footnotesize
 \bg\label{jgaddar}
  e^{-\varphi}\sqrt{-G}~{G}^{ZZ}{G}^{x_1x_1} & = & - {1\over 2}~g_sN_f~\kappa_1 r_h^2\left(1+\log~r_h\right)
  \left[{9b^2 \over e^{2Z}}\left({1\over \alpha^2_{\theta_1}} - {2\alpha^2_{\theta_1} \over \alpha^4_{\theta_2}}\right)
  + {6\alpha^2_{\theta_1} \over \alpha^4_{\theta_2}}\right] {g_sM^2\over N} \nonumber\\
  & = & - 3 g_sN_f \left({g_sM^2\over N}\right)\left({\alpha^2_{\theta_1} \over \alpha^4_{\theta_2}}\right) \kappa_1 r_h^2
  \left(1 + \log~r_h\right), \nd}
  where $\kappa_1 = {\kappa_o\over g_s^{3/2}}$ and $\kappa_o$
  is the same constant that appeared earlier in \eqref{kohlapuri}, and in the last line we have used the large $Z$ limit \eqref{mircalla} to eliminate the $e^{-2Z}$ piece. The above result differs clearly from
  \eqref{kohlapuri}, which was computed for $r_h$ as in \eqref{insaafka}. Here we expect $r_h$ to be
  small $-$ as shown in {\bf Appendix} \ref{saldana}  $-$ and so \eqref{jgaddar} will finally be proportional to $r_h^2~ \log~r_h$.

Having done the first part of the computation  in \eqref{asidana}, let us now  investigate the second part which is the
ratio $E'_T(Z)/E_T(Z)$. The functional form for $E_T(Z)$ is given in \eqref{kumbaya} and is expressed in terms of
coefficients $C_{\pm}$ which could in principle be complex. The ratio then can be written as:
\bg\label{alana}
{E_T'(Z) \over E_T(Z)} = {C_+ e^{-gZ} \left({\alpha\over Z} - Q\right)
+ C_-e^{-{\overline{g}}Z} \left({\alpha\over Z} - {\overline{Q}}\right) \over C_+ e^{-gZ} + C_-e^{-\overline{g}Z}}, \nd
where we have introduced three functions $\alpha, g$ and $Q$ that are in general complex. In fact what we require
is that the $\alpha, g$ and $Q$ functions remain complex for large $Z$ and small $r_h$. The precise forms for these functions are:
\bg\label{dhaidhai}
\alpha \equiv {1\over 2} - i\mathbb{I}, ~~~~~ g \equiv 1 +  {i\over \sqrt{\bf A}}, ~~~~~
Q \equiv g + Z ~{dg\over dZ} + {1\over 2\mathbb{P}_2} {d\mathbb{P}_2 \over dZ},  \nd
where $\mathbb{I}$ and ${\bf A}$ are defined in \eqref{prelim} and \eqref{ADdeaf} respectively. Note that in the limit of large $N$, small $g_sN_f$,  and small $r_h$, ${\bf A}$ is  large number, implying a small (but non-zero) complex piece in $g$.  On the other hand, $\mathbb{I}$ is a large number being proportional to $g_sN$ and inversely proportional to the horizon radius $r_h$. However to avoid contradictions, we will not take any limit at this stage and
 continue with the operations with exact expressions. This gives us:
 \bg\label{evans}
 {E_T'(Z) \over E_T(Z)} = {C_+\left[\mathbb{P}_o - i \left({\mathbb{I}\over Z} + \mathbb{Q}_o\right)\right]  +
 C_-\left[\mathbb{P}_o - i \left({\mathbb{I}\over Z} - \mathbb{Q}_o\right)\right] {\rm exp}\left({2iZ\over \sqrt{\bf A}}\right)
 \over C_+ + C_-  {\rm exp}\left({2iZ\over \sqrt{\bf A}}\right)}, \nd
  where now there are three distinct sources of imaginary pieces from \eqref{evans}: they can come from the $C_\pm$ coefficients, the exponential term $e^{iZ/\sqrt{\bf A}}$ and the bracketed terms in \eqref{evans}. The bracketed terms are defined with respect to two new functions $\mathbb{P}_o$ and $\mathbb{Q}_o$, which may be written as:
  \bg\label{graglam}
 \mathbb{P}_o = -1 + {1\over 2 Z} + {3\over 2 \mathbb{P}_2}, ~~~~~~
 \mathbb{Q}_o = {1\over \sqrt{\bf A}}\left[1 - {9Z\over 2{\bf A} \mathbb{P}^2_2}\left(1 + {3\over 2\mathbb{P}_2}\right)\right]. \nd
The limit that we are looking for now, and as mentioned earlier, is the large $Z$, large $N$ and small $r_h$ limit  where $Z$ becomes large as \eqref{mircalla}. Essentially then it is the large $N$ and large $\vert\log~r_h \vert$ limit. In this limit $\mathbb{P}_2$ can be expressed using $Z$ as \eqref{tarazu}, which would tell us that it is a small number\footnote{$\mathbb{P}_2 \approx {9b^2 r_h^2\over 2N^{2/3}} ~{\rm exp}\left(-{8\pi\over g_sN_f}\right)$.}. Plugging the values of ${\bf A}$ from \eqref{kirnstein}, and $Z$ from \eqref{mircalla} now implies that $\mathbb{Q}_o$ may be approximated by:
\bg\label{glamdanger}
 \mathbb{Q}_o ~\approx~ {3b^2 r_h^2\over N^{2/3}}\left[\log\left({r_h^2\over N^{2/3}}\right) - {8\pi\over 3g_sN_f}\right]
 {\rm exp}\left(-{8\pi\over 3g_sN_f}\right) ~\equiv ~ 3b^2 ~{}^{\rm lim}_{x \to 0} \left[ x ~\log~x\right] , \nd
 where on the RHS we have shown the behavior of the function as it approaches zero, by ignoring a constant
 additive factor as the term in the bracket on the LHS of \eqref{glamdanger} will always dominate. We have also defined $x$ and then $Z$ as a function of $x$ in the following way:
\bg\label{milla}
 x  \equiv   {r_h^2\over N^{2/3}}{\rm exp}\left(-{8\pi\over 3g_sN_f}\right), ~~~~ Z  \equiv  -{1\over 2} ~\log~x, \nd
 where the latter should be viewed as an alternative expression for \eqref{mircalla}.
 For large $N$, small $r_h$ and  $g_sN_f \to 0$, it is easy to see that $x$ vanishes whereas $Z$ becomes very large.
 However $\mathbb{Q}_o$ will always go to zero in this limit.
  What we now want to claim, in this limit, is that:
 \bg\label{abdanger}
 \Big\vert {\mathbb{I} \over Z} \Big\vert ~ \gg ~ \vert \mathbb{Q}_o \vert, \nd
 which is easy to justify from the form on $Z$ in \eqref{milla} and the fact that multiplying $x$ with any powers of
 $\log~x$ will always approach zero in the above limit.

 The dominance of $\mathbb{I}/Z$ over $\mathbb{Q}_o$ is a huge simplification for us because this will not only render the expression \eqref{evans} manageable without worrying about contributions from the exponential pieces, but also remove the ambiguity of its dependence on the constants $C_\pm$ whose values have not been explicitly determined. In fact after plugging in all the values from \eqref{graglam} and \eqref{prelim} in \eqref{evans} and using the limiting conditions \eqref{glamdanger} and \eqref{abdanger}, it is easy to see that:
 \bg\label{jovovich}
 {E_T'(Z)\over E_T(Z)} = -1 + {1\over 2Z_{\rm uv}} + {3\over \mathbb{P}_2}
 - {im_0 \omega \over 4Z_{\rm uv} m_{0^{++}}}\sqrt{6b^2 +1 \over 9b^2 + 1} + {\cal {O}}\left({1\over N^2}\right), \nd
 where $m_{0^{++}}$ is the mass of the lightest glueball expressed in terms of scale $m_0$ and is given in
 \eqref{doclaura}.  In fact this is all we need, because the imaginary part of \eqref{jovovich} can then take the following form:
 \bg\label{elizshue}
 {\bf Im}\left[{E_T'(Z)\over E_T(Z)}\right] = - {\omega \sqrt{4\pi g_sN}\over 4 r_h Z_{\rm uv}}\sqrt{6b^2 +1 \over
 9b^2 + 1}, \nd
 where we have used \eqref{doclaura} to write it in this form.
 One may note its linear dependence on $\omega$, the frequency parameter that we encountered earlier. It is also inversely proportional to the horizon radius $r_h$, a fact that will be useful soon.

 The logic behind the above series of computations should be clear now. What we are looking for is the retarded Green's function in the zero momentum limit. This is now easy to extract from \eqref{asidana}, and can be written as:
 \bg\label{melleo}
 \mathbb{G}_{x_1x_1}^{({\rm R})}(\omega, q = 0) \equiv
 \Omega_2 T_{D6}\left[e^{-\varphi}\sqrt{-G}~{G}^{ZZ}
{G}^{x_1x_1}\left(\frac{\partial_{Z}E_{T}(Z)}{E_{T}(Z)}\right)\right]_{Z_{\rm uv}}, \nd
which precisely contains the two pieces of computations that we performed above, namely the coefficient of $E_T'/E_T$ in \eqref{jgaddar} and the ratio $E'_T/E_T$ itself in \eqref{jovovich}. One additional input was the imaginary piece in \eqref{jovovich} that we extracted in \eqref{elizshue}. The reason for this extra bit of work is apparent: the spectral function is exactly the imaginary piece of the retarded Green's function, i.e:
\bg\label{donmckay}
\rho(T, \omega) \equiv -2 {\bf Im}~\mathbb{G}^{({\rm R})}(\omega, q = 0), \nd
 where $T$ is the temperature that will be related to the horizon radius $r_h$. Since \eqref{jgaddar} is all real, the imaginary piece in the retarded Green's function can only come from \eqref{jovovich}. Any other contributions to the imaginary piece will be suppressed by higher powers of $1/N$ so does not concern us here. Putting everything together then gives us the required expression for the spectral function:
 \bg\label{prunej}
 {\rho(T, \omega)\over \omega} = {3\over 4}~g_sN_f \left({g_sM^2\over N}\right) \mathbb{F}_a(N, g_s, Z_{\rm uv})
 \mathbb{F}_b(b, \alpha_{\theta_i}) ~br_h~\log~r_h, \nd
where expectedly this is proportional to $g_sN_f$ and $g_sM^2/N$. It is also proportional to $r_h$ (and also
$\log~r_h$), so at zero temperature $\rho(0, \omega) = 0$. We can use \eqref{a}, or the footnote \ref{internet}, to express the combination $br_h$ in terms of the resolution parameter as $a(r_h) \equiv br_h + {\cal O}\left({g_sM^2\over N}\right)$. This way, the pre-factor multiplying $\log~ r_h$ in \eqref{prunej} is not explicitly but only implicitly dependent on $r_h$ and it brings out the resolution in the gravity dual rather succinctly. The two other functions appearing in \eqref{prunej} may be defined in the following way:
\bg\label{tagchink}
\mathbb{F}_a(N, g_s, Z_{\rm uv}) = {N^{1/10}\sqrt{4\pi g_sN}\over g_s^{3/2} Z_{\rm uv}}, ~~~~~
\mathbb{F}_b(b, \alpha_{\theta_i}) = {\beta \over n_o} \left({\alpha^2_{\theta_1} \over \alpha^4_{\theta_2}}\right)
\sqrt{6b^2 +1 \over 9b^2 + 1}, \nd
where $n_o$ is a numerical constant defined after \eqref{1000}, and $\beta$ is defined in footnote \ref{internet}. Note that if we use the strong string coupling result, as opposed to the weak string coupling analysis
presented here (both a strong 't~Hooft coupling of course), $\beta$ can be defined from \eqref{a} with $\beta = c_1 = c_2$. The coefficient
$c_1$ appears in
\eqref{dansor} and $c_2$ is bounded by \eqref{em18}. Following this logic, what we now need is the $g_sN_f$ independent pieces to define $\beta$. Thus if we take the negative definite constant piece of $c_1$ from \eqref{dansor} and use this to define both $c_2$ and $\beta$ then we can ignore higher order $g_sN_f$ dependences. Thus essentially, from both strong and weak type IIA couplings,
$\beta$ will be another constant to ${\cal O}(g_sN_f)$, which in turn would make $\mathbb{F}_b$ to be another constant\footnote{Recall that the parameters $\alpha_{\theta_1}$ and $\alpha_{\theta_2}$ are constants.}, that we shall call
${\bf f}_b$. However the worrisome feature is the other
function in \eqref{tagchink}, i.e the function $\mathbb{F}_a$ that depends on $N, g_s$ and $Z_{\rm uv}$. Both
$N$ and $Z_{\rm uv}$, with $Z_{\rm uv}$ defined in \eqref{mircalla}, go to infinity whereas $g_s$ approaches zero.
If we define $\zeta_1 \equiv g_s, \zeta_2 \equiv 1/N$ and $\zeta_3 \equiv 1/Z_{\rm uv}$, then we can choose the behavior of each of these parameters such that:
\bg\label{teenmeye}
\lim_{\zeta_i \to 0} \mathbb{F}_a(\zeta_1, \zeta_2, \zeta_3) \equiv {\bf f}_a, \nd
 with a constant ${\bf f}_a$.\footnote{In the MQGP limit wherein $g_s\stackrel{<}{\sim}1$, one can argue that ${\bf f}_a$ will be a finite non-zero constant as follows. As $r_h<r_0$ or $|\log r_h|>|\log r_0|$ ($r_0$ being the $r$ where the $D3$-branes have been entirely cascaded away, and noting min$(r)=r_h$), hence instead of choosing $r_h$ to satisfy \eqref{pauldom}, assume $|\log r_h| = \frac{N^{1/3}}{\kappa}\left(\frac{1}{f}\right), 0<f<1$ and $\kappa = n_b g_s M^2/3$ from \eqref{pauldom}. As $Z_{UV} \sim |\log r_h| + \log N^{1/3} \sim   \frac{N^{1/3}}{\kappa}\left(\frac{1}{f}\right)$, so $\mathbb{F}_a \sim \frac{N^{3/5} f \kappa}{N^{1/3}} = N^{4/15} f \kappa$. If $g_s\sim {\cal O}(1)$ then $N\sim10^2$ is sufficient to consider large 't-Hooft coupling $g_s N$, one can choose $f: N^{4/15} f \kappa \sim {\cal O}(1)$.}.
 As $T \to 0$, $r_h$ vanishes, and from the expression of the spectral function in
 \eqref{prunej}, this also
 vanishes. Therefore we can finally put everything together and argue that:
\bg\label{jun17122}
\lim_{\omega \rightarrow 0}\frac{\rho(\omega)}{\omega} \equiv \lim_{\omega \rightarrow 0}\Biggl[\frac{\rho(T, \omega)}{\omega} - \frac{\rho(T=0, \omega)}{\omega}\Biggr]  \propto \frac{1}{3} - c_s^2,
\nd
where we have used \eqref{cmukha} to express the RHS in terms of the sound speed. Of course, as mentioned above, \eqref{cmukha} is a strong coupling result, so the comparison has to be done with $c_1$ and $c_2$ being proportional to $g_sN_f$ and not constants (as opposed to the weak string but strong 't~Hooft coupling answer). Taking all these into consideration we see a clear {\it linear} dependence on (${1\over 3} - c_s^2$) at strong 't~Hooft coupling, perfectly consistent with the results of sections \ref{nf0} and \ref{nfn0}.

Few comments are in order now. Our analysis is based on small $r_h$ as derived in {\bf Appendix} \ref{saldana}, so the natural question is what happens when $r_h$ is of order 1, i.e the one given in \eqref{insaafka}. When the horizon radius is of order 1, it means we are at the point where new degrees of freedom are about to enter, i.e we are in Region 2 of \cite{metrics}. Therefore unless we know the detailed metric configuration of Region 2 and beyond, we cannot perform the analysis as clearly as we have done here because of our definition the radial coordinate as
$r = r_h e^Z$. When $r_h$ is small we are still in Region 1 of \cite{metrics} and so precise computations may be performed (as shown here).

Secondly $r_h$ itself is bounded below by \eqref{tree}. This bound is of course to prevent any appearances of unphysical imaginary pieces in the computations. Clearly for the range of  $Z$ that we are concerned here, this poses no constraints. Thus happily all the results lead to the following conclusion:
\bg\label{203pm}
\lim_{\omega\rightarrow 0} ~{\rho(\omega)\over \omega} ~ \propto~ {1\over 3} - c_s^2. \nd

\subsection{The strong string coupling limit and pure classical supergravity \label{gsnf1}}

Most of the analysis section \ref{spectral} is done with $g_s \to 0$ and with large $M$. This differs a bit from what we did
in section \ref{nfn0} where $g_s = {\cal O}(1)$, so that natural question is whether
we can work through the analysis of sections \ref{spec2} $-$ \ref{onshala} assuming ($g_s, N_f$)
$\sim{\cal O}(1)$ and $N\gg1$ as part of the MQGP Limit of \cite{MQGP}\footnote{See footnote \ref{7khoon}.}.
This is an unusual large $N$ limit but still warrants the use of pure classical supergravity. To see this, one notes that by including terms higher order in $g_s N_f$  in the RR and NS-NS three-form fluxes than those considered in \cite{MQGP} and the NLO terms in the angular part of the metric, one sees that in the IR in the MQGP limit, there occurs an IR color-flavor enhancement of the length scale as compared to a Planckian length scale in the Klebanov-Strassler (KS) model \cite{KS} for large $M$, thereby showing that stringy corrections will be suppressed. To see this more explicitly, we summarize the main ideas of \cite{NPB,EPJC-2} here. Using \cite{metrics} let us define an effective number of color in the following way:
\bg\label{Neff}
N_{\rm eff}(r) = N\left[ 1 + \frac{3 g_s M_{\rm eff}^2}{2\pi N}\left(\log ~r + \frac{3 g_s N_f^{\rm eff}}{2\pi}\log^2 r\right)\right], \nd
where $M_{\rm eff}$ and $N_f^{\rm eff}$ are the effective number of bi-fundamental and fundamental flavors
respectively that are defined for our background in the following way:
\bg\label{mnf}
& & N^{\rm eff}_f(r)  \equiv N_f\left(1 + \sum_{m, n} k_{mn}N_f^m M^n\right) \nonumber\\
& & M_{\rm eff}(r)  \equiv  M\left[1 + {3g_sN_f\over 2\pi}\left(\log~r + \sum_{m, n} f_{mn} N_f^m M^n\right)\right],
\nd
where ($m, n$) indices are summed from ($m, n$) $=$ ($0, 0$) onwards, and henceforth to avoid clutter we will
use Einstein summation convention. The coefficients $k_{mn} \equiv k_{mn}(r, g_s)$ and $f_{mn} \equiv f_{mn}(r, g_s)$ and therefore the effective flavors are constructed from the higher orders $g_sN_f$ and ${g_sM^2\over N}$ corrections \cite{metrics}. Combining these together,
it was argued in \cite{NPB,EPJC-2} that  the length scale in the IR at $r = \Lambda$ will be dominated by:
\bg\label{length-IR}
L^4 \equiv 4\pi\alpha^{'2} \left(g_sN_f\right)^3\left({3g_sM\over 2\pi}\right)^2 \Bigl(f_{mn}(\Lambda) N_f^m M^n\Bigr)^2
\Bigl(k_{pq}(\Lambda) N_f^pM^q\Bigr)\log~\Lambda . \nd
In the IR, relative to KS geometry, we thus see that (\ref{length-IR})  implies  the abovementioned color-flavor enhancement of the length scale. Therefore in the IR, even for $g_s = 0.45, M=3$ and $N_f = 2$, upon inclusion of of $n, m >1$  terms in
$M_{\rm eff}$ and $N_f^{\rm eff}$ in (\ref{mnf}), the characteristic length scale in the MQGP limit \cite{MQGP}
involving $g_s \le 1$ satisfy:
\bg\label{length}
L ~ \gg ~ L_{\rm KS}, \nd
where $L_{\rm KS}$ is the characteristic length scale for the Klebanov-Strassler model \cite{KS} in the far IR.  Because of this enhancement, the stringy corrections are suppressed implying that one can still trust classical supergravity.

It is however interesting to note that in the IR,  one {\it can}  obtain $g^2_{YM} = {\cal O}(1)$ even for $g_s\rightarrow 0$, provided $N_f \ne 0)$. To see this let us first consider vanishing $N_f$.
The NSVZ RG flow equation for the SU$(M)$ gauge group that survives at the end of the Seiberg duality cascade,
 gives us:
 \bg\label{nsvz}
 {\partial \over \partial \log~\Lambda} \left({8\pi^2\over g^2_{YM}}\right) = 3M, \nd
 where the RHS appears from the integral of the NS two-form field over a vanishing two cycle ${\bf S}^2$ in the type IIB side. This is of course the same two-cycle discussed at the beginning of section \ref{nf0}, parametrized by ($\theta_2, \phi_2$), on which we have $M$ wrapped  D5-branes. The question is whether \eqref{nsvz} can allow
 $g^2_{YM} = {\cal O}(1)$.

Solving the equation \eqref{nsvz} gives the inverse YM coupling in terms of $M$ and $\log~r$, for $r = \Lambda$. It is easy to see that, with $M = {\cal O}(1)$ this is only possible if $\Lambda$ is proportional to the UV cutoff itself. Clearly since we want to concentrate on far IR physics, such a choice is not feasible. Additionally, since near the UV cutoff we expect the theory to become scale invariant, $M$ automatically vanishes there.

On the other hand, when $N_f \ne 0$, the above conclusion can change because the dilaton on the gravity side
is no longer a constant. Recall that, with $N_f$ flavors, the dilaton takes the following form \cite{ouyang, metrics}:
\bg\label{rahasya}
e^{-\phi} = \frac{1}{g_s}\left[1 - \frac{g_s N_f}{8\pi}\log \left(r^6 + 9 r^4a^2\right) - \frac{g_s N_f}{2\pi}\log\left(\sin\frac{\theta_1}{2}\sin\frac{\theta_2}{2}\right)\right], \nd
where $a^2$ is the resolution parameter that we encountered earlier. Using the fact that we have an almost vanishing resolution parameter, and the angular coordinates ($\theta_1, \theta_2$) are parametrized by \eqref{harper}, the inverse of the YM coupling now satisfy:
\bg\label{nyelle}
{1\over g^2_{YM}} \propto M\left[1 - {g_s N_f\over 2\pi} ~\log\left(\alpha_{\theta_1} \alpha_{\theta_2}\right) - {3g_sN_f\over 4\pi}~\log~r
+ {g_s N_f\over 4\pi} ~\log~N\right] \Bigl(\log ~r + {\cal O}(g_sN_f)\Bigr), \nonumber\\ \nd
at the scale $r = \Lambda$ measured with respect to the cutoff scale $\Lambda_{\infty}$. What we are looking for now is a $\Lambda$ in the IR whereat $g^2_{SU(M)} =  {\cal O}(1)$.  The scenario is more subtle now because of the additional ${\cal O}(g_sN_f)$ pieces appearing in \eqref{nyelle}. These pieces come from carefully looking at the
NS B-field threading the vanishing two-sphere on which we have the wrapped D5-branes. The B-field
is more non-trivial than what we had above, and is given by:
\bg\label{jfonda}
{\bf B}_2\vert_{S^2} = 3g_s M~\log~r\Bigl[1 + \mathbb{Q}(r, \theta_1, \theta_2) g_s N_f + {\cal O}(g_s^2N_f^2)\Bigr]
\sin~\theta_2~d\theta_2 \wedge d\phi_2, \nd
where the first term is precisely what we had on the RHS of \eqref{nsvz} for the case with vanishing $N_f$, and the second term involves the $g_sN_f$ corrections. These correction terms have been worked out in \cite{metrics}, and may be expressed as:
\bg\label{morningafter}
\mathbb{Q}(r, \theta_i) \equiv {9\over 8\pi} \log~r - {1\over 4\pi} \left(2 + {1\over \log~r}\right)
\log\left(\sin~{\theta_1\over 2}  \sin~{\theta_2\over 2}\right) - {1\over 4\pi}  \cot~\theta_2 \cot~{\theta_2\over 2},
\nonumber\\ \nd
where we have removed any dependence on the resolution parameter when writing \eqref{morningafter} from
\cite{metrics}\footnote{There is one subtlety that we are putting under the rug.  A part of the B-field in \eqref{jfonda}
goes as  ${9\over 4\pi} \left(g_sM\right) \left(g_sN_f\right) \log~r~\log~|a|$, where $a$ is the resolution parameter. This blows up in the limit $a \to 0$, so one might be worried that $\mathbb{Q}$ given in \eqref{morningafter} is not well defined in this limit. This is however not the case because the derivation of the B-field in \cite{metrics} was done with non-zero resolution parameter, and for zero resolution parameter we have to do the analysis separately. The result then is of course independent of the $\log~|a|$ piece, and is as given in \eqref{morningafter}.}.
In fact the ${\cal O}(g_sN_f)$ term alluded to in \eqref{nyelle} comes precisely from
 $\mathbb{Q}$ in \eqref{morningafter}. There is however one subtlety associated with the angular variables
 $\theta_i$ and $\phi_2$. Since the {\it integral} of the ${\bf B}_2$ field over the two-sphere parametrized by ($\theta_2, \phi_2$) contributes to the YM coupling $g^2_{YM}$, one needs to be careful while imposing \eqref{harper}.
 One way would be to impose \eqref{harper} to $\theta_1$ in \eqref{jfonda} and then integrate over $\theta_2$.  In that case an additional $N$ dependence would appear from the second term in \eqref{morningafter}. Alternatively we could also insert the value of $\theta_2$ from \eqref{harper} after integration over the two-sphere. The latter would imply that the integration of ${\bf B}_2$ field over the two-sphere is concentrated mostly near the regime defined in \eqref{harper}. After the dust settles, the equation that we need to solve to determine
$\Lambda$ can be derived from \eqref{nyelle} as:
\bg\label{pchopra}
\mathbb{A}~ \log^2\left({\Lambda_{~~}\over \Lambda_\infty}\right) + \mathbb{B}~ \log\left({\Lambda_{~~}\over \Lambda_\infty}\right) + \mathbb{C} = 0, \nd
which is a quadratic equation to the first order in $g_sN_f$. To higher orders in $g_sN_f$ the equation starts becoming more complicated. The various coefficients of \eqref{pchopra} are defined
as\footnote{We have used the following values of the integrals governing the ${\bf B}_2$ field using the $\theta_i$ values given in \eqref{harper}:
\bg
&& \int d\theta_2~\sin~\theta_2~ \log\left(\sin~{\theta_2\over 2}\right) = {\cos~\theta_2\over 2} + (1 - \cos~\theta_2) \log\left(\sin~{\theta_2\over 2}\right) \approx {1\over 2} \nonumber\\
&& \int d\theta_2~\cos~\theta_2 ~\cot~{\theta_2\over 2} = \cos~\theta_2 + 2 \log\left(\sin~{\theta_2\over 2}\right) \approx 1 +
2\log~\alpha_{\theta_2} - {3\over 5} \log~N \nonumber \nd}:
\bg\label{phyllis}
&& \mathbb{B} = 1 - {g_sN_f\over 4\pi} \Bigl[1 + \log\left(\alpha^3_{\theta_1} \alpha^4_{\theta_2}\right)\Bigr] +
{g_sN_f\over 2\pi} \log~N \nonumber\\
&& \mathbb{A} = {3 g_sN_f \over 8\pi}, ~~~~ \mathbb{C} = 1 - {g_sN_f\over 8\pi} \left(\log~\alpha_{\theta_2} -
{1\over 5}~\log~N\right). \nd
Let us pause a bit to see what are the dominating terms in the above set of coefficients. We want $g_s \to 0$, and small $N_f$, but we also want very large $N$. Let us therefore take the following limiting values for $g_s, N_f$ and
$N$:
\bg\label{kolchak}
g_s \to \epsilon, ~~~N_f = {\cal O}(1), ~~~ N \to {\rm exp}\left({\alpha_{{}_N}\over \epsilon^b}\right), \nd
where $\alpha_N$ could be a large number and $1 < b < 2$. This clearly shows that the $g_sN_f \log~N$ term in $\mathbb{B}$ dominates and $\mathbb{B}^2 \gg 4\mathbb{A}\mathbb{C}$. Using this criteria, and solving
\eqref{pchopra} immediately gives us:
\bg\label{japtag}
\Lambda ~ = ~ {\Lambda_\infty\over N^{4/3}} ~ << ~\Lambda_\infty, \nd
implying that $\Lambda$ can be in the deep IR.
Hence, one can obtain an ${\cal O}(1) g_{YM}$ in the IR without requiring an O(1) $g_s$,  in the presence of flavors but not in their absence. In the IR, of course $N_f\neq0$.

Before ending this section, let us make a few observations. First, if we also take $M$ to be very large, then the first term of
$\mathbb{C}$ in \eqref{phyllis} will be suppressed by $1/M$. This of course will not change the conclusion
of \eqref{japtag}. Secondly,
in Section {\bf 6}, (\ref{tree}) will be replaced by the observation that for large $Z$ the argument of the square root in (\ref{runimara}) is obviously positive and for small $Z$:
\bg\label{pool}
{1\over \mathbb{P}_2^2}  ~\sim ~ \frac{1}{(\log ~N - 3\log ~r_h)^2} ~ \ll ~ 1, \nd
where $\mathbb{P}_2$ is given in \eqref{ginlynn};
as long as  $\log N,|\log r_h|\gg1$. This is obviously true from our earlier considerations. Therefore, the argument of the square root in $P(Z)$ in (\ref{runimara})  is always positive.
	
\section{Conclusions and discussions \label{konthi}}

The aim of this work is to study bulk viscosity, a universal physical quantity defining conformal anomaly of a system, in accessible computational limits of the 't~Hooft coupling constant $\lambda=g_{YM}^2 M$. Therefore, the discussion covers very different methodologies relevant for different strengths of the coupling. They all, however, describe the IR limit of the SU$(M)$ gauge theory. The discussion is meant to be exhaustive enough since we not only focus on presenting new results, but also broadly show the context of our studies and elaborate on tools needed for computation of the coefficient. 

It is fair to state that the limits considered in this work $-$ kinetic theory and strings $-$ have traditionally been tackled by two distinct communities. Hence one of the goals pursed in this enterprise is to facilitate a rapprochement between two groups of practitioners that have more often than not remained distinct. Building on the unity of physics,  different groups can learn from each other.

In light of this, one of the main tasks of the analysis was to express the bulk viscosity as a function of the speed of sound within well-established first-principle SU$(M)$ gauge theory in large $M$ limit. Our efforts were put into clarifying possible differences in the parametric form of the ratio $\zeta/\eta$, obtained at different coupling constants. At weak coupling, kinetic theory was used, which is currently the most common and efficient approach to calculate transport coefficients. In our studies we provided justification of the effective kinetic theory using a fundamental diagrammatic approach. At strong ('t Hooft) coupling, the UV complete type IIB holographic dual (and its M theory uplift when addressing also the strong string coupling limit) of large-$N$ thermal QCD was employed. The intermediate coupling behavior, most relevant for the quark gluon plasma produced experimentally in the heavy ion collisions, was also briefly discussed. We mainly summarized known challenges related to the first-principles extraction of bulk viscosity.

To discuss the weak coupling limit we encapsulated the analysis of bulk viscosity of QCD done extensively within the kinetic theory in Ref.~\cite{Arnold:2006fz}. When the interaction is governed by the 't~Hooft coupling $\lambda=g_{YM}^2 M$ and $M \to \infty$, the behavior of bulk viscosity is controlled by gluons only as the quark contributions are suppressed by a factor $1/M$.\footnote{The number of colors is $N+M$ in the UV and $M$ in the IR; both are kept very large in sections \ref{kinetic} and \ref{nf0} and $N_f$ (along with string coupling) could be taken to be O(1) in sections \ref{nfn0} and \ref{spectral} keeping $N$ to be very large as part of the "MQGP" limit.} The parametric form of the bulk viscosity as a function of the speed of sound is $\zeta/s \propto 1/3-c_s^2$, while the ratio $\zeta/\eta \propto (1/3-c_s^2)^2$. Then, starting from the Kubo formula, we performed a multi-loop analysis which enabled us to determine which scattering processes contribute to the collision kernel of the Boltzmann equation and provided a power counting in the weak 't~Hooft coupling and high temperature. Collecting all evaluated diagrams we have shown a schematic procedure how to derive an integral equation which may be thought of as a diagrammatic representation of the Boltzmann equation. The integral equation is formed by infinite number of planar diagrams with propagators and vertices being dressed. Both number conserving and number changing processes have to be included in the complete bulk viscosity examination. For the vertices a separate integral equation, governed mainly by the soft physics and capturing the LPM effect, has to be solved. The integral equations can be also obtained using $n$PI formalism and the findings were also briefly summarized.

Within the intermediate coupling region, we shortly presented a state of the knowledge on the bulk viscosity studies. Although the prescription of calculation of bulk viscosity is given by the Kubo formula, it is difficult to reliably establish the hydrodynamic limit of the spectral function and determine which physical phenomena may be responsible for its shape. Therefore we can only conclude that all compiled findings do not allow one for quantitative determination of the bulk viscosity behavior in this region starting from first principles and new methods and/or perspectives are needed.

After analyzing the weak and the intermediate 't~Hooft coupling regimes, we go to the next stage, i.e the strong 't~Hooft coupling regime. Clearly neither pQCD, nor lattice results can help us here. A new paradigm is needed and is given by the so-called gauge/gravity duality. This is a refined form of the famed AdS/CFT duality, constructed precisely to tackle strongly coupled gauge theories that are non-conformal. In section \ref{nf0} we study a SU$(M)$ gauge theory in the IR at high temperature (i.e the temperature above the deconfinement temperature) and at strong 't~Hooft coupling. We take large $M$, but keep the string coupling $g_s$ very small, such that $\lambda = g_sM$ is still very large. To avoid additional complications, we take no flavor degress of freedom.

 In such a setup, the computation of bulk viscosity boils down to the computation of metric fluctuations in the corresponding gravity dual. In section \ref{bvcomp} we study the equations governing the fluctuations using two steps: one, in section \ref{amandas}, we relax some of the constraints and study a toy example which in turn provides a nice solvable system; and two, in section \ref{galgad}, we do a more precise and careful computations of the fluctuation equations. Knowing the precise fluctuations help us to compute both the sound speed as well as the ratio of the bulk to the shear viscosities. In section \ref{sound} we perform the aforementioned computations and show that the ratio of the bulk to shear viscosities is indeed bounded below by the deviation of the square of the sound speed from its conformal value but more interesting, is independent of the cut-off.

 It is believed that QGP is an example of a strongly coupled system at finite temperature wherein unlike as considered in most gravity duals, the gauge coupling and hence the string coupling, is of ${\cal O}(1)$. Motivated by the same and with the idea of also including the flavor degrees of freedom as well as
the UV region, in section \ref{nfn0}, we calculate holographically at finite string coupling, the deviation of the square of the speed of sound from its conformal value, the attenuation constant and the ratio of the bulk and shear viscosities and find a Buchel-like bound for the latter. Finite string coupling necessitates addressing these issues from the M-theory uplift of the type IIB construct of \cite{metrics} which was obtained by  the M-theory uplift of the SYZ type IIA dual in \cite{MQGP}.  This also enjoys the additional benefit of not having to keep track of the NS5-degrees of freedom that one needs to while working with a single T-dual of the type IIB configuration of \cite{metrics}. Based on \cite{klebanov quasinormal, EPJC-2}, an equation of motion (EOM) for a combination of scalar modes of metric perturbations invariant under infinitesimal diffeomorphisms, is constructed. Upon investigating this EOM near the horizon, it is realized that for a non-zero bare resolution parameter, the horizon turns out to be an irregular singular point. Demanding the same of an ansatz for the solution to the same, in section \ref{lostgls}, the dispersion relation for the quasinormal modes obtains not only the conformal values of the speed of sound and the attenuation constant but also their respective non-conformal corrections. Interestingly, for the case of a vanishing bare resolution parameter, by looking at the solution to the EOM near the asymptotic boundary, in section \ref{zerbare} one realizes that one can not consistently impose Dirichlet boundary condtion (at the asymptotic boundary); like section \ref{nf0}, non-normalizable modes are required to propagate. In section \ref{suhaag}, with a non-zero bare resolution parameter, we first show that the KSS bound on the shear-viscosity-to-entropy-density is not violated having incorporated the non-conformal corrections. We then obtain the bulk-viscosity-to-entropy-density ratio and the deviation of the square of the speed of sound from its conformal value, and confirm that the conformal value of both vanish and they are both hence determined entirely by the non-conformality of the theory. One of the main results of this section is a crisp bound: $\frac{\zeta}{\eta}\geq\frac{91}{5}\left(\frac{1}{3} - c_s^2\right)$ with(out) the flavor degrees of freedom.

In section \ref{spectral}, we approach the issue of obtaining the deviation of the square of the speed of sound from its conformal value, from two-point correlators involving gauge field fluctuations on the world-volume of flavor D6-branes using the prescription of \cite{Minkowskian-correlators}. To begin with one considers the weak-string-coupling strong-'tHooft-coupling limit. The fluctuations are considered over a background value of the gauge field $-$ worked out in section \ref{spec2} $-$ assumed to be having only a temporal component and radial dependence. In the zero-momentum limit, interestingly and as shown in section \ref{spec3}, there is only a single second order equation in a gauge-invariant perturbation field $-$ the `electric field' $-$ which needs to be and is solved for
(in section \ref{onshala}). Finally the subtracted (zero temperature from the non-zero temperature) spectral function per unit frequency in the vanishing-frequency limit yields that the same is proportional to the linear power of the deviation of the square of the speed of sound from its conformal value, thereby validating the same as obtained in the previous sections \ref{nf0} and \ref{nfn0}. We conclude section \ref{spectral} with some remarks (in section \ref{gsnf1}) arguing that this result remains unchanged even in the strong-string-coupling stong-'tHooft-coupling, or the true MQGP limit of \cite{metrics}.

Let us briefly discuss some future directions. It would be rather interesting to probe better the regime of intermediate 't Hooft coupling whereat the number of colors is large, the gauge coupling is small but the 't Hooft coupling is finite, i.e., neither small (weak coupling regime) nor large (strong coupling regime). As discussed, techniques based on QCD do not offer currently a reliable way to explore this region. The ansatz proposed for the spectral function parametrization does not capture properly a high frequency tail and the QCD sum rule cannot be directly applied to constrain bulk viscosity. Since it is not clear how to handle the issues with the QCD tools, the region can be alternatively explored within the supergravity framework. One could invoke higher derivative corrections in the supergravity action which would hence back-react on the background. The same in the context of ${\cal N}=4$ SYM has been studied recently in \cite{solana}. For the present case there are two ways to go about it. One, we could start from the type IIB background of \cite{metrics} and consider corrections to the metric and fluxes in powers of $\alpha'^3$ and solve the modified equations of motion up to ${\cal O}(\alpha'^3)$. Two, we could use the MQGP limit (with $g_s \sim {\cal O}(1)$, large $N$ but finite $g_sM$) and start with $D=11$ supergravity action up to sextic power in the eleven-dimensional Planck length \cite{grimm}, and construct solutions to the EOMs as Planck-length perturbation of the M-theory uplift of \cite{MQGP}. Clearly the latter is a bit more practical because of the reduced number of fields in M-theory. Following this, one can then include metric perturbations and solve for their EOMs and hence see the effects of the inclusion of the aforementioned higher derivative terms on some spectral functions. It would be interesting to evaluate the non-zero frequency contribution to the spectral function per unit frequency and compare with previous studies on this topic as in \cite{dileptonSC} (which had excluded higher derivative corrections) in ${\cal N}=4$ SYM.

Another possible future direction could be to look at simultaneously turning on gauge and vector modes of metric perturbations \cite{D.Tong_Cracow_13} and then see the modification in the spectral function of gauge fluctuations considered in Section \ref{spectral}. The same in the context of type IIB for evaluating electrical and thermal conductivities, was considered in \cite{EPJC-2}.

\section*{Acknowledgements}

We would like to thank Simon Caron-Huot, Dimitri Kharzeev and Guy Moore for helpful discussions.
The work of AC is supported in part by the program Mobility Plus of the Polish Ministry of Science and Higher Education. The work of KD, CG and SJ is supported in part by the Natural Sciences and Engineering Research Council of Canada. CG gratefully acknowledges support from the Canada Council for the Arts through its Killam Research Fellowship program.
The work of
KS is supported by a Senior Research Fellowship from the Ministry of Human Resource and Development, Government of India.
AM would like to thank McGill University for the wonderful hospitality during the completion of this work. AM is supported in part
by the Indian Institute of Technology, Roorkee, India and the department of Physics, McGill University, Canada.

\newpage

\appendix
\section{A Gauge-Invariant Combination of Scalar Modes of Metric Perturbations \label{seceqaa}}

 The black M3-brane metric of \cite{MQGP}, after dimensional reduction to five dimensions, can be extracted from the eleven-dimensional metric \eqref{Mtheorymet}.
Consider a linear perturbation of the above metric as:
\begin{equation}\begin{split}
g_{\mu\nu} = G_{\mu\nu}+h_{\mu\nu},
\end{split}
\end{equation}
where $G_{\mu\nu}$ is the unperturbed background metric. Assuming the perturbations to propagate in the above background with momentum along the $x$-direction, it can be defined as following Fourier decomposed form:
\begin{equation}\begin{split}
h_{\mu\nu}(t,x,u)=\int \frac{d^4q}{(2\pi)^4}e^{-iwt+iqx}h_{\mu\nu}(w,q,r),
\end{split}
\end{equation}
where $u \equiv {r_h\over r}$ as defined earlier.
For the scalar channel the nonzero independent perturbations may be split into the following five components:
\bg\label{wonderw}
h_{tt}(u),~~~h_{tx}(u),~~~h_{xx}(u),~~~h_{yy}(u)=h_{zz}(u), \nd
where we have considered all the cross-terms to vanish, i.e we choose  $h_{r\mu}=0$ (for all $\mu$) gauge.
As is the convention in the literature, \cite{KSS}, \cite{klebanov quasinormal}, \cite{Herzog-vs}, we define new variables for the above perturbations:
\bg\label{porte}
&& H_{tt} \equiv h^t_t= G^{tt}h_{tt}, ~~~~~~ H_{tx} \equiv h^{x}_t= G^{xx}h_{tx} \nonumber\\
&&H_{xx} \equiv h^x_x= G^{xx}h_{xx}, ~~~~ H_{yy} \equiv h^y_y= G^{yy}h_{yy}, ~~~~ H_{zz} \equiv h^z_z= G^{zz}h_{zz},
\nd
where, as mentioned above,  the Lorentz indices for the fluctuations follow the conventions of \cite{KSS, klebanov quasinormal, Herzog-vs}; and the
unperturbed M-theory inverse metric components may be expressed in the following way:
\bg\label{3minot}
G^{tt} = \frac{e^{\frac{2\varphi}{3}}}{g_{tt}}, ~~~~~~ G^{xx} = G^{yy} = G^{zz}=\frac{e^{\frac{2\varphi}{3}}}{g_{\mathbb{R}^3}}, \nd
wherein $g_{tt}, g_{\mathbb{R}^3}$ are as given in equation (\ref{metdef}) and $\varphi$ is the type IIA dilaton.
We will also find useful to construct the following linear combinations of the
fluctuation modes $H_{xx}$ and $H_{yy}$:
\bg\label{acedoubles}
H_s(u)  \equiv  H_{xx}(u) + 2 H_{yy}(u), \nd
whose equation of motion will be discussed in section \ref{Hs}.
The EOMs for the other fluctuation modes are expectedly correlated to each other which will be illustrated in the
following. We will start with the fluctuation mode $H_{tt}$, then go to the mode $H_s$, and finally discuss the remaining modes.

\subsection{The equation of motion for the fluctuation mode $H_{tt}$ \label{Htt}}

To elaborate the implications of the above discussion, let us discuss the EOM
for $H_{tt}$.  This can be expressed in terms of the other fluctuation modes in the following way:
\bg\label{HTT}
H_{tt}^{\prime\prime}(u) = {{\bf A}_1 H_{tx}(u) + {\bf A}_2H^\prime_{tt}(u) + {\bf A}_3 H_{tt}(u) + {\bf A}_4 H_{xx}(u) + {\bf A}_5 H_{yy}(u)\over r_h^4 (u^4-3)^2(u^4-1)}, \nd
which at first glance seems to be well defined in the regimes $r_h > 0$ and $u \ge 1.32$. The precise regimes of interest however is not important for the kind of details that we are aiming for here. This will be illustrated later. Note also that  ${\bf A}_i$ are not constants but certain nested functions whose forms may be given in the following
way:
\bg\label{pmason}
&&{\bf A}_1 = {{\cal D}_1 - r_h^2 {\cal D}_2 \over u^2}, ~~~~ {\bf A}_2 = {r_h^2 u(u^4-3) (u^4-1){\cal D}_3 + r_h^4 {\cal D}_4 \over 4N u} \nonumber\\
&& {\bf A}_{(4, 5)} = {a^2 {\cal D}_{(7, 9)} - r_h^2 {\cal D}_{(8, 10)} \over (2, 1) u^2}, ~~~~ {\bf A}_3  = {a^2 {\cal D}_5 - r_h^2(u^4-1) {\cal D}_6 \over 2 u^2},
\nd
where the denominator of the form ($a, b$) is to be understood as being identified with the subscript bracket ${\bf A}_{(a, b)}$ so that individual relations for ${\bf A}_a$ and
${\bf A}_b$ may be constructed. The nested function ${\cal D}_1$ takes the following form:
\bg\label{cald1}
{\cal D}_1 & \equiv &  a^2 {g_s} \pi  q w\Big[6 \left(u^4-15\right) {\cal B}'(u) u^5+8 \left(4 u^8-9 u^4-9\right) {\cal B}(u) \nonumber\\
&+& \left(u^4-3\right) \left(9
   \left(u^4-1\right) {\cal B}''(u) u^2+8 N \left(4 u^4+3\right)\right)\Big], \nd
  where ${\cal B}(u)$ is defined earlier in \eqref{Bdef} and $N$ is the usual number of colors. The information of the $N_f$ flavors are in the definition of ${\cal B}(u)$. In the
  same vein ${\cal D}_2$ takes the following form:
{\footnotesize
\bg\label{cald2}
{\cal D}_2 \equiv
{g_s} \pi  q w \left[2
   \left(u^4-15\right) {\cal B}'(u) u^3+16 \left(u^4-3\right) {\cal B}(u) u^2+\left(u^4-3\right) \left(16 N u^2+3 \left(u^4-1\right) {\cal B}''(u)\right)\right], \nonumber\\ \nd}
 which is also expressed in terms of ${\cal B}(u)$ in a somewhat similar form as in \eqref{cald1} above. Together they would determine the coefficient ${\bf A}_1$
 in \eqref{HTT}. The next coefficient ${\bf A}_2$ is now determined in terms of ${\cal D}_3$ and ${\cal D}_4$. The former is simple:
 \bg\label{cald3}
 {\cal D}_3 \equiv  ua^2\left[ 4N\left(u^4+3\right)+9 u \left(u^4-1\right) {\cal B}'(u)\right], \nd
and  expressed in terms of ${\cal B}(u)$, whereas the latter is more involved and may be expressed in the following way:
 {\footnotesize
 \bg\label{cald4}
 {\cal D}_4 \equiv
 6 u \left(2 u^{12}-17 u^8+30 u^4-15\right) {\cal B}'(u)+\left(u^4-3\right) \left[3 u^2 {\cal B}''(u) \left(u^4-1\right)^2+4 N \left(u^8+2
   u^4+9\right)\right]. \nonumber\\  \nd}
 \noindent The first four coefficients ${\cal D}_i$  as in \eqref{cald1}, \eqref{cald2}, \eqref{cald3} and \eqref{cald4} seem to illustrate a pattern: any ${\cal D}_k$ may be expressed
 in terms of powers of $u$ and powers of derivatives of ${\cal B}(u)$. In other words the pattern seems to be:
 \bg\label{pattern}
 {\cal D}_k \equiv \sum_{n, m} c_{knm} u^n {\cal B}^{(m)}(u), \nd
 where $c_{knm}$ coefficients are independent of $u$, but functions of $N$, $a^2$ etc., and ${\cal B}^{(2)}(u) \equiv {\cal B}^{\prime}(u)$ for example. One may
 easily read up the values for $c_{1nm}, c_{2nm}, c_{3nm}$ and $c_{4nm}$ from \eqref{cald1}, \eqref{cald2}, \eqref{cald3} and \eqref{cald4} respectively.
 Working out the coefficients $c_{5nm}$ lead us to express ${\cal D}_5$ in the following way:
  {\footnotesize
 \bg\label{cald5}
 && { \cal D}_5 \equiv 6 {g_s} \pi  q^2 \left(u^8-16 u^4+15\right)
   {\cal B}'(u) u^5+8 {g_s} \pi  q^2 \left(4 u^{12}-13 u^8+9\right) {\cal B}(u)\nonumber\\
   && + \left(u^4-3\right) \left[8 {g_s} N \pi  \left(4 u^8-u^4-3\right) q^2+9 {g_s} \pi  u^2
   \left(u^4-1\right)^2 {\cal B}''(u) q^2+2 {r_h}^2 u^2 \left(5 u^8-42 u^4+33\right)\right], \nonumber\\ \nd}
 \noindent where the ${\cal B}(u)$ independent terms appear, in our notation, as ${\cal B}^{(0)}(u)$  and one may verify the uniqueness of the proposed form \eqref{pattern}. This  is also evident from the next coefficient, namely ${\cal D}_6$ which may be determined from $c_{6nm}$ as:
  {\footnotesize
 \bg\label{cald6}
 {\cal D}_6 \equiv {g_s} \pi  q^2 \left[2 \left(u^4-15\right) {\cal B}'(u) u^3+16 \left(u^4-3\right) {\cal B}(u) u^2+\left(u^4-3\right) \left(16 N u^2+3
   \left(u^4-1\right) {\cal B}''(u)\right)\right],  \nonumber\\ \nd}
  \noindent which as one may easily check follows \eqref{pattern}. The other set of coefficients, namely $c_{7nm}$, combines in a way to reproduce the next coefficient ${\cal D}_7$
  appearing in ${\bf A}_4$ as:
  {\footnotesize
  \bg\label{cald7}
  && {\cal D}_7 \equiv  6 {g_s} \pi  \left(u^4-15\right)
   w^2 {\cal B}'(u) u^5+8 {g_s} \pi  \left(4 u^8-9 u^4-9\right) w^2 {\cal B}(u)\nonumber\\
   & & \hskip -0.4in -\left(u^4-3\right) \left[2 {r_h}^2 \left(u^8-12 u^4+3\right) u^2-9 {g_s} \pi
   \left(u^4-1\right) w^2 {\cal B}''(u) u^2-8 {g_s} N \pi  \left(4 u^4+3\right) w^2\right]. \nd}
   \noindent We see that the structure is somewhat similar to \eqref{cald5}, i.e the coefficient ${\cal D}_5$ in the sense that we have
   ${\cal B}^{(0)}$, ${\cal B}^{(1)}$, ${\cal B}^{(2)}$
   and ${\cal B}^{(3)}$ terms  distributed in an identical way (although the precise $c_{knm}$ coefficients differ) as the derivations of these terms involve
   similar manipulations of the Einstein's equations. This is evident from the form of the next coefficients, namely ${\cal D}_8$ which may be expressed as:
   {\footnotesize
   \bg\label{cald8}
   {\cal D}_8 \equiv   {g_s} \pi  w^2 \left[(2 \left(u^4-15\right) {\cal B}'(u) u^3+16 \left(u^4-3\right) {\cal B}(u) u^2+\left(u^4-3\right) \left(16 N u^2+3
   \left(u^4-1\right) {\cal B}''(u)\right)\right], \nonumber\\ \nd}
   \noindent which is structurally similar to ${\cal D}_6$ in \eqref{cald6}. On the other hand, the last two coefficients require a slightly different
   analysis and therefore we expect them to differ
   from  the above ${\cal D}_k$ coefficients. This becomes clear from the expression of ${\cal D}_9$ which is written as:
   {\footnotesize
   \bg\label{cald9}
   {\cal D}_9 &\equiv &  6 {g_s} \pi
   \left(u^4-15\right) \left(\left(u^4-1\right) q^2+w^2\right) {\cal B}'(u) u^5
   +8 {g_s} \pi  \left(u^4-3\right) \left(\left(u^8+8 u^4-3\right) q^2+\left(4 u^4+3\right)
   w^2\right) {\cal B}(u)\nonumber\\
   & + & \left(u^4-3\right)\Big[-2 {r_h}^2 \left(u^8-12 u^4+3\right) u^2+9 {g_s} \pi  \left(u^4-1\right) \left(\left(u^4-1\right) q^2+w^2\right)
   {\cal B}''(u) u^2 \nonumber\\
   &+&8 {g_s} N \pi  \left(\left(u^8+8 u^4-3\right) q^2+\left(4 u^4+3\right) w^2\right)\Big] , \nd}
   that takes the form, although similar to \eqref{pattern}, different from the other ${\cal D}_k$ coefficients.
   The final coefficient, ${\cal D}_{10}$,  may be presented in the following to
   illustrate the same point:
  {\footnotesize
  \bg\label{cald10}
  {\cal D}_{10} &\equiv &
   {g_s} \pi  \Big[2 \left(u^4-15\right) \left(\left(u^4-1\right) q^2+w^2\right) {\cal B}'(u) u^3 + 8 \left(u^4-3\right)
   \left(\left(u^4+1\right) q^2+2 w^2\right) {\cal B}(u) u^2\nonumber\\
   & + & \left(u^4-3\right) \left(8 N \left(\left(u^4+1\right) q^2+2 w^2\right) u^2+3 \left(u^4-1\right)
   \left(\left(u^4-1\right) q^2+w^2\right) {\cal B}''(u)\right)\Big]. \nd}
This completes our analysis of the EOM \eqref{HTT} for  $H_{tt}(u)$. Our next step is to analyze the EOM for  $H_s(u)$ defined above in \eqref{acedoubles}.

\subsection{The equation of motion for the combined mode $H_s$ \label{Hs}}

The functional form for $H_s(u)$, as evident from \eqref{acedoubles}, can be expressed as
certain linear combination of $H_{xx}$ and $H_{yy}$. As in \eqref{HTT}, we can express the EOM for $H_s(u)$ in the following way:
 \bg\label{UCSB}
H_{s}^{\prime}(u) = {{\bf B}_1 H_{tx}(u) + {\bf B}_2H^\prime_{tt}(u) + {\bf B}_3 H_{tt}(u) + {\bf B}_4 H_{xx}(u) + {\bf B}_5 H_{yy}(u)\over r_h^4 (u^4-3)^2(u^4-1)}, \nd
 where we see that both the denominator and the numerator have the same set of factors as in the denominator and the numerator of \eqref{HTT}. The only thing
 that would differ are the actual values of ${\bf B}_k$. The functional forms for ${\bf B}_k$ may be expressed  in terms of certain nested functions in the following way:
 \bg\label{statestreet}
 &&{\bf B}_1 = a^2 {\cal F}_1 - r_h^2 {\cal F}_2, ~~~~~~ {\bf B}_4 =  a^2 {\cal F}_7 +  r_h^2 {\cal F}_8, ~~~~~~ {\bf B}_5 = 2a^2 {\cal F}_9 + r_h^2 {\cal F}_{10}\\
 && {\bf B}_3 = a^2 (u^4 -1){\cal F}_5 + r_h^2 (u^4 -1) {\cal F}_6, ~~~ {\bf B}_2 = {r_h^4(u^4-1)\over 2N} {\cal F}_3 - r_h^4(u^4-3)(u^4-1) {\cal F}_4, \nonumber \nd
where the ${\cal F}_k$ functions may be compared with the ${\cal D}_k$ functions in \eqref{pmason}.  In fact one may even express the functional forms for ${\cal F}_k$
as a power series in $u$ and derivatives of ${\cal B}(u)$, much like \eqref{pattern}, but now with coefficients $g_{knm}$ instead of $c_{knm}$. The coefficients
$g_{knm}$ are independent of $u$, and may be determined easily as before by analyzing the corresponding Einstein's equations. For example
finding $g_{1nm}$ and
$g_{2nm}$ immediately reproduces the functional forms for ${\cal F}_1$  and ${\cal F}_2$ in the following way:
\bg\label{calf12}
{\cal F}_1  = 2 u^2 {\cal F}_2 \equiv  6 \pi {g_s} q u^3 w \left[3 u \left(u^4-1\right) {\cal B}'(u)-4 \left(u^4-3\right) {\cal B}(u)-4 N \left(u^4-3\right)\right], \nonumber\\ \nd
 which as expected takes the form \eqref{pattern}. One may also easily see the pattern repeating for the next two coefficients, namely ${\cal F}_3$ and ${\cal F}_4$,
 in the following way:
 \bg\label{calf34}
 {\cal F}_3 = {{\cal F}_4 \over u^5 {\cal B}^\prime(u)} = 3(u^4 - 1). \nd
 We can now go to the other set of coefficients where we can see how we could relate to the ${\cal D}_k$ coefficients studied above.  A priori there shouldn't be
 any apparent connections, but the functional forms for ${\cal F}_5$ and ${\cal F}_6$ are similar to what we had earlier. For example:
 {\footnotesize
 \bg\label{calf5}
 {\cal F}_5 \equiv  9 \pi  {g_s} q^2 \left(u^4-1\right) u^4 {\cal B}'(u)-12 \pi  {g_s} q^2 \left(u^4-3\right) u^3 {\cal B}(u)+\left(u^4-3\right) u
   \left[{r_h}^2 \left(5 u^4-3\right)-12 \pi  {g_s} N q^2 u^2\right],  \nonumber\\ \nd}
 \noindent  which should be somewhat reminiscent of \eqref{cald5}. Similarly the functional form for ${\cal F}_6$, expressed here as:
  \bg\label{calf6}
  {\cal F}_6 \equiv  \pi  {g_s} q^2 u \left[-3 u
   \left(u^4-1\right) {\cal B}'(u)+4 \left(u^4-3\right) {\cal B}(u)+4 N \left(u^4-3\right)\right], \nd
should be reminiscent of ${\cal D}_6$ in \eqref{cald6}. Of course all of these could also be expressed as \eqref{pattern} with $c_{knm}$ replaced by appropriate
$g_{knm}$ as we mentioned earlier. Interestingly comparing \eqref{calf6} with \eqref{calf12}, we see that they are related via the following
relation:
\bg\label{spmass}
{{\cal F}_6\over q} + {{\cal F}_1\over 6wu^2} = 0. \nd
Thus knowing ${\cal F}_1$ would determine the functional forms for ${\cal F}_2$ as well as ${\cal F}_6$. In fact one can show that ${\cal F}_1$  or
${\cal F}_6$ can also fix the functional
forms for two other coefficients, namely ${\cal F}_8$ and ${\cal F}_{10}$, in the following way:
\bg\label{calf810}
{{\cal F}_6 \over q^2} = {{\cal F}_8\over w^2} = {{\cal F}_{10}\over 2[q^2(u^4-1) + w^2]}. \nd
The remaining two coefficients, namely ${\cal F}_7$ and ${\cal F}_9$, are however more complicated and are not anyway related to ${\cal F}_1$ in a simple way. For
example the functional form for ${\cal F}_7$ may be expressed as:
\bg\label{calf7}
{\cal F}_7 &\equiv & 9 \pi  {g_s} \left(u^4-1\right) u^4 w^2 {\cal B}'(u)-12 \pi  {g_s}
   \left(u^4-3\right) u^3 w^2 {\cal B}(u) \nonumber\\
  &&~~~~~ -  \left(u^4-3\right) u \left[12 \pi  {g_s} N u^2 w^2+{r_h}^2 \left(u^8+2 u^4-3\right)\right], \nd
which of course follows the pattern similar to \eqref{pattern}, but cannot be decomposed in terms of any of the above ${\cal F}_k$ coefficients. A similar thing
may also be said for the coefficient ${\cal F}_9$, written as:
\bg\label{calf9}
{\cal F}_9 &\equiv & 9 \pi  {g_s} \left(u^4-1\right)
   u^4 {\cal B}'(u) \left[q^2 \left(u^4-1\right)+w^2\right]
   -12 \pi  {g_s} \left(u^4-3\right) u^3 {\cal B}(u) \left[q^2 \left(u^4-1\right)+w^2\right] \nonumber\\
   &&~~~-\left(u^4-3\right) u
   \left[12 \pi  {g_s} N u^2 \left(q^2 \left(u^4-1\right)+w^2\right)+{r_h}^2 \left(u^8+2 u^4-3\right)\right], \nd
which evidently takes a more non-trivial form. Thus together with \eqref{calf12}, \eqref{calf34}, \eqref{calf5}, \eqref{calf6}, \eqref{calf810}, \eqref{calf7} and \eqref{calf9},
the EOM for $H_s(u)$ may be succinctly presented in terms of the other fluctuation modes.

\subsection{The equations of motion for the remaining fluctuation modes \label{remmo}}

Knowing the functional form for $H_s(u)'$ in \eqref{UCSB} in terms of $H_{xx}(u), H_{yy}(u), H_{tt}(u)$ and
$H_{tx}(u)$ tells us that we can express the EOM for $H'_{tx}(u)$ in the following way:
\bg\label{hardrock}
 H^\prime_{tx}(u) = \frac{4 q u^3 H_{tx}(u)+w \left[2 u^3 {H_{xx}}(u)+4 u^3 H_{yy}(u)-\left(u^4-1\right) {H_s}'(u)\right]}{q \left(u^4-1\right)}, \nd
whose form is, not surprisingly, similar to  \eqref{HTT} for the $H_{tt}(u)$ component. In fact since both \eqref{HTT} and \eqref{UCSB} are expressed in terms of  $H_{ab}(u)$ and $H'_{tt}(u)$, where $a, b$ take values in ($t, x, y$),
\eqref{hardrock} may also be expressed in terms of $H_{ab}(u)$ and $H'_{tt}(u)$. This pattern follows for the
next component $H'_{yy}(u)$ as:
\bg\label{moodies}
  H^\prime_{yy}(u) = -\frac{q \left(u^4-1\right) H_{tt}'(u)+2 q u^3 H_{tt}(u)+w H_{tx}'(u)}{2 q \left(u^4-1\right)},
  \nd
 implying that solutions may be found once we know the background values.  Finally, combining the above set of
 equations with the defining equation for $H_s(u)$, namely \eqref{acedoubles}, gives us a way to formulate the EOM
 for $H_{xx}(u)$ as:
 \bg\label{holresort}
  H^\prime_{xx} = {H_s}'(u)-2 H_{yy}'(u). \nd
Basically this is all we need to construct gauge invariant perturbation modes. For us, following
\cite{klebanov quasinormal}, a specific combination of the above set of perturbations is useful to quantify the
required perturbation as:

{\footnotesize
\bg\label{pegagus}
Z_s(u)=H_{yy} \left(q^2 + {q^2 u^4\over \pi^2 T^2}-w^2-\frac{{\cal B}^\prime(u) q^2 u^5 N_f g_s^2 }{2 N}\right)+q^2 \left(u^4 - 1\right)
   H_{tt}+2 q w H_{tx}+w^2 H_{xx}. \nonumber\\ \nd}
This is the perturbation \eqref{gaugu} that we described earlier.
Our aim now is to write down the set of four equations, namely \eqref{HTT}, \eqref{hardrock}, \eqref{moodies} and
\eqref{holresort},  as a single second-order equation in terms of the gauge invariant variable $Z_s(u)$ in the following way (see \eqref{Z-EOM}):
\bg\label{restoOPA}
Z_s^{\prime\prime}(u) = m(u) Z_s^{\prime}(u) + l(u) Z_s(u), \nd
with the two coefficients $m(u)$ and $l(u)$ of $Z_s^{\prime}(u)$ and $Z_s(u)$ respectively to be determined. Now the system of equations, \eqref{HTT}, \eqref{hardrock}, \eqref{moodies} and
\eqref{holresort}, is written in such a way that on the RHS of these equations there are {\it no} derivatives of the variables $H_{ab}$ except only single derivatives on $H_{tt}$. Hence the double derivatives of each variable: $H_{xx}, H_{yy}, H_{tx}$ and $H_{tt}$ contain double derivatives only of $H_{tt}$. This means the expression for
$Z_s^{\prime\prime}(u)$ and $Z_s^{\prime}(u)$,
as evaluated using \eqref{HTT}, \eqref{hardrock}, \eqref{moodies} and
\eqref{holresort}, can have a single derivative term acting only on $H_{tt}$, with no double derivatives on any variables. Since the expression of $Z_s(u)$ in \eqref{pegagus} has no $H^{\prime}_{tt}$, one can easily determine $m(u)$ by taking the ratio of the coefficient of $H^{\prime}_{tt}$ from $Z_s^{\prime\prime}(u)$ to the coefficient of
$H^{\prime}_{tt}$ from $Z_s^{\prime}(u)$. Equation \eqref{novharpx} is a precise reproduction of this fact.
Once $m(u)$ is determined,
$l(u)$ can also be obtained by equating the coefficient of $H_{tt}$ from $Z_s^{\prime\prime}(u)$ to the sum of coefficients of $H_{tt}$ from $Z_s^{\prime}(u)$
and $Z_s(u)$. In \eqref{lu1} we quoted the functional form for $l(u)$ for $u \to 1$. The generic form for $l(u)$ is straightforward but technically challenging, and is therefore left as an exercise for the reader.

After the dust settles, one may verify that the EOM  \eqref{restoOPA} is satisfied by the gauge-invariant choice of the perturbation $Z_s(u)$ in \eqref{pegagus}.

\newpage

\section{A derivation of the on-shell action and the Green's function \label{peridotv} }

The four-dimensional action that we considered in \eqref{lodgers} uses the pull-back metric  ${\cal G}_{\mu\nu}$
constructed out of
the type IIA metric, the NS B-field and the world-volume gauge field background. When the gauge field fluctuation, whose Fourier component is written as ${\cal A}_\mu$ in \eqref{doctor}, is also taken into account, the four-dimensional action takes the following form:
\bg\label{lodgers3}
\mathbb{S}_4  =  \frac{\Omega_2T_{D6}}{2}\int d^{4}x\Biggl[ e^{-\varphi}\sqrt{-{\cal G}}
\left(({\cal G}^{tZ})^2{\cal A}_{t}\partial_{Z}{\cal A}_{t}- \sum_{a = 0}^3 {\cal G}^{x_ax_a}
{\cal G}^{ZZ}{\cal A}_{x_a}\partial_{Z}{\cal A}_{x_a}
\right)\Biggr]^{Z_{\rm uv}}_{Z_{h}},  \nonumber\\
\nd
where $x_0 \equiv t$, $T_{D6}$ is the tension of the probe D6-brane,  and $\Omega_2$ is the volume of the two-sphere that we had in \eqref{D6DBI3}. The presence of $Z$ derivative in the integrand, despite integrating out the $Z$ variable, is from a total derivative term as may be inferred from \eqref{lodgers}. This also explains the two limits of
$Z$ in \eqref{lodgers3} Note that we took $Z_h$ to be the lower limit of $Z$ to be consistent with the lower bound \eqref{tree}.
However what we seek here is in fact the on-shell action and the Green's function at the boundary $Z = Z_{\rm uv}$, so the near-horizon
geometry is not too relevant for us. At the boundary
$F_{tZ}=-F_{Zt}=0$, so we must set ${\cal G}^{tZ}=0$ and replace $\sqrt{-{\cal G}}$ by $\sqrt{-G}$. Incorporating these changes, the boundary value of the on-shell action may now be re-written from \eqref{pawmil} as:
\bg\label{pawmil2}
\mathbb{S}_4 =-\frac{\Omega_2T_{D6}}{2}\int d^{4}x \left[e^{-\varphi}\sqrt{-{G}}{G}^{ZZ}
\left(\sum_{a = 0}^3 {G}^{x_a x_a}{\cal A}_{x_a}(Z, -k)\partial_{Z}{\cal A}_{x_a}(Z, k)\right) \right]_{Z_{\rm uv}}
\nd
where we have suppressed the $\omega$ dependence, and will have to resort back to ${\cal G}_{\mu\nu}$ component if we want to take $Z_h$, i.e the lower limit of $Z$. Recall also that we have used EOM to get to the boundary action \eqref{pawmil2}, so it makes sense to use the EOM further to simplify the above action. For example we can use \eqref{laura1} to rewrite $G^{tt}$ in the following way:
\bg\label{second}
G^{tt}(\partial_{Z}{\cal A}_{t})=\frac{q}{\omega}G^{xx}(\partial_{Z}{\cal A}_{x}),
\nd
for non-vanishing $\omega$. Plugging \eqref{second} in \eqref{pawmil2} then gives us the following action:
\bg\label{third}
\mathbb{S}_4 =-\frac{\Omega_2T_{D6}}{2\omega}\int d^{4}x \left[e^{-\varphi}\sqrt{-{G}}{G}^{ZZ}
\left(\sum_{a = 1}^3 {G}^{x_a x_a}\mathbb{E}_{x_a}(Z, -k)\partial_{Z}{\cal A}_{x_a}(Z, k)\right) \right]_{Z_{\rm uv}},
\nd
which is almost similar to the action \eqref{pawmil2} except with three major differences: one, the appearance of
${1\over \omega}$ as an overall factor; two, the sum over $a$ now being from 1 to 3;  and three, the appearance of  three new variables $\mathbb{E}_{x_a}$ for $a = 1, 2, 3$. The new variables are defined in the following way:
\bg\label{new}
\mathbb{E}_{x_1} \equiv q {\cal A}_{t} + \omega {\cal A}_{x}, ~~~~~~\mathbb{E}_{x_2} \equiv \omega {\cal A}_{y}, ~~~~~~\mathbb{E}_{x_3} \equiv \omega {\cal A}_{z}, \nd
which are clearly borne out from \eqref{second} and explains the appearance of the ${1\over \omega}$ suppression of the full action. We could also use \eqref{new} to express ${\cal A}_y$ and ${\cal A}_z$ in terms of
$\mathbb{E}_{x_2}$ and $\mathbb{E}_{x_3}$ respectively, but we won't do this right way.  Instead let us use the first
equation in \eqref{new} to write:
\bg\label{four}
\partial_{Z}E_{x_1}  =   q\partial_{Z}{\cal A}_{t} + \omega \partial_{Z}{\cal A}_{x}
 =  \left(\omega + \frac{q^2}{\omega}\frac{G^{xx}}{G^{tt}}\right)\partial_{Z}{\cal A}_{x},
\nd
where to get the second equality in the above we have used equation \eqref{second}. To complete the picture we will need the ratio of the two metric components. Using the fact that $r \equiv r_h e^Z$, we can easily argue that
$\frac{G^{xx}}{G^{tt}}=-\left(1-e^{-4Z}\right)$. Plugging this in \eqref{four} gives us:
\bg\label{fifth}
\partial_{Z}{\cal A}_{x} = \frac{\omega (\partial_{Z}E_{x_1})}{\left(\omega^2-\frac{q^2}{1-e^{-4Z}}\right)}.
\nd
This is all we need, because the derivatives on the other components are straightforward replacements of
$\mathbb{E}_{x_a}$ with $a = 2, 3$. Therefore combining \eqref{fifth} with \eqref{new} and plugging this in
\eqref{third} gives us the final action:
\bg\label{final}
\mathbb{S}_4 &= & -\frac{T_{D6}\Omega_2}{2}\int dx^4 \Biggl[e^{-\varphi}\sqrt{G}G^{xx}\Biggl\{\frac{E_{x_1}(Z, -k)\partial_{Z}E_{x_1}(Z, k)}{\omega^2-\frac{q^2}{1-e^{-4Z}}} \nonumber\\
&&+  \frac{E_{x_2}(Z,-k)}{\omega^2}\partial_{Z}E_{x_2}(Z,k)+\frac{E_{x_3}(Z,-k)}{\omega^2}\partial_{Z}E_{x_3}(Z,k)\Biggr\}\Biggr]_{Z_{\rm uv}},
\nd
which is the action given earlier in \eqref{duckling}. The $k_a^2$ appearing in \eqref{duckling} are the poles in
\eqref{final} and may be identified with the variables of \eqref{final} as:
\bg\label{final2}
k_1^2 = {q^2\over 1 - e^{-4Z}}, ~~~~~~ k_2^2 = k_3^2 = 0. \nd
Since we are only interested in the $x_1$ part of the fluctuation, the values of $k_2^2$ and $k_3^2$ are not very useful for us. Of course one may perform a more generic study, but we will not do so here. For the simplest case, the next step would be to define \eqref{stephanie} and then re-write the action as in \eqref{asidana}. From here the story follows as depicted in section \ref{onshala}.

\newpage

\section{Effective number of three-brane charges with background three-forms and the horizon radius \label{saldana}}

The horizon radius that we computed in \eqref{insaafka} was typically a ${\cal O}(1)$ number that was written as
$r_h = 1 - \epsilon^2$ by demanding the vanishing of \eqref{tarazu}. The small parameter $\epsilon$ is defined as:
\bg\label{altamira}
\epsilon \equiv {b\sqrt{3}\over \sqrt{2} N^{1/3} {\rm exp}\left({4\pi\over 3g_sN_f}\right)}, \nd
with both $b$, the resolution parameter, and $g_sN_f$ small. This choice of the horizon radius is not very useful for us because this would imply that $r_h$ is placed right at the point where new degrees of freedom would appear  to UV complete the system. With the definition of the radial coordinate $r$ as $r = r_he^Z$, this means $Z$ only measures geometry beyond $r_h$, i.e the geometry of Regions 2 and 3. Question then is how to place $r_h$ deep inside
Region 1 where the background is well known. However we cannot make $r_h$ arbitrarily small, as there exists a lower bound on $r_h$  given in \eqref{tree}. If $r_h^{(o)}$ denotes the lower bound, then:
\bg\label{tree2}
0 ~ < ~ r^{(o)}_h ~ \le ~ \left({{\bf C} \over 8\pi \alpha^4_{\theta_1}}\right)^{1/3}, \nd
with ${\bf C}$ being an integration constant that appeared in \eqref{AtEOM-1}, and we expect the horizon radius to satisfy $r_h > r^{(0)}_h$.
Such a lower bound is necessary otherwise an expression like $P(Z)$ in \eqref{runimara} will develop unphysical imaginary piece.

To find an appropriate $r_h$ it would be easier to do the analysis in the type IIB side instead of the mirror type IIA side. Such an analysis won't change the expression for $r_h$ as the mirror transformation {\it a la} SYZ \cite{syz}
keeps the radial coordinate unchanged. To proceed then, let us define an {\it effective} number of three-brane charge
as:
\bg\label{infinitecrisis}
N_{\rm eff}(r) = \int_{\mathbb{M}_5} {\cal F}_5 + \int_{\mathbb{M}_5} {\cal B}_2 \wedge {\cal F}_3, \nd
where ${\cal B}_2, {\cal F}_3$ and ${\cal F}_5$ are given, for $N_f = 0$ and in the Baryonic branch, in \eqref{heidi}. The five-dimensional internal space $\mathbb{M}_5$,
with coordinates ($\theta_i, \phi_i, \psi$), is basically the resolved warped-defomed conifold of \eqref{goeskalu}, or its simplified avatar given in \eqref{met}. What we now need is the
functional form for ${\cal B}_2$ and ${\cal F}_3$ with non-zero $N_f$ and non-zero axio-dilaton $\tau$. This may be worked out in the following way:

{\footnotesize
\bg\label{B2wedgeF3}
 {\cal B}_2\wedge {\cal F}_3 & =& {\cal B}_2\wedge \widetilde{\cal F}_3 + {\bf Re}~\tau \left({\cal B}_2\wedge
{\cal H}_3\right)\nonumber\\
&= &   {\bf Re}~\tau\left[\left(b_2 d_1 - a_2 c_1 +  b_1 d_2\right) d\phi_2 - a_1 c_2 d\phi_1\right]
 {dr}\land {d\theta_1}\land d\theta_2\land {d\psi}\\
&+&\left[ (a_0 {b_2}+ e_0{c_2}+ f_0{d_2}) {d\theta_1}\land {d\theta_2}
+ d_0\left(b_2 d\theta_2 - a_2 d\theta_1\right) \wedge dr \right] \wedge d\phi_1 \wedge d\phi_2 \wedge d\psi \nonumber\\
& + &  {\bf Re}~\tau\left[(-a_2c_1 + b_2d_1+b_1d_2)\cos~ \theta_1 + a_1(b_2 + c_2 \cos~\theta_2)\right]
 {dr}\land {d\theta_1}\land {d\theta_2}\land {d\phi_1}\land {d\phi_2}, \nonumber
\nd}
where $\widetilde{\cal F}_3$ is the standard combination of ${\cal F}_3$ and $-{\bf Re}~\tau ~{\cal H}_3$, and is used because of its appearance directly from the type IIB EOM.
The various other coefficients appearing above may be defined in the following way:

{\footnotesize
\bg\label{infinitywar}
& & a_0\equiv M \sin ~\theta_1 \left(\frac{3 {g_s} {N_f} \log ~r}{2 \pi }+1\right)
   \left(\frac{9 a^2 {g_s} {N_f} (2-3 \log ~r)}{4 \pi  r^2}+1\right)\nonumber\\
   & & b_0\equiv M \sin ~\theta_2 \left(\frac{3 {g_s} {N_f} \log ~r}{2 \pi }+1\right)
   \left(\frac{9 a^2 {g_s} {N_f} (2-3 \log ~r)}{4 \pi  r^2}+1\right)
   \left(\frac{81 a^2 {g_s} {N_f} \log ~r}{2 \left(9 a^2 {g_s} {N_f}
   (2-3 \log ~r)+4 \pi  r^2\right)}+1\right)\nonumber\\
   & & c_0\equiv  \frac{3 {g_s} M {N_f} \cot \left(\frac{\theta_2}{2}\right)
   \left(\frac{18 a^2 \log ~r}{r^2}+1\right)}{4 \pi  r}\nonumber\\
   & & d_0\equiv \frac{3 {g_s} M {N_f} \cot \left(\frac{\theta_1}{2}\right)
   \left(\frac{18 a^2 \log ~r}{r^2}+1\right) \left(\frac{36 a^2 \log ~r}{18 a^2 r
   \log ~r+r^3}+1\right)}{4 \pi  r}\nonumber\\
   & & e_0\equiv  \frac{3 {g_s} M {N_f} \sin ~\theta_1~\sin ~\theta_2~ \cot
   \left( \frac{\theta_2}{2}\right) \left(1-\frac{18 a^2 \log ~r}{r^2}\right)}{8
   \pi }\nonumber\\
   & & f_0\equiv \frac{3 {g_s} M {N_f} \sin ~\theta_1~ \cot
 \left( \frac{\theta_1}{2}\right) \sin ~\theta_2 \left(1-\frac{18 a^2
   \log ~ r}{r^2}\right) \left(\frac{36 a^2 \log ~ r}{r^2-18 a^2 \log
   ~r}+1\right)}{8 \pi } \nonumber\\
   & & a_1\equiv \frac{3 {g_s} M \left(1-\frac{3 a^2}{r^2}\right) \sin ~\theta_1
   \left(\frac{9 {g_s} {N_f} \log ~r}{4 \pi }+\frac{{g_s} {N_f} \log
   \left(\sin \left(\frac{\theta_1}{2}\right) \sin
   \left(\frac{\theta_2}{2}\right)\right)}{2 \pi }+1\right)}{r}\nonumber\\
   & & b_1\equiv \frac{3 {g_s} M \left(1-\frac{3 a^2}{r^2}\right) \sin ~\theta_1
   \left(\frac{9 {g_s} {N_f} \log (r)}{4 \pi }+\frac{{g_s} {N_f} \log
   \left(\sin \left(\frac{\theta_1}{2}\right) \sin
   \left(\frac{\theta_2}{2}\right)\right)}{2 \pi }+1\right)}{r}\nonumber\\
   & & c_1\equiv \frac{3 {g_s}^2 M {N_f} \cot \left(\frac{\theta_1}{2}\right)
   \left(\frac{36 a^2 \log~ r}{r}+1\right) \left(\frac{72 a^2 \log ~r}{36 a^2 \log~
   r+r}+1\right)}{8 \pi  r}\nonumber\\
   & & d_1\equiv \frac{3 {g_s}^2 M {N_f} \cot \left(\frac{\theta_2}{2}\right)
   \left(\frac{36 a^2 \log ~r}{r}+1\right)}{16 \pi }\nonumber\\
   & & e_1\equiv \frac{3 {g_s}^2 M {N_f} \cot \left(\frac{\theta_2}{2}\right)
   \left(\frac{36 a^2 \log ~r}{r}+1\right)}{16 \pi }\nonumber\\
   & & f_1\equiv \frac{3 {g_s}^2 M {N_f} \cot \left(\frac{\theta_1}{2}\right)
   \left(\frac{36 a^2 \log ~r}{r}+1\right) \left(\frac{72 a^2 \log ~r}{36 a^2 \log
   ~r+r}+1\right)}{16 \pi }\nonumber\\
   & & a_2\equiv 3 {g_s} M \left(1-\frac{3 a^2}{r^2}\right) \sin ~\theta_1
   \left(\frac{{g_s} {N_f} (2 \log ~r+1) \log \left(\sin
   \left(\frac{\theta_1}{2}\right) \sin
   \left(\frac{\theta_2}{2}\right)\right)}{4 \pi }+\frac{9 {g_s} {N_f}
   \log ^2 r}{4 \pi }+\log ~r\right)\nonumber\\
   & & b_2\equiv 3 {g_s} M \left(1-\frac{3 a^2}{r^2}\right) \sin ~\theta_2 \left(\frac{3 a^2
   {g_s}}{r^2-3 a^2}+1\right) \left(\frac{{g_s} {N_f} (2 \log ~r+1) \log
   \left(\sin \left(\frac{\theta_1}{2}\right) \sin
   \left(\frac{\theta_2}{2}\right)\right)}{4 \pi }+\frac{9 {g_s} {N_f}
   \log ^2 r}{4 \pi }+\log ~r\right)\nonumber\\
   & & c_2\equiv \frac{3 {g_s}^2 M {N_f} \log ~r \cot \left(\frac{\theta_2}{2}\right)
   \left(\frac{36 a^2 \log ~r}{r}+1\right)}{8 \pi }\nonumber\\
   & & d_2\equiv \frac{3 {g_s}^2 M {N_f} \log ~r \cot \left(\frac{\theta_1}{2}\right)
   \left(\frac{36 a^2 \log ~r}{r}+1\right) \left(\frac{72 a^2 \log ~r}{36 a^2 \log
  ~r+r}+1\right)}{8 \pi }.
   \nd}
To estimate the value of the horizon radius $r_h$, lets us fix a point on the radial direction $r = r_0$ where the effective number of three-brane charges vanish, i.e $N_{\rm eff}(r_0) = 0$. We will take the other internal coordinates, namely ($\theta_i, \phi_i, \psi$) to have the span:
\bg\label{celined}
\theta_1\in \left[\frac{\alpha_{\theta_1}}{N^{\frac{1}{5}}},\pi\right],~~~~ \theta_2\in \left[\frac{\alpha_{\theta_2}}{N^{\frac{3}{10}}},\pi \right], ~~~~\phi_{1,2}\in \left[0,{2\pi}\right], ~~~~ \psi\in \left[0,4\pi\right], \nd
where the effective lower limits of the $\theta_i$ angular terms have been described earlier. If we now collectively denote the lower limits of all the angular variables, namely ($\theta_i, \phi_i, \psi$) as $\mathbb{R}_-$ and the upper limits of all the angular variables as $\mathbb{R}_+$; and also use the fact that at fixed $r$, $dr = 0$, then the effective number of three brane charges take the following form:
\bg\label{mirelle}
N_{\rm eff}(r) = N + \int_{\mathbb{R}_-}^{\mathbb{R}_+}
(a_0 {b_2}+e_0{c_2} + f_0{d_2} ) {d\theta_1}\land {d\theta_2}\land {d\phi_1}\land {d\phi_2}\land {d\psi}, \nd
thus simplifying the expression \eqref{infinitecrisis}  tremendously. Here $N$ denotes the integral over ${\cal F}_5$, and is therefore related to the integer D3-branes in the dual gauge theory side at the Higgsing scale. The second term
combined with $N$ then denotes the effective number of cascading D3-brane charges at the scale $r = r_0$.
The functional forms for ($a_0, e_0, f_0, b_2, c_2, d_2$) can be extracted from \eqref{infinitywar}. Combined together leads to the following expression for $N_{\rm eff}$:

 {\footnotesize
 \bg\label{grantmunro}
 N_{\rm eff} &=& N + {3g_sM^2 \log~r\over 10 r^4} \bigg\{18\pi r (g_sN_f)^2 \log~N \sum_{k = 0}^1\left(18a^2 (-1)^k\log~r + r^2\right)\left({108a^2 \log~r\over 2k+1} + r\right) \\
 &+& 5 \left(3 a^2 ({g_s}-1)+r^2\right) (3 {g_s} {N_f} \log ~r+2\pi ) (9 {g_s} {N_f} \log ~r+4 \pi )
 \left[9 a^2 {g_s} {N_f} \log\left({e^2\over r^3}\right) +4 \pi  r^2\right]\biggr\}, \nonumber\\
 & = & N\left[1 + 6\pi\log~r\left(3 g_s N_f \log~r + 2\pi\right)\left(9 g_s N_f \log~r + 4\pi\right){g_sM^2\over N}  \right] +
 {\cal O}\left[{g_sM^2\over N}\left(g_sN_f\right)^2 \log~N\right]\nonumber, \nd}
 where we have only kept terms linear in ${g_sM^2\over N}$, linear and quadratic in $g_sN_f$, and ignored higher order terms. Of course one may question the logic of suppressing a term linear in $\log~N$. Such a term typically  comes with $(g_sN_f)^2$ and with either $a^2$ or with higher powers of $r = r_0$.  Since we will be assuming
 $r_0 << 1$, we can safely ignore the $\log~N$ piece. Note that the assumption of small $r_0$ is crucial here.  This implies the domination of $g_sN_f\vert\log~r_0 \vert $ over other constant pieces in \eqref{grantmunro}.  Implementing this\footnote{Otherwise one will have to solve a cubic equation in $\log~r_0$ from \eqref{grantmunro}. This will have one real solution that we can identify with the horizon radius $r_h$.} ,
 and putting $N_{\rm eff} = 0$, gives the following estimate for $r_0$ that we will identify with the horizon radius
 $r_h$ as:
\bg\label{pauldom}
r_h~ \sim ~ r_0 ~ \sim ~ {\rm exp}\left[- {1\over n_b \left(g_sN_f\right)^{2/3} \left({g_sM^2\over N}\right)^{1/3}}\right],
\nd
where $n_b \equiv 3\left(6\pi\right)^{1/3}$. Since both $g_sN_f$ and ${g_sM^2\over N}$ are very small quantities, the horizon radius is indeed deep inside Region 1.
Note that this estimate has to be bigger than the lower bound $r_h^{(o)}$ which in turn has a range \eqref{tree2}.

\newpage

\section{Equivalence between various alternative approaches of computing the bulk to shear ratio
 \label{UBC}}

The analysis that we performed in sections \ref{nf0} and \ref{nfn0} gave us precise formulae for the sound speed as well as the ratio of bulk to shear viscosities. The results of section \ref{nf0} was derived using 
Wilsonian method with zero flavors, whereas the results of section \ref{nfn0} were derived using a formalism similar to the renormalized perturbation theory with non-zero flavors. In the regime where the parameters of 
both the sections could be identified, the results for the sound speed as well as the ratio of bulk to shear viscosities match. For example using the expressions for $T$ and $s$ in \eqref{T} and \eqref{entoden} 
respectively, one can express ${1\over 3} - c_s^2$ as:

{\footnotesize
\bg\label{tagchink}
{1\over 3} - c_s^2 &=& {1\over 3} - {d\log~t\over d\log~s} = {1\over 3} - {s\over T}
\left({\partial T/\partial r_h \over \partial s/\partial r_h}\right)\\
& = & {g_sM^2\over 4\pi N}\left[C_{11}(1) + {\hat C}_{23} + 2{\hat C}_{01} + {\pi\sqrt{6}\over 2}\left(2c_1 + 
c_2\right) -\left(3{\hat C}_{23} - \pi\sqrt{6} c_2 + {3g_sN_f\over 2\pi}\right)\log~r_h\right], \nonumber \nd}
where $C_{kj}(u)$ is defined in \eqref{jkay85} and ${\hat C}_{kj}$ is $C_{kj}(u)$ for $u \to 1, \theta_i \to 1$.
Happily, like the quasinormal mode analysis, the aforementioned thermodynamical computation also 
yields:
\bg\label{hkpgkm}
\frac{1}{3} - c_s^2 \equiv {\cal O}\left(\frac{g_sM^2}{N}, (g_sN_f)\frac{g_sM^2}{N}\right). \nd
This is a good start but we want to show that the match is exact. From a quasinormal mode analysis of scalar modes of metric perturbations in the hydrodynamical limit,  which also is used in 
\eqref{vedpra}, one obtains from \eqref{cs} the following relation:
\bg\label{kkkbkbb}
{1\over 3} - c_s^2 = {g_sM^2\over N} \left({c_1 + c_2 \log~r_h\over 3\sqrt{6}} - {C_{21}(1)\over 16\pi}\right), 
\nd
where we again see the expected $g_sM^2/N$ factor emerging, as well as the ($c_1, c_2$) dependence. 
The $c_1$ factor, as defined in \eqref{dansor}, has an undetermined piece $\sigma$ that we did not compute in section \ref{nfn0}.  On the other hand $\vert c_2\vert$ is bounded above by \eqref{em18}, and 
for consistency we require $c_2 = -\vert c_2\vert$. If we now demand the equality between 
\eqref{hkpgkm} and \eqref{kkkbkbb}, then the following bound on $\vert\sigma\vert$ emerges:     
\bg
\label{|sigma|}
\vert \sigma\vert  > \frac{{2}\sqrt{3} g_s N_f\left(19 \log ~N + 552 \vert \log ~r_h\vert \right)}{64\pi^2}, \nd
leading to a consistent framework. Turning the idea around, bounding $\vert\sigma\vert$ below by 
\eqref{|sigma|},
one obtains an identical result for $\frac{1}{3} - c_s^2$ from a quasinormal mode analysis in the hydrodynamical limit and from thermodynamics.

Having shown the consistency of two different approaches to determine the sound speed above, and the consistency of the ratio of the bulk to shear viscosities for the two approaches instigated in 
sections \ref{nf0} and \ref{nfn0}, it is time to compare the ratio with other results from the literature. Unfortunately many of the results in the literature have been determined using  {\it bottom-up} approaches, so a direct comparison would be  futile to perform as most of the bottom-up approaches cannot be embedded in string theory. In other cases where embeddings {\it appear} feasible,  these backgrounds do 
not solve all the supergravity equations of motion\footnote{In most cases problems appear with Bianchi
identities.}. 
An alternative way out is to take our background, and look for limiting scenario where it would appear to resemble a class of background coming from the bottom-up approaches. As a concrete example, let us consider the model studied in \cite{EO}, where the ratio of 
bulk to shear viscosities has been discussed. Question is, what simplifications should we impose to our background so that it resembles the model of \cite{EO}?

First we need to dispose away the angular coordinates ($\psi, \phi_i, \theta_j$), and secondly decouple the internal space $M_6(\theta_{1,2}, \phi_{1, 2}, \psi, x_{10})$ from the five-dimensional space-time 
$M_5(t, x_{1, 2, 3}, u)$. The former can be easily done by choosing an appropriate slicing (see \cite{metrics}) whereas the latter can be achieved by imposing \eqref{harper}. With both these in place,  
one sees that one can write the resulting five-dimensional metric in the form of equation (4) of [21] in 
\cite{EO} using the following identification:
\bg\label{teenangul}
 e^{2A}\equiv  {e^{-2\varphi\over 3}\over \sqrt{h}}, \nd
 where $\varphi \equiv \varphi(r, \langle\theta_{1, 2}\rangle)$ is the type IIA string coupling and 
 $h \equiv h(r, \langle\theta_{1, 2}\rangle)$ is the warp-factor; and both are measured for 
 $\langle\theta_{1, 2}\rangle$. The scalar field
in the 5D action (1) of the aforementioned \cite{EO}'s reference [21], in our context could correspond to one of the metric components $g_{mn}(r,\langle\theta_{1,2}\rangle)$, where $(m,n)\in \theta_{1,2}, \phi_{1,2},\psi, x_{10}$. Additional simplifications are imposed by ignoring most components of internal fluxes, that would appear as scalar fields in five-dimensions, and taking a {\it single} 
scalar field and then, like [21] of \cite{EO}, work in a gauge wherein the five-dimensional radial coordinate 
$r$ is set equal to this scalar field.
 
Ignoring issues like compatibility with equations of motion, flux-quantizations etc, we can now make some precise connections with the results of \cite{EO}. According to \cite{EO}, the ratio of bulk to shear viscosities, is expressed as:
\bg\label{chatangul}
{\zeta\over \eta} = \left(\frac{1}{3 A'(r_h)}\right)^2, \nd
where $A(r_h)$ is the warp-factor near $r = r_h$ which we can now identify with our dilaton $\varphi$ 
and $h$ as in \eqref{teenangul} and substitute them in \eqref{chatangul}. In fact we can do slightly better by finding the ratio for both LO in $N$, 
i.e ${\cal O}\left({1\over N^0}\right)$, as well as for NLO in $N$ , i.e ${\cal O}\left({1\over N}\right)$, in the same setting. In other words:
\bg\label{chemer}
{\zeta\over \eta} = \left({\zeta\over \eta}\right)^{(0)} + \left({\zeta\over \eta}\right)^{(1)}.\nd
 The LO result is easy to figure out by plugging in the values for the type IIA dilaton and the warp-factor at the slice. The dilaton appears from the mirror transform of the type IIB dilaton, and may be read up from 
\cite{MQGP}, including the form of the warp-factor. After the dust settles, the LO result can be presented as:
\bg\label{goddleng}
\left({\zeta\over \eta}\right)^{(0)} = r_h^2 + {r_h^2 - 3a^2\over \log~N}\left[{3(1 + 2\log~r_h)\over \log~N} -
{8\pi\over (g_sN_f)\log~N} + 2\right] + {\cal O}\left({a^4\over r^2}\right), \nd
where $a^2$ is the resolution parameter which, as discussed in \cite{metrics, MQGP}, is a function of the horizon radius $r_h$. 
One easily sees that by choosing the resolution parameter to be:
\bg\label{lilprince}
a(r_h) = \frac{r_h}{\sqrt{3}}\left[1 + \frac{\log^2N}{12|\log~ r_h|} + {\cal O}\left(\frac{\log^3 N}{|\log^2r_h|}\right)\right], \nd
the leading order value \eqref{goddleng} vanishes\footnote{From \eqref{pauldom} we should keep in mind that  $\frac{\log^2 N}{12|\log~ r_h|}\sim\frac{\log^2 N}{N^{\frac{1}{3}}}$.}.
More interestingly, the choice \eqref{lilprince} connects some of the discussions regarding Regions 2 and 3 that we had in section \ref{nf0}. 
Given that $r>\sqrt{3}a$ is treated as large $r$ when dealing with resolved conifolds, this ties in very nicely with having chosen in \cite{meson Yadav+Misra+Sil, EPJC-2} the ${\rm D5}-\overline{\rm D5}$ 
separation ${R}_{\rm D5/\overline{\rm D5}}\sim\sqrt{3}a>r_h$ as the boundary of Regions 2 i.e the IR-UV interpolating region,  and 3, i.e the UV region (see Fig. \ref{allregions}). This additionally implies $r_h< \sqrt{3}a$, that appears from 
\eqref{lilprince},  is indeed in the IR and distances exceeding $\sqrt{3}a$ are in the UV.  All in all:
\bg\label{omoja}
\left(\frac{\zeta}{\eta}\right)^{(0)}=0, \nd
fits consistently with not only our whole IR picture, but also elucidates how the scales are separated. Additionally,
our approach reproduces the NLO result rather succinctly as: 
\bg
\label{jdhingsha}
\left(\frac{\zeta}{\eta} \right)^{(1)} = {3g_s M^2\over 4\pi N} \left(g_sN_f\right) a^2~\log~r_h ~ \propto~ 
{1\over 3} - c_s^2, \nd
where we have used \eqref{cs} to tie the ratio with the sound speed $c_s$. In the regime of interest discussed above, this matches precisely with the class of models using bottom-up approaches, for example 
\cite{EO}. Our analysis, in particular the ones we performed in section \ref{nfn0} onwards, reveals the power of the UV complete set-up in not only matching up with results from other models (once appropriate simplifications are inserted in), but also reveals how various parameters of the system, namely colors 
($N, M$), flavors $N_f$, resolution parameter $a^2$ as well as the horizon radius $r_h$ conspire to 
produce answers that may be consistently compared to experimental data (see for example 
\cite{meson Yadav+Misra+Sil}).

%



{}
 \end{document}